\documentclass[cm,final, 11pt]{article}
\pdfoutput=1
\usepackage{MRnotesBB}
\usepackage[section]{placeins}
\usetikzlibrary{arrows.meta}

\newcommand{\prn}[1]{\left ( #1 \right )}

\providecommand{\MRtodo}[1]{\todo[color=red!15,bordercolor=red!40]{MR: #1}}

\newcommand{\Gret}{G\subm{R}}
\newcommand{\Gadv}{G\subm{A}}
\newcommand{\cmtrace}{\cos(2\pi\sigma)}
\newcommand{\phGR}{\arg \Gret}
\newcommand{\hbw}{\frac{\beta\mkern2mu\omega}{2}}
\newcommand{\bw}{\beta\mkern2mu\omega}
\newcommand{\delphR}{\delta\phi\subM{R}}

\newcommand{\tkray}{t_{\vu{k}}}
\newcommand{\tzeta}{t_\zeta}

\newcommand{\cmexp}{\sigma}

\newcommand{\rp}{r_{+}}
\newcommand{\bdy}{\mathrm{bdy}}
\newcommand{\hor}{\mathrm{hor}}

\newcommand{\freq}{\mathfrak{w}}

\newcommand{\wkb}{\mathrm{WKB}}
\newcommand{\SCwkb}{\Sigma\subM{WKB}} 
\newcommand{\lwkb}{\lambda\subM{WKB}} 
\newcommand{\lgeod}{\lambda\subm{geod}} 
\newcommand{\Aper}{V_{A_s}} 
\newcommand{\RAper}{\Re\!\Aper}
\newcommand{\IAper}{\Im\!\Aper}
\newcommand{\Pphys}{P\subm{phys}} 
\newcommand{\Iwkb}{\mathcal{I}\subM{WKB}} 
\newcommand{\RIwkb}{\Re\!\ms\Iwkb}
\newcommand{\IIwkb}{\Im\!\ms\Iwkb}
\newcommand{\rbull}{r_\bullet} 
\newcommand{\rstar}{r_\star} 

\newcommand{\zcrit}{\zeta\subm{crit}}

\newcommand{\zpr}{\mathtt{z}}
\newcommand{\rpr}{\mathtt{r}}
\newcommand{\rpbull}{\rpr_\bullet}
\newcommand{\rptp}{\rpr\subm{tp}}
\newcommand{\rpstar}{\rpr_\star}
\newcommand{\upr}{\mathtt{u}} 

\newcommand{\trs}{\widehat{\zeta}} 
\newcommand{\rtr}{\widehat{\rpr}}

\newcommand{\Half}[1]{\frac{#1}{2}}

\title{Expiscating the black hole singularity in holography}

\author{Hewei Frederic Jia${}^{a}$,  Mukund Rangamani$^{b}$}

\affiliation[a]{Institute for Advanced Study, Tsinghua University, Beijing,
100084, China}
\affiliation[b]{Center for Quantum Mathematics and Physics (QMAP)\\ Department
of Physics \& Astronomy, University of California, Davis, CA 95616 USA}

\emailAdd{heweifred@gmail.com}
\emailAdd{mukund@physics.ucdavis.edu}

\abstract{
We argue that the leading exponential rate in the high frequency asymptotics of holographic thermal two-point functions is independent of the conformal dimension and spin (held fixed) of the operator. 
We analyze the leading universal WKB spectral curve governing the asymptotics at non-vanishing momentum scaling with frequency. 
We show that its classical action governs time domain signatures of the bouncing geodesic singularity, highly damped quasinormal modes, and the asymptotics of multi-stress tensor OPE coefficients. 
Beyond the leading order, we examine linear and dimension-dependent sublinear scalings of momentum with frequency. 
In both regimes, we derive the connection matrix between the boundary and the horizon using exact-WKB techniques. 
The results show a factorization of the spectral function into a perturbative part controlled by the OPE and a non-perturbative part governed by WKB transport. 
For linear scaling at fixed $k/\omega$ in the timelike regime, the turning points controlling physical transport remain separated from the curvature singularity. Nevertheless, a singularity-anchored representative of the same transport homology class determines the leading WKB action. 
In the sublinear regime, the relevant turning points approach the singularity. 
We match an exact solution of the near-singularity scaled equation to the exterior WKB solution to determine the connection matrix and spectral function. 
Our results clarify how the black hole curvature singularity is encoded in dual field theory observables at leading order in the planar expansion.
}
\newpage

\begin{document}
\maketitle


\part{Motivation}
\label{prt:motive}

\section{Introduction}
\label{sec:intro}

Ever since the inception of the holographic AdS/CFT correspondence nearly three
decades ago, an abiding question has been: \emph{How does the Lorentzian
  spacetime geometry, and its attendant features, such as global causal structure,
emerge in the dual field theoretic description?} As a necessary prerequisite one
could first ascertain the imprints of spacetime geometric features in terms of
field theory observables. Such an encoding dictionary ought to be constructed by
starting in the regime where semiclassical geometry can be trusted. There have
indeed been many investigations on this front, arguing for the encoding of bulk geometric
data into dual field theory quantities such as correlation functions,
entanglement entropy, etc. We will be interested in the encoding of spacetime
geometric features into correlation functions of gauge invariant operators. Of
particular focus will be the imprint of the black hole singularity as a finite
time divergence in real-time thermal correlation
functions~\cite{Fidkowski:2003nf}. In addition, we will also comment on
bulk-cone singularities in boundary correlators~\cite{Hubeny:2006yu}, for, as we
shall see explicitly, there are intimate connections between the two as we vary
parameters.

The black hole singularity is, of course, a fascinating feature of classical
spacetime geometry. A clear-cut encoding of its imprint in the dual field theory
in the planar strong coupling limit, would provide a target for understanding
its resolution by non-planar and finite coupling effects. Recall from the
holographic dictionary that the former correspond to Planckian, while the latter
correspond to stringy corrections to the spacetime geometry.

\begin{figure}[htp!]
  \centering
  \includegraphics[width=0.85\textwidth]{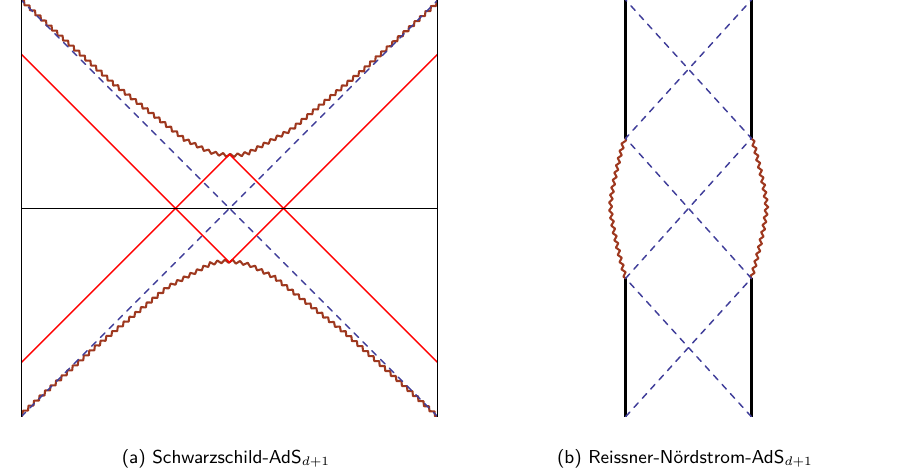}
  \caption{
    The eternal \SAdS{d+1} and \RNAdS{d+1} Penrose diagrams, with the bending of the curvature singularities. For the former we also denote the enveloping pair of `bouncing null geodesics', but refrain from indicating the analogous one for the charged case. As we shall argue later, the relevant geodesics are complex ones that stay within one thermofield double chamber of the Penrose diagram.
  }
  \label{fig:ads-SRN-pd}
\end{figure}

The key observation in~\cite{Fidkowski:2003nf} was that radial spacelike
geodesics connecting two asymptotic points on either side of the eternal
\SAdS{d+1} geometry are repelled by the black hole singularity. Furthermore, the
family of such spacelike geodesics, which is characterized by a single conserved
quantum number---the energy conjugate to the timelike Killing field---has a
limiting behavior at high energy. This limit is an envelope characterized by a
pair of null geodesics, colloquially referred to as the `bouncing null
geodesic', owing to the fact that the pair appears to meet and bounce off the
singularity. What is important here is that the presence of this limiting pair of null
geodesics signals a critical time $t_c$ beyond which there isn't a spacelike
geodesic connecting the two asymptotic boundaries (in the real Lorentzian
geometry), as illustrated in~\cref{fig:ads-SRN-pd}.

Let us assume for the moment that two-sided correlation functions of gauge
invariant operators $\mathcal{O}_\Delta$ of the dual field theory can be
approximated eikonally by a geodesic connecting the insertion points on the
boundaries. This is justified for high-dimensional operators $\Delta \gg 1$
using the Hamilton-Jacobi approximation of the wave
equation~\cite{Festuccia:2005pi}. In that case, the correlator
$\expval{\mathcal{O}_\Delta(t,\mathbf{0})\, \mathcal{O}_\Delta(0,\mathbf{0}) }_\beta$
for operators at the same spatial location (chosen for simplicity to be the
origin), would diverge as $(t - t_c)^{-2\ms\Delta}$. This would constitute a clear
strong signal of the black hole singularity. As noted in~\cite{Fidkowski:2003nf}
this conclusion is contingent on the family of radial spacelike geodesics lying
on the path of steepest descent for the eikonal approximation. This
unfortunately fails to be the case, and the signal, while present, is on a
secondary sheet. Furthermore, the asymptotic behavior of quasinormal modes at
high overtone number is intimately
related to the presence of the geodesic singularity in the
correlator~\cite{Festuccia:2005pi,Festuccia:2008zx}.

Recent investigations both from thermal OPE and WKB analyses have clarified the
picture significantly. Firstly,~\cite{Ceplak:2024bja} observe that the
asymptotics of multi-stress tensor OPE contributions alone predict a time domain
singularity for thermal correlators, albeit in a domain where the correlator is
guaranteed to be analytic by the KMS property.\footnote{The large order behavior of the  stress-tensor sector of the OPE in holography was analyzed in detail in the earlier work~\cite{Parisini:2023nbd}. The authors also note the connection between the asymptotics and the bouncing null geodesic.} However, the multi-stress tensors
do not by themselves comprise the complete contribution to the thermal OPE;
there are non-trivial contributions from the double-twist operators
$\mathcal{O}_\Delta\, (\nabla^2)^n\, \partial^\ell \mathcal{O}_\Delta$. It is
their presence that alleviates the naive multi-stress tensor induced singularity
and ensures that the correlator is analytic in the fundamental thermal strip
$\Im(t) \in (-\frac{\beta}{2}, \frac{\beta}{2})$. In a subsequent work they also
argued that the timelike singularity of charged black holes similarly leaves an
imprint on the boundary correlator~\cite{Ceplak:2025dds} (see
  also~\cite{Dodelson:2025jff, AliAhmad:2026wem,Chakravarty:2026fdm}
for an analysis of timelike singularities of charged/rotating black holes). This case is a priori
puzzling: there is no spacelike geodesic that connects the two sides of the
thermofield state which gets close to the singularity,
cf.~\cite{Brecher:2004gn}. Instead, in the Lorentzian section there is a
timelike geodesic that connects two asymptotic boundaries, one to the future of
the other in the maximally extended Penrose diagram!

On a different front, the asymptotic behaviour of thermal Green's functions in
frequency space was directly shown to control the location of the singularity in
time domain~\cite{Afkhami-Jeddi:2025wra,Jia:2025jbi,Giombi:2026kdz,Jia:2024zes}.
These works build on old
observations~\cite{Teaney:2006nc,Kovtun:2006pf,Caron-Huot:2009ypo} and show that
the spectral function (minus twice the imaginary part of the retarded Green's
function in our conventions) has exponentially small terms related to the bouncing geodesic.
Specifically, the deviation from the vacuum spectral function behaves at high frequency as
$\delta\rho\subm{np} \sim \omega^\gamma\, e^{- t_c \, \omega}$. The exponential damping term
then controls the location of the time domain singularity. At the same time, we
have the observations of~\cite{Festuccia:2005pi} relating the asymptotic
quasinormal mode spectrum (specifically the direction at which the mode attains
large $\abs{\omega}$) to the bouncing geodesic time. The connection to the
asymptotic behavior can be sharpened by exploiting the thermal product formula
of~\cite{Dodelson:2023vrw} as argued in~\cite{Dodelson:2025jff}.
The thermal OPE expansion can be mapped to the quasinormal mode expansion~\cite{Arnaudo:2026axe, Barrat:2026jfg} and demonstrates the connection between the
highly damped quasinormal modes and the bouncing singularity.\footnote{The connection, as we note in~\cite{Jia:2026ryl} and explain below, can also be inferred from the analysis of~\cite{Motl:2003cd}; the asymptotic behavior is determined by the composite monodromy between the horizon and the boundary. Recent works such as ~\cite{Hatsuda:2026ghx,Hatsuda:2026qaw} use exact WKB techniques to analyze the asymptotic highly damped quasinormal modes of asymptotically flat black holes.}
The recent work~\cite{Grassi:2026ekz,Grassi:2026wip} analyzes the resurgent structure of the momentum space OPE and argues the Borel singularities are related to the black hole singularity signature in some regimes.
Other works exploring the connection between analytic structure of thermal correlators and
bouncing geodesics include~\cite{Arnaudo:2026der,Arnaudo:2026tcy,Grozdanov:2026cut},
who also explore non-asymptotically AdS black holes. In addition, we have evidence that
quantum mechanical models possess such singularities in complex time
domain~\cite{Dodelson:2024atp,Dodelson:2026gak,Buric:2026qsp}. Recent work~\cite{Dodelson:2026hjm} extends the analysis broadly to large $N$ matrix quantum mechanical models.

Most of the aforementioned analyses focus on stationary (eternal) black holes.
Perturbations around the stationary solution are expected to have a strong impact in the vicinity of the curvature singularity. For neutral black holes with a spacelike singularity, the near-singularity evolution is governed by the chaotic BKL map~\cite{Belinsky:1970ew}. The imprint of the Kasner epochs as one approaches the singularity has recently been argued to be visible in the dual CFT correlators~\cite{Hartnoll:2026vhu,Xiao:2026pir,Ceplak:2026ard,Coviello:2026hdc,Oh:2026dky}.

Finally, as mentioned at the outset, the canonical approach would be to understand the manner in which semiclassical geometry is encoded in the dual field theory observables. However, one also ought to ask how stringy and quantum corrections modify this picture. Recent work~\cite{Komargodski:2026oob,Batra:2026gbq,Caron-Huot:2026unp,Liu:2026luf} sheds light on the latter question. Specifically,~\cite{Batra:2026gbq} analyzes a geodesic particle in low-dimensional JT gravity and demonstrates how its backreaction at finite $G_N$ (which scales inversely with $N$) smooths out the imprint. By studying analytic properties of thermal CFT correlators at finite $N$~\cite{Komargodski:2026oob} places bounds on the amplitude of the correlator on the second sheet.\footnote{Similar arguments can be adapted to the case of bulk-cone singularities, cf.~\cite{Batra:2026otu,Almheiri:2026cpv}.} In some examples of black holes with solvable string worldsheet theory, such as the geometry of near-extremal NS5 branes, finite string length corrections also resolve the singularity~\cite{Buric:2026sit} (see also~\cite{Itzhaki:2018rld}).

\looseness=-1
This is a fascinating collection of results, and while it points to an
interesting picture for what is the holographic encoding of the black hole
singularity, one cannot help but wonder what is the unifying picture that ties
all of these ends together. Moreover, they raise a number of questions. First,
how sensitive is the location of the singularity to the choice of local gauge
invariant operator whose thermal correlator we measure?\footnote{One
  can also ask how other observables perceive the singularity. There have been
  explorations using D-brane probes, cf.~\cite{Horowitz:2009wm}. We will, however,
limit our focus to correlation functions of gauge invariant observables.}
For example, do  operators
such as conserved currents or the energy-momentum tensor (which can lay claim
to being universal) have the same behaviour as scalar primaries? Second, how are
the OPE observations tied more seamlessly to the spectral function asymptotics?
Are the multi stress tensor (or current) exchanges, whose asymptotics picked out
the bouncing geodesic singularity, suitably universal? Third, why is it that the
time singularity of eternal charged black holes seems to require timelike
geodesics, which have no obvious interpretation in the thermofield double
picture? And finally, how robust is the result to considering spatially separated operators? The final question, which informs our analysis, is interesting, as even classically
the nature of the geodesics changes.

\subsection{Spatial resolution on the boundary}
\label{sec:whymom}

We propose a simple picture that ties together these questions, building up a
hierarchy of universal features associated with the holographic encoding of the
black hole singularity. We work strictly in the planar limit anchoring ourselves
on the semiclassical geometry, which we exploit extensively to arrive at our results.

Before we outline the general philosophy, we should
highlight the fact that we examine the behavior of observables at
non-zero spatial separation. This is along the lines suggested by our earlier analysis
in~\cite{Jia:2025jbi,Jia:2026ryl}, and in contrast to most discussions
hitherto. To be specific, we explore the asymptotics of real-time thermal
observables such as the spectral function in a limit of high frequency and large
momentum (or angular momentum).

Our analysis focuses on two distinct scenarios: first we take momentum to scale linearly with frequency and define
\begin{equation}
\omega, k = \abs{\vb{k}} \gg 1\,, \quad  \text{with }\quad
\frac{k}{\omega} = \zeta \,,\ \text{fixed}\,.
\end{equation}
We largely analyze the case where $\zeta \in (0,1)$, so momentum is real and timelike.\footnote{The literature largely focuses on the regime of fixed finite $k$ while taking $\omega$ large, with~\cite{Festuccia:2008zx} being an exception, for they also explore the near lightlike limit.}
We will refer to this limit as the \emph{linear-scaling limit}. In addition, we also motivate and briefly examine the behaviour at purely imaginary momenta $\zeta \in \ii \mathbb{R}$. With Wick rotated momenta spacelike geodesics are repelled from the singularity (akin to the radial geodesics), while with real momenta they fall in directly; the difference arises from the sign of the centrifugal barrier.\footnote{This analysis and the core-scaling regime discussed below were inspired in the course of discussions with Steve Shenker.}

The second scenario we explore is one where we scale the momentum by a dimension dependent sub-linear factor in the high frequency limit. We refer to this as the \emph{core-scaling limit}. Here we take
\begin{equation}
\omega, k = \abs{\vb{k}} \gg 1\,, \quad  \text{with }\quad
k \, \omega^{-\frac{d-2}{2\,(d-1)}} \propto \trs\,,\ \text{fixed}\,.
\end{equation}

The two-point function with fixed spatial separation $\vb{x}$ has branch cuts at the edge of the fundamental thermal strip, at $t = \pm \norm{\vb{x}} \pm i \frac{\beta}{2}$. At zero spatial separation the light-cone cuts merge, while at fixed momentum these cuts get smeared across the edge of the strip. However, keeping momentum finite in the scaling limit of high frequency allows us some spatial resolution. With real spatial momentum spacelike geodesics no longer access the region close to the singularity. The situation is better with purely imaginary momentum, but even here the turning point gets further from the singularity (and closer to the horizon) as momentum is increased. These statements suggest that the behavior at vanishing momentum is special.

Motivated by these observations, we examine thermal correlators of holographic CFTs using WKB techniques for the wave equations of fields in the black hole backgrounds. With non-zero momentum in our linear-scaling limit ($\zeta \neq 0$) the WKB turning points separate from the singularity.  This allows us some control, both technical (the turning points are standard Airy like turning points), and conceptual (disambiguating what is the signal of the singularity).

\begin{figure}[tp!]
  \centering
  \includegraphics[width=0.9\textwidth]{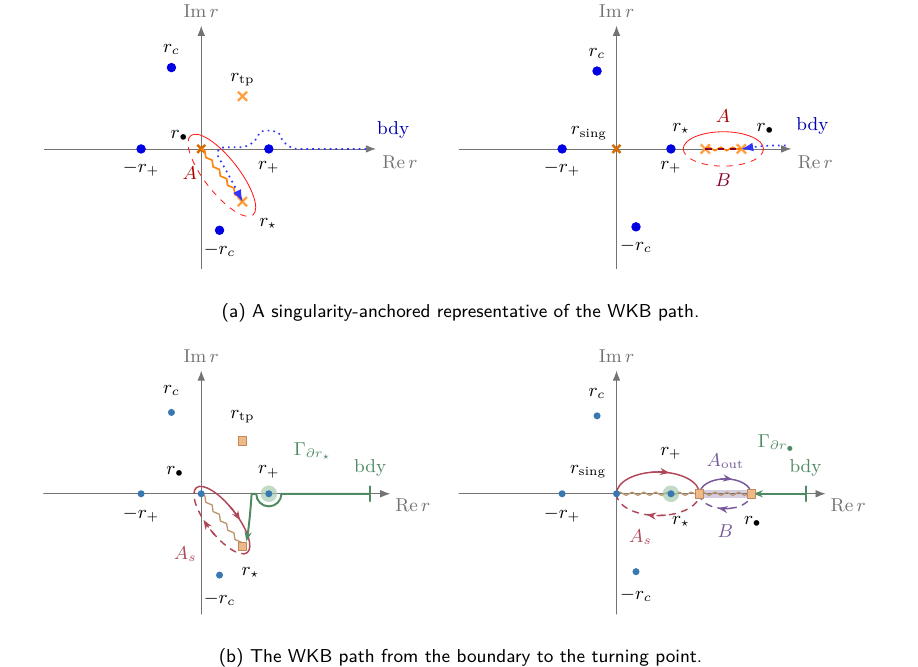}
  \caption{
    A schematic illustration of possible WKB trajectories in neutral black hole
    spacetimes in $d = 4$. The top left panel displays a physically motivated heuristic WKB path, that passes through the vicinity of the singularity before detouring to the physical turning point $r_\star$. The bottom left panel displays the true WKB trajectory, displaying direct transport from the boundary to the turning point $r_\star$. This is the path relevant for computing the boundary retarded Green's function. The former is a particular representative in the same homology class as the latter. Both of these left panels display situations where the turning points are complex (they may be inside or outside the horizon). The right panels, on the other hand, show situations where they are real and lie outside the horizon, which occurs at sufficiently high angular momentum for global solutions, and corresponds to bound states associated with periodic orbits.
  }
  \label{fig:globalads5-chamber-schematic}
\end{figure}

First, we show that a \emph{naive} WKB analysis can be employed to motivate that the leading behavior is always controlled  by a WKB trajectory or \emph{transport path}\footnote{Since we will motivate the use of exact-WKB to analyze the full connection matrix, we will often refer to the trajectories as the WKB  transport path. The terminology is meant to indicate the transport of a wavefunction basis from the AdS asymptopia to the outer horizon.} that passes through the singularity. However, the true WKB trajectory, which we examine subsequently, connects the boundary to a (complex) turning point, one which like the spatial geodesic turning points alluded to above, need not be close to the singularity. Indeed, we explain that the naive analysis picks out a particular singularity-anchored representative of the true WKB transport path within the homology class. We have attempted to illustrate the possibilities  in~\cref{fig:globalads5-chamber-schematic}, where we also indicate the possibility that in some cases the turning points lie well outside the horizon (they correspond to bound orbits).

However, there are two situations in which the singularity becomes accessible. The first is by an analog of quantum tunneling; at complex frequencies one picks up contributions from the vicinity of the singularity (in some, but not all cases). A more direct probe at real frequencies is the core-scaling limit. In this limit the WKB turning points lie within a small neighborhood of the singularity. The size of this neighborhood vanishes as $\omega \to \infty$, so for all intents and purposes one may view the transport path as reaching all the way to the singularity. The scaling behaviour extracted from the WKB trajectory gives a prediction for how the time domain singular locus at $t_c$ behaves as a function of spatial separation. Technically, the WKB analysis is interesting as the core region near the black hole singularity has to be handled differently in this scaling regime. This regime also encompasses the zero momentum case, where all the turning points merge into the black hole singularity.

Our discussion has been split into four key parts: the universal properties of wave equations are described in~\cref{prt:thermal-WKBgen}, the naive analysis of the singularity's imprint is in~\cref{prt:WKB1}, an exact-WKB analysis in the linear-scaling regime is explained in~\cref{prt:WKB2}, and finally the core-scaling limit analysis is presented in~\cref{prt:WKB3}. To orient the reader we summarize the salient results of our analysis in the next subsection.

\subsection{Summary of results}
\label{sec:summary}

We establish the following statements in the course of our discussion. These are organized thematically in the various parts of the paper.

\paragraph{A. Universality of leading WKB asymptotics (\cref{prt:thermal-WKBgen}):}
\begin{enumerate}[wide,left=0pt]
  \item[(A1).] The leading exponential asymptotics of the spectral function in the
    timelike high frequency limit are controlled by a universal WKB spectral curve,
    and are agnostic of the operator in question, cf.~\cref{sec:ode}. In particular, it holds for any
    dimension, spin or statistics. This statement follows from an elementary
    Hamilton-Jacobi analysis (or the characteristics of
    the wave operator).
  \item[(A2).] The WKB action computed from the spectral curve and the location
    of the time domain singularity share a simple relation, cf.~\cref{sec:spgeods}: the latter
    is a Legendre transform of the former with respect to the momentum ray
    parameter $\zeta$. In particular, one can derive a universal singular
    locus for fixed spatial separation directly from our universal spectral
    curve. The origins of this Legendre relation lie in our working with
    large momentum, which in saddle point terms translates to fixed spatial
    position.
  \item[(A3).] The spectral curve only knows about the background state. In
    geometric terms given the black hole geometric dual to the equilibrium
    thermodynamic ensemble, one has a unique spectral curve. In particular,
    it is absolutely agnostic to the quantum numbers of the operator in
    question, as long as they are held fixed in our high frequency
    limit.\footnote{This is qualitatively different from the original
      analysis in~\cite{Fidkowski:2003nf} and~\cite{Festuccia:2005pi}, who
      motivate the geodesic connection by invoking eikonal behavior for high
      dimension operators. The analysis
      of~\cite{Amado:2008hw} argues for something closer to what we have in
    mind.}
\end{enumerate}
\paragraph{B. A singularity-anchored WKB action (\cref{prt:WKB1}):}
\begin{enumerate}[wide,left=0pt]
  \item[(B1).] The WKB action which dictates the asymptotics can be determined naively by
    free propagation from the boundary to the vicinity of a branch point at
    a real radial position (the contour is originally ingoing radially from the boundary).
    This branch point $\rbull$ is typically the black
    hole singularity, but not universally so. Exceptions are seen in the
    case of global black holes where it may lie outside the horizon, or in
    charged black holes where it may lie inside the inner horizon, but not
    at the curvature singularity. Having made it there, the corrections to
    the free propagation arise from a migration to a distinguished complex
    turning point $\rstar$.\footnote{In our examples this complex turning
      point is distinguished by $\frac{3}{2}\pi < \arg \rstar < 2\pi$, which
    we observe empirically, but do not prove.} The action receives contributions
    both from the free propagation, and from the complex excursion. In
    addition, propagation in the classically admissible region requires
    traversing across horizons, which contribute additional imaginary (half)
    thermal periods (with a sign $\eta_j $). Schematically, the WKB action is then given by (cf.~\cref{sec:Iwkb-conj})
    \begin{equation}\label{eq:naiveIwkb}
    \Iwkb = \underbrace{2 P_{\mathrm{free}}}_{\text{real phase}}
    + \,\ii \sum_{j \, \in \,\mathrm{horizon}}\, \underbrace{\frac{1}{2}\,\eta_j\, \beta_j}_{\text{thermal period}}
    + \underbrace{V_{\mathrm{complex}}}_{\text{complex period}} \,.
    \end{equation}
    We refer to this as the singularity-anchored representative of the WKB path.  
  \item[(B2).] We demonstrate that the above WKB period controls the asymptotics of
    the spectral function, the behaviour of highly damped quasinormal modes,
    the location of OPE singularities arising from the exchange of just
    conserved current operators, and of course, the complex time-domain
    singularities of the two-sided thermofield double correlator outside the
    fundamental strip, see~\cref{sec:examples} for  evidence in various backgrounds.
  \item[(B3).] The technical gadget that ties all of these
    observables together in the analysis is the trace of composite monodromy
    between the boundary and the physical (outer) horizon.
    In the course of our analysis, we also argue that the phase of the
    retarded Green's function $\exp(\ii \phGR) = \frac{\Gret}{\abs{\Gret}}$
    is an exquisitely sensitive diagnostic of the asymptotics. It is even
    more robust than the composite monodromy trace, to which it can be
    straightforwardly related.\footnote{A closely related object (the thermal phase
      shift relative to the vacuum) was analyzed in recent
    literature~\cite{Jia:2026pmv,Araya:2026shz}.}
  \item[(B4).] For CFTs on compact spaces (e.g., on spatial
    $\mathbf{S}^{d - 1}$) in the highly spinning regime the WKB trajectory
    stays outside the black hole and ends up determining the bulk-cone
    singularity~\cite{Hubeny:2006yu} loci. This point has been appreciated
    in recent
    literature~\cite{Dodelson:2023nnr,Jia:2026pmv,Araya:2026shz}, and we
    shall recover it as part of our general analysis.
\end{enumerate}

\paragraph{C. The exact-WKB analysis (\cref{prt:WKB2}):}
\begin{enumerate}[wide,left=0pt]
  \item[(C1).] To better understand the naive WKB answer~\eqref{eq:naiveIwkb}  we undertake an exact-WKB analysis for scalar wave equations. We  work out the exact-WKB connection matrix for transport between the AdS boundary and the horizon (cf.~\cref{sec:ewkb-invite}) in the linear-scaling regime.\footnote{ The composite monodromy trace can be examined with an exact-WKB analysis~\cite{KawaiTakei2005}, as we explained  in~\cite{Jia:2026ryl}. } This picks out the true WKB contour on the spectral curve, which traverses between the boundary and the turning points before getting to the horizon. At no point does it get anywhere close to the black hole singularity.
  \item[(C2).] The exact-WKB connection formulae allow us to show that the spectral function factorizes cleanly into an OPE controlled part and a non-perturbative part (see~\cref{sec:ewkb-connection-imp}). The latter depends on the transport from the boundary to the horizon, while the former is a near-boundary feature consistent with earlier analysis. We derive an explicit transseries for the spectral function from this analysis, cf.~\cref{sec:ewkb-examples}.
  \item[(C3).] The naive WKB formula~\eqref{eq:naiveIwkb} can be obtained through a simple contour deformation of the true WKB path. In other words, the singularity-anchored representative is a particular choice one can make staying within the given homology class of paths. Not only is it not singled out in any way, but one can also make the path pass through this locus even when the singularity is a regular point on the WKB spectral curve.
  \item[(C4).] The full quantum WKB problem does carry knowledge of the singularity through the presence of tunneling type trajectories, which can be exposed by working at non-trivial complex frequency (cf.~\cref{sec:qperiod-ddp}).
\end{enumerate}

\paragraph{D. Zooming into the singularity (\cref{prt:WKB3}):}
\begin{enumerate}[wide,left=0pt]
  \item[(D1).] To access the vicinity of the black hole singularity without consideration of representatives within a homology class anchored at the specific locus, we motivate and study the core-scaling limit~\eqref{eq:triple-scale}. In this limit the turning points lie arbitrarily close to the singularity providing direct access to the high curvature region.
  \item[(D2).] The exact-WKB analysis in this core-scaling regime has some novel features. We analyze the problem therefore in two different coordinate charts. One involves resolving the near-singularity core region, so that the turning points remain separated from the core, and the other involves using a WKB action chart. In both, we show that the exact-WKB analysis involves identifying the core connection matrix. This is characterized by a core Stokes multiplier, a reflection coefficient for scattering off the singularity, which replaces the canonical Airy Stokes factor of a simple turning point.
  \item[(D3).] The core-scaling limit is anisotropic in the vicinity of the singularity. Our analysis which was motivated by the location of the turning points can equivalently be understood as exploiting the Kasner scaling. This is indeed the motivation for the contemporaneous paper~\cite{Valach:2026xjk}, who obtain similar results as us.
\end{enumerate}

\subsection{Outline of the paper}
\label{sec:outline}

The outline of the paper is as follows. We give a quick overview of spacelike geodesics in various black hole backgrounds in~\cref{sec:spgeod-mom} for orientation. In~\cref{prt:thermal-WKBgen}
we present the core argument for leading order WKB universality and the WKB spectral curve in~\cref{sec:wkbspectral}, and review some salient results about thermal Green's functions in holography in~\cref{sec:thermal-spectral-functions}.
A factorization property of the thermal spectral function for  integer dimension scalar primaries is proved in~\cref{sec:rho-Z-factors}, generalizing an observation from~\cite{Jia:2026ryl}.

In~\cref{prt:WKB1} we begin our WKB analysis with a heuristic discussion. In~\cref{sec:specasym} we motivate a singularity-anchored WKB path. This path is forced to pass through this specific spacetime locus, irrespective of whether this locus is a branch point of the WKB spectral curve or not. Using the WKB action thus obtained we relate spectral function asymptotics in~\cref{sec:timedomain} to the time-domain singular loci. In~\cref{sec:examples} we then provide numerical evidence for our WKB action. It agrees  extremely well with direct numerical integration of wave equations in neutral and charged black hole backgrounds in various dimensions ($d=4,6,8$).
In~\cref{sec:ope-wkb} we connect our WKB analysis to the OPE decomposition of the expectation value of the commutator.

In~\cref{prt:WKB2} we perform an exact-WKB analysis of the scalar wave equation (uncharged) for simplicity. We show in~\cref{sec:ewkb-invite} the boundary to horizon connection matrix has a three-step exact-WKB treatment: translation of Frobenius basis at boundary and horizon into appropriate WKB basis, and transport of WKB basis from boundary to horizon.
Our analysis predicts an exact factorization of the spectral function, see~\cref{sec:ewkb-connection-imp}. We exemplify the exact-WKB problem with explicit examples in~\cref{sec:ewkb-examples}, tying back to the heuristic discussion of~\cref{sec:examples}. Finally, we explain the success of the naive WKB approach of~\cref{prt:WKB1} in~\cref{sec:ewkb-bhsing} as a contour deformation statement, and argue that quantum tunneling at complex frequencies is  a natural probe of the singularity.

We motivate and study a core-scaling regime of parameters in~\cref{prt:WKB3} in order to provide a robust probe of the near-singularity region. The idea is to be able to separate the turning points from the singularity in~\cref{sec:tripscale-core}. This  regime leads to a more involved WKB analysis: a core region where we do not necessarily have a small parameter, and therefore must treat it exactly, followed by matching to an exterior region where the exact-WKB results of~\cref{prt:WKB2} apply. We analyze this in a chart that resolves the core region obtaining the non-perturbative contributions to the spectral function in~\cref{sec:core-examples}. To complement this discussion, we also explain how the leading WKB action chart can be adapted to obtain the same results in~\cref{sec:core-wkb-borel}.

We end with a brief discussion in~\cref{sec:discuss}. Some technical details are collected in the appendices. \Cref{app:Q0Q2examples} collects the  opers for scalar fields we analyze;~\cref{app:imaginary-mom} analyzes imaginary momentum behavior;~\cref{app:determinants} derives the core Stokes multipliers as functions of the core-scaled momentum. A companion website \href{https://mukundrangamani.github.io/stokes-atlas/}{Stokes atlas visualizers} collects the Stokes graphs for the examples we study in this manuscript in an interactive visualizer format.

\section{Spacelike geodesics in black hole spacetimes}
\label{sec:spgeod-mom}

Consider static asymptotically AdS black hole spacetimes (e.g., Schwarzschild, Reissner-Nordstr\"{o}m),
whose line element is given as
\begin{equation}
\label{eq:staticbh}
ds^2 = -r^2\,f(r)\, dt^2 + \frac{dr^2}{r^2\,f(r)} + r^2\, d\Sigma_K^2\,.
\end{equation}
Here $d\Sigma_K^2$ is the line element on a $d - 1$ dimensional constant
curvature spatial manifold with $K \in \set{0,\pm1}$, corresponding to planar,
global (spherical), and hyperbolic spatial geometries, respectively. The
function $f(r)$ depends on the geometry in question
\begin{equation}
f(r) = \frac{1}{r^2} \pqty{\frac{r^2}{\lads^2} + K - \frac{M}{r^{d-2}}
+ \frac{Q_c^2}{r^{2\,(d-2)}}} \,.
\end{equation}
This is the familiar two parameter solution of the Einstein-Maxwell equations with $Q_c$
parameterizing the charge, and $M$ parameterizing the mass. We will later
rewrite these in terms of physical horizon radii, but for now record that $\rp$
will denote the location of the outer horizon with $f(r_+) = 0$.
The factorization of $r^2$
in~\eqref{eq:staticbh} is a convenient choice specifically adapted to planar
horizon topologies.

\begin{figure}[bh!]
  \centering
  \includegraphics[width=0.9\linewidth]{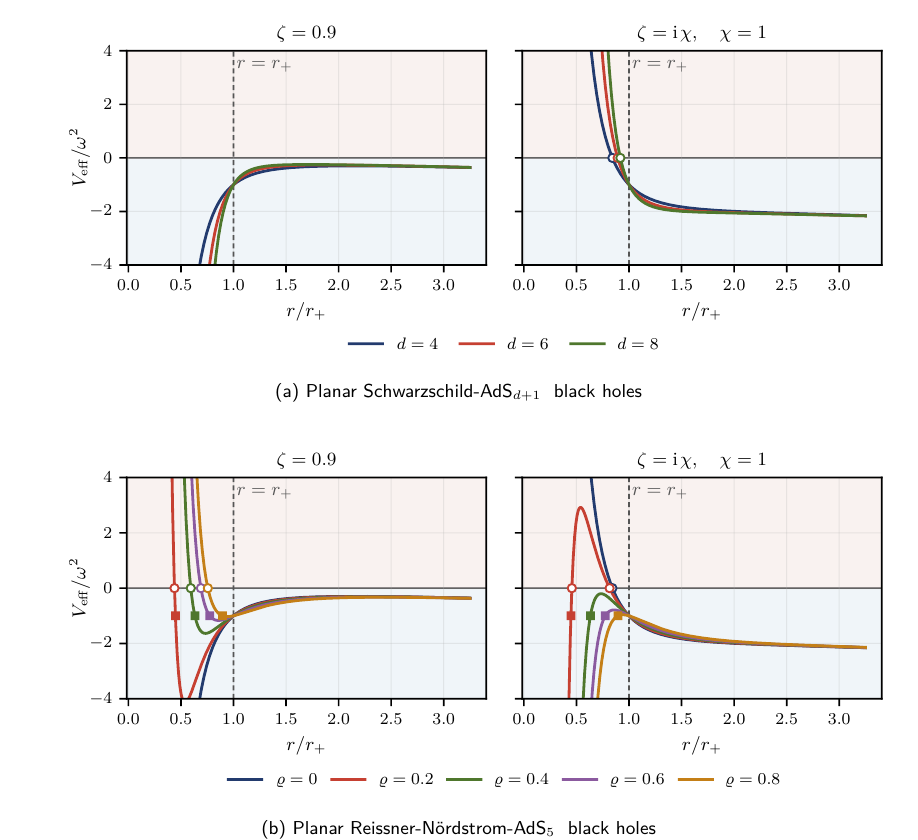}
  \caption{
    Effective potentials scaled down by $\omega^2$ for planar \SAdS{d+1} black
    holes (top) for $d = 4,6,8$ and \RNAdS{5} black holes (bottom). The real radial turning points are indicated by the open circles and the dashed vertical line is the outer horizon. Positive and negative effective potential domains are indicated with color coding. The left panels display the situation for real timelike momentum $\zeta = 0.9$, while the right panels show the case of
    purely imaginary momentum $\zeta = \ii \chi = \ii$ (both use $\omega=8$). The colored filled squares in the charged panels mark the location of the  inner horizon.
  }
  \label{fig:spacelike-geodesic-potential-planar}
\end{figure}

\begin{figure}[th!]
  \centering
  \includegraphics[width=0.9\linewidth]{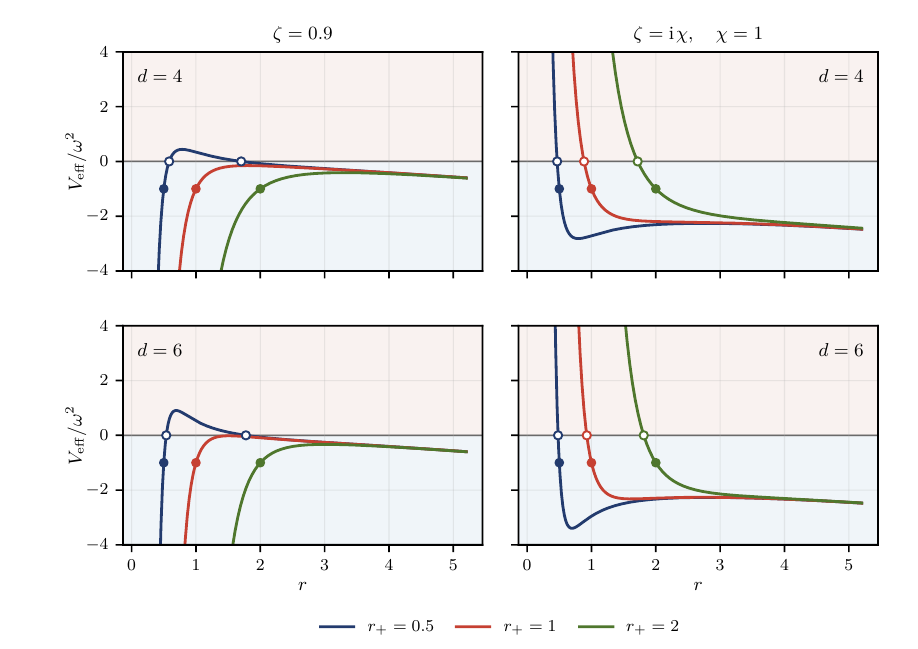}
  \caption{
    Global \SAdS{d+1} effective potential for $d = 4,6$ with $\omega=8$.  The conventions are as
    in~\cref{fig:spacelike-geodesic-potential-planar}. Since we have
    set $\lads = 1$ we mark the location of the horizon with filled circles.
    Note there are closed circular orbits outside the horizon for a range
    $\zeta_{\mathrm{crit}} < \zeta < 1$.
  }
  \label{fig:spacelike-geodesic-potential-global}
\end{figure}

To understand the relevance of different choices  (real and
imaginary momentum), consider spacelike geodesics in the
geometry~\eqref{eq:staticbh}. Using the Killing vectors to define conserved
quantities,
\begin{equation}
\omega = -p_t = r^2\,f\, \dot{t} \,, \qquad k= p_x = r^2\, \dot{x} \,.
\end{equation}
Here $x$ is a spatial coordinate along the boundary. For planar horizon
topologies one can use the norm of the spatial vector $x = \abs{\vb{x}}$, while
for spherical black holes it can be a Killing angular direction (say restricted
to the equatorial plane). Either way, we obtain a radial equation
\begin{equation}
\dot{r}^2 + V_{\mathrm{eff}}(r) = 0\,, \qquad \frac{1}{\omega^2}\, V_{\mathrm{eff}}
= r^2\,f(r) \pqty{\frac{\zeta^2}{r^2} -\frac{1}{\omega^2}}
- 1\,,
\end{equation}
with
\begin{equation}
\zeta = \frac{k}{\omega} \,.
\end{equation}

The plots of the effective
potential~\cref{fig:spacelike-geodesic-potential-planar,fig:spacelike-geodesic-potential-global}
suffice to make the point eloquently. Let us highlight some salient features
\begin{itemize}[wide,left=0pt]
  \item For real timelike momenta, $0 < \zeta < 1$, spacelike geodesics crash
    into the singularity of neutral black holes as noted
    in~\cite{Jia:2025jbi}. However, for charged black holes, they encounter
    a repulsive barrier inside the inner horizon. Exceptions to this statement arise for global black holes for large angular momentum.
  \item On the other hand, when the momentum is purely imaginary
    $\zeta = \ii \chi$ with $\chi \in \mathbb{R}_+$, the situation flips:
    neutral and weakly charged black holes have turning points close to the
    singularity, while highly charged black holes have the geodesic turning
    point between the inner and outer horizons.
  \item The situation for global black holes is similar, with a new
    wrinkle: the presence of bound orbits for a sufficiently large timelike
    momentum. In particular, for a given horizon size $r_+/\lads$ there is a
    range of real timelike momenta $\zeta_{\mathrm{crit}} < \zeta < 1$ where
    we have trapped orbits.
\end{itemize}

We highlight these features at the outset using the spacelike geodesics, as they
make it amply clear that accessing the vicinity of the singularity
using them requires one to work with complex momenta $\zeta \in \ii \mathbb{R}$.
Of course, this means one has to take an excursion into the complex domain of the spatial
section of the boundary since $\vb{x} \in -\ii \mathbb{R}$. On the other hand,
it is also clear that the momentum needs to be sufficiently small to
get close to the curvature singularity. For real $\zeta$ the turning points are still complex, but close to the origin, while for imaginary momentum they lie on the real radial line. We will quantify later that $\zeta \to 0$ will suffice for arbitrary
$\arg \zeta$.

We do not wish to focus on the geodesic analysis itself, but rather
examine the WKB limit of the wave equation. These simple features will end up
having a corresponding image and thereby determine whether the high frequency asymptotics of the thermal correlators are `cognizant' of the black hole singularity.

\newpage
\part{Thermal Green's functions: WKB universality \& general properties}
\label{prt:thermal-WKBgen}

\section{The WKB spectral curve and geodesics}
\label{sec:wkbspectral}

Let us turn to the main theme of our discussion: the WKB analysis of
thermal correlators. We first make a crucial observation motivating the
study of the high frequency limit at fixed quantum numbers (conformal dimension,
spin) of the dual CFT operators. Consider a primary operator of the CFT
transforming in a particular representation of $\mathrm{SO}(d,2)$. By the GKPW
dictionary~\cite{Gubser:1998bc,Witten:1998qj}, in the bulk spacetime it will be
dual to some tensor field carrying the same quantum numbers. We now examine the
consequences of the bulk-boundary dictionary for such operators.

Given a tensor\footnote{The discussion holds not just for bosonic fields, but
  also for fermionic operators. However, we will write formulae with tensor
indices to avoid cluttering notation with spinor indices.} field
$\mathcal{F}_{A_1 \cdots A_n}$ the wave operator acting on it is schematically
of the form $(g^{AB} \nabla_A \nabla_B -m^2) \mathcal{F}_{C_1 \cdots C_n} = 0$ where
$\nabla_A$ is the covariant derivative for the geometry~\eqref{eq:staticbh}
defined with respect to the Levi-Civita connection. If the tensor corresponds to a gauge field or a linearized metric perturbation, then there are
gauge constraints to include atop the dynamical wave equation.

The spatial manifold being a maximally symmetric space admits
$\frac{1}{2}\, d\,(d - 1)$ Killing vector fields, and the
geometry~\eqref{eq:staticbh} has an explicit timelike Killing vector
$\partial_t$. We can use these isometries to pass to the Fourier domain, as we did
in~\cref{sec:spgeod-mom}. Let $\vb{x}$ collectively denote the coordinates of
$\Sigma_{d - 1}$. We take $Y_\Sigma(\vb{x})$ to be an eigenfunction of the corresponding
Laplace operator acting on tensors $-\nabla_\Sigma^2 Y_\Sigma = k^2\,Y_\Sigma$, which defines $k^2$. One can then
decompose the field as
$\mathcal{F}_{C_1 \cdots C_n} (r) e^{-\ii \omega \ms t} \, Y_\Sigma(\vb{x})$ and
write the wave equation as a second order radial ODE with parameters $\omega$
and $k$.\footnote{Our notation is biased toward planar horizon topologies, for
  which $k = \abs{\vb{k}}$. For spherical black holes one has
  $k^2 = \ell \, (\ell + d - 2)$ for scalar spherical harmonics, and more
  generally a quadratic function of $\ell$ for tensor harmonics. The limit we
  describe in~\eqref{eq:zetadef} is to be viewed as the high angular momentum
limit, $\ell \to \infty$.}

We will be interested in the high frequency limit at large spatial momentum,
viz., $\omega, \;k \to \infty$ keeping their ratio
\begin{equation}
\label{eq:zetadef}
\zeta = \frac{k}{\omega} \, \quad \text{fixed}\,.
\end{equation}
This defines our linear-scaling limit. While taking this limit, we
keep all other quantum numbers, such as conformal
dimension $\Delta$, spin $J$, charge $q$, etc., of the field (and therefore the
dual operator) fixed. We are therefore considering the high frequency
limit at non-zero  momentum (linear or angular) of a generic operator
with finite quantum numbers under $\mathrm{SO}(d,2)$ and any global (R) symmetry
group. Towards the end of our discussion (cf.~\cref{prt:WKB3}) we motivate a second core-scaling limit, where $\zeta$ is also scaled towards zero at a specific rate.

During the course of our discussion we primarily examine the case where the momentum is real and timelike, viz., $0 < \zeta < 1$. However, given the repulsion of spacelike geodesics with imaginary momentum from the interior of the spacetime, we also explore purely imaginary momentum $\zeta = \ii \chi$ with
$\chi \in \mathbb{R}_+$ in~\cref{app:imaginary-mom}.
The leading order WKB action turns out to have a smooth $\zeta \to 0^+$ limit. However, obtaining the subleading corrections of observables is more intricate, and will only be accessed in the core-scaling limit. Furthermore, taking the lightlike limit $\zeta \to 1$  is also more involved. This is better analyzed by setting $\zeta =1$ and has already been examined thoroughly in~\cite{Festuccia:2008zx}.

We will write various equations in terms of $\zeta$ in the course of our
discussion, and revert to $\chi$ when discussing specific features of the
imaginary momentum case. One could of course imagine working with a generic
phase, but these two choices serve to illustrate some of the features we wish to highlight.

\subsection{Principal symbols and the WKB curve}
\label{sec:ode}

We will begin with a statement of universality of the wave equations, which will
be interpreted subsequently in terms of leading non-perturbative corrections
to the thermal correlation functions.

\paragraph{Universality of the principal symbol:}
Our first observation is that all of these wave equations have a universal
\emph{principal symbol}. We recall that the symbol associated with a
differential operator is the polynomial obtained by replacing the derivatives by
variables $\partial_{x_A} \mapsto \xi_A$. The principal symbol is the leading
homogeneous component. For the second order wave operators of interest, this is
the set of terms with \emph{exactly} two derivatives in the radial, temporal and
spatial coordinates. The statement of universality is that the principal symbol
is $g^{AB} \, \xi_A \, \xi_B$ with $\xi_A$ being an element of the cotangent
space, irrespective of the tensor in question.

Operationally, this observation can be rephrased by saying that every wave
equation of interest once decomposed into a radial ODE can be brought to a
Schr\"odinger form,\footnote{While the background is time-reversal invariant,
  the fields in question may only satisfy $CT$ (C: charge-conjugation) invariance,
  and thus we can have odd powers of $\omega$ in the expansion. A charged scalar
  in \RNAdS{d+1} provides an example where odd powers of frequency appear through
the gauge coupling.}
\begin{equation}
\label{eq:schreq}
\psi''(r) - \omega^2\,\bqty{ Q_{0}(r) + \sum_{n \geq 1} \omega^{-n}\, Q_{n}(r) } \psi(r) = 0\,,
\end{equation}
for some function $\psi(r)$.\footnote{The precise relation between $\psi(r)$ and
  $\mathcal{F}_{C_1 \cdots C_n} (r)$ can be deduced on a case by case basis, but
is not germane for our discussion.} The observation regarding the universality
of the principal symbol implies that the function $Q_0$ is universal. In our
linear-scaling limit it is given  by
\begin{equation}
\label{eq:Q0wkb}
Q_{0}(r) =- \frac{1- \zeta^2\, f(r)}{r^4\, f(r)^2}
\end{equation}
The pieces with $Q_{n}(r)$ with $n \geq 1$ capture the details of the spin,
mass, and other parameters. The range of $n$ in the subleading parts is
typically bounded $0 \leq n \leq n_{\mathrm{max}}$ (e.g., $n_{\mathrm{max}} = 2$
for scalars). However, there are examples where the sum does not terminate
($n_{\mathrm{max}} \to \infty$), as exemplified by the scalar polarization of
gravitons in the black hole backgrounds.\footnote{In the case of scalar
  polarizations of gravitons (or for the graviton-photon system both the scalar
  and the vector sectors) the physical linear combinations comprising the gauge
  invariant modes involve rational or algebraic functions of the spatial momentum,
  which is being scaled at the same rate. This leads to rational functions in the
effective potential whose expansion does not terminate.} The statement may be
easily verified for various operators of interest.\footnote{A concise
  compilation of the wave equations for scalars, gauge fields, and tensors can be
found in the Appendices of~\cite{Loganayagam:2022teu}.}

This universality is a simple consequence of the eikonal approximation. Consider
a Hamilton-Jacobi ansatz for the wave equation in the high timelike frequency
limit. Let us strip off the polarization and write a field agnostic
Hamilton-Jacobi functional as
\begin{equation}
\label{eq:HJansatz}
S\subM{HJ} =  -\omega\, t + k\, x + W(r)\,.
\end{equation}
In writing this we have absorbed the factor of $\ii$ to reside in the Fourier
waveform for simplicity, i.e., we use $\exp(\ii S\subM{HJ})$. The Hamilton-Jacobi equation is just the null condition
$g^{AB}\, p_A \, p_B = 0$, with $p_A$ being the momenta, i.e., the
aforementioned cotangent vectors. Explicitly, we have for the radial momentum
\begin{equation}
p_r^2 = W'(r)^2 = \omega^2\, \frac{1- \zeta^2\, f(r)}{r^4\, f(r)^2} \,,
\end{equation}
which is the familiar WKB form of the Schr\"odinger equation.

One may obtain the WKB/geodesic form of the Schr\"odinger equation by taking
other parameters to be large. Indeed, as discussed in depth
in~\cite{Festuccia:2005pi} the Hamilton-Jacobi ansatz also allows contact with
eikonal limit of massive geodesics for large dimensions $\Delta \gg 1$ of the probe operator. As emphasized hitherto we will keep $\Delta$ fixed and finite and let $\omega$ play the role of the WKB parameter. Similar observations were made in~\cite{Amado:2008hw}.

\paragraph{The WKB spectral curve:}
Let us recast our observations in the language of exact-WKB analysis. Good
accounts of the technology can be found in~\cite{Iwaki:2014vad,Iwaki_2015}, and
a recent review of the material is~\cite{Iwaki:2025tmq} (a brief summary can
also be found in~\cite[Section 4]{Jia:2026ryl}).

Given a Schr\"odinger equation of the form~\eqref{eq:schreq} one defines the WKB
spectral curve as $\SCwkb = \set{(r,\nu) \mid \nu^2 = Q_0(r)}$.\footnote{For
  wave equations in black hole backgrounds, it will be efficacious to redefine the
  radial coordinate into $z(r)$, with the $r$ plane being a branched cover over the
$z$-plane. We will implement this in due course in explicit examples.}
For our wave-equation~\eqref{eq:schreq} the spectral curve can be read off directly. It is endowed with a quadratic differential $Q_0(r)\, dr^2$ which is
just our principal symbol. The formal WKB solutions to~\eqref{eq:schreq} are
then
\begin{equation}
\psi_\pm(r,\omega) = \frac{1}{\sqrt{P_{\mathrm{odd}}(r,\omega)}}\, \exp(\pm \int_{r_0}^r\, dr' P_{\mathrm{odd}}(r',\omega)) \,.
\end{equation}
For potentials that are even functions of $\omega$, $P_{\mathrm{odd}}$ is the
odd-part of the all-orders WKB momentum,\footnote{On the other hand, for
  potentials that are not even in $\omega$, the odd-part of the momentum
  $P_{\mathrm{odd}}$ is the antisymmetric combination
  $\frac{1}{2} (P^{(+)} - P^{(-)})$ built atop the leading solution
$P^{(\pm)} = \pm\,\omega\, \sqrt{Q_0} + \order{1}$.} i.e.,
\begin{equation}\label{eq:Podddef}
P_{\mathrm{odd}} = \sum_{m = 0}^{\infty} P_{2m - 1}\, \omega^{1 - 2m}\,.
\end{equation}

The all-orders momentum is conventionally said to define the WKB differential
\begin{equation}
\lambda = P_{\mathrm{odd}} \,dr   \,.
\end{equation}
Its leading piece $P_{-1}$ is the familiar WKB momentum. We will later fix the branch.
Of interest in the WKB analysis are period integrals of $\lambda$ along contours
$\gamma \in \SCwkb$. Following standard exact-WKB terminology, we refer to such
periods as
\begin{equation}
V_\gamma = \oint_\gamma\, \lambda
\end{equation}
as the Voros period, and their exponential
\begin{equation}
X_\gamma = e^{V_\gamma}\,,
\end{equation}
as the Voros symbol, following the classic analysis
in~\cite{Voros1983ReturnQuarticOscillator}.

Usually, one is interested in the all-orders WKB differential, whose
leading term $P_{-1}(r)\ms dr$ is the meromorphic differential
characterizing the null Hamilton-Jacobi radial momentum. Define therefore,
\begin{equation}
\label{eq:lambdaWKB}
\lwkb(\zeta) = \sqrt{-Q_{0}(r)}\, dr
= \frac{\sqrt{1-\zeta^2\, f(r)}}{r^2\, f(r)} \, dr \,.
\end{equation}
This differential depends on $\zeta$ as we have explicitly indicated.\footnote{In general, we will assume it understood that the leading WKB differential is evaluated at non-zero $\zeta$ and simply write $\lwkb$. However, when we need to specify its behavior at particular values, say $\zeta =0$, we shall do so explicitly as $\lwkb(0)$.} Thus,
\begin{equation}
\lambda = -\ii \omega\, \lwkb(\zeta)  + \order{\omega^{-1}}\, dr\,.
\end{equation}
We have assumed for simplicity that the probes are also time-reversal invariant, so no odd powers of frequency appear in~\eqref{eq:schreq}.

The leading order WKB action is then simply given by the period integrals of
this differential around appropriately chosen contours $\gamma$, viz.,
\begin{equation}
\label{eq:Vmone}
V^{\mathrm{cl}}_\gamma(\zeta) = \oint_\gamma\, \lwkb(\zeta) + \cdots\,.
\end{equation}
We are not presently distinguishing the all orders symbol from the leading one. The Voros symbols then take the form of an asymptotic series
\begin{equation}
X_\gamma = \exp(-\ii\omega\, \oint_\gamma\, \lwkb(\zeta))\,
\sum_{m = 0}^\infty\, X_{\gamma, m}(\zeta)\, \omega^{-m}\,.
\end{equation}
As explained in~\cite{Jia:2026ryl} this can be Borel-resummed to obtain an exact
transseries (with suitable median symbols at saddle configurations). We will denote the Borel resummed Voros symbol for a cycle $\gamma$ as
\begin{equation}
\mathcal{X}_\gamma^{(\vartheta)} = \mathcal{S}_\vartheta[X_\gamma]\,,
\end{equation}
with $\mathcal{S}_\vartheta$ denoting the Borel summation in the direction $\vartheta$. If necessary, the summation in the specified direction should be defined using the median prescription~\cite{Delabaere:1997srq}.
Of import in our analysis is the simpler statement: the leading piece is governed by the
classical WKB action.

Let us take stock: we started with an observation of the universality of the
principal symbol. However, the principal symbol defines the WKB quadratic
differential and thus controls the leading WKB solution. Therefore, the
wavefunction for any tensor valued field in the high-frequency scaling limit is
simply governed by the leading WKB differential $\lwkb$. While the bulk field's
wavefunction itself is not an observable from the dual CFT perspective, it
nevertheless controls the CFT Green's functions. We will therefore find that
$V_\gamma$ for suitable $\gamma$ will capture the universal non-perturbative
piece of thermal two-point functions. In the course of our analysis we will determine the cycle $\gamma$. Initially, we shall do so by a heuristic argument, which we then will justify using exact-WKB methods.

\subsection{Null geodesics in the black hole geometry}
\label{sec:spgeods}

Before we proceed with boundary observables, let us take stock of one more bulk
geometric feature. Consider a null geodesic in the
background~\eqref{eq:staticbh}, which is characterized as the curve $x^A(\tau)$
for some affine parameter $\tau$. Using the conservation
associated with Killing fields,
\begin{equation}
\dv{t}{\tau} = \frac{\omega}{r^2\,f} \,, \qquad \dv{\vb{x}}{\tau}
= \frac{\vb{k}}{r^2}\,,
\end{equation}
the radial null geodesic equation reads
\begin{equation}
\pqty{\dv{r}{\tau}}^2 = \omega^2 \, (1 - \zeta^2\, f(r)) \,.
\end{equation}
Along any trajectory $\gamma$, the time and spatial (NB: $x = \abs{\vb{x}}$)
separations are
\begin{equation}
\label{eq:txsep}
\begin{aligned}
\Delta t & = \int_\gamma\, \frac{dr}{r^2\,f \, \sqrt{1-\zeta^2\, f(r)}}\,, \\
\Delta x & = \int_\gamma\, \frac{\zeta\, dr}{r^2\, \sqrt{1-\zeta^2\, f(r)}}\,,
\end{aligned}
\end{equation}

At a superficial level the geodesic problem appears different from the WKB
analysis of~\cref{sec:ode}. The only obvious point of contact is that the
geodesic turning points coincide with those of the leading WKB potential.

To facilitate a comparison, define a second differential, now associated with
the geodesic
\begin{equation}
\label{eq:lamgeod}
\lgeod(\zeta) = \frac{dr}{r^2\,f \, \sqrt{1-\zeta^2\, f(r)}}\,.
\end{equation}
It is then easy to check that the two one-forms, which are defined on the same
algebraic curve, are related as
\begin{equation}
\lgeod(\zeta) = \lwkb(\zeta) - \zeta\, \dv{\lwkb(\zeta)}{\zeta} \,.
\end{equation}
Furthermore, they are obviously equal at $\zeta = 0$, which lies at the edge of
our domain of interest.

Given that the differentials are defined on the same algebraic curve, it follows
that their periods are related. We can map the geodesic time and spatial separations
back onto Hamilton-Jacobi periods, and conclude that the spatial and temporal
periods are given as
\begin{equation}
\label{eq:periodrels1}
\begin{aligned}
(\Delta x)_\gamma & = -\dv{\zeta}V^{\mathrm{cl}}_\gamma(\zeta) \,, \\
(\Delta t)_\gamma & = V^{\mathrm{cl}}_\gamma(\zeta) - \zeta\, \dv{\zeta} V^{\mathrm{cl}}_\gamma(\zeta) \,.
\end{aligned}
\end{equation}

Having identified how the WKB spectral curve can be repurposed to compute
periods (or open integrals) that relate to temporal and spatial differences
along geodesics, we are now in a position to relate to the observations
in~\cite{Fidkowski:2003nf}. The second relation in~\eqref{eq:periodrels1} says
that the geodesic time separation is a Legendre transform of the classical WKB
period with respect to the parameter $\zeta$. This follows immediately from our
Hamilton-Jacobi ansatz, which gives a phase
\begin{equation}
\label{eq:HJphase}
\Phi_\gamma = \omega \,\bqty{ V^{\mathrm{cl}}_\gamma(\zeta) - (\Delta t )_\gamma + \zeta \, (\Delta x)_\gamma}
\end{equation}
The stationary phase approximation at fixed spatial separation
$(\Delta x)_\gamma$ is precisely what fixes~\eqref{eq:periodrels1}. The key
difference from~\cite{Fidkowski:2003nf} is that they considered zero spatial
separation and thus obtained $\Delta t$ directly. We will later carefully
specify the distinction between these calculations once we have identified the
relevant cycles.

\subsection{An interpretation of the spectral curve}
\label{sec:swinterp}

We now turn to a simple characterization of our leading WKB spectral curve, by
interpreting it in terms of Seiberg-Witten (SW) data. This is per se not all too
surprising considering the modern interpretation arising from analysis of
$\mathcal{N} = 2$ class $\mathcal{S}$ gauge theories in four
dimensions~\cite{Gaiotto:2009hg,Gaiotto:2012rg}. Equivalently, this perspective
can also be motivated from the Fuchsian structure of black hole wave equations,
as originally used in obtaining quasinormal modes~\cite{Aminov:2020yma}.\footnote{The
  singular point analysis for the wave equations of interest can be found
in~\cite{Loganayagam:2022teu}.}

We will be interested in algebraic curves $\Sigma$ equipped with a meromorphic
one-form $\lambda_{\mathrm{SW}}$ whose periods define special coordinates. An
invariant characterization is best done in terms of the analytic structure of
the differential. One characterizes such curves by the following data:
\begin{itemize}[wide,left=0pt]
  \item The holomorphic differential itself (so-called first kind data).
    Typically, $\lambda_{\mathrm{SW}}$ is not holomorphic by itself, but
    rather its derivatives with respect to some parameters $u_i$,
    $\partial_{u_i} \lambda_{\mathrm{SW}}$, are. The periods of these
    differentials comprise the period matrix of the curve.
  \item The second kind of data are meromorphic differentials with vanishing
    residues and higher order polar-parts. These may be absent in a given
    setting.
  \item Finally, the third kind of data are simple poles with non-zero residues.
    They fix the monodromy at the punctures.
  \item In addition, the quadratic differential may have vanishing loci, which
    are the branch points of the spectral cover.
\end{itemize}

The WKB spectral curves come equipped with a distinguished one-form
$\lwkb(\zeta)$. Its derivative,
\begin{equation}
\label{eq:holdiff}
\pdv{\zeta^2} \lwkb(\zeta) \,,
\end{equation}
is a holomorphic differential. To assert this one has to verify regularity at
all points above $r = 0, \infty$, but this can be done straightforwardly. This
implies that the spatial separation $(\Delta x)_\gamma$ is related to the period
(around $\gamma$) of a holomorphic differential (modulo a $2\,\zeta$
multiplicative factor). The zeros of $f(r)$ which are locations of complex
horizon roots are poles of $\lwkb$. The residues at these poles are proportional to complex
horizon (inverse) temperatures, and capture the indicial exponents at the regular singular
points of the wave equation. They prescribe the rigid local monodromies of the
problem and are clearly independent of $\zeta$. The only other piece of data is
the location of the turning points, which do depend on $\zeta$. We do not
encounter second-kind data for spectral curves associated with AdS black holes.
As is clear from the structure of $\lwkb$ the $\zeta \to 0$ limit is special,
since then the turning points merge with the singularity.

In general, for the non-extremal AdS black holes considered here, the radial
opers are Fuchsian. This is the reason for the absence of second-kind puncture
data. Therefore, the principal WKB differential carries only first-kind moduli
and third-kind horizon residues. However, the second-kind puncture data can
arise in confluent limits; for instance, at extremality, where the degenerate
horizon becomes an irregular singular point.

In the case of genus-1 curves, which will be of interest for several (but not
all) examples, we can pick a homology basis of A and B cycles, and consider in
SW terminology
\begin{equation}
a(\zeta^2) = \frac{1}{2\pi \ii}\, \oint_A\, \lwkb \,, \qquad a_D(\zeta^2)
= \frac{1}{2\pi \ii}\, \oint_B\, \lwkb
\end{equation}
In $\mathcal{N} = 2$ gauge theories, $a_D(a)$ is related to the derivative of
the prepotential. We recognize that the spatial and temporal separations
precisely map to the periods of~\eqref{eq:holdiff} --- in particular,
$\pdv{\zeta^2} a$, as noted above, is $\frac{\ii}{4\pi\ms\zeta}\, (\Delta x)_A$. We shall
reinterpret these relations using the AGT dictionary~\cite{Alday:2009aq} in terms of semiclassical
Virasoro blocks below.

\section{Thermal spectral functions in holography}
\label{sec:thermal-spectral-functions}

Thus far we have discussed some general features of black hole wave equations
identifying the universal spectral curve which governs WKB dynamics. The natural
next step is to answer what this observation is good for. Therefore, we first
turn to a quick review of the holographic computation of boundary correlation
functions.

Consider a gauge invariant primary operator $\mathcal{O}$ in a holographic CFT
with conformal dimension $\Delta_{\mathcal{O}} = \frac{d}{2} + \nu$. We suppress the other quantum numbers, such as spin $J$, charge $q$, etc., for brevity. The dual
tensor field satisfies wave equations of the form discussed in~\cref{sec:ode}.
We will be interested in the retarded Green's function
\begin{equation}
\label{eq:Gret}
\Gret(t,\vb{x}) = -\ii \Theta(t)\, \expval{[\mathcal{O}(x), \mathcal{O}(0)]}_\beta \,,
\end{equation}
and the two-sided (Wightman) correlation function
\begin{equation}
\label{eq:CWdef}
C(t,\vb{x}) = \expval{\mathcal{O}\pqty{t - \frac{\ii}{2}\,\beta,\vb{x}}\,
\mathcal{O}\pqty{0,\vb{0}}}_\beta\,.
\end{equation}

A related object that will prove slightly simpler to analyze is the thermal
spectral function, typically defined in frequency space as the Fourier transform
of the commutator
\begin{equation}
\label{eq:rhospec}
\rho(\omega, \vb{k}) = \int\, dt \, \int d^{d - 1} \vb{x}\, e^{\ii \omega\ms t} \, F_{\vb{k}}(\vb{x})\, \expval{[\mathcal{O}(x), \mathcal{O}(0)]}_\beta \,.
\end{equation}
Here $F_{\vb{k}}(\vb{x})$ is the appropriate tensor harmonic basis function on
the $(d - 1)$-dimensional spatial manifold $\Sigma_K$, and our CFT now lives on
$\mathbb{R} \times \Sigma_K$. For example, scalar operators on
$\mathbb{R}^{d - 1}$ have a conventional plane wave expansion with
$F_{\vb{k}}(\vb{x}) = e^{-\ii \vb{k} \cdot \vb{x}}$ while the selfsame operators
on $\vb{S}^{d - 1}$ would be expanded in the spherical harmonic basis. The
spectral function is related to the imaginary part of the retarded Green's
function, and more straightforwardly to the two-sided correlator via
\begin{equation}
\label{eq:Comega}
C(\omega,k) = \frac{\rho(\omega, k)}{2\,\sinh(\hbw)} \,.
\end{equation}

In holography the Green's functions are computed by solving the wave equation
for the dual field in an appropriate black hole background. For static canonical
or grand canonical ensembles with abelian global symmetry chemical potential we
can use the geometries described in~\eqref{eq:staticbh}. This makes the wave
equations described in~\cref{sec:ode} of immediate relevance. The retarded
Green's function is obtained by solving the wave equation with ingoing boundary
conditions at the future horizon. The most straightforward recipe for computing
it is to take the ingoing solution and extract the normalizable (vev) and
non-normalizable (source) fall-offs near the boundary. After discarding contact
terms, the appropriately normalized vev/source ratio gives the Green's function
of interest~\cite{Son:2002sd}.\footnote{This prescription can be given a clean
  justification in the gravitational Schwinger-Keldysh framework
  (see~\cite{Glorioso:2018mmw,Jana:2020vyx}). The fact that the retarded Green's
  function can be computed by remaining within the domain of outer communication
of the black hole will be of relevance in our discussion later.} One can
therefore phrase the determination of the Green's function as a connection
problem between the horizon and the boundary for the bulk wave equation. For
unfathomable reasons, all the bulk wave equations of interest are Fuchsian, with
regular singular points at the AdS boundary, the complex roots of the metric
function $f(r)$, and potentially at the singularity.\footnote{Some equations,
  such as scalar polarizations of gravitons, have additional apparent singular
  points, but these are innocuous as the wavefunctions are single-valued at these
loci.}

In general the wave-equations do not admit closed form solutions (except in
$d = 2$). So one therefore does not have a closed form expression for the
aforesaid connection coefficients. Nevertheless, thanks to the Fuchsian nature
of these equations, one can exploit results from two-dimensional CFTs, or
equivalently, from Seiberg-Witten theory (using the AGT
correspondence~\cite{Alday:2009aq}) to give exact connection
formulae~\cite{Bonelli:2022ten} (see~\cite{Lisovyy:2022flm} for related
developments). We will use a variant of the connection formula obtained
in~\cite{Jia:2024zes} to write down immediately an expression for the spectral
density.\footnote{One can view the formula derived in~\cite{Bonelli:2022ten} as
  a semiclassical correlation function in an auxiliary 2d CFT in the $t$-channel,
  while the expression we use here is more akin to the $s$-channel correlator for
  four point functions (i.e., four regular singular points). Technically, we are
  also using the generalization discussed in~\cite{Jia:2026ryl} since our
discussion is not limited to a specific spacetime dimension.}

\subsection{Thermal Green's function from an auxiliary 2d CFT}

Let us start with the expression for the retarded Green's function, which can be
shown to take the form
\begin{equation}
\label{eq:exact-Gret}
\begin{split}
\Gret(\omega,k) & = \mathsf{N}_\nu\, \mathscr{W}(\omega, k)\, \frac{\Gamma\prn{\frac{1+\nu}{2}- i\frac{\bw}{4\pi}\pm\cmexp}}{\Gamma\prn{\frac{1-\nu}{2}- i\frac{\bw}{4\pi}\pm\cmexp}} \,.
\end{split}
\end{equation}
In writing this expression, we have adopted a convenient shorthand: one takes
the product of terms on the r.h.s.~over all the sign choices. The normalization
factor $\mathsf{N}_\nu$ only depends on the conformal dimension
\begin{equation}\label{eq:Nnu-op-norm}
\mathsf{N}_\nu = \frac{\Gamma(1-\nu)}{\Gamma(1+\nu)}\,,
\end{equation}
and is harmless.
The quantity $\mathscr{W}(\omega, k)$ is the exponential of the derivative
of the semiclassical Virasoro block for the correlation function in the chosen
channel
\begin{equation}
\mathscr{W}(\omega, k) = e^{-2\, \partial_{\nu} \mathcal{W}}\,,
\end{equation}
with $\mathcal{W}$ being the comb channel correlator
\begin{equation}
\label{eq:Wcomb}
\mathcal{W} = V_{\mathrm{bdy}}\,
\begin{tikzpicture}[
    baseline={(0,0)},
    x=0.85cm,
    y=1cm,
    line cap=round,
    line width=0.8pt
  ]
  \draw (0,0) -- (4,0);
  \draw (5.5,0) -- (9.5,0);

  \foreach \x/\V in {
    1.3/{$V_{\mathrm{hor}}$}, 2.7/{$V_{\mathrm{hor}}^{(1)}$}
  }{
    \draw (\x,0) -- (\x,1.35); \node[above] at (\x,1.35) {\V};
  }

  \draw (7.5,0) -- (7.5,1.35);
  \node[above, xshift=3pt] at (7.5,1.35)
  {$V_{\mathrm{hor}}^{(\mathsf{m})}$};

  \node[anchor=north, yshift=-3pt] at (2.0,0) {$\sigma$};
  \node[anchor=north, yshift=-3pt] at (3.35,0) {$\sigma^{(1)}$};
  \node[anchor=north, yshift=-3pt] at (6.5,0)
  {$\sigma^{(\mathsf{m})}$};

  \node at (4.75,0) {$\cdots$};
\end{tikzpicture} \,V_{\mathrm{sing}}\,.
\end{equation}
As noted above, one views the operators in this auxiliary CFT as being inserted
at the punctures $r_i$ determined by the background geometry. We have a total of
$\mathsf{m} + 3$ punctures, where we are assuming that the black hole function
$f(r)$ has $\mathsf{m} + 1$ roots.

The spectral function can now be straightforwardly determined using
\begin{equation}
\label{eq:spectral-retarded-relation}
\rho(\omega, k) = -2\Im\, \Gret(\omega,k)
= \ii \pqty{\Gret(\omega, k) - \Gret(-\omega, k)}\,,
\end{equation}
where we used $\Gret(\omega,k)^* = \Gadv(\omega,k) = \Gret(-\omega,k)$ for
static ensembles (and assume real frequency and momentum). The upshot is that the spectral
function takes the following simple form
\begin{equation}
\label{eq:adsbh-rho-exact}
\rho(\omega, k) =  \frac{1}{\pi\,\nu\,\Gamma(\nu)^2} \, \sinh(\hbw)\, \mathscr{W}(\omega, k)\, \Gamma\pqty{\frac{1+\nu}{2}\pm \ii \frac{\bw}{4\pi} \pm\sigma}\,.
\end{equation}

As is well known, the semiclassical block does not have a closed form
expression. In the conventional phrasing the differential equation has a set of
accessory parameters which are related to $\partial_{r_i} \mathcal{W}$ by the
Zamolodchikov relation. These relations can be used to solve for the
$\sigma^{(i)}$, of which $\sigma^{(0)} \equiv \sigma$ has been singled out. We
will find it more convenient to view $\sigma^{(i)}$ as
composite monodromy exponents between the punctures. Explicitly, let $\mathcal{C}^{(j)}$
be a monodromy contour encircling the punctures starting from the boundary
$V_{\mathrm{bdy}}$ and $V_{\mathrm{hor}}^{(i)}$ for
$i \in \set{0,1, \cdots, j}$. Then
\begin{equation}
\Tr M_{\mathcal{C}^{(j)}} = - 2\, \cos(2\pi \sigma^{(j)})\,.
\end{equation}
%

\subsection{The channel momentum and the composite monodromy}

In the channel chosen for the decomposition in~\eqref{eq:Wcomb} we have singled
out one of the parameters $\sigma$. It is distinguished as the channel momentum
flowing upon fusion of the physical outer horizon and the boundary. Once one has
worked out the consequences of the exact-WKB analysis,  in a local domain one can
identify this quantity with a period integral\footnote{In the rest of our discussion, $\Aper$ is a classical period, but we refrain from indicating it as such with a superscript, viz.,
$\Aper^{\mathrm{cl}} \equiv \Aper$. }
\begin{equation}
\sigma = -\frac{\omega}{2\pi} \, \oint_{A_s}\, \lwkb\,, \qquad 2\pi \ii \sigma
= -\ii\frac{\omega}{\rp}\, \Aper\,.
\end{equation}
This is immediate from our discussion in~\cref{sec:swinterp} and the AGT
dictionary. However, one should view this as a local period coordinate, but it is
not yet the exact definition of $\sigma$. In other words, the period is
essentially local fusion data (valid for large frequencies) and does not by itself contain information about the transport data.

The precise definition of $\sigma$, and one that has an invariant meaning is the
composite monodromy invariant
\begin{equation}
\cmtrace = -\frac{1}{2}\, \Tr(M_{\hor}\, M_{\bdy}) \,.
\end{equation}
We are here considering monodromies around the singular points corresponding to the
physical outer horizon and the asymptotic boundary. This is precisely what the
connection problem computing the Green's function ordered. Note that this
composite horizon-boundary monodromy is a non-trivial function of $\omega$, $k$
and other quantum numbers of the operator $\mathcal{O}$, which is, in part,
clear from the definition as the channel momentum.

A few caveats are in order: this composite monodromy is clearly independent of any
local basis choice, but $\sigma$ is only defined up to signs and integer shifts
using it. Further, suppose we have identified the appropriate $A$-cycle and had
access to $\Aper$. Then, the composite monodromy is naively given by the Voros
symbol $X_A$, and thus, is predicted to have an exponential envelope atop an
oscillatory function. The exponential growth/damping rate is set by $\RAper$, while
the oscillations are characterized by $\IAper$.

However, this is not a complete characterization of the asymptotics. To work out
the proper large frequency asymptotics, which will turn out to be amenable to a
closed form analysis, one can employ exact-WKB methods. As described
in~\cite{Jia:2026ryl} the composite trace is a Laurent polynomial in the Voros
symbol $X_A$ for some selected $A$-cycle, its conjugate $X_A^*$, and some local
horizon monodromy factors (which serve to supply thermal factors). We will see
this more precisely below.

This observation is important for the following reason: since the composite
monodromy trace oscillates under the exponential envelope, one cannot invert it
to obtain the channel momentum pointwise. At real frequencies~\cite{Jia:2026ryl}
used the intuition that $\cmtrace$ ought to be real, and argued that the median
Stokes resummation serves to ensure this. Nevertheless, the oscillations
indicate that the monodromy passes from being hyperbolic, $\abs{\cmtrace} > 1$,
to elliptic, $\abs{\cmtrace} < 1$. The median resummation precisely accounts for
this, and shows that the true envelope gets dressed by horizon monodromy
factors. In fact, we will find that the envelope will actually follow $\IAper$
and not the real part as naively predicted.

\subsection{The phase of the retarded Green's function}
\label{sec:RRA}

In fact, we now argue that for time-reversal invariant systems there is actually
a very simple relation between the Green's function and the composite monodromy.
To see this, let us recall that the monodromy around a regular singular point of
a Fuchsian differential equation is given by a diagonal matrix with entries set
by the indicial exponent (assuming non-resonant indicial exponents). For the wave equations of interest (which are
Fuchsian), we want the composite monodromy between the horizon and the boundary.
The indicial exponents at the horizon are $\frac{1}{2}\pm \ii \frac{\bw}{4\pi}$,
corresponding to ingoing/outgoing modes, respectively. At the AdS asymptopia the
indicial exponents are $\frac{1}{2}\pm \frac{\nu}{2}$. The non-normalizable source mode is $(r^{-2})^{\frac{1-\nu}{2}}$ and normalizable response (or expectation value) is $(r^{-2})^{\frac{1+\nu}{2}}$ in an  asymptotic expansion with coordinate $r^{-2}$. These results follow
from a local analysis around a non-degenerate Killing horizon and the AdS
asymptopia.

Consider a local solution near the horizon, which will be in terms of ingoing
and outgoing modes. The solution near the AdS boundary will be expressed in
terms of the sources and responses as usual. Pick a convention where the
amplitudes of the ingoing/outgoing waves when expanded in the asymptotic
solution basis take the form
\begin{equation}
\begin{aligned}
\psi_{\mathrm{in}}^{\Gamma}  & = A_{i}\, \psi_{\mathrm{source}} + B_{i} \, \psi_{\mathrm{vev}} \\
\psi_{\mathrm{out}}^{\Gamma} & = A_{o}\, \psi_{\mathrm{source}} + B_{o} \, \psi_{\mathrm{vev}}\,.
\end{aligned}
\end{equation}
The near-horizon basis has been normalized with suitable gamma functions of the
indicial exponents, viz.,
\begin{equation}
\label{eq:Ghorbasis}
\psi_{\mathrm{in}}^{\Gamma} = \frac{\psi_{\mathrm{in}}}
{\Gamma\pqty{1-\ii\frac{\beta\omega}{2\pi}}}\,, \qquad \psi_{\mathrm{out}}^{\Gamma}
= \frac{\psi_{\mathrm{out}}}
{\Gamma\pqty{1+\ii\frac{\beta\omega}{2\pi}}}\,.
\end{equation}

The composite monodromy of interest is
\begin{equation}
\cmtrace = -\frac{1}{2}\, \Tr(M_{\mathrm{bdy}}\, M_{\mathrm{hor}}) \,.
\end{equation}
Working with the asymptotic basis, note that
\begin{equation}
\label{eq:horizon-monodromy}
M_{\mathrm{bdy}} = -\mathrm{diag}(e^{-\ii\pi\ms \nu}, e^{\ii\pi\ms \nu}) \,, \qquad M_{\mathrm{hor}}
= -\mathscr{C}_{\mathrm{bdy}, \mathrm{hor}}^{-1}\ \mathrm{diag}\pqty{
e^{\hbw}, e^{-\hbw}}\
\mathscr{C}_{\mathrm{bdy}, \mathrm{hor}}\,,
\end{equation}
where $\mathscr{C}_{\mathrm{bdy}, \mathrm{hor}}$ is the connection matrix
between the asymptotic and horizon bases. But this matrix has entries set by the connection coefficients $\set{A_i,B_i, A_o,B_o}$. Let us explicitly write out the connection matrix mapping the horizon basis onto the boundary basis as
\begin{equation}
\label{eq:Connect-hor-bdy}
\mathscr{C}_{\mathrm{bdy},\mathrm{hor}} =
\begin{pmatrix}
  A_i & B_i \\
  A_o & B_o
\end{pmatrix}.
\end{equation}

All we need to do is to work out the trace, and account for the normalization
through the Wronskian (which results in the hyperbolic functions below).\footnote{We use the standard definition  $\operatorname{\text{Wr}}(f,g) = f \, g' - g \, f'$.\label{fn:Wronskian}} This
leads to the following expression
\begin{equation}
\label{eq:monoeq1}
\cmtrace = -\frac{\cosh(\hbw - \ii \pi \ms\nu)- \mathcal{R}_{\mathrm{RA}}\,
\cosh(\hbw + \ii \pi\ms \nu)}{1-\mathcal{R}_{\mathrm{RA}}}\,,
\end{equation}
with
\begin{equation}
\mathcal{R}_{\mathrm{RA}} = \frac{B_i}{A_i}\,\frac{A_o}{B_o}
= \frac{\Gret(\omega,k)}{\Gadv(\omega,k)}\,.
\end{equation}
In the final step we have used the standard prescription for the boundary
observables~\cite{Son:2002sd}: the normalized response/source ratio of the
ingoing mode defines the contact-term-subtracted retarded Green's function
$\Gret$, while the same prescription for the outgoing mode defines the advanced
Green's function $\Gadv$. One could equivalently have arrived at this answer
from~\eqref{eq:Gret}. For time-reversal symmetric states dual to static black
hole geometries, $\Gadv(\omega,k) = \Gret^*(\omega,k)$, which upon
using~\eqref{eq:Gret}, leads to a compact form of~\eqref{eq:monoeq1}
\begin{equation}
\label{eq:monoratio}
\frac{\Gret(\omega,k)}{\Gadv(\omega,k)} = \exp(2\ii \phGR)
= \frac{\cosh(\hbw - \ii \pi \ms \nu ) + \cmtrace}{\cosh(\hbw + \ii\pi \ms\nu ) + \cmtrace }\,.
\end{equation}
Specifically, this implies that the phase carried by the retarded Green's
function is simply related to the composite monodromy trace, which is a
slightly simpler statement than that derived for the spectral
density~\eqref{eq:adsbh-rho-exact}. For one, the block factor cancels out in the
ratio of the retarded and advanced Green's functions, and for another it is
actually straightforward to numerically compute it for any differential
equation. We will exploit this relation to derive the asymptotics of $\cmtrace$
at high frequencies.

\paragraph{The phase and quasinormal modes:}
There is an interesting corollary of our observation~\eqref{eq:monoratio}.
Recall that the poles of the retarded Green's function are the locations of the
quasinormal modes. Similarly, the poles of the advanced Green's function are the
anti-quasinormal modes. However, $\mathcal{R}_{RA}$ is a pure phase on the real
$\omega$ axis for reasons mentioned above. Near a retarded Green's function pole
$\omega\subm{QNM}$ we would write
\begin{equation}
\Gret(\omega) = \frac{\Res\subM{QNM}}{\omega - \omega\subM{QNM}} \;\;
\Longrightarrow \;\; \exp(\ii \phGR)
= \frac{\Res\subM{QNM}}{\abs{\Res\subM{QNM}}}\, \frac{\abs{\omega-\omega\subM{QNM}}}{\omega-\omega\subM{QNM}}\,.
\end{equation}
Since the phase is bounded (here we use the time-reversal symmetry), it doesn't
blow up. However, as we vary frequency along the real-axis, a quasinormal pole  at
point in the lower half-plane ends up resulting in a rapid phase variation. The
variation, i.e., the relative phase change per unit frequency, is stronger the
closer the pole is to the real axis.

As a cross-check notice that we would predict the quasinormal frequencies to be
the zero loci determined by
\begin{equation}
\label{eq:qnmloc}
\cosh(\hbw + \ii \pi\ms \nu ) + \cmtrace = 0 \,.
\end{equation}
The statement holds provided that there are no accidental zeros of $\Gadv$ and that the   branch of $\sigma$ is chosen appropriately. This can be checked to give the same condition derived from the poles
of~\eqref{eq:exact-Gret} obtained in~\cite{Jia:2024zes}.

We will see that the phase indeed oscillates around the baseline value
$-\pi \nu \pmod{\pi}$ at large frequency in explicit examples. This is not in
any way because the modes themselves have small imaginary part, but rather an
effect of the monodromy. The oscillation occurs under an exponential envelope
and the WKB action can directly be used to predict the spacing of the quasinormal
mode lattice, as we demonstrate in due course.

\section{Factorization of the spectral function}\label{sec:rho-Z-factors}

Before we turn to the main theme of the present paper, we take a small detour to
prove a statement conjectured for $\Delta_{\mathcal{O}} \in \mathbb{Z}$
in~\cite{Jia:2026ryl}. The result we seek is an exact factorization formula for
the spectral function into a perturbative piece controlled by the OPE, and a
non-perturbative piece. We now establish this for $\Delta - \frac{d}{2} = \nu \in \mathbb{Z}_{>0}$. For even $d$ this includes integer dimension operators away from the BF bound.  Later in~\cref{sec:rho-factorization} we will generalize our result to arbitrary $\Delta_{\mathcal{O}}$ from the exact-WKB analysis, but the present discussion closes a small gap more straightforwardly.

For this discussion we will be more specific: consider the Green's function of a scalar primary operator. The dual (neutral) scalar
field  is quantized with standard (Dirichlet) boundary
conditions at the AdS asymptopia. The background is a static, non-extremal black hole with spatial cross-sections of the horizon being a maximally symmetric space.

For this analysis it will be convenient to work with the inverse-radius
coordinate
\begin{equation}\label{eq:zcoord}
z = \frac{\rp}{r}\,.
\end{equation}
We set
\begin{equation}
2\ms \nu = 2\ms \Delta_{\mathcal{O}} - d \in \mathbb{Z}_{> 0}\,.
\end{equation}
We expand the Schr\"odinger form of the scalar wave equation in the
neighbourhood of the boundary, writing for the scalar oper
\begin{equation}
\psi''(z) + t(z)\, \psi(z)
= 0\,, \qquad t(z)
= \frac{\frac14-\nu^2}{z^2}
+ \sum_{j = 0}^{\infty}t_j\,z^j\,.
\end{equation}
The normalizable and non-normalizable solutions then have a local Frobenius
expansion albeit with a possible logarithmic term because the indicial
exponents differ by an integer. We have
\begin{equation}
\begin{aligned}
\psi_{\mathrm{vev}}(z)    & = z^{\frac12 + \nu}(1 + \cdots)\,, \\
\psi_{\mathrm{source}}(z) & = z^{\frac12 - \nu}\sum_{n = 0}^{\infty}a_n\,z^n + \mathcal L_{\nu}^{(z)}\,\psi_{\mathrm{vev}}(z)\log z\,, \\
a_0                       & = 1\,, \qquad a_{2\nu} = 0\,.
\end{aligned}
\end{equation}
This local solution is determined by standard techniques to be given by
recursion relations
\begin{equation}
n(n - 2\nu)\,a_n+ \sum_{j = 0}^{n - 2}\,t_j \,a_{n - 2 - j}
= 0\,, \qquad 1
\leq n
< 2\nu\,,
\end{equation}
and
\begin{equation}
\mathcal L_{\nu}^{(z)} = -\frac{1}{2\nu} \sum_{j = 0}^{2\nu - 2}
t_j \, a_{2\nu - 2 - j}\,.
\end{equation}
The absence of the logarithmic branch for integer dimensions is equivalent to
the locus $z = 0$ being an apparent singular point. For this to occur, we need
$\mathcal L_{\nu}^{(z)} = 0$, which then defines the apparent singularity
condition at the boundary. We are now in a position to make the following claim for integral $\nu$.

\begin{proposition}[Exact resonant factorization]
  For $\nu \in \mathbb{Z}_{> 0}$, the spectral function factorizes exactly
  as\footnote{We have decorated $\sigma$ with a $z$ subscript to make clear the
  coordinate choice.}
  \begin{equation}\label{eq:rho-factor}
  \rho(\omega,k) = \rho_{\mathrm{log}}(\omega,k)\, \rho_{\mathrm{mon}}(\omega,k)\,,
  \end{equation}
  where
  \begin{equation}\label{eq:rho-monodromy}
  \rho_{\mathrm{log}} = -\pi \, \mathcal{L}_{\nu}^{(z)}\,, \qquad \rho_{\mathrm{mon}}
  = \frac{2\sinh(\hbw)}
  {\cosh(\hbw)+ \cos(2\pi\sigma_z)}\,.
  \end{equation}

\end{proposition}

\begin{proof}
  We have already introduced the gamma normalized horizon basis earlier
  in~\eqref{eq:Ghorbasis}. To fix normalizations use the Wronskian for the
  asymptotic basis, which is
  $\operatorname{\text{Wr}}[\psi_{\mathrm{source}}, \psi_{\mathrm{vev}}] = 2\ms \nu$. Then it follows
  from our connection matrix that
  \begin{equation}
  \psi_{\mathrm{in}}(z) \sim (1 - z)^{\frac12 - \ii\frac{\beta\omega}{4\pi}}\,, \qquad \psi_{\mathrm{out}}(z)
  \sim (1 - z)^{\frac12 + \ii\frac{\beta\omega}{4\pi}}\,,
  \end{equation}
  Using the standard prescription for real-time correlators~\cite{Son:2002sd}
  \begin{equation}
  \Gret = - \frac{B_i}{A_i}\,, \qquad \Gadv
  = - \frac{B_o}{A_o}\,,
  \end{equation}
  Here we assumed that after subtracting off local contact terms, the physical correlators are defined by stripping off a positive factor $2\ms\nu\, \mathcal{N}(\nu)\, \rp^{2\nu}$. This leads to the above simple expression for the Green's function.
  Using the Wronskian of the normalized horizon basis
  solutions~\eqref{eq:Ghorbasis}, we find
  \begin{equation}
  A_i \, B_o-A_o\, B_i = -\frac{\ii}{2\pi\nu}\, \sinh(\hbw)\,, \qquad \rho(\omega,k)
  = \frac{\sinh(\hbw)}{2\pi\ms \nu\, A_i\,A_o}\,.
  \end{equation}
  The main change from~\eqref{eq:horizon-monodromy} is the Jordan normal form of
  the boundary monodromy due to the logarithmic branch
  \begin{equation}
  M_{\mathrm{bdy}}^{(z)} = (-1)^{2\nu + 1}
  \begin{pmatrix}
    1  & 2\pi\ii\mathcal L_{\nu}^{(z)} \\
    0 & 1
  \end{pmatrix},
  \end{equation}
  An explicit computation, along with the use of the determinant of the
  connection matrix derived from the Wronskians gives immediately
  \begin{equation}
  \cos(2\pi\sigma_z) = (-1)^{2\nu + 1} \bqty{
    \cosh(\hbw)
    +4\pi^2\nu\mathcal L_{\nu}^{(z)}A_iA_o
  }\,.
  \end{equation}
  The central thesis of the proposition now immediately follows by direct
  substitution of $A_i\,A_o$.  We immediately find the factorized result as
  quoted upon setting $(-1)^{2\nu} =1$ for $\nu \in \mathbb{Z}_{>0}$.
\end{proof}

Some important points are in order: the argument above uses neither the
representation of the spectral function using the Virasoro block
technology~\eqref{eq:adsbh-rho-exact}, nor does one assume any
cancellation of poles related to the monodromy. Moreover, it is applicable in a
dimension agnostic manner, for at no point in the above discussion did we have
to fix any particular value of $d$. Thus, this argument extends the
conjecture of~\cite{Jia:2026ryl}. There only the $d = 4$ statement
was verified explicitly by computing the blocks entering~\eqref{eq:adsbh-rho-exact} numerically.

Turning this around,
the above factorization formula suggests that the generic semiclassical comb
channel Virasoro block satisfies some non-trivial identities to ensure the
factorization occurs for $\nu \in \mathbb{Z}_{> 0}$.

\paragraph{Remark (double-cover form).}
In fact, for comparison with~\cite{Jia:2026ryl} we need a small modification.
Their earlier result was derived in a projective coordinate
\begin{equation}\label{eq:zproj}
\zpr = z^2 =  \frac{r_+^2}{r^2}\,.
\end{equation}
Now at the boundary $\zpr = 0$, we define the logarithmic branch coefficient
with an explicit sign
\begin{equation}
\begin{aligned}
\psi_{\mathrm{source}}^{(\zpr)}(\zpr)
& = \zpr^{\frac{1-\nu}{2}}(1 + \cdots) + \mathcal L_{\nu}^{(\zpr)}\,
\psi_{\mathrm{vev}}^{(\zpr)}(\zpr)\log \zpr\,, \\
\psi_{\mathrm{vev}}^{(\zpr)}(\zpr)
& = \zpr^{\frac{1+\nu}{2}}(1 + \cdots)\,.
\end{aligned}
\end{equation}
With this choice, the coefficients are related as
\begin{equation}
\mathcal L_{\nu}^{(\zpr)}
= \frac{1}{2}\, \mathcal L_{\nu}^{(z)}\,.
\end{equation}
In the double-cover coordinate the monodromy is not $\sigma_z$ but $\sigma_\zpr$.
However, the two traces are simply related by
\begin{equation}
\cos(2\pi\sigma_z) = 2(-1)^{\nu}\cos(2\pi\sigma_\zpr)+\cosh(\hbw)\,.
\end{equation}
Consequently, the factors in the spectral function now are
\begin{equation}
\rho_{\mathrm{log}}
= -2\pi\,\mathcal{L}_{\nu}^{(\zpr)}\,, \qquad
\rho_{\mathrm{mon}} = \frac{\sinh(\hbw)}
{\cosh(\hbw)+(-1)^{\nu}\cos(2\pi\sigma_\zpr)}\,.
\end{equation}
The results of~\cite{Jia:2026ryl} were presented in this form. We will have further occasion to work with the projective $\zpr$ coordinate in our WKB analysis.

\newpage

\part{A singularity-anchored WKB analysis}
\label{prt:WKB1}
\section{WKB heuristics and asymptotics of the spectral function}
\label{sec:specasym}
The high frequency behavior (at fixed $\zeta$) of the Green's function can be
deduced once we control how the composite monodromy parameter $\sigma$ behaves.
This is immediate from the simplification of the spectral
function~\eqref{eq:adsbh-rho-exact}. For the case of planar \SAdS{5} black holes
the basic procedure was laid out in~\cite{Jia:2026ryl}. There it was argued that
the exact-WKB analysis is well suited to ascertain the asymptotics of the
composite monodromy trace $\Tr(M_{\mathrm{bdy},\mathrm{hor}})$ which is all one
needs for our spectral function asymptotics. Moreover, the authors used the
complementary form of the trace: in suitable coordinates the wave equation has
singular points at the horizon, boundary, complex horizon and singularity. The
composite trace between horizon and boundary could be equivalently phrased in
terms of the composite trace between the singularity and the complex horizon(s).
We will tackle the problem here directly between the boundary and the horizon.

One advantage of planar \SAdS{5} was that there were only two turning points at
fixed $\zeta$. In general dimensions or for other black holes, there are more
solutions to the equation $\zeta^2\, f(r) = 1$. In practice, this implies that
we have to be circumspect in our identification of the principal $A_s$-cycle of
interest. Moreover, as in the case of planar \SAdS{5} black holes the
asymptotics along the ray $\Im(\omega) = 0$ requires regulation using median
symbols. All these are easily achieved given the WKB spectral curve and its
associated Stokes graph. Logically, using this information would be the correct way to proceed as we do in~\cref{prt:WKB2}. However, it is instructive to attempt a simpler analysis first, which will turn out to provide a \emph{naive}, but correct, answer for the leading WKB result.

We assume at the outset that the appropriate $A_s$-cycle has been identified
(and suitably median regulated). More pertinently, we  hypothesize that the cycle
encircles the singularity. Doing so we will be able to write down an abstract
formula for the asymptotics and the location of time domain singularities
directly in terms of the apposite periods. The logic behind our construction is
simple: imagine the WKB wavefunction transported from the boundary into the
spacetime. Barring any obstruction (i.e., Stokes boundary) the wavefunction can
freely make its way down to the singularity, and then detour to the nearest
complex turning point; see~\cref{fig:globalads5-chamber-schematic} for an illustration.
In this process one will pick up a thermal residue from
the simple pole across horizon crossings. The formula we write down will be
based on this simple `free propagation + complex turning point' picture. While
heuristic, a numerical extraction of the asymptotics in~\cref{sec:examples} will serve to justify the final result.

Numerical ratification of our hypothesis notwithstanding, this picture should rightly be regarded as preliminary and heuristic. A priori it does not explain why the free propagation all the way down to the singularity is
relevant for causal observables such as the retarded Green's function.
Moreover, in some examples the locus of the singularity is a regular point of the leading WKB curve and ought not to play a distinguished role.

We examine carefully the assumptions we make in the present analysis
in~\cref{prt:WKB2}, and will find that they hold, provided we are willing to
assume a certain analytic structure for the bulk wave operator. Furthermore, even though exact-WKB trajectories do not generally pass close to the singularity (for non-vanishing $\zeta$), one can deform the contour to always find a singularity-anchored representative. The discussion of this part should be viewed as providing a particular WKB representation of the black hole singularity.

\subsection{Leading near-boundary WKB analysis}
\label{sec:nearbdywkb}

The leading order WKB potential~\eqref{eq:Q0wkb} being insensitive to the
operator dimension, the associated WKB wavefunction does not know about the
Frobenius basis of solutions asymptotically.\footnote{Since we are
  working with operators whose dimensions are held fixed,  $\nu$ dependent
pieces are subleading. Dropping them results in the wave equation not being aware of the correct AdS asymptotic fall-off conditions.} Said differently, using the WKB wavefunctions
alone one is unable to fix the non-normalizable and normalizable modes that are
central to defining the boundary Green's function.

It will therefore be useful to examine the universal WKB wave equation, while
keeping track of the $\nu$ dependence so that we can ensure that the correct AdS
asymptotics are retained. Effectively, we are asking how the universal WKB basis
maps onto the Frobenius basis, i.e., the standard source/vev or
non-normalizable/normalizable basis of wavefunctions near the AdS boundary. This
is a one-off exercise and allows us to convert the WKB answers into boundary
correlation functions.

Working in the projective coordinate $\zpr = \frac{\rp^2}{r^2}$ defined in~\eqref{eq:zproj} and only retaining terms to
leading order near the boundary (which is at $\zpr = 0$) one has\footnote{See~\cref{app:Q0Q2examples} for explicit examples of the wave equations for scalar operators that we study in the course of our discussion.}
\begin{equation}
\psi''(\zpr) + \bqty{
  \frac{1-\nu^2}{4\,\zpr^2}  +
\frac{(1-\zeta^2)\,\omega^2 + \order{\omega^0}}{4\,\rp^2\,\zpr} + \order{\zpr^0}}\psi(\zpr)
= 0\,.
\end{equation}
Clearly, the WKB parameter is the dimensionless frequency\footnote{Since we work
  in general $d$ we will rescale the physical frequency by $\rp$. In $d = 4$ a
  rescaling by $2\rp$ as employed in~\cite{Jia:2026ryl} is convenient, so readers
comparing the two papers should be alert to this modification.}
\begin{equation}
\freq = \frac{\omega}{\rp}
\end{equation}
This is straightforwardly solved in terms of Bessel functions
\begin{equation}\label{eq:bdyJbasis}
\psi(\zpr) = \sqrt{\zpr}\, \pqty{c_1\,  J_{-\nu}\pqty{\xi} + c_2 \,J_{\nu}\pqty{\xi} }\,, \qquad \xi
\equiv \sqrt{1-\zeta^2}\, \freq \,\sqrt{\zpr}\,.
\end{equation}
The two local solutions near the boundary pick up the source,
$\psi_{\mathrm{source}} \propto J_{-\nu}$, and the expectation value,
$\psi_{\mathrm{vev}} \propto J_{\nu}$, respectively.\footnote{This basis assumes
  $\nu \notin \mathbb{Z}_{\geq 0}$. For integer $\nu$
we need to pick a suitable non-degenerate Bessel basis.} We have just completed the
usual asymptotic fall-offs into a leading WKB basis in our high frequency
scaling regime. The WKB wavefunctions can be matched to the near-horizon basis
$\psi_{\mathrm{in}}$ and $\psi_{\mathrm{out}}$, respectively, by noting that the
ingoing wave is $e^{i\xi}$ in the above notation. This fixes the WKB connection
coefficients
\begin{equation}\label{eq:ads-bdy-wkb}
\begin{aligned}
A_i & = \mathscr{N}_\nu\; \Gamma(1 + \nu)\, \pqty{\frac{(1-\zeta^2)\,\freq^2}{4}}^{-\frac{\nu}{2}}\, e^{\ii\frac{\pi}{2}\ms \nu}\,, & \qquad A_o & = A_i^*\,, \\
B_i & = -\mathscr{N}_\nu\; \Gamma(1 - \nu)\, \pqty{\frac{(1-\zeta^2)\,\freq^2}{4}}^{\frac{\nu}{2}}\, e^{-\ii\frac{\pi}{2}\ms \nu}\,, & \qquad B_o & = B_i^*\,.
\end{aligned}
\end{equation}
Here $\mathscr{N}_\nu$ is some local normalization factor (dependent on $\nu$) that will cancel out in the ratio. A convenient choice turns out to be
$\mathscr{N}_\nu = \frac{1}{\sqrt{\pi}\, \nu}$. From here we can read off the WKB form of the
retarded Green's function, which is dominated by the vacuum contribution (since it is a near-boundary analysis). Define therefore the vacuum Green's function computed in the empty \AdS{} geometry
\begin{equation}\label{eq:GR-vac}
\Gret^{\mathrm{vac}}
=\frac{\Gamma(1-\nu)}{\Gamma(1+\nu)}
\left(\frac{(1-\zeta^2)\, \freq^2}{4}\right)^\nu\, e^{-i\pi\nu}\,.
\end{equation}
In the black hole geometry one then finds the expected result, viz.,
\begin{equation}
\Gret(\omega,k) \sim \Gret^{\mathrm{vac}}(\omega,k)\,, \qquad \freq
\to \infty\,,
\end{equation}
using which we can deduce that
\begin{equation}
\label{eq:leadingWKB-GR}
\abs{\Gret(\omega,k)} \sim \abs{\frac{\Gamma(1-\nu)}{\Gamma(1+\nu)}}\, \pqty{\frac{\omega^2 -k^2}{4\,\rp^2}}^{\nu}\,, \qquad \phGR(\omega,k)
\sim - \pi \ms\nu\,,
\end{equation}
along the real frequency ray. There is a physically invariant statement that one
can make using the observation that the gamma ratio is positive for
$0 < \nu < 1$. As we change $\nu$ the sign changes are absorbed by demanding
\begin{equation}
\phGR^{(0)} = -\pi\ms \nu \pmod{\pi}\,,
\end{equation}
with the correct branch determined by the positivity of the spectral function.
We will use these results later in extracting the asymptotics of the Green's
functions and spectral density. In fact, it will suffice to fix $\nu \in (0,1)$
and examine asymptotics of the various quantities of interest.

\subsection{Subleading correction and the composite monodromy}
\label{sec:subleadsigma}

Since we have the leading piece from WKB, we can make progress by
rearranging~\eqref{eq:monoeq1}. Define
\begin{equation}
\label{eq:lead-phiGR}
\delphR \equiv \phGR - \phGR^{(0)}
= \phGR + \pi\ms \nu\,, \qquad (0 < \nu < 1 )\,.
\end{equation}
Then using the definition we have
\begin{equation}
\cmtrace = -\frac{1}{2\, \sin(\phGR)}\,\bqty{
  e^{\hbw}\, \sin(\phGR + \pi \nu) +
e^{-\hbw}\, \sin(\phGR - \pi \nu)} \,.
\end{equation}
Upon using~\eqref{eq:lead-phiGR} we can express the composite monodromy as
\begin{equation}
\label{eq:mono-phase-rel}
\cmtrace = \frac{e^{\hbw}}{2\, \sin(\pi\nu)}\, \, \delphR
- \cos(\pi \nu)\, e^{-\hbw}
+ \order{e^{\hbw}\, \delphR^2, e^{-\hbw}\, \delphR} \,.
\end{equation}
Notice that as promised earlier we have picked up a thermal factor $e^{\hbw}$,
which is not obviously present from the exponentiation of the naive period
integral. In fact, all the information about the periods is now buried in the
deviation of $\phGR$ from its UV value $-\pi\ms \nu$, which is where the WKB
corrections and thus the asymptotics of the spectral function are hidden. More
precisely, the branch convention choice is $\phGR \sim - \pi \nu \, \pmod \pi$
from the leading WKB analysis, so we will seek corrections to $\delphR$.

\subsection{Periods and conventions}\label{sec:perconv}

The turning points of the WKB spectral curve are located at the roots of
$\zeta^2 \, f(r) = 1$. Depending on the background of interest and on
$\arg \zeta$ one may have a real turning point at some physical location
$r > 0$, or only complex turning points. Denote the set of all turning points by
$R\subm{tp} = \set{r\subm{tp} \mid \zeta^2\, f(r\subm{tp}) = 1}$. Further, let
$\rbull$ be the outermost positive real turning point, if present, and zero otherwise:
$\rbull = \max\bigl(\set{0} \cup (r_{\mathrm{tp}} \in R\subm{tp} \cap r_{\mathrm{tp}} \in \mathbb{R}_{>0}) \bigr)$.
This is the nearest point to the boundary that lies on the original WKB contour running
along the ray $\arg r = 0$ (since the transport originates at the boundary).
Furthermore, let $\rstar$ be a distinguished (complex or real)
turning point of the WKB spectral curve.

Given this, define a closed cycle, which we declare for now to be the
$A_s$-cycle, with the subscript suggestive of the cycle being associated with
the singularity. This cycle encircles the cut running from $r = \rbull$ to
$r = \rstar$.\footnote{The case where $\rbull = 0$ requires some care. Depending
  on the spacetime dimension $d$, $r = 0$ may end up being a regular point of the
  WKB spectral curve. If so, it is not a branch point of a cut, and one needs a
  more careful argument. We will see how to deal with these situations
in~\cref{sec:pschwads}.} We will also be interested in another cycle, a relative
contour running from the asymptotic boundary to $\rbull$. Denote this as
$\Gamma_{\partial \rbull}$. The integral of the WKB differential along this
cycle will determine the WKB action for the physical trajectory, which we
declare to be the physical phase, i.e.,
\begin{equation}\label{eq:Pphys-def}
\Pphys \equiv \rp \ \mathrm{PV} \int_{\rbull}^\infty\, \lwkb \,.
\end{equation}
The integral is on a sheet where $\lwkb >0$ outside the horizon. While the transport path is directed inward from the boundary, $\Pphys$ is defined as the positive outward-oriented phase used in the action computation. Define further, the $A_s$-cycle period (NB: this is still a classical period)
\begin{equation}
\Aper = \rp\, \oint_{A_s}\, \lwkb \,,
\end{equation}
and the contour integral (taken counterclockwise) around the physical outer horizon, viz., the positive
root of $f(r)$ set to be at $r = \rp$
\begin{equation}
\oint_{\rp} \lwkb = \frac{\ii}{2}\,\beta\,, \qquad \beta
= \frac{1}{T}
= \frac{4\pi}{\rp^2\, f'(\rp)}\,.
\end{equation}
One can similarly define period integrals around other roots of $f(r)$. These
may be the inner (Cauchy) horizon of charged \RNAdS{d+1} black holes or other
complex horizons, which also will contribute residues $\beta_k$. For
definiteness, let $\beta_{\mathrm{inner}}$ denote the inverse temperature of
the Cauchy horizon, and let $\beta_{c,(j)}$ denote the analog for complex
horizons with appropriate phase choices (which will be fixed below). We then
claim that the asymptotics of the monodromy is controlled by the period $\Aper$,
horizon contributions $\beta_k$, and the physical WKB action $\Pphys$.

\subsection{WKB actions and the phase of the retarded propagator}\label{sec:Iwkb-conj}

Our aim is to control the deviation of $\phGR + \pi \nu$ away from the origin and derive non-perturbative corrections to it.
Let us look at $\lwkb$ which contains information about the
turning points as discussed above. On the physical sheet we are in the
classically allowed region, so the asymptotic WKB wavefunctions are happily
transported to a real turning point (if it exists) or extend all the way to
$r = 0$, the black hole singularity. This is the action integral along
$\Gamma_{\partial \rbull}$. Any deviations come from exponentially small
corrections owing to the presence of additional (real/complex) turning points in
the set $R\subm{tp}$, and the associated Stokes crossings.

Insofar as the ratio of the asymptotic fall-off coefficients $A_i$ and $B_i$ is
concerned, the Stokes crossings carry a weight which is set by the relative
action from the boundary to the turning point along the crossing paths. The WKB
weight is $\exp(i\, \freq\, \Iwkb)$ with
\begin{equation}
\label{eq:Iwkb-abstract}
\Iwkb = 2\,\pqty{ \Pphys +
  \ii\pi\, \rp\,\sum_{k}\, \eta_k\, \Res_k\, \lwkb +
\rp \int\limits_{\rbull}^{r_\mathrm{tp}}\, \lwkb }\,,
\end{equation}
for a turning point $r\subm{tp} \in R\subm{tp}$. Here, we are assuming that the
path to the turning point runs past some poles of $\lwkb$, which contribute to
the residues with signs $\eta_k$ depending on the direction of traversal (above
or below the pole). The integrals between the turning points can be converted to
periods enclosing the cuts, leading to
\begin{equation}
\int_{\rbull}^{r\subm{tp}}\, \lwkb = \frac{1}{2} \, \oint\limits_{\Gamma_{\rbull\, r_{\mathrm{tp}}}}\, \lwkb\,.
\end{equation}

In all examples we study we will find that this turning point cycle is precisely
the $A_s$-cycle singled out above, with $\rstar$ as the distinguished complex
turning point. For neutral black holes $\rbull$ will end up being the
singularity for planar horizons in $d \equiv 0 \pmod{4}$. For global neutral black holes the location depends on $\zeta$, while charged black holes have a real turning point.
Assuming that we have correctly identified the $\rstar$ turning point, we
find that the accumulated phase for the WKB trajectory accounting for the Stokes
crossings upon using the half-residue for a crossing of the horizons is
\begin{equation}
\label{eq:Iwkb}
\Iwkb = 2\, \Pphys + \ii \sum_k\, \eta_k\, \frac{\rp\,\beta_k}{2} + \Aper \,.
\end{equation}

Since the physical WKB phase is real, it follows that the oscillations are
controlled by
\begin{equation}
\RIwkb = 2\, \Pphys + \RAper\,,
\end{equation}
while the exponential envelope is controlled by
\begin{equation}
\IIwkb = \sum_k\, \eta_k\, \frac{\rp\,\beta_k}{2} + \IAper\,.
\end{equation}
In writing this we assume that all the horizon crossings contribute real half-periods, viz., $\Im \beta_k =0$.

Accounting for these phases we conclude that
\begin{equation}
\label{eq:argGR-asymptotics}
\phGR + \pi \nu \sim \kappa\, e^{-\IIwkb\, \freq}\, \cos(\RIwkb\,\freq + \varphi)\,,
\end{equation}
with an overall normalization and additional phase shift to be derived. Using
the leading order Bessel analysis one can fix the $\order{1}$ factors $\kappa$
and $\varphi$. To do so, note that a simple turning-point Stokes crossing
shifts the boundary phase by
$-2\, \sin(\pi \nu)\, \Im(S\, e^{\ii \freq\ms \Iwkb})$. The factor
$2\sin(\pi\nu)$ is the Bessel connection factor derived hitherto from the near
boundary analysis. The Stokes factor $S$ for crossing a Stokes line emanating
from a simple turning point can be determined by a local Airy analysis. In our
conventions it supplies a factor of $S = -\ii$. This then fixes
\begin{equation}
\kappa = 2\, \sin(\pi\nu)\,, \qquad \varphi = 0\,,
\end{equation}
leading to the result
\begin{equation}\label{eq:delph-asymptotics}
\delphR = \phGR + \pi \nu
\sim 2\, \sin(\pi\nu)\, e^{- \freq \ms \IIwkb}\, \cos(\RIwkb \ms \freq)\,.
\end{equation}

Armed with this result we can finally state the asymptotics of the composite
monodromy trace. Plugging into~\eqref{eq:mono-phase-rel} we arrive at
\begin{equation}
\label{eq:sigma-asymptotics}
\begin{aligned}
\cmtrace & \sim \exp(\bqty{\frac{\beta\,\rp}{2}  - \IIwkb}\ms \freq)\, \bqty{\cos(\RIwkb\ms \freq)
+ \order{\freq^{-1}} + \cdots} \\
& \qquad \qquad - \cos(\pi \nu)\, \exp(-\frac{\beta\,\rp}{2}\,\freq) \,,
\end{aligned}
\end{equation}
in the limit $\freq \to \infty$. The composite monodromy trace has an
exponential envelope with oscillations. In particular, away from the zeros of
the oscillatory factor it lies in the hyperbolic region. Near the oscillatory
nodes the subleading terms and the decaying KMS term decide whether it crosses
into the elliptic domain. We  highlight the presence of the physical and complex horizon temperatures through $\beta$ and $\beta_k$ in the final answer.

We will provide a sharper argument for~\eqref{eq:Iwkb}
in~\cref{prt:WKB2}. What we can check directly is that it agrees with explicit numerical evaluation of $\delphR$ or $\cos(2\pi \sigma)$ very well in the large $\freq$ regime. We shall see this for a variety of examples, planar-\SAdS{d+1} black holes in
$d = 4,6,8$, global \SAdS{d+1} black holes in $d = 4,6$, and the planar \RNAdS{5}
black holes, in~\cref{sec:examples}. It is also interesting to the note that these features arise in the study of high-energy reflection coefficients of asymptotically flat black holes, cf.~\cite{Basha:2018bvi}. 

\subsection{Asymptotics of the spectral function}
\label{sec:rho-asym}

We now have sufficient data to work out the asymptotics of the spectral
function. Since
\begin{equation}
\rho(\omega,k) = -2\, \Im(\Gret(\omega,k))
= -2\, \abs{\Gret(\omega,k)}\, \sin(\arg\Gret(\omega,k))\,,
\end{equation}
we may directly use the leading WKB answer~\eqref{eq:leadingWKB-GR} along with
the deduced correction to the phase~\eqref{eq:argGR-asymptotics} to immediately
obtain
\begin{equation}
\label{eq:specfn-asymnu}
\begin{aligned}
\rho(\omega,k) & = \frac{2\pi}{\nu\,\Gamma(\nu)^2}\, \pqty{\frac{\omega^2 -k^2}{4\,\rp^2}}^{\nu}\, \bqty{1- 2\, e^{-\freq\, \IIwkb} \,\cos(\pi\nu) \,
\cos(\RIwkb\, \freq) } + \cdots
\end{aligned}
\end{equation}
We have picked the correct branch for $\phGR$, cf.~the discussion
around~\eqref{eq:mono-phase-rel}. The exponential damping scale and the phase of the oscillations follow from the asymptotics of $\delphR$. We shall show later in~\cref{prt:WKB2} that exact-WKB analysis shows that the amplitude and absolute phase are off. The exact expression requires the replacement $\cos(\pi\nu) \,\cos(\RIwkb\, \freq) \to \cos(\RIwkb\, \freq - \pi\nu)$, which can be derived from~\eqref{eq:ewkb-Gret}.

Let us contrast this analysis with the special case of
$\nu \in \mathbb{Z}_{> 0}$ examined in~\cite{Jia:2026ryl}. Since the WKB
spectral curve is agnostic about $\Delta_{\mathcal{O}}$ or $\nu$, not only is
this a legitimate specialization we can use, but also it provides a useful
cross-check of our formalism. Strictly speaking we are not entirely agnostic
about $\nu$, but the only place $\nu$ enters is the leading WKB analysis when we
had to keep track of the asymptotic fall-offs. While integer dimensions require
logarithmic terms in holographic renormalization, notice that the final answer
for the spectral function, say~\eqref{eq:specfn-asymnu} has a smooth limit as
$\nu$ is taken to be an integer. The $\sin(\pi \nu)$ factors arising from the
Wronskian which feed into the holographic renormalization precisely ensure the
smoothness of the limit.

For $\nu \in \mathbb{Z}_{> 0}$ as in~\cite{Jia:2026ryl} we can then work out
asymptotics directly using~\eqref{eq:adsbh-rho-exact}, which can be checked to
simplify in this case (using the reflection identity of the gamma functions) to
\begin{equation}
\label{eq:rho-Intnu}
\rho(\omega,k) = P_{2\nu}\pqty{\ii\frac{\bw}{4\pi},\sigma}\, \mathscr{W}(\omega,k)\, \frac{\sinh(\hbw)}{\cosh(\hbw) + (-1)^\nu\, \cmtrace}\,,
\end{equation}
with $P_{2\nu}(x,y)$ being a degree $2\nu$ polynomial of its two arguments. As a
cross-check of our analysis we can recover~\eqref{eq:specfn-asymnu} for
$\nu \in \mathbb{Z}_{> 0}$ by using~\eqref{eq:sigma-asymptotics} directly
in~\eqref{eq:rho-Intnu}.

\paragraph{A prediction for the semiclassical Virasoro block:}
There is an interesting corollary of this observation regarding the
semiclassical block. It was demonstrated in~\cite{Jia:2026ryl} that the
conformal block $\mathscr{W}(\omega,k)$ combines with the prefactor $P_{2\nu}$
to reproduce the power law behavior, $(\omega^2 - k^2)^{\nu}$, of the vacuum
spectral function, for planar \SAdS{5}. In~\cref{sec:rho-Z-factors} we demonstrated exact factorization using logarithmic monodromy for $\nu \in \mathbb{Z}_{>0}$. The leading high-frequency piece from there already gives the vacuum power law, with subleading thermal corrections.  However, if as we might suspect,
$\nu \in \mathbb{Z}_{> 0}$ is just a special case of a more general result,
it must be the case that the factor of $\pqty{\omega^2 - k^2}^{\nu}$, even for
non-integral values of $\nu$, must imply that the semiclassical blocks ought to simplify.

We can attempt to extract the asymptotics of the channel momentum as long as we are
away from the regime where the oscillatory factor
in~\eqref{eq:sigma-asymptotics} starts to be important. In other words, from the
region where the composite monodromy trace is in the hyperbolic class, we could
posit that
\begin{equation}
\sigma \sim \frac{\ii}{2\pi}\, \freq\, \pqty{\frac{\beta\,\rp}{2}  - \IIwkb}\,.
\end{equation}
The spectral function~\eqref{eq:adsbh-rho-exact} can then be simplified using
Stirling's approximation.\footnote{We can use
  \begin{equation*}
  \abs{\Gamma(x + i\, y)}^2 = 2\pi\, \abs{y}^{2\,x - 1} \, e^{-\pi \,\abs{y}}\, \pqty{1+ \order{y^{-1}}}\,, \qquad \abs{y}
  \to \infty\,.
  \end{equation*}
} This approximation implies that the leading contribution to the spectral
function behaves as
\begin{equation}
\rho \sim \frac{2\pi}{\nu\,\Gamma(\nu)^2} 
\frac{\freq^{2\,\nu}}{(2\pi)^{2\,\nu}} \, \bqty{\pqty{ \frac{\beta \, \rp}{2}}^2 - \pqty{\frac{\beta\,\rp}{2}  - \IIwkb}^2}^{\nu}\, \mathscr{W}(\omega,k) \,.
\end{equation}
Combining this crude prediction with the leading WKB
result~\eqref{eq:leadingWKB-GR} we would require that
\begin{equation}
\mathscr{W} \sim \pqty{\frac{\pi^2 \, (1-\zeta^2)}{
\frac{(\beta \, \rp)^2}{4} - \pqty{\frac{\beta\,\rp}{2}  - \IIwkb}^2}}^{\nu}
+ \cdots \,.
\end{equation}
Thus, we end up with a simple prediction for the semiclassical block,
generalizing the result of~\cite{Jia:2026ryl} for
$\nu \notin \mathbb{Z}_{\geq 0}$.

\section{Analytic structure in the time domain \& the black hole singularity}
\label{sec:timedomain}

Given the high frequency asymptotics of the spectral function we can immediately
deduce that of the two-sided thermofield double correlation function, by
stripping off the statistical factor~\eqref{eq:Comega}. We want to examine the
analytic structure of this observable in the time domain, at fixed $\zeta$.

Recall that the KMS invariance of the thermal observable guarantees analyticity
in the fundamental thermal strip
$\Im(t) \in (-\frac{\beta}{2}, \frac{\beta}{2})$. For fixed spatial separation
one has a branch cut emanating from the $\Re(t) = \pm \abs{\vb{x}}$ (for
translationally invariant ensembles) at $\Im(t) = \pm \frac{\beta}{2}$ at the
edge of the thermal strip. Working at fixed spatial momentum entails that the
cut gets smeared; the function obtained in the fundamental strip need not
have a periodic analytic continuation outside of it, cf.~\cite{Dodelson:2025jff}.

At fixed $\zeta$, we can consider two observables in the time domain: the
frequency Fourier transform along a fixed momentum direction $\vu{k}$, or an
angularly averaged observable by smearing over the directions. Using the Fourier
kernel $e^{-\ii (\omega \ms t - k\, \vu{k} \cdot \vb{x})}$, we learn that the
transform is along a fixed ray $t - \zeta\, \vu{k} \cdot \vb{x}$. We can then
independently integrate over the angular components and use a saddle point to
extract the time domain singularities. To distinguish the two cases, introduce
\begin{equation}
\tkray = t - \zeta\, \vu{k} \cdot \vb{x}\,, \qquad \tzeta
= t - \zeta\, \abs{\vb{x}}
\end{equation}
Hence, the observables in the time domain are
\begin{equation}
\label{eq:C-fixed-vs-smeared}
\begin{aligned}
C_\zeta(\tkray,\hat{\vb{k}}) & = \int_{-\infty}^\infty\, d\omega\, e^{-\ii \omega\ms \tkray} \, C(\omega, \zeta\, \omega) \\
C_\zeta(\tzeta,\abs{\vb{x}}) & = \int_{-\infty}^\infty\, d\omega\, e^{-\ii \omega\ms t} \, \int_{\mathbf{S}^{d - 2}}\, d\Omega_{\vu{k}}\, C(\omega, \zeta\, \omega, \hat{\vb{k}})\, e^{\ii \zeta\, \omega\, \hat{\vb{k}} \cdot \vb{x}}\,.
\end{aligned}
\end{equation}
Thus, owing to the spatial smearing the time-domain thermofield double
correlator can (and will) have singularities outside the thermal strip. Their
locations will be controlled by our asymptotics of the spectral
function~\cite{Afkhami-Jeddi:2025wra,Dodelson:2025jff,Jia:2025jbi,Jia:2026ryl}.

\subsection{Time domain singular loci}
\label{sec:timesing}

Given our observation that $C(\omega) \sim \omega^{2\nu}\, e^{-\gamma\,\omega}$,
we deduce from the Fourier integral (using the time reflection symmetry of
static ensembles)
\begin{equation}
\int_0^\infty\, d\omega\, \cos(\omega\, t)\, \omega^{2\nu}\, e^{-\gamma\,\omega} = \frac{\Gamma(2\nu+1)}{2}\,\bqty{ \pqty{\frac{1}{\gamma+ \ii t}}^{2\nu +1} +
\pqty{\frac{1}{\gamma- \ii t}}^{2\nu +1}} \,.
\end{equation}
This pins the location of the complex time singularities to
$t_c = \pm \ii \gamma$. We can read off the value of $\gamma$
from~\eqref{eq:specfn-asymnu} to be related to $\IIwkb$ (modulo thermal factors).

For the fixed momentum ray observable, $C_\zeta(\tkray,\hat{\vb{k}})$, we only
perform a single Fourier transform. The thermofield double correlator then has
singularities at
\begin{equation}
\label{eq:tc-angle}
\begin{aligned}
t_{\hat{\vb{k}},\, c} & = \ii \frac{\beta}{2} + \frac{1}{\rp}\, \Iwkb\,,  & \qquad t_{\hat{\vb{k}},\, c} & = \ii \frac{\beta}{2} - \frac{1}{\rp}\,\Iwkb^*\,, \\
t_{\hat{\vb{k}},\, c} & = -\ii \frac{\beta}{2} - \frac{1}{\rp}\, \Iwkb\,, & \qquad t_{\hat{\vb{k}},\, c} & = -\ii \frac{\beta}{2} + \frac{1}{\rp}\,\Iwkb^*\,.
\end{aligned}
\end{equation}
Since the WKB action is independent of $\nu$ the location of the singularity is
universal. The order of the singularity for this fixed direction observable,
however, depends on $\nu$, and justifiably so, since the latter determines the
scaling exponents.

For the correlator smeared over the spatial directions,
$C_\zeta(\tzeta,\abs{\vb{x}})$, we also need to integrate over the angles.
Happily, this  is a standard Bessel integral
\begin{equation}
\int_{\mathbf{S}^{d - 2}}\, d\Omega_{\hat{\vb{k}}}\, e^{\ii \zeta\, \omega\, \hat{\vb{k}} \cdot \vb{x}} = (2\pi)^{\frac{d-1}{2}}\, (\zeta\, \omega\, \abs{\vb{x}})^{\frac{3-d}{2}}\, J_{\frac{d-3}{2}}(\zeta\, \omega\, \abs{\vb{x}})\,.
\end{equation}
This has the effect of shifting the order of the singularity from $2\nu + 1$ to
$2\nu - \frac{d-4}{2}$ using the large $\omega$ asymptotics of the Bessel
function (assuming $\omega \, \abs{\vb{x}} \gg 1$). The singular loci are now at
\begin{equation}
\label{eq:tc-position}
\begin{aligned}
t_c & = \pm \, \zeta\, \abs{\vb{x}} + \ii \frac{\beta}{2} + \frac{1}{\rp} \,\Iwkb\,, & \qquad t_c & = \pm \, \zeta\, \abs{\vb{x}} + \ii \frac{\beta}{2} - \frac{1}{\rp}\, \Iwkb^*\,, \\
t_c & = \pm \, \zeta\, \abs{\vb{x}} - \ii \frac{\beta}{2} - \frac{1}{\rp} \,\Iwkb\,, & \qquad t_c & = \pm \, \zeta\, \abs{\vb{x}} - \ii \frac{\beta}{2} + \frac{1}{\rp}\, \Iwkb^*\,.
\end{aligned}
\end{equation}
Notice that in both cases the thermal statistical factor ensures that the
thermofield double correlator is analytic in the thermal strip
$\abs{\Im(t)} < \frac{\beta}{2}$. This would not be the case for the spectral
function itself, which can end up with singularities within this domain.

The two observables we discussed above are not the correlator at fixed spatial
position. The angle averaged observable $C_\zeta(\tzeta, \abs{\vb{x}})$ is still
restricted to a fixed momentum ray. We can attempt to determine the fixed position
answer by completing the integral over $\zeta$ in a saddle point approximation.
This would give,
\begin{equation}
\rp\,\abs{\vb{x}} = -\dv{\zeta} \Iwkb \,, \qquad t_c
= \ii\frac{\beta}{2}
+ \frac{1}{\rp} \pqty{\Iwkb - \zeta\,\dv{\zeta} \Iwkb} \,.
\end{equation}
Operationally, one must solve for the complex $\zeta$ value for fixed $\abs{\vb{x}}$
and then use that to evaluate the location of the singularity in the complex
time plane. This has of course to be carried out with due care, for we are
obtaining the WKB actions anchored on $\zeta \in (0,1)$. But this is precisely
the connection between the WKB analysis and the geodesic discussion
of~\cite{Fidkowski:2003nf,Festuccia:2005pi}. Given the WKB action
$\mathcal{I}_{\wkb}$ the full eikonal phase is obtained using the
Hamilton-Jacobi ansatz~\eqref{eq:HJphase}. The implications for the
black hole singularity's imprint will be explored in~\cref{sec:examples}, cf.~\eqref{eq:psads-tc-x}, and connect to the OPE in~\cref{sec:ope-wkb}.

Recall that our WKB action has contributions from the physical phase shift from
the classically allowed region $\infty \to \rbull$, and a contribution from the
excursion to the complex turning point $\rbull \to \rstar$. In between, we pass
through the horizon crossings, which only show up in the time separation, since
they are $\zeta$ independent. We will check below that this general picture does
reproduce the predicted answer for the location of the bouncing singularity in the $\zeta \to 0$ limit. In $d = 4$ for
neutral black holes we only have a single horizon crossing, while for charged
black holes the turning point lies inside the inner horizon and thus picks up
the contribution from crossing it as well (this reproduces the observation
of~\cite{Ceplak:2025dds}).

\subsection{Connection to asymptotic quasinormal modes}
\label{sec:asymqnm}

We argued earlier that the location of the quasinormal modes, the poles of the
retarded Green's function, are given by the solution to~\eqref{eq:qnmloc}.
We can now use the asymptotics of the composite monodromy trace to obtain a prediction
for the asymptotic behaviour of the quasinormal spectrum at fixed $\zeta$, which is an analytically continued fixed momentum ray slice. This
discussion generalizes the original study of asymptotic quasinormal modes of
flat spacetime black holes using monodromy techniques in~\cite{Motl:2003cd}. The
corresponding analysis for \AdS{} black holes was carried out
in~\cite{Natario:2004jd} and further key results derived using the semiclassical
WKB analysis in~\cite{Festuccia:2008zx,Arnaudo:2026tcy}. Note that most of these
analyses were at fixed spatial/angular momentum with the only large parameter being
frequency (in~\cite{Festuccia:2008zx} the high momentum limit was also analyzed). We are largely interested in keeping $\zeta$ fixed. The earlier
results comprise the limiting case at the timelike end $\zeta \to 0$ (direction
of approach being set by $\arg \zeta$). The limiting behaviour on the lightlike
end $\zeta \to 1$ (with real momenta) gives the results
of~\cite{Festuccia:2008zx} who demonstrated the presence of weakly damped modes.

\begin{figure}[h!]
  \centering
  \includegraphics[width=\linewidth]{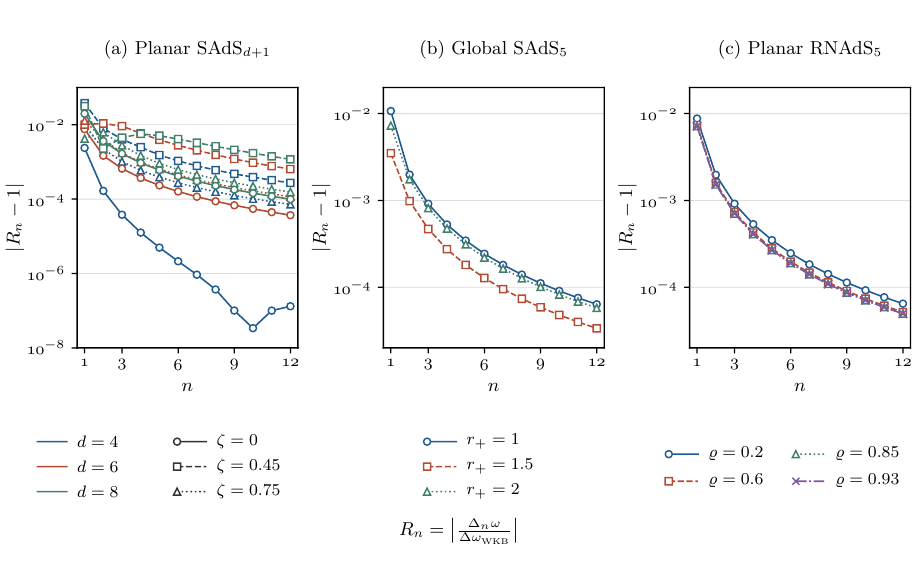}
  \caption{A numerical check of our WKB prediction for the asymptotic quasinormal mode spacing for various examples.
    We plot $\abs{R_n-1}$ on a logarithmic scale against the overtone number $n$. $R_n$ is the ratio of the difference between successive quasinormal mode locations to our WKB spacing prediction, cf.~\eqref{eq:Rnqnm}.
    The three panels show:
    (a) Planar \SAdS{d+1} for $d=4,6,8$ (color coded), each at
    $\zeta=0,0.45,0.75$ (with different markers). (b) Global \SAdS{5} at
    $\rp=1,1.5,2$, with $\zeta=0.75$. (c) Planar \RNAdS{5}:
  $\varrho=r_-^2/\rp^2=0.2,0.6,0.85,0.93$, with $\zeta=0.75$. The data are nearly collinear at higher values of the charge. In all cases we used a neutral scalar wave equation with $\nu =3/4$, and held $\zeta$ fixed. }
  \label{fig:qnm-lattice-spacing-summary}
\end{figure}

For now, we will obtain a general prediction, and subsequently compare against
the explicit predictions when we discuss explicit examples. However, it is
nevertheless tempting to present the result for the asymptotic spacing upfront,
which the reader can therefore find in~\cref{fig:qnm-lattice-spacing-summary}
for the cases we examine in~\cref{sec:examples}.

To obtain the lattice spacing, we proceed as follows. From the
asymptotics~\eqref{eq:sigma-asymptotics} we obtain upon setting
\begin{equation}
\Iwkb = \beta\, r_+\, I \,, \qquad I = I_1 + \ii I_2\,,
\end{equation}
the condition
\begin{equation}
e^{-\hbw} \, \cmtrace \sim \frac{1}{2} \pqty{e^{\ii I\ms \bw} + e^{-\ii I^* \ms \bw}}\,.
\end{equation}
The quasinormal mode condition~\eqref{eq:qnmloc} can be solved by balancing one
of these terms against the corresponding growing piece of
$\cosh(\hbw + \ii \pi \ms \nu)$ to obtain (depending on $\arg \omega$) the
families
\begin{equation}
\begin{aligned}
i\, I\, (\bw_n)_+ = \log\pqty{-e^{\ii \pi \ms \nu}}
+ 2\pi\ii \ms n\,, \qquad
- i\, I^*\, (\bw_n)_-
= \log\pqty{-e^{\ii \pi \ms \nu}} + 2\pi\ii \ms n\,.
\end{aligned}
\end{equation}
Here $n$ is an integer, and since we are seeking asymptotic behavior, $n \gg 1$.
The relations above predict the asymptotic spacing is set by
\begin{equation}\label{eq:asym-qnm-pred}
\Delta(\bw)_+ = \bw_{n + 1} - \bw_{n} \sim \frac{2\pi}{I} + \cdots \,,
\end{equation}
with the ellipses denoting subleading corrections at large $n$. Similarly, on
the negative branch the spacing is $-2\pi (I^*)^{-1}$.

In~\cref{fig:qnm-lattice-spacing-summary} we show
the effectiveness of our prediction for the asymptotic spacing~\eqref{eq:asym-qnm-pred} at
$\zeta \neq 0$ for various examples of asymptotically AdS black holes. In fact, we also  verify that these spacings reproduce known results in the $\zeta \to 0$ limit. The plot was obtained by computing quasinormal modes by directly solving the wave equation (chosen example was a scalar with $\nu = 0.75$). The plot shows on a logarithmic scale, the relative deviation $\abs{R_n - 1}$ across overtone numbers, which we defined as
\begin{equation}\label{eq:Rnqnm}
R_n= \frac{\omega_n - \omega_{n-1}}{\omega\subM{WKB}} =
\Iwkb\, \frac{\omega_n - \omega_{n-1}}{2\pi\ \rp}\,.
\end{equation}
The relative deviation captured by $\abs{R_n -1}$ is already small for overtone numbers $n \geq 8$ or thereabouts. We present this result here mostly to indicate that our singularity-anchored approach actually captures the correct asymptotics. In the next section we study the examples used above in greater detail, exposing  the predicted oscillations in the retarded Green's function phase.

\section{Black hole WKB spectral curves}
\label{sec:examples}

We now turn to a detailed examination of the WKB spectral curves for various examples.
The aim here is to provide a clear picture of the turning points, the singular loci, etc., of the leading WKB curve, and justify our leading WKB action numerically. This analysis will be a useful precursor to the more careful exact-WKB treatment for the same set of examples in~\cref{prt:WKB2}.
We shall start with planar \SAdS{d+1} black holes, then turn to the global
(spherically symmetric) neutral solutions, before finally describing the case of
planar \RNAdS{d+1} black holes. In the explicit evaluation of periods we will
restrict to even boundary dimensions $d \geq 4$.\footnote{For odd dimensional
boundaries the bulk black hole spacetimes do not admit an $r \to -r$ involution.  Since we exploit this at several points of our analysis we have refrained from an explicit analysis for $d$ odd. We thank M.~Kulaxizi for a discussion on this point.}

\subsection{Planar Schwarzschild-AdS black holes}
\label{sec:pschwads}

The metric function for the planar black holes is
\begin{equation}
f(r) = 1 - \frac{\rp^d}{r^d} \,,
\end{equation}
which has $d$ roots at $r = \rp \, e^{\frac{2\pi \ii m}{d}}$ with
$m \in \set{0,1,\cdots, d-1}$, and as noted above $d$ is even. The inverse
temperature of the dual CFT is
\begin{equation}
\label{eq:beta-planar}
\beta = \frac{4\pi}{d\,\rp}
\end{equation}
The WKB spectral curve in the radial coordinate variable is defined by the
quadratic differential
\begin{equation}
Q_0(r) = - r^{d - 4}\, \frac{r^d\, (1- \zeta^2) + r_+^d \, \zeta^2 }{ (r^d - r_+^d)^2}
\end{equation}
As indicated in~\cref{sec:wkbspectral} we will consider both $\zeta \in (0,1)$
and $\zeta = \ii\chi$ with $\chi \in \mathbb{R}_+$, though general formulae will
be written in terms of $\zeta$ for simplicity.

One can exploit the $\mathbb{Z}_2$ symmetry $r \to -r$ visible in the curve and
work with the following projective coordinate:
\begin{equation}\label{eq:rproj}
\rpr = \frac{1}{\zpr} = \frac{r^2}{\rp^2} \,.
\end{equation}
Locally, this $\rpr$ coordinate will be more convenient than $\zpr$. The boundary is at $\rpr = \infty$ and the curvature singularity is at $\rpr = 0$, while keeping the
horizons at $\rpr_m = e^{-\frac{4\pi i m}{d}}$ with
$m \in \set{0,1,\cdots, \tfrac{d}{2} -1}$. The WKB spectral curve is then
characterized by the differential\footnote{When we write the WKB differential in dimensionless coordinates like $\rpr$ we refrain from including factors of $\rp$. These will instead be absorbed into the frequency, so that the WKB action is built out of dimensionless variables. }
\begin{equation}
\lwkb(\rpr) = \rpr^{\frac{d}{4} - \frac{3}{2}}\, \frac{\sqrt{\rpr^{\frac{d}{2}}\, (1-\zeta^2)+\zeta^2}}{2\,\pqty{\rpr^{\frac{d}{2}} -1}}\, d\rpr\,.
\end{equation}
We analyze the two cases distinguished by $\arg \zeta$ separately.

\begin{figure}[p]
  \centering
  \includegraphics[width=\linewidth]{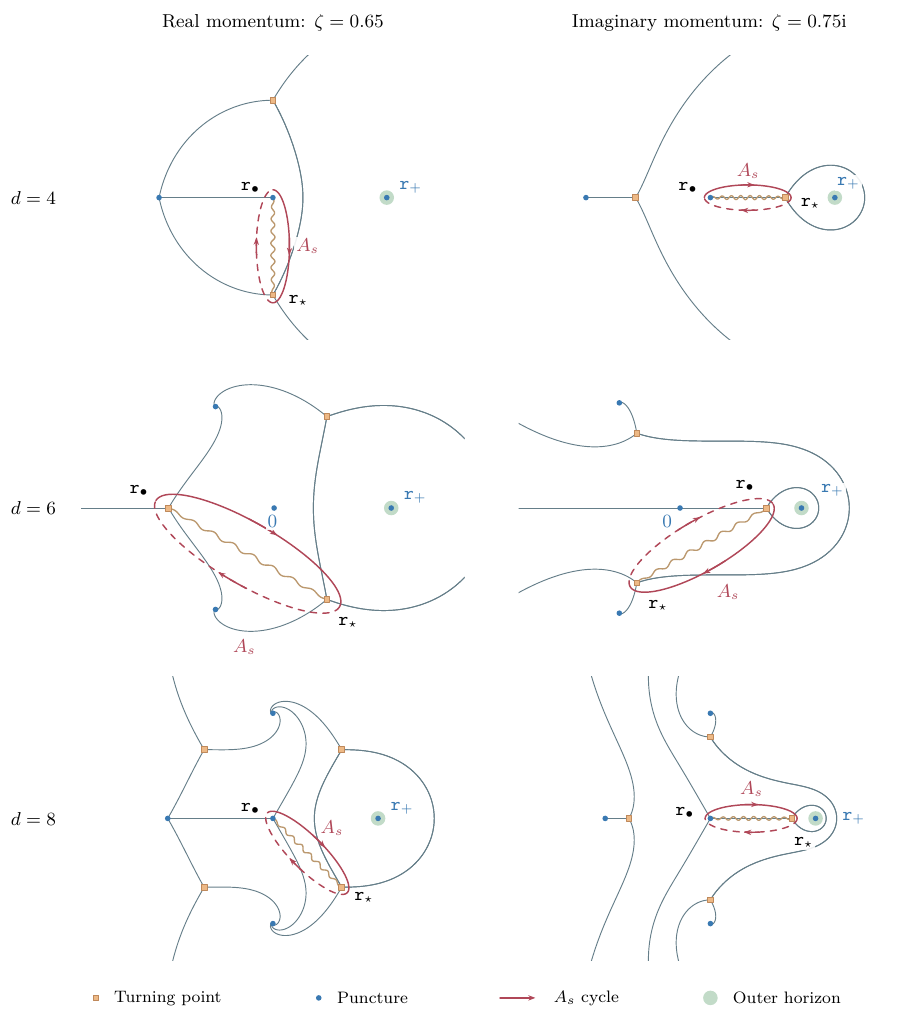}
  \caption{
    Stokes graphs of the planar \SAdS{d+1} leading WKB spectral curve in
    the $\rpr$ coordinate, where the curvature singularity is at $\rpr=0$.
    We plot these for $d=4,6,8$ comparing for each case real timelike momentum (at $\zeta=0.65$, left) and purely imaginary momentum ($\zeta=0.75\ii$, right). The graphs are for real frequency, and are critical saddle graphs.  The WKB turning points are marked as orange squares, while the blue circles are the punctures (horizons, singularity, \AdS{} boundary). The red $A_s$ cycle is depicted as the closed cycle encircling the cut
    running from $\rpstar$ to $\rpbull$. We also highlight (green shading) the physical outer horizon. Note that
    the $d=4$ and $d=6$ curves have genus one.
  The transport path from the boundary to the horizon along the $\Re \rpr$ axis crosses Stokes lines, which emanate from the turning  point. These are temporarily ignored for the heuristic analysis of this section, but will be fully  accounted for in~\cref{prt:WKB2}. The crossings do not affect the leading action; they only matter for subleading corrections.}
  \label{fig:planar-sads-stokes-cycles}
\end{figure}

The turning points of the WKB potential with $\arg \zeta = 0$ are located at
\begin{equation}
\label{eq:planar-uturn}
\rpr_{j} = \pqty{\frac{\zeta^2}{1-\zeta^2}}^{\frac{2}{d}}\, \exp(\frac{2\pi \ii}{d}\, (2\, j + 1)) \,, \qquad j
\in \set{0,1,\cdots , \tfrac{d}{2}-1}\,.
\end{equation}
Thus, we have $\frac{d}{2}$ turning points. In addition, the AdS boundary at
$\rpr = \infty$, is a branch point of the above differential, as is clear from the
explicit $\rpr^{\frac{3}{2}}$ factor. However, the status of $\rpr = 0$, the curvature
singularity, depends on $d\pmod{4}$ (equivalently on the odd/even parity of
$\frac{d}{2}$).

\paragraph{A distinction between $d = 4\,n$ and $d = 4n + 2$:}
For $d = 4\,n$ the locus $\rpr =0$ is a branch point,
but it is invisible to the leading WKB
differential for $d = 4\,n + 2$. The total number of branch points in either
case is even. When $\rpr = 0$ is regular, for $\zeta \in (0,1)$ it will transpire that there is a real turning point at $\rpr < 0$.
The full set of singular points of the WKB curve are these branch points
along with the complex horizon roots at $\rpr^{\frac{d}{2}} = 1$. The turning point
$\rpr_{\frac{d}{2}-1}$ of~\eqref{eq:planar-uturn} will be our distinguished complex turning point $\rpstar$ for all the
cases we examine. The inward directed WKB trajectory starting from $\rpr = \infty$ traversing along
the ray $\arg \rpr = 0$ will either make it to $\rpr = 0$ when $d = 4\,n$, or sail past
it to $\rpr_{\frac{d-2}{4}}$ when $d = 4\,n + 2$. In the latter case,
$\rpr_{\frac{d-2}{4}}$ is the one real turning point (the total number of roots is
odd). Therefore, depending on $d \equiv 0,2\pmod{4}$ one of these will be the
real endpoint $\rpbull$. The Stokes graphs for the cases we examined are
presented in~\cref{fig:planar-sads-stokes-cycles} in the $\rpr$-coordinate.\footnote{As the caption indicates the analysis we undertake will first simply ignore Stokes crossings on the transport path. These are of course relevant, and will be accounted for properly in~\cref{prt:WKB2}.}

The determination of the distinguished turning point was made by taking
inspiration from the $d = 4$ analysis carried out in~\cite{Jia:2026ryl}. Having made the choice, one checks that the leading contribution to the composite monodromy
trace between the boundary and horizon arises precisely from this cycle. The $A_s$-cycle is then argued to be the closed cycle encircling the cut between the singularity $\rpr = 0$ (this is $\rpbull$ for these examples) and $\rpstar$.
This pattern continues for other values of $d = 4\,n$. We shall check below that this works numerically in $d = 4, 8$. For $d = 6$ on the other hand
$\rpbull = - \zeta^\frac{2}{3}\, (1 - \zeta^2)^{-\frac{1}{3}}$. We analyze
this case separately to show that our identification of the distinguished
turning point is reasonable. Despite this difference we will see below that the
WKB action will have a uniform behaviour in $d$.

One important point of clarification is in order. We see that for $d = 4\,n + 2$
the curvature singularity is not a branch point of the spectral curve. However,
consider the wave equations governed by the linear differential operator whose
(radial) kinetic term is $\dv{r}(r^{d + 1}\, f(r)\, \dv{r})$. Since the
potential pieces are generically regular as $r \to 0$, this part already makes
clear that $r = 0$ is a regular singular point of the wave equation. Thus while
the black hole singularity is genuinely a singular point for physical wavefunctions,
the corresponding piece of the wave operator is subleading in the large
frequency limit, and therefore disappears from the leading WKB curve. When we
compute the periods, we will see how it is that the WKB action has a smooth
dimension agnostic behavior. Nevertheless, the significance of the curvature
singularity will become clear when we consider scaling the momentum to zero,
whence the turning points start to approach the singularity
(cf.~\cref{prt:WKB3}).

The period integrals in all dimensions can be written uniformly in terms of a
function $\Pi_d(\zeta)$. It is defined as\footnote{To evaluate the period integrals it helps to redefine the coordinate once more. Setting
$\mathtt{t}(\rpr) = \frac{\sqrt{1-\zeta^2}}{\zeta}\, \rpr^{\frac{d}{4}}$, the turning points collapse onto the set $\mathtt{t}^2= - 1$. Aided by this transformation the period integrals collapse to hypergeometric integrals, and can be computed directly.}
\begin{equation}
\Pi_d(\zeta) = \frac{1}{\pi}\,\cos\!\prn{\frac{\pi}{d}}\, \operatorname{B}\!\prn{\frac{1}{2}-\frac{1}{d},\frac{3}{2}}\,\zeta^{2 - \frac{2}{d}}\, {}_2F_1\!\prn{
  \frac{1}{2}-\frac{1}{d},
  1-\frac{1}{d};
  2-\frac{1}{d};
  \zeta^2
}. \label{eq:Pi-d-general-hypergeometric}
\end{equation}
Here $\operatorname{B}(x,y)$ is the Euler beta function.  We fixed normalizations by requiring the function to satisfy $\Pi_d(0) = 0$ and $\Pi_d(1) = 1$.
This result can be analytically continued from $\arg \zeta = 0$ to $\arg \zeta = \frac{\pi}{2}$, enabling us to write the WKB action for imaginary momentum using the same
function.

\begin{figure}[ht]
  \centering
  \includegraphics[width=0.9\linewidth]{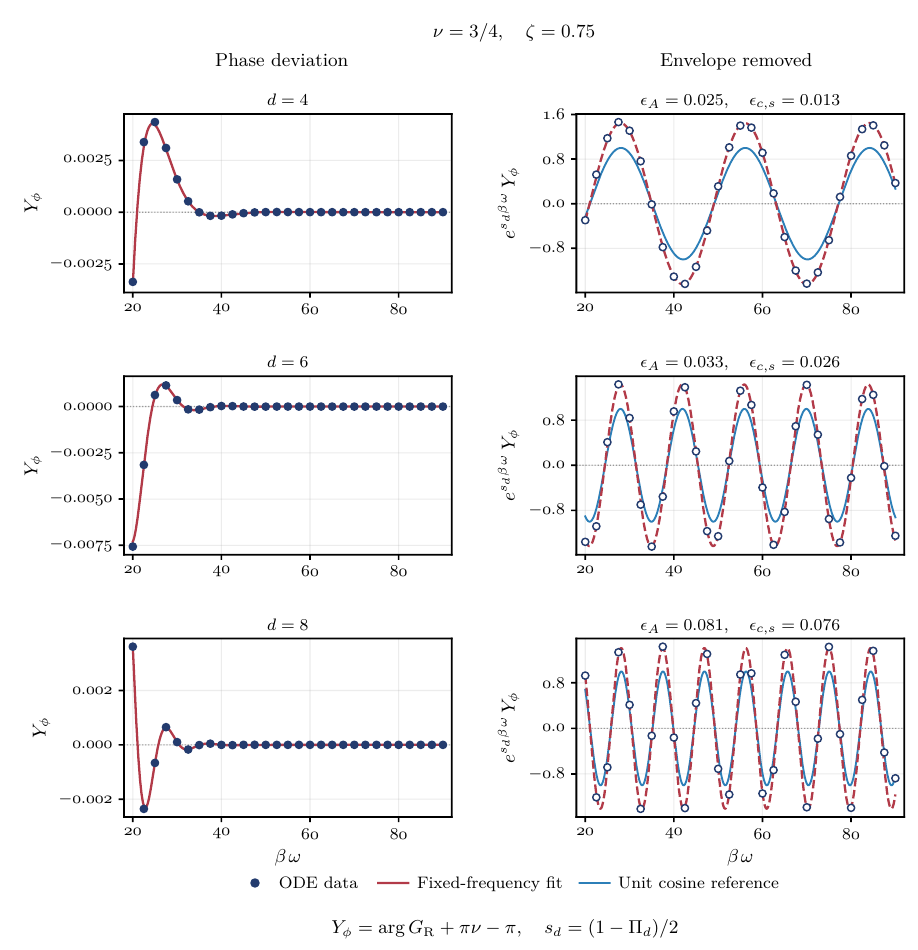}
  \caption{
    Oscillations of the retarded Green's function phase for
    $\nu=3/4$ and $\zeta=0.75$. The left panels show the
    vacuum-subtracted phase $Y_\phi=\arg\Gret+\pi\nu-\pi$, while the right panels remove the exponential envelope using the predicted factor:
    $s_d=(1-\Pi_d(\zeta))/2$.  The oscillation frequency is also fixed at the predicted value  $\xi_d=(1-\Pi_d)/(2\tan(\pi/d))$. Both the filled and empty markers correspond
    to the same ODE data before and after rescaling.
    The red curves show the fixed-frequency fit to a cosine plus sine ansatz, viz., $h_c\cos(\xi_d\bw)+ h_s\sin(\xi_d\bw)$. The blue curve on the right panel is a unit cosine phase reference plot.
  The residuals $\epsilon_A$ and $\epsilon_{c,s}$ refer to a fitted cosine amplitude  and a fitted cosine--sine pair, respectively, in the rescaled data. The fitted amplitude in the latter case agrees well with the predicted answer $2\sin(\pi\nu)=\sqrt{2}$.}
  \label{fig:planar-phase-oscillations}
\end{figure}

\begin{figure}[ht!]
  \centering
  \includegraphics[width=0.9\linewidth]{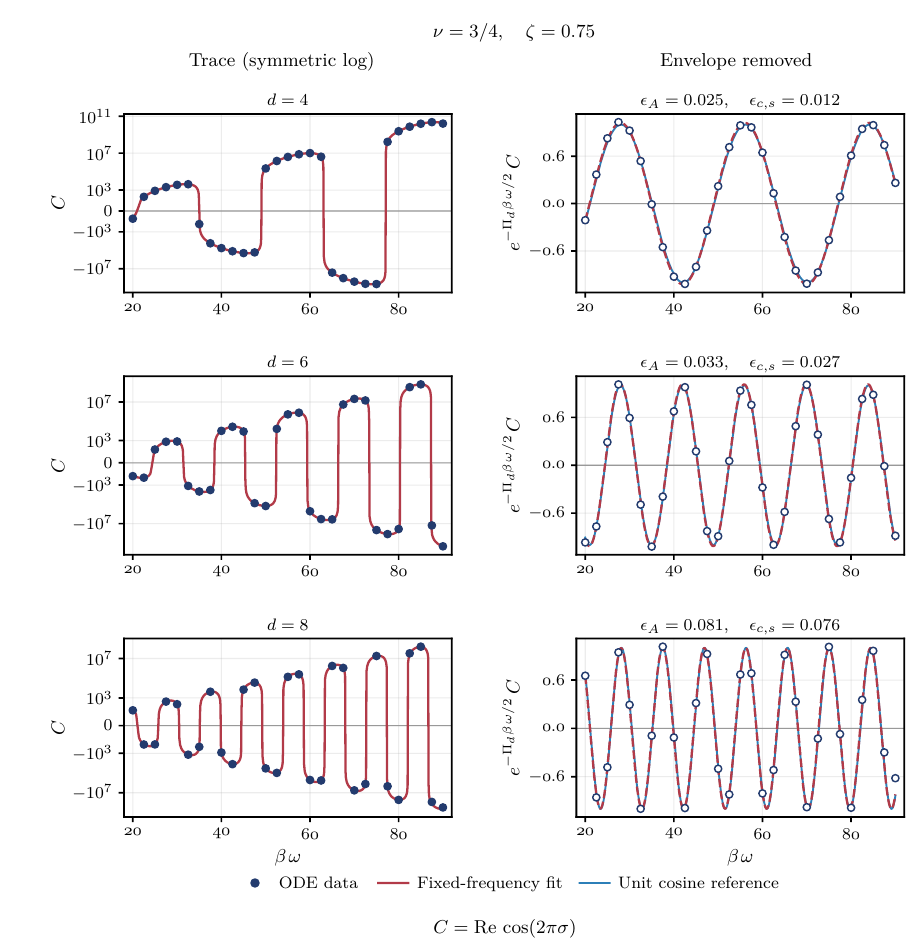}
  \caption{The composite monodromy trace for $\nu=3/4$ and $\zeta=0.75$.
    The left panels show $C=\Re\cmtrace$ on a (symmetric) logarithmic scale, while the right panels remove the exponential growth envelope by a factor
    $e^{-\Pi_d(\zeta)\,\hbw}$. The conventions for the plots are similar to those used
    in~\cref{fig:planar-phase-oscillations}.
  }
  \label{fig:planar-sigma-oscillations}
\end{figure}

\paragraph{Periods, composite monodromy, and the WKB action:}

Before working out the period integrals for the cycles we identified above, we will ascertain the composite monodromy trace and retarded Green's function's phase numerically. This is done by solving the differential equation numerically. Concretely, we fix the ingoing mode at the horizon and extract the source and expectation value from an expansion near the boundary. Knowing the latter the phase can be computed directly. The monodromy itself may also be evaluated by evaluating the connection coefficients explicitly (or by integrating in from both sides and matching at a suitable midpoint).

Having determined the quantity of interest numerically, we attempt a fit to an ansatz that is inspired by our expectation for the WKB action~\eqref{eq:Iwkb-abstract}.
For instance, the phase of the retarded Green's function would behave as
\begin{equation}
\delphR \sim e^{-s_d\, \bw}\, \pqty{ h_c\, \cos(\xi_d\, \bw)  + h_s\, \sin(\xi_d\, \bw) }\,.
\end{equation}
Numerically the expected behavior is clearly ratified, cf.~\cref{fig:planar-phase-oscillations}. The fit  assumes the analytically obtained rate of the exponential envelope and the oscillatory phase.
In practice, we fix the functional form of $\xi_d$ and $s_d$ to the predicted values (which we derive below), and only fit the coefficients $h_c$ and $h_s$. Specifically, we first establish analytically from the period computations that
\begin{equation}
s_d = \frac{1}{2}\, (1 - \Pi_d) \,, \qquad \xi_d
= \frac{1-\Pi_d}{2\, \tan(\frac{\pi}{d})} \,,
\end{equation}
and use it to work out the optimal value of the amplitudes.  We also examine the composite trace, in~\cref{fig:planar-sigma-oscillations}. In
this case we attempt to fit the numerically computed composite monodromy trace
to the ansatz
\begin{equation}
\label{eq:planar-sigma-dasym}
\cmtrace \sim e^{\Pi_d\, \hbw}\, \bqty{h_c\,
  \cos(\frac{1-\Pi_d}{2\, \tan(\frac{\pi}{d})}\, \bw)
  + h_s\, \sin(\frac{1-\Pi_d}{2\, \tan(\frac{\pi}{d})}\, \bw)
}\,.
\end{equation}

\begin{figure}[bhp!]
  \centering
  \includegraphics[width=0.85\linewidth]{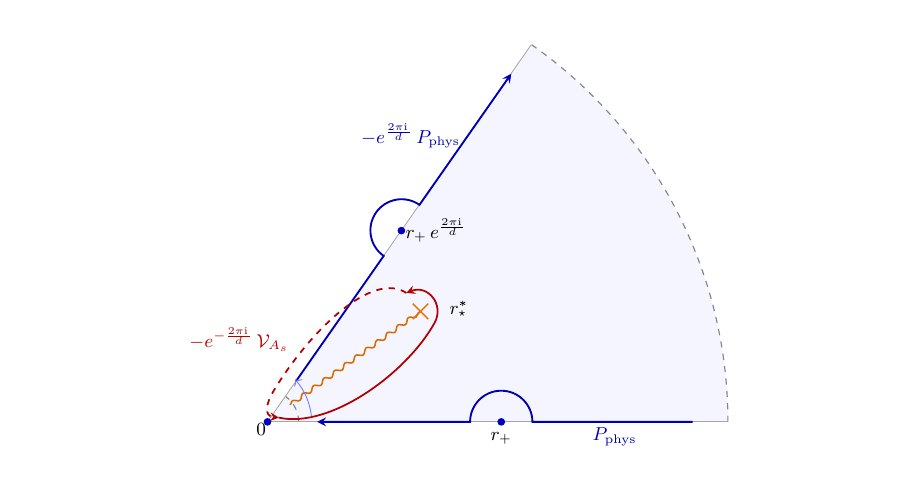}
  \caption{
    The contour deformation argument underlying the derivation of~\cref{eq:Psumrule}. The angle between the two boundary rays is  $2\pi/d$, and the two indicated horizon roots are at $r=\rp$ and $r=\rp e^{2\pi\ii/d}$. The two boundary integrals contribute to the
    principal-value term with a relative phase and the poles at the horizon supply (halved) residues, which assemble into the thermal factors. Importantly, the deformation also picks up the phase-rotated period integral around the cut. We have depicted the domain containing the conjugate turning point $r_\star^*$ for simplicity.
  }
  \label{fig:planar-Pphys-deformation}
\end{figure}

Taking the data used to generate either~\cref{fig:planar-phase-oscillations}
or~\cref{fig:planar-sigma-oscillations} we strip off the exponential using the predicted growth. We then attempt to fit the coefficients $h_c$ and $h_s$. Adjusting
just one coefficient $h_c$ with $h_s = 0$ already gives a clean fit through the
data points as is clear in the right panels of the corresponding figures. There is
a modest improvement upon inclusion of an additional phase offset with the coefficient $h_s$. The quoted error in the residual $\epsilon$ is determined as the unweighted least-squares residuals at the sampled frequencies for the oscillatory function.

Based on the results, we predict that the WKB action~\eqref{eq:Iwkb} should be
given by
\begin{equation}
\label{eq:planar-Iwkb}
\Iwkb = \frac{\beta\,\rp}{2} \, (1 - \Pi_d)\, \pqty{\cot(\frac{\pi}{d}) + \ii} \,.
\end{equation}
The astute reader should be concerned that at no point in this discussion did we
single out the $d = 4n$ case from the $d = 4n + 2$ case. The periods, composite
monodromy, and therefore the WKB action appear insensitive to the nuance we
discussed at the outset. We will now sketch the derivation and in the process
explain why these qualitative differences wash out at the end.
We assemble the pieces involved in the determination of $\Iwkb$ below.

\paragraph{Derivation of the WKB action for $d = 4n$:}
In this case we assume that the $A_s$-cycle encircles $\rpbull = 0$ and
$\rpstar = \rpr_{\frac{d}{2}-1}$, cf.~\eqref{eq:planar-uturn}. The period integral around this cycle can be computed by elementary methods and is given by
\begin{equation}
\label{eq:planar-A-period}
\Aper = \frac{2\pi}{d}\, \pqty{\tan\frac{\pi}{d} - \ii}\, \Pi_d(\zeta)\,.
\end{equation}

For the integral from the boundary to the point of departure from the $\Re \rpr$
trajectory we need the principal value integral from $\rpr = \infty$ to
$\rpbull = 0$, i.e.,
\begin{equation}
\Pphys = \rp\, \mathrm{PV}\, \int_0^\infty\, \lwkb\,.
\end{equation}
We will evaluate this directly in the original radial coordinate as indicated in the equation.  To complete the integral we consider the principal value integral along two
rays. The first is oriented inward from the boundary to the singularity
with one horizon crossing (picking up a half thermal period). The
second is oriented along an outgoing ray from the singularity to the boundary
through the complex horizon immediately adjacent to the distinguished conjugate turning
point $\rstar^*$. The two contours are at $\arg(r) = 0$ and
$\arg (r) = \frac{2\pi}{d}$ as depicted in~\cref{fig:planar-Pphys-deformation}.
There is no contribution from the Jordan arc at large $r$, nor is there a
contribution from the small arc crossing between the rays. The contribution from
the rays is balanced by the contribution from the $A_s$-cycle period and the
horizon crossings, leading to a sum-rule
\begin{equation}
\label{eq:Psumrule}
\pqty{1- e^{-\frac{2\pi \ii}{d}}}\, \Pphys - e^{-\frac{2\pi \ii}{d}} \, \Aper - \frac{\ii\pi}{d}\, \pqty{1 + e^{-\frac{2\pi\ii}{d}}} = 0\,.
\end{equation}
This exploits the relation $\overline{\Aper} =- e^{-\frac{2\pi \ii}{d}}\, \Aper$. Taking the real part we arrive at
\begin{equation}
\pqty{1-\cos(\frac{2\pi}{d})}\, \Pphys + \RAper = \frac{\pi}{d}\, \sin (\frac{2\pi}{d})\,.
\end{equation}
Finally, using the period we determined~\eqref{eq:planar-A-period} and the knowledge of~\eqref{eq:Pi-d-general-hypergeometric}, we obtain the relation
\begin{equation}
\frac{2\pi}{d}\, \Pi_d(\zeta) = -\IAper = \cot \frac{\pi}{d}\, \RAper\,.
\end{equation}
Therefore, we predict that the composite monodromy should have an exponential
envelope with exponent
\begin{equation}
\freq \pqty{\frac{\beta\,\rp}{2} - \IIwkb} = \frac{2\pi}{d}\, \freq\, \Pi_d(\zeta)\,,
\end{equation}
and an oscillatory factor with frequency
\begin{equation}
\freq\,\pqty{
2\, \Pphys + \RAper}
= \frac{2\pi}{d}\, \cot(\frac{\pi}{d})\, (1 - \Pi_d(\zeta))\, \freq.
\end{equation}
This completes the derivation of the WKB action $\Iwkb$ given
in~\eqref{eq:planar-Iwkb} for these cases.

\paragraph{Derivation of WKB action for $d = 4n + 2$:}
Let us first give an abstract argument for two puzzling features of the above result.
First, there is a physical turning point $\rpbull = \rpr_{\frac{d-2}{4}}$ that the contours ignore, and second, $\rpr =0$ being regular cannot be the end-point of a cut.
Consider the action integrals of the one-form $\lwkb$ between some reference point
$\rpr_{\mathrm{ref}}$ and the endpoints of
the WKB trajectory $\rpr = \infty$ and $\rpr = \rpstar$. Assume we have two such
reference points, one at the physical turning point $\rpr= \rpbull$ and another at
$\rpr = 0$.
As long as both reference points involve the same number of
horizon crossings, it follows that the difference of the WKB actions is
\begin{equation}
\Iwkb(\rpr_{\mathrm{ref}} = \rpbull) - \Iwkb(\rpr_{\mathrm{ref}} = 0) = 2\, \mathrm{PV}\, \int_0^{\rpbull} \, \lwkb + 2 \, \bqty{ \int_{\rpbull}^{\rpstar}\ -  \int_0^{\rpstar} } \lwkb = 0\,.
\end{equation}
The first term is the real line integral between the two reference points, and
the second term comes from the periods (or the open cycle from the
regular $\rpr = 0$ locus). As long as the two reference points can be
connected without passing through a singularity of $\lwkb$ the difference of
them vanishes as indicated in the equation. But it is easy to check that there
is no singular locus of the WKB differential between $\rpr =0$ and $\rpbull$. Therefore, the result follows.

One can verify this explicitly for the planar \SAdS{7} black hole. We use $\beta r_+ = \frac{2\pi}{3}$ and evaluate the period integral between the turning points at $\rpbull = - \zeta^\frac{2}{3}\, (1 - \zeta^2)^{-\frac{1}{3}}$ and
$\rpstar =  \zeta^\frac{2}{3}\, (1 - \zeta^2)^{-\frac{1}{3}}\, e^{-\frac{\ii\pi}{3}}$, and the principal value part for the real trajectory from infinity. One finds
\begin{equation}
d = 6: \quad \Aper
= \frac{\pi}{3}\, \bqty{\cot(\frac{\pi}{6}) -i}\, \Pi_6(\zeta) \,, \qquad \Pphys
= \frac{\pi}{3}\, \cos(\frac{\pi}{6}) \pqty{1 - 2\, \Pi_6(\zeta) } \,.
\end{equation}
Assembling the pieces we arrive at~\eqref{eq:planar-Iwkb} as expected.

\paragraph{The time domain singular loci:}
The WKB action~\eqref{eq:planar-Iwkb} controls the location of the singularities
in the complex time domain as described in~\eqref{eq:tc-angle}
and~\eqref{eq:tc-position} depending on the observable we choose. Notice that
since $\Pi_d(0) = 0$, in the limit where spatial momenta are held finite we
recover the contribution from the bouncing null geodesic. In $d = 4$ this
recovers the result originally derived in~\cite{Fidkowski:2003nf} and the higher
dimensional answers agree with the geodesic lengths computed
in~\cite{Ceplak:2024bja}. This should not be a surprise: in the limit
$\zeta \to 0$ the turning points coalesce into the singularity and there is no
complex excursion. In this limit we therefore have an explicit Lorentzian
geodesic controlling the high frequency behavior. Explicitly,
\begin{equation}
\lim_{\zeta \to 0} t_c = \frac{\beta}{2} \cot(\frac{\pi}{d}) + \ii \beta \,,
\end{equation}
and its reflection images across the $\Re(t)$ and $\Im(t)$ axes.

For non-zero $\zeta$ we have to take a detour to the complex turning point,
which entails that we pick up the relative period between the boundary and the
turning point. While we have only explicitly checked our answers for
$d = 4, 6, 8$, it should be clear that the picture holds in all even dimensions
and therefore the result for the singular loci is valid more generally.

We can attempt to ascertain the fixed $\abs{\vb{x}}$ singular locus. Working at
small $\zeta$ where we can hope to retain analytic control, we have
\begin{equation}
\Pi_d(\zeta) = \alpha_d\, \zeta^{2 - \frac{2}{d}}
+ \order{\zeta^{4- \frac{2}{d}}} \,, \qquad \alpha_d
\equiv \frac{\cos(\frac{\pi}{d})}{\pi}\, \mathop{B}\pqty{\frac{1}{2} -\frac{1}{d},\frac{3}{2}} \,.
\end{equation}
Then the saddle solution for $\zeta$ for fixed $\abs{\vb{x}}$ is
\begin{equation}
\zeta_*(\abs{\vb{x}}) \sim \pqty{\frac{d\,\abs{\vb{x}}}{(d-1)\,\alpha_d\, \beta\, \pqty{\cot (\frac{\pi}{d}) +\ii}}}^{\frac{d}{d-2}}\,.
\end{equation}
This predicts a singular locus
\begin{equation}\label{eq:psads-tc-x}
t_c(\abs{\vb{x}}) = \frac{\beta}{2} \cot(\frac{\pi}{d})
+ \ii \beta
+ \frac{d-2}{2\,d}\,\alpha_d\,  \beta\, \pqty{\cot (\frac{\pi}{d}) +\ii}\, \zeta_*(\abs{\vb{x}})^{\frac{2(d-1)}{d}}
\end{equation}
We will later see in~\cref{sec:ope-wkb} that this picture fits very nicely with the thermal OPE analysis.

\paragraph{Asymptotic quasinormal spectrum:}
From our general result in~\cref{sec:asymqnm} we predict that the asymptotic
spacing of the quasinormal modes for planar \SAdS{d+1} black holes is
\begin{equation}
\Delta(\bw)_+ = \frac{4\pi}{1-\Pi_d(\zeta)}\, e^{-\ii\frac{\pi}{d}}\sin(\frac{\pi}{d}) \,.
\end{equation}
As we take $\zeta \to 0$ we find that this simplifies to
\begin{equation}
\Delta \omega = 4\pi\ms T\,e^{-\ii\frac{\pi}{d}}\sin(\frac{\pi}{d})\,,
\end{equation}
which is in precise agreement with~\cite{Natario:2004jd}
and~\cite{Festuccia:2008zx}. In the lightlike limit on the other hand we see
that the spacing diverges as $\Pi_d(\zeta) \to 1$ as $\zeta \to 1$. This is
signalling that the turning point is approaching the boundary and requires a
more refined analysis to extract the subleading scaling obtained
by~\cite{Festuccia:2008zx}.

\subsection{Global \SAdS{} black holes}
\label{sec:globalSAdS}

The metric function for global \SAdS{d+1} black holes has
\begin{equation}
f(r) = \frac{1}{r^2} \, \pqty{\frac{r^2}{\lads^2} + 1 - \frac{\rp^{d-2}}{r^{d-2}}\, \pqty{\frac{\rp^2}{\lads^2} +1}} \,,
\end{equation}
where we have made explicit the \AdS{} length scale $\lads$. The inverse
temperature is
\begin{equation}
\beta = \frac{4\pi\, \rp\,\lads^2}{(d-2)\, \lads^2 + d\, \rp^2} \,.
\end{equation}
We will continue to set $\lads = 1$ and work with $\rp$ setting the scale as
before for convenience, and restrict to $d$ being even.

The WKB spectral curve is characterized by expanding in spherical harmonics with
spin $\ell$ and setting $\ell\,(\ell +d-2) = \zeta^2\, \omega^2$. For general $d$ it is given by
\begin{equation}
\lwkb(r) = \frac{r^{\frac{d}{2} -2}\, \sqrt{(1-\zeta^2)\, r^{d} - \zeta^2\, r^{d-2} + \zeta^2\, \rp^{d-2} \, (1+\rp^2)}}{r^{d-2} \,(1+ r^2) - \rp^{d-2}\, (1+\rp^2)}\, dr
\end{equation}
The spectral curve is a hyperelliptic curve of genus $\frac{d}{2} - 1$. Since
the curve has an $r \to -r$ involution we can project it down to a curve of
genus $\floor{\frac{d}{4}}$ using the projection $\rpr = r^2$, as in the planar
case. In particular, the $d = 4$ curve is elliptic in the $r$ plane, while the
$d = 6$ curve is of genus-$2$ in the $r$-plane. However, the latter descends to a genus-$1$ curve in the $\rpr$-plane.

\subsubsection{General features for timelike momentum}
\label{sec:genfeatures}

There are curious features of these backgrounds that bear highlighting. Firstly,
in the radial variable the singular points of the WKB differential are the zeros
of $f(r)$ alone, viz., the physical horizon and its complex partners. The
curvature singularity at $r = 0$ is not distinguished in any way for any even
$d$. As in the planar case it should be emphasized that this is a statement
pertaining to the leading universal WKB differential -- the full wave equations
retain $r = 0$ as a regular singular point, for similar reasons as discussed
hitherto.

Projecting to the $\rpr$-plane, one can end up introducing a branch cut at
$r = \rpr = 0$. The details of the behavior seen in the leading WKB differential depend on the dimension.  For $d=4$ the origin is a simple pole, but when $d = 4\,n$ for $n \geq 2$ it is a zero. In this case, the curvature
singularity retains a role in the leading order WKB as a turning point. When
$d = 4\,n + 2$, on the other hand, even after the projection to the $\rpr$-plane
the black hole singularity is entirely absent from the spectral curve. So the
WKB trajectory picks out some locus that is different from $r = 0$ to migrate
into the complex plane. All of this is similar to the planar \SAdS{d+1} black
holes discussed above.

Secondly, there is also a question of what is the physical WKB turning point
closest to the boundary. We shall see that there can be turning points for
$\Re(r) > 0$, in fact, even $\Re(r) > \rp$ for fixed $\zeta$. This means that
the WKB trajectory for the boundary conditions of interest does not get inside
the black hole, but rather turns around from a real turning point
$r\subm{tp} > \rp$. In these cases we do not get direct access to the black hole
singularity from the spectral function or thermofield double correlator
asymptotics, but rather to the bulk cone singularities of~\cite{Hubeny:2006yu}, which have been explored in~\cite{Dodelson:2023nnr,Araya:2026shz}.  We exemplify the general features with global \SAdS{5} and \SAdS{7} black holes.

\paragraph{The global \SAdS{5} black hole:}
Let us start by writing down the curves in the $\rpr$-plane for $d = 4$. The
differential is given by
\begin{equation}
\lwkb = \frac{\sqrt{(1-\zeta^2)\, \rpr^2 - \zeta^2\, \rpr + \zeta^2\,\rp^2\, (1+\rp^2)}}{
2\,\sqrt{\rpr}\, (\rpr-\rp^2)\, (\rpr +1 +\rp^2)}\, d\rpr \,.
\end{equation}
The singularities of the WKB curve are at $\rpr = \rp^2$, the outer horizon, and at
$\rpr = - (1 + \rp^2)$, the complex horizon root. $\rpr = 0$, the black hole
singularity is clearly a branch point (arising from the change in the basis
differential from $dr$ to $d\rpr$). The turning points are at
\begin{equation}
\rptp^{\pm} = \frac{\zeta^2 \pm \sqrt{\zeta^4 - 4\, \zeta^2 \, (1-\zeta^2)\, \rp^2\, (1+\rp^2)} }{2\, (1-\zeta^2)}\,.
\end{equation}

\begin{figure}[bht!]
  \centering
  \begin{tikzpicture}[x=1bp,y=1bp]
    \node[anchor=south west,inner sep=0pt,outer sep=0pt] at (0,0) {\includegraphics[width=0.9\linewidth]{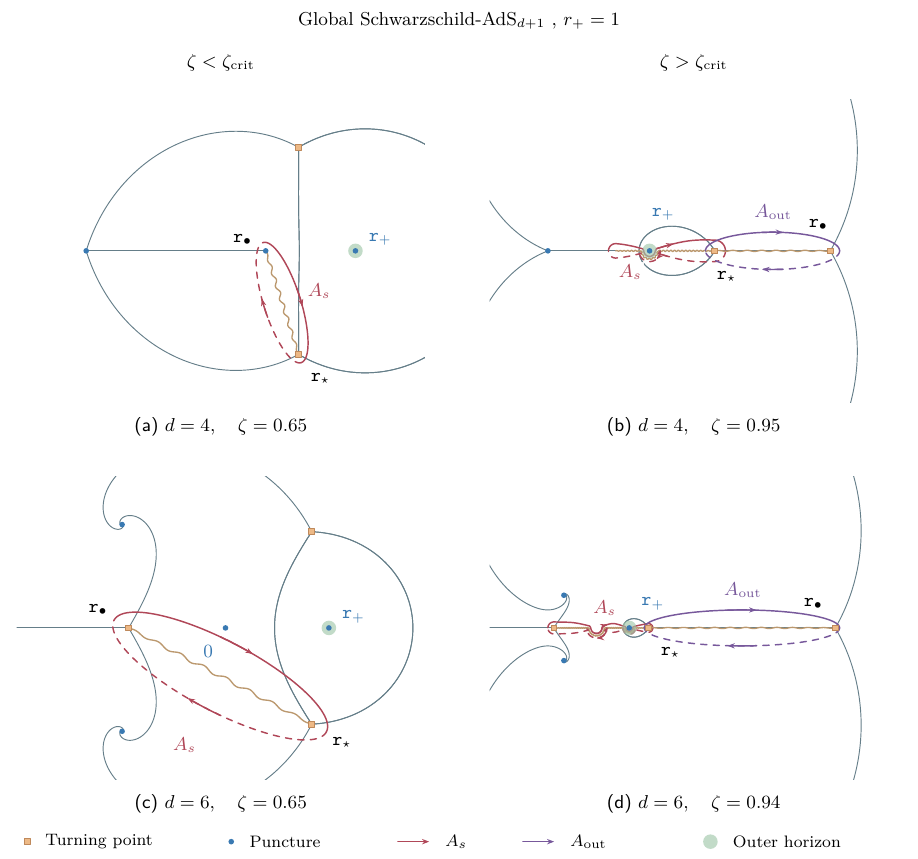}};
    \node[inner sep=0pt,outer sep=0pt] at (8,365) {\phantomsubcaption\phantomsection\label{fig:gsads5-in}};
    \node[inner sep=0pt,outer sep=0pt] at (235,365) {\phantomsubcaption\phantomsection\label{fig:gsads5-out}};
    \node[inner sep=0pt,outer sep=0pt] at (8,184) {\phantomsubcaption\phantomsection\label{fig:gsads7-in}};
    \node[inner sep=0pt,outer sep=0pt] at (235,184) {\phantomsubcaption\phantomsection\label{fig:gsads7-out}};
  \end{tikzpicture}
  \caption{
    Stokes graphs for global \SAdS{d+1} with $\rp=1$, in the projected
    $\rpr$ coordinate for $d=4$ and $d=6$. The left and right sides are for
    $\zeta<\zcrit$ and $\zeta>\zcrit$, respectively. In the former the WKB trajectory reaches inside the black hole, and in the latter the centrifugal barrier prevents it from doing so. The conventions are as specified in~\cref{fig:planar-sads-stokes-cycles}, and the chosen parameters are indicated  below the plots.
  }
  \label{fig:global-SAdS-stokes}
\end{figure}
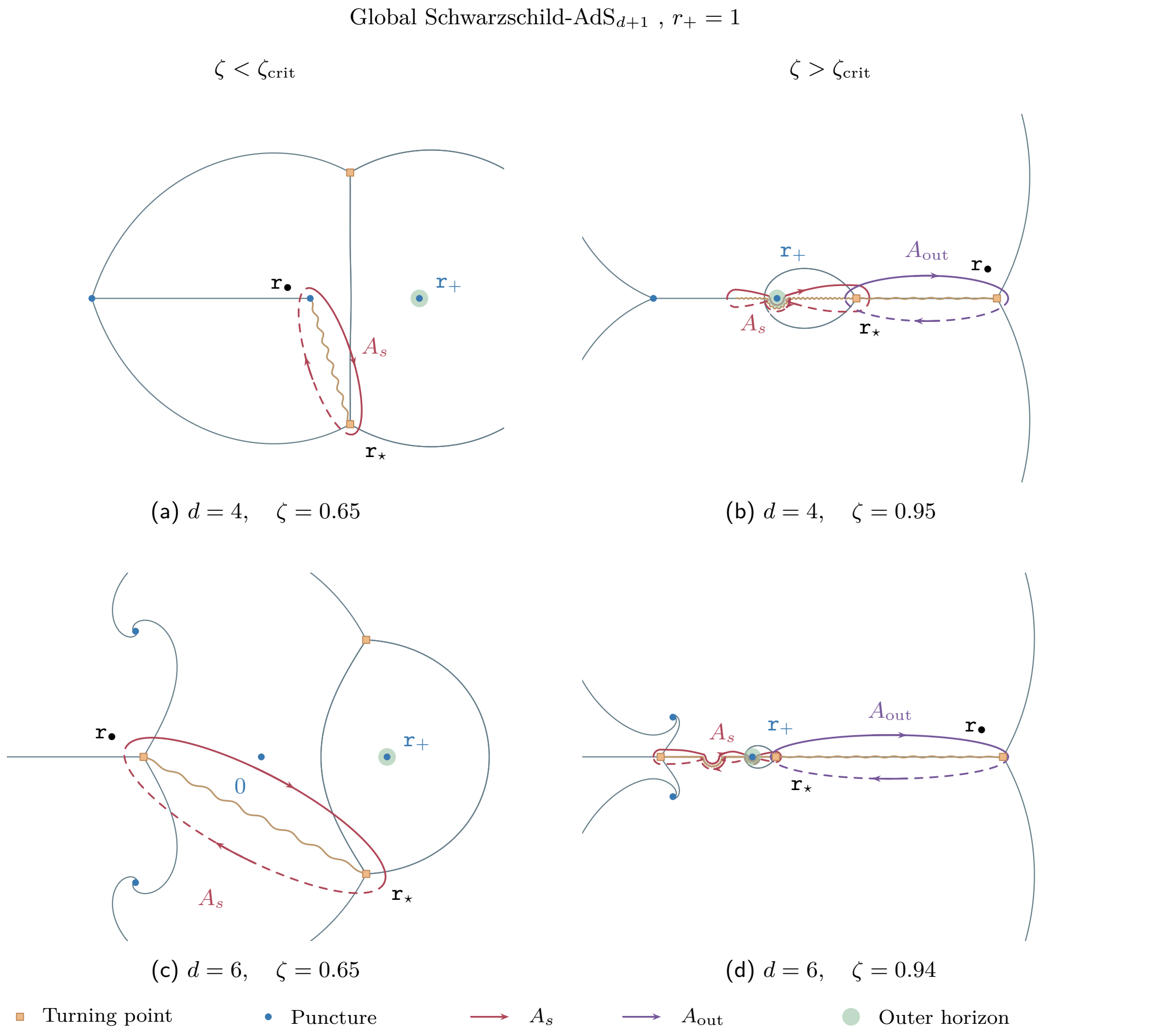

As presaged above, the turning points $\rptp^{\pm}$ can actually be real and lie
outside the horizon. One can check that the roots are real when
\begin{equation}
\zeta \geq \zcrit = \sqrt{1-\frac{1}{(1+2\,\rp^2)^2}}\,.
\end{equation}
Moreover, in this range both the turning points are outside the horizon
since $\frac{\zeta^2}{2\, (1-\zeta^2)} > \rp^2$. Therefore, the WKB trajectory
does not get anywhere near the horizon or the singularity. It is intuitively
clear why this happens from the geodesic perspective. The global black hole
supports circular orbits and the WKB trajectory here is the much more familiar
unbound orbit in the geometry. The deviation comes from the region where the
centrifugal barrier permits local bound states in the potential. The Stokes
graphs for the two cases make this amply clear, as can be seen
from~\cref{fig:gsads5-in} and~\cref{fig:gsads5-out}.

\paragraph{The global \SAdS{7} black hole:}
In the case of \SAdS{7} the WKB differential in the $\rpr$ plane takes the form
\begin{equation}
\lwkb = \frac{\sqrt{(1-\zeta^2)\, \rpr^3  - \zeta^2\, \rpr^2
+ \zeta^2\, \rp^4\, (1+\rp^2) }}{2\,(\rpr - \rp^2)\, \bqty{\rpr^2 + (1+\rp^2)\, (\rpr+\rp^2)}}\, d\rpr\,.
\end{equation}
As indicated at the outset, the black hole singularity at $\rpr = 0$ is not in any way
distinguished on this spectral curve. There are three turning points, and one of
them clearly has to be real. It can be checked that this lies at $\rpr < 0$ for all
$\zeta \in (0,1)$ and $\rp > 0$. Two other roots are complex, but turn real
when
\begin{equation}
\rp^4\, (1 + \rp^2) < \frac{4\, \zeta^4}{27\, (1-\zeta^2)^2} \;\;
\Longrightarrow \;\; \zeta
> \zcrit
= \frac{3\,\rp\, (1+\rp^2)^{\frac{1}{4}}}{\pqty{9 \,\rp^2\, \sqrt{1+\rp^2} + 2\,\sqrt{3}}^{\frac{1}{2}}}\,.
\end{equation}
So when $\zeta < \zcrit$ the WKB solution is transported from the
boundary to $\rpbull < 0$ through the outer horizon and the singularity along the
real $\rpr$-axis, before migrating to the complex $\rpstar$ turning point. However,
for $\zeta > \zcrit$ the turning points lie outside the horizon and the probe
gets caught in the local potential well. The Stokes graphs for the two cases are
displayed in~\cref{fig:gsads7-in} and~\cref{fig:gsads7-out}.

\subsubsection{The WKB analysis}
\label{sec:wkbglobal}

\begin{figure}[t]
  \centering
  \includegraphics[width=0.9\linewidth]{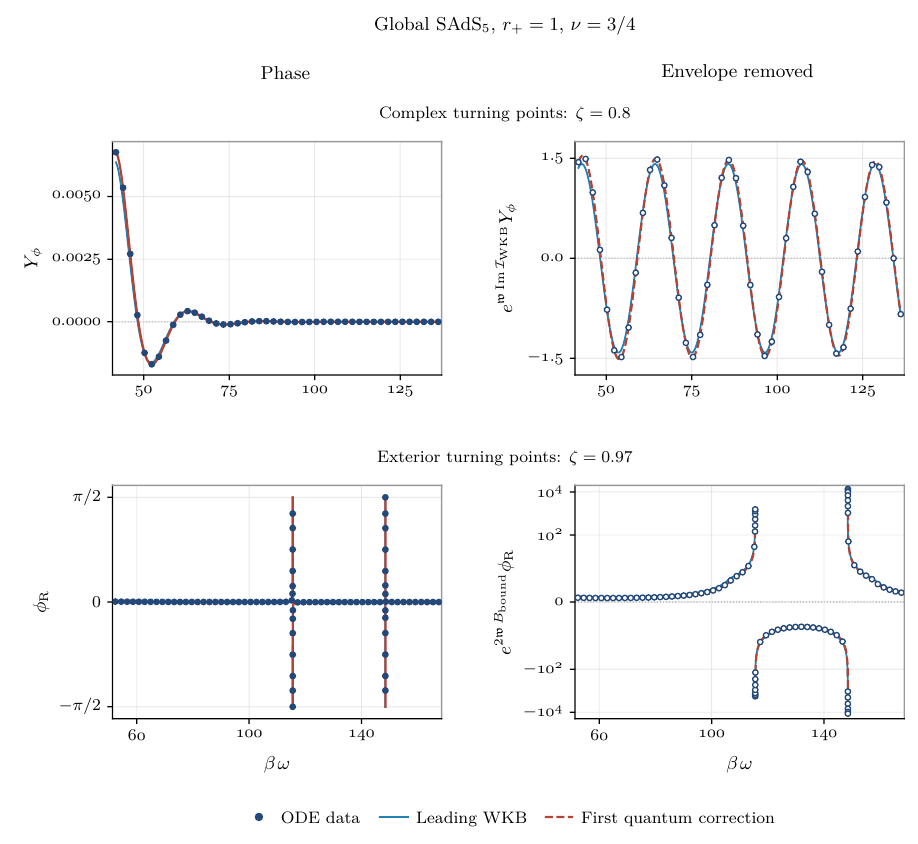}
  \caption{ The phase of the retarded Green's function for global \SAdS{5} with
    $\rp =1$ and $\nu = 0.75$ in the same set of conventions as~\cref{fig:planar-phase-oscillations}. For complex turning points, we picked
    $\zeta=0.8$ and show $Y_\phi=\arg\Gret+\pi\nu-\pi$, while for the exterior roots
    we chose $\zeta=0.97$ and plot $\phi_{\rm R}=\tfrac12\operatorname{Arg}(G_{\rm R}/G_{\rm A})$. In the case of exterior turning points (bottom panels) the phase which is defined modulo $\pi$ is vanishing except near poles and zeros of the retarded Green's function, in the vicinity of which it rapidly oscillates. In the plot shown, the first spike is a pole and the second is a zero; see~\cref{sec:boundstates} for an explanation.
  }
  \label{fig:globalads5-root-phase}
\end{figure}

\begin{figure}[ht!]
  \centering
  \includegraphics[width=0.9\linewidth]{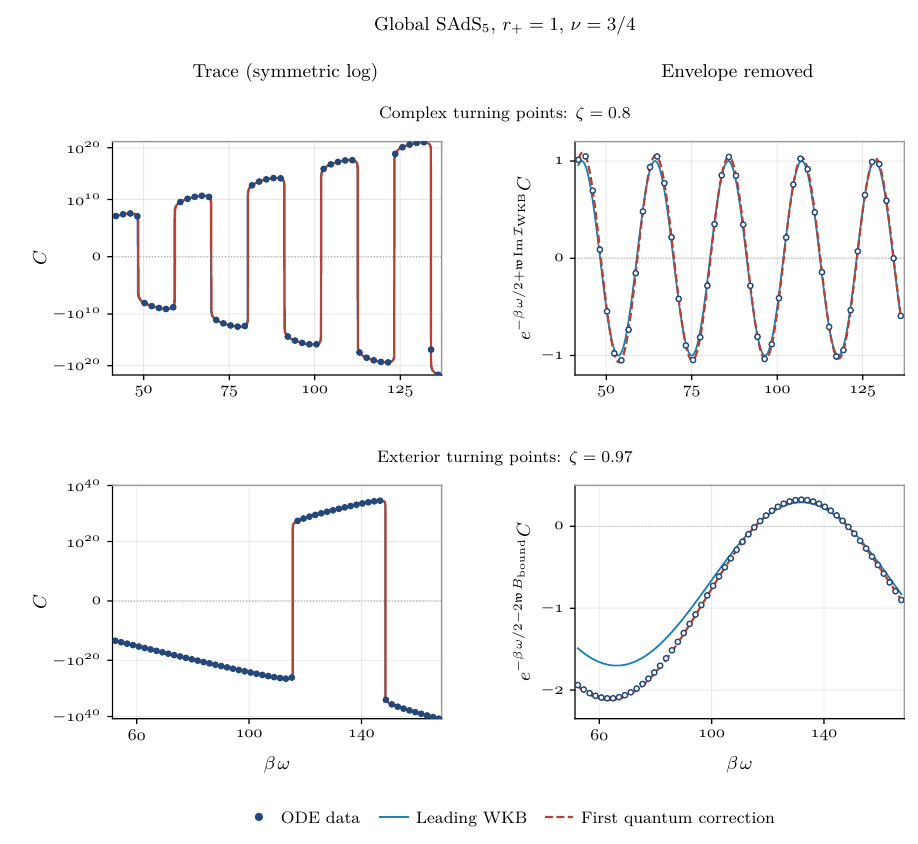}
  \caption{The composite monodromy trace for the background discussed in~\cref{fig:globalads5-root-phase}. The conventions are as in~\cref{fig:planar-sigma-oscillations} with the bare trace on the left and its exponential stripped oscillations on the right. The new feature is the bound states (lower row) for which we strip off the factor $e^{-\hbw -2\freq \,B\subm{bound}}$. }
  \label{fig:globalads5-root-trace}
\end{figure}

We now discuss the two cases separated by the critical value
$\zeta_{\mathrm{crit}}$ separately, as the physics probed by the high frequency
spectral function is quite different. The behavior of the retarded Green's
function's phase and the composite monodromy trace is displayed
in~\cref{fig:globalads5-root-phase} and~\cref{fig:globalads5-root-trace},
respectively for both cases, for both $d = 4$ and $d = 6$. In what follows we
will write down explicit formulae for $d = 4$. The $d = 6$ case is similar; the
spectral curve is still elliptic of genus-$1$ in the projective $\rpr$-coordinate.
The periods can be evaluated in terms of elliptic integrals, but the
expressions are unilluminating, and so we refrain from displaying them.

\paragraph{A. Complex turning points:}
When $\zeta < \zcrit(\rp)$ the turning points $\rptp^{\pm}$ are both complex
conjugates of each other. The $A_s$-cycle turns out to be the one encircling the
cut running from $\rpr = 0$ to $\rptp^-$, with the cycle oriented such that $\Re \Aper > 0$ and $\Im \Aper < 0$.
Given these identifications, we  have $\rptp^+ = \rpstar^*$. With these conventions the
$A_s$-cycle period is
\begin{equation}
\Aper = 2\,\rp\, \sqrt{\frac{1-\zeta^2}{\rpstar^*}} \bqty{
  \frac{\rp^2\,\Pi\pqty{\frac{\rpstar}{\rp^2}, k}
  + (1+\rp^2)\, \Pi\pqty{-\frac{\rpstar}{1+\rp^2} , k}}
  {(1+2\,\rp^2)\,(1-\zeta^2)}
  - K(k)
}\,, \qquad k^2
= \frac{\rpstar}{\rpstar^*} \,.
\end{equation}
Here $K(k)$ and $\Pi(\alpha,k)$ are elliptic integrals in standard notation. We
have made a sign convention choice such that $\RAper > 0$ and $\Im(\Aper) < 0$.

The rest of the WKB action requires $\Pphys$, which may be obtained
using a contour deformation argument analogous to~\eqref{eq:Psumrule}. One finds
\begin{equation}
\Pphys = \frac{\pi\,\rp\,\sqrt{1+ \rp^2}}{2\,(1+2\, \rp^2)} - \RAper\,.
\end{equation}
The WKB action~\eqref{eq:Iwkb} straightforwardly assembles, and is given by
\begin{equation}
\Iwkb = \frac{\pi\,\rp\,\sqrt{1+ \rp^2}}{1+2\, \rp^2}
- \RAper
+ \ii \pqty{ \frac{\pi\,\rp^2}{1+2\,\rp^2} + \IAper}\,.
\end{equation}

Our fits in the plots use the exponential envelope and oscillation frequency
predicted from this action. Specifically, the left panels
of~\cref{fig:globalads5-root-phase} and~\cref{fig:globalads5-root-trace} show
the behaviour of the retarded Green's function phase and the composite trace
monodromy, respectively. The bare data for the former show a clean exponential
convergence. Stripping off the exponential with the predicted factor $\IIwkb$ we
expose the oscillations underneath. These fit quite nicely in our two parameter
ansatz with a frequency set by $\RIwkb$, as is clear
from~\cref{fig:globalads5-root-phase}. The corresponding plots for the composite
trace~\cref{fig:globalads5-root-trace} show worse fits, largely because of
numerical sensitivity.

From here we can read off the location of the time-domain singularity, since the
thermofield double correlator has its singularities determined solely by
$I_\wkb$.

\paragraph{B. Real turning points:}
Now we have two real turning points for $d = 4$ and these are at
$\rptp^+ > \rptp^- > \rp^2$. Following our stated convention, we now set
$\rptp^+ = \rpbull$ and $\rptp^- = \rpstar$. The WKB action is assembled as the
phase accumulated along the real trajectory from the boundary to $\rpbull$ and
the transition through the classically forbidden region to access $\rpstar$. The
WKB period comprises the phase picked up along the classically admissible
trajectory, and the period around the cut running between the two turning
points. To wit, let
\begin{equation}\label{eq:gsads-outsidePB}
P\subm{tp} = \rp\, \int_{\rpbull}^\infty\, \lwkb \,, \qquad
B\subm{bound} = -\ii \rp \, \int_{\rpstar}^{\rpbull}\, \lwkb \,,
\end{equation}
so that
\begin{equation}\label{eq:gsads-Iwkb}
\frac{1}{2}\, \Iwkb = P\subm{tp} + \ii B\subm{bound} \,,
\end{equation}
where we have fixed a phase convention for the classically forbidden region
contribution. The subscripts are meant to be evocative: the real part is the
turning point action, and the imaginary part sets the bound state decay width.
We will verify this directly below.

The integrals themselves can again be evaluated in terms of elliptic
integrals, but since the explicit formulae do not offer much intuition we have refrained from quoting them.

\begin{figure}[t]
  \centering
  \includegraphics[width=\linewidth]{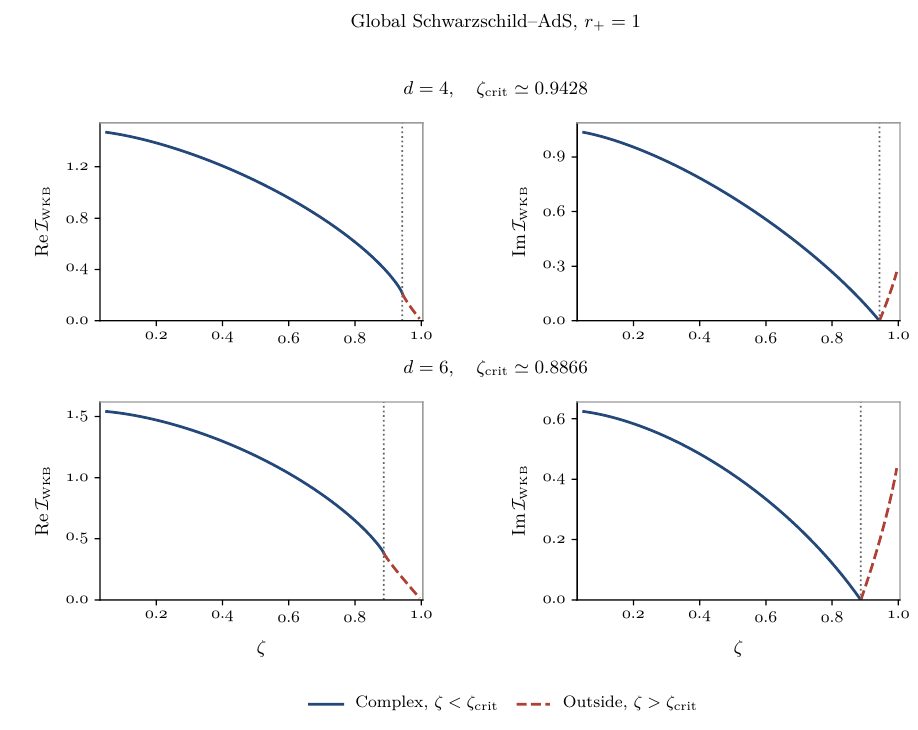}
  \caption{
    The behavior of the WKB action for global \SAdS{d+1} for $d=4,6$
    at $\rp=1$ as we scan over $\zeta$ through the critical value $\zeta\subm{crit}$. The real part is smooth, but the imaginary part switches direction at $\zeta\subm{crit}$.
  }
  \label{fig:global-sads-iwkb-zeta-rp1}
\end{figure}

\paragraph{The WKB action:}
We can examine the WKB action explicitly as a function of $\zeta$ for fixed
$\rp$ by numerically evaluating the integrals. The result is displayed
in~\cref{fig:global-sads-iwkb-zeta-rp1}. As the figure illustrates the real
part is smooth across $\zeta_{\mathrm{crit}}$, but the imaginary part
has a derivative discontinuity. The principal branch picked out by the phase of the retarded
Green's function is defined in the two cases to be
\begin{equation}
\label{eq:global-phase-convention}
\phGR \to
\begin{dcases}
- \pi \ms\nu \,, & \quad \zeta < \zcrit    \\
0\,,             & \quad \zeta > \zcrit\,.
\end{dcases}
\end{equation}

The easiest way to obtain this is to work out the boundary Green's function, which modulo a real normalization factor is given as
\begin{equation}\label{eq:GR-bound}
\Gret
\propto
-\frac{
  \cos\pqty{\freq P\subm{tp}+\frac{\pi\ms\nu}{2}}
  -\frac{\ii}{4}\, e^{-2\ms\freq\ms B\subm{bound}}
\sin\pqty{\freq \, P\subm{tp}+\frac{\pi\ms\nu}{2}}}
{
  \cos\pqty{\freq \, P\subm{tp}-\frac{\pi\ms\nu}{2}}
  -\frac{\ii}{4} \, e^{-2\ms\freq \ms B\subm{bound}}
  \sin\pqty{\freq \, P\subm{tp}-\frac{\pi\ms\nu}{2}}
}\,.
\end{equation}
This is obtained as the ratio of response and source modes, both now computed as WKB waves in the region between the boundary and the turning points.

\subsubsection{Bound state spectrum}
\label{sec:boundstates}
\begin{figure}[ht!]
  \centering
  \includegraphics[width=0.9\linewidth]{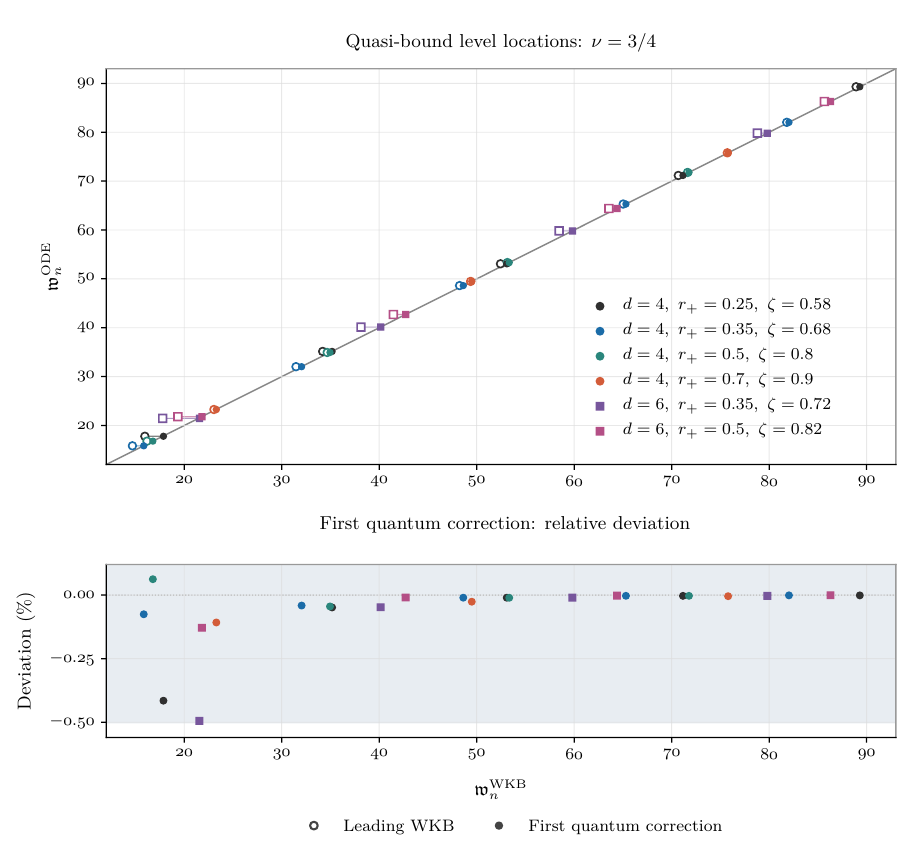}
  \caption{The (quasi) bound states of global \SAdS{d+1} for $d=4,6$ for a scalar field with $\nu= 0.75$.  The top plot shows the numerically computed bound state energy  compared against the WKB prediction at leading (open marker) and first subleading order (closed marker).
    The lower panel shows only the signed corrected discrepancy,
  $100(\freq_n^{\rm ODE}/\freq_n^{\rm WKB}-1)$, on a linear scale.}
  \label{fig:global-outside-bound-states}
\end{figure}

The presence of turning points outside the black hole indicates that the
effective potential has a local well that can trap bound states. In our WKB
analysis we should recover the bound state energies by a standard
Bohr-Sommerfeld quantization condition. The leading part of the bound state
energy should be fixed by the turning point action between the outermost turning
point and the AdS boundary, the quantity $P\subm{tp}$ defined in~\eqref{eq:gsads-outsidePB}. The decay
width is controlled by the trapping in the tunneling barrier to the horizon, and
hence by $B\subm{bound}$. Using the near boundary WKB
analysis to fix the asymptotic data, we conclude that the bound state spectra
are fixed as
\begin{equation}\label{eq:bohrsm-bound}
P\subm{tp}\, \Re\freq_n = \pi \,\pqty{n + \frac{\nu+1}{2}}
+ \order{\freq_n^{-1}} \,, \qquad \Im \freq_n
\sim - \frac{1}{4\, P\subm{tp}}\, e^{-2\ms B\subm{bound}\, \Re\freq_n} \,.
\end{equation}
This follows from~\eqref{eq:GR-bound} and is a fixed $\zeta$ result. Ignoring the decay width of the bound state, one obtains the standard quantization by demanding that we have a normalizable ingoing solution, i.e., when the denominator vanishes. The decay width is then responsible for pushing the pole slightly below the real axis.
A useful consistency check is the limit $\rp \to 0$ when we should recover the
harmonic trap bound state spectra of global \AdS{d+1}. Recall that the
eigenenergies are $\omega\subm{AdS} = 2\ms n + \ell + \Delta_{\mathcal{O}}$. To compare we should convert the fixed $\zeta$ answer to a fixed $\ell$ expression, which shifts $ P\subm{tp} \to  P\subm{tp} - \zeta\,  \partial_\zeta P\subm{tp}$ in the denominator of $\Im \freq_n$. Using
the fact that $\zeta \omega \sim \ell + \frac{d}{2} - 1$ (asymptotically), one can check that the factors assemble to reproduce the expected answer.

Notice that the Green's function vanishes at  frequencies where the response coefficient for fixed source mode does. This can be read off from the numerator of~\eqref{eq:GR-bound} and predicts zeros at $\freq\, P\subm{tp} = \pi\,\pqty{m + \frac{1-\nu}{2}}$. The swap $\frac{1+\nu}{2} \leftrightarrow \frac{1-\nu}{2}$ follows from the exchange of operator and shadow operator dimension which is explicit in the Green's function formula~\eqref{eq:exact-Gret}. Near a zero or pole of $\Gret$ the phase receives a dominant contribution from the subleading damping term.

These bound states show up as near real frequency poles of the retarded Green's
function. So one can directly solve the connection problem and deduce their
location numerically. We do so for $d = 4,6$ for a range of black hole sizes and
a few different values of $\zeta$ determined by the chosen $\rp$ (since the
potential well exists only for $\zeta > \zeta\subm{crit}$). Scanning the
first few levels $n = 0, \cdots, n\subm{max}$ with $n\subm{max} = 3$ or $4$ we
see that the eigenenergies are recovered quite reasonably (numerical errors of $\freq_n^{\rm ODE}/\freq_n^{\rm WKB}-1$ are
at most~$5 \times 10^{-4}$). A comparison between the ODE determined
real part of the energy and the WKB prediction from our period calculation is
displayed in~\cref{fig:global-outside-bound-states}. The leading order result obtained from the Bohr-Sommerfeld quantization formula~\eqref{eq:bohrsm-bound} deviates from the true bound state energy determined from direct solution of the differential equation. In plotting the result we therefore also included the first correction to the WKB integral between $\rbull$ and the boundary.

\subsection{Planar \RNAdS{d+1} black holes}
\label{sec:planarRN}

As our final example we will examine planar charged black holes. We will find it
convenient to write the metric function by setting $M = \rp^d\,(1 + Q^2)$ and replacing $Q_c = Q\, \rp^{d-1}$ to make the charge parameter dimensionless, viz.,
\begin{equation}
\label{eq:RNmetricfunction}
f(r) = 1
- \frac{r_+^d\, (1+Q^2)}{r^d}
+ Q^2\, \pqty{\frac{r_+}{r}}^{2\,d - 2} \,.
\end{equation}
Then in the projective variable $\rpr$, cf.~\cref{eq:rproj} we have the WKB one form
\begin{equation}
\lwkb = \frac{\rpr^{\frac{d}{2}-2}\, \sqrt{\rpr^{d-1}  - \zeta^2\,F(\rpr) }}{F(\rpr)} \, \frac{d\rpr}{2}\,, \qquad
F(\rpr) = \rpr^{d - 1} - (1 + Q^2)\, \rpr^{\frac{d}{2} - 1} + Q^2\,.
\end{equation}
There are an odd number ($d - 1$) of turning points from the numerator, which
along with the point $\rpr = \infty$ make up the $d$ branch points. The spectral
curve is therefore hyperelliptic of genus $\frac{d}{2} - 1$, and the black hole
singularity remains invisible to the WKB curve. This case is different from the
\SAdS{d+1} examples in that the curvature singularity is at worst an apparent singular
point of the differential equation, cf.~\cite{Loganayagam:2022teu}. This statement holds in the $r$ coordinate chart. In projective coordinates (such as $\rpr$) one could encounter a branch-cut at that locus.

\begin{figure}[ht!]
  \centering
  \includegraphics[width=\linewidth]{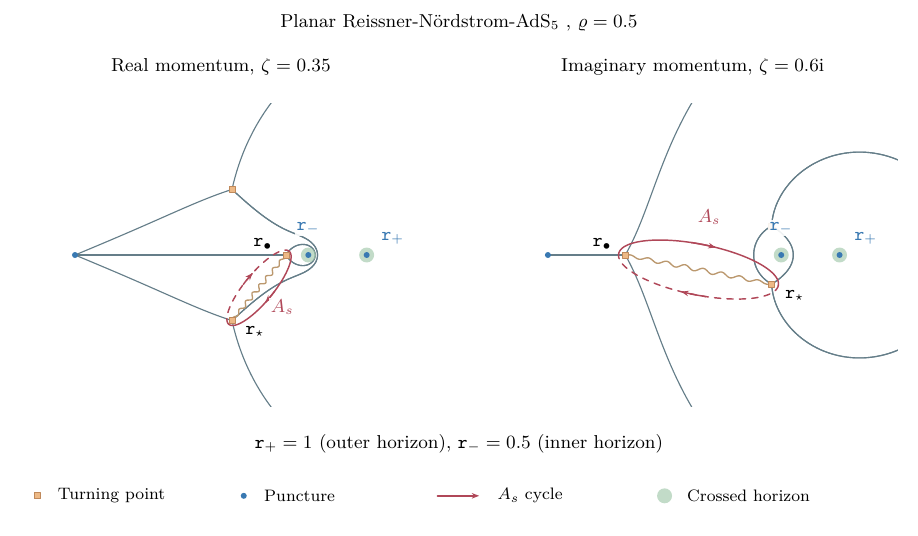}
  \caption{
    Stokes graphs and $A_s$ cycles for planar \RNAdS{5} at
    $\varrho=0.5$, in the projected $\rpr$ coordinate following conventions used for~\cref{fig:planar-sads-stokes-cycles}, for real momentum $\zeta=0.35$ (left) and imaginary momentum $\zeta=0.6\ms\ii$ (right).
  }
  \label{fig:rnads5-stokes-cycles}
\end{figure}

Of the finite odd number of turning points, one is located on the $\Re \rpr$
axis at $\rpbull > 0$, but inside the inner horizon. This follows from the fact that the
polynomial inside the square root changes sign between $\rpr = 0$ and the inner
horizon (where $F(\rpr)$ vanishes). This behavior is dimension agnostic, so we will
examine only the tractable case $d = 4$ in detail below.

\begin{figure}[bh!]
  \centering
  \includegraphics[width=0.9\linewidth]{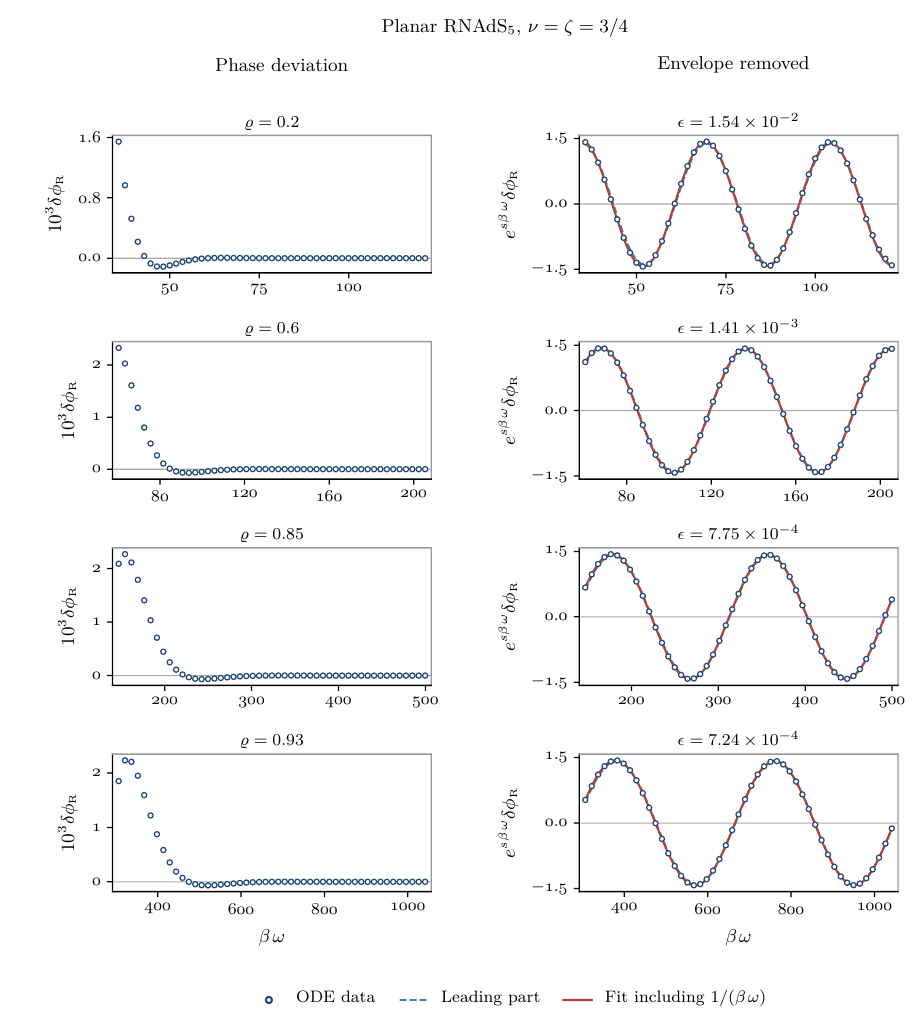}
  \caption{
    The phase of the retarded Green's function in planar \RNAdS{5}, with
    $\nu= 0.75$ and $\zeta=3/4$. The left shows a rescaled bare phase, while the right strips off an exponential with $s=\Im\Iwkb/(\beta\rp)$. Conventions are as  indicated in~\cref{fig:planar-phase-oscillations}.
  }
  \label{fig:rnads5-lateral-normalized-oscillations}
\end{figure}

In $d = 4$ the formulae are simple if we introduce
\begin{equation}
\label{eq:Qvarrho}
Q^2 = \varrho\, (1 + \varrho) \,, \qquad \varrho
= \frac{r_-^2}{\rp^2} \,, \qquad 0
< \varrho
< 1\,.
\end{equation}
The singular points of the differential are at $\rpr = 1$, $\rpr = \varrho$ and
$\rpr = -(1 + \varrho)$, respectively, corresponding to the outer, inner, and the
third horizon root (which is at real $\rpr$). The turning points are roots of
\begin{equation}\label{eq:rn-rturnp}
(1 - \zeta^2)\, \rpr^3 + \zeta^2\, (1 + \varrho + \varrho^2)\, \rpr - \zeta^2\, \varrho\, (1 + \varrho) = 0\,.
\end{equation}
Denote the distinguished real one which lies between $\rpr = 0$ and $\rpr = \varrho$
as $\rpbull$. This structure is summarized in the Stokes
graph~\cref{fig:rnads5-stokes-cycles}.

\begin{figure}[th!]
  \centering
  \includegraphics[width=0.9\linewidth]{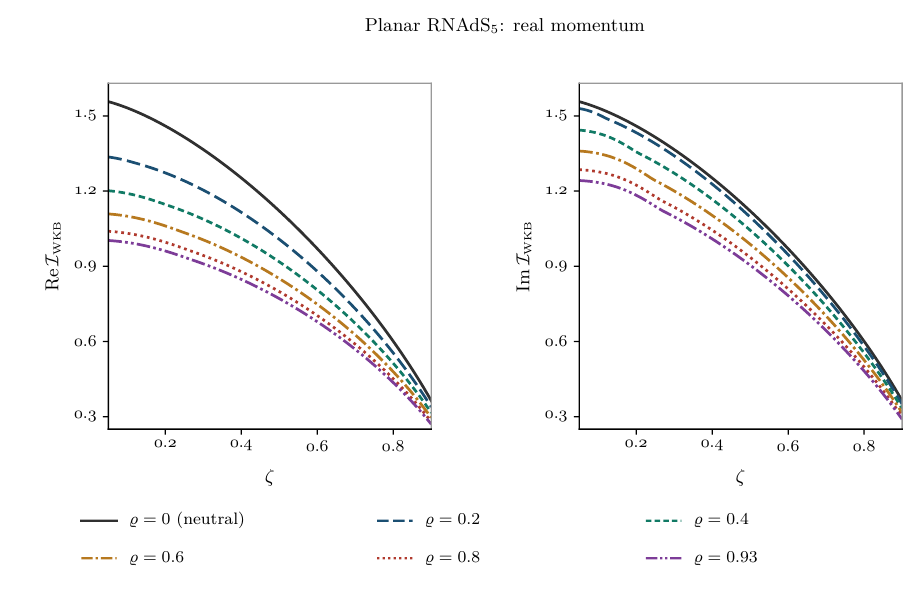}
  \caption{
    The WKB action~\eqref{eq:rnads5-relative-action-explicit} for
    planar \RNAdS{5} at real momentum.
    The panels show $\Re\Iwkb$ and $\Im\Iwkb$ for the six charge
    parameters, with the solid curve being the neutral answer.
  }
  \label{fig:rnads5-iwkb-zeta}
\end{figure}

The $A$-cycle of interest, as depicted in~\cref{fig:rnads5-stokes-cycles},
envelopes the cut from $\rpbull$ to the complex turning point with
$\pi < \arg \rpr < \frac{3\ms\pi}{2}$. The period $\Aper$ can be computed in terms of
elliptic integrals as in the global \SAdS{5} case. We will refrain from writing
out the result, which isn't particularly illuminating. To write the action
in a clean form, let us record the inverse temperatures at the three horizon
loci. They are
\begin{equation}\label{eq:rnads5-residue-horizon-periods}
\begin{aligned}
\beta
& = \frac{2\,\pi\,}{r_+\, (1-\varrho) \, (2+\varrho)} \,,
\\
\beta_{\mathrm{inner}}
& = \frac{2\pi\, \varrho^{3/2}}{r_+\, (1-\varrho)\,(1+2\varrho)}\,,
\\
\beta_c
& = \frac{2\pi(1+\varrho)^{3/2}}{r_+\, (2+\varrho)\, (1+2\varrho)} \,.
\end{aligned}
\end{equation}

The WKB action can be assembled as before by computing the physical propagation
factor $\Pphys$ by the contour deformation. However, in this case since $\rpbull$
is inside the inner horizon, the outgoing ray in the analog
of~\cref{fig:planar-Pphys-deformation} goes out along the third complex horizon
root. The analog of the sum-rule~\eqref{eq:Psumrule} which one derives here is the identity
\begin{equation}
2\, \Pphys + \Aper = \frac{r_+\,\beta_c}{2} - \Aper^*\,.
\end{equation}
In addition, the action also picks up an imaginary contribution from jumping not
only the outer horizon, but also from the jump across the inner horizon pole.
One then assembles the WKB action to obtain
\begin{equation}
\Iwkb = \frac{r_+}{2} \,\bqty{ \beta_c + \ii (\beta - \beta_{\mathrm{inner}})}
- \Aper^* \,.
\label{eq:rnads5-relative-action-explicit}
\end{equation}
We verify that this action reproduces the predicted behaviour of the retarded
Green's function's phase $\phGR$
in~\cref{fig:rnads5-lateral-normalized-oscillations}.

Finally, given the WKB action we can investigate the behaviour as we vary the
charge starting from the neutral limit $\varrho = 0$ to the close-to-extremal
limit ($\varrho = 0.93$). The result displayed in~\cref{fig:rnads5-iwkb-zeta}
shows a smooth behaviour, with no emergent feature as we move towards
extremality. This is per se, a bit surprising since the approach to extremality modifies the infra-red dynamics. This occurs even classically~\cite{Liu:2009dm} without including the quantum effects present in the near-extremal regime~\cite{Ghosh:2019rcj,Iliesiu:2020qvm}. However, the features we are examining correspond to the UV, and are less affected by these effects.

\section{Commutator OPE and WKB periods}
\label{sec:ope-wkb}

The analysis of large order thermal OPE coefficients initiated in~\cite{Ceplak:2024bja} from the bulk perspective provided a concrete link between thermal OPE and the black hole. Furthermore, now we have a clear picture of how to map the quasinormal mode expansion of thermal correlators to their OPE expansion~\cite{Arnaudo:2026axe,Batra:2026gbq}. One would therefore anticipate that the leading WKB asymptotics motivated above ought to be directly connected to the thermal OPE coefficients. We will now demonstrate that this is indeed the case. For this discussion, we will focus on scalar primaries of dimension $\Delta = \frac{d}{2} + \nu$. Our discussion will use the position space OPE; the momentum space counterpart has also recently been explored in~\cite{Grassi:2026ekz}.

\subsection{Singularity in commutator OPE}
\label{sec:commutator-large-order-ope}

The thermal observables discussed hitherto have been the retarded Green's function and its relations, such as the spectral function, or the thermofield double Wightman function. For the present OPE discussion, we shall consider the expectation value of the commutator itself (viz., eschew the time ordering step function). Consider, therefore,
\begin{equation}
\mathcal{G}(t,\vb{x}) = \expval{\comm{\mathcal{O}(t,\vb{x})}{\mathcal{O}(0,\vb{0})}}_{\beta}\,,
\qquad
\Gret(t,\vb{x}) = -\ii\, \Theta(t)\,\mathcal{G}(t,\vb{x}) \,.
\end{equation}
The commutator expectation value when Fourier transformed to momentum space is the spectral function. Let us examine the Lorentzian OPE of the commutator with $t, \abs{\vb{x}}$ small, but keeping the ratio
\begin{equation}
\kappa = \frac{x}{t} \in (0,1)\,, \qquad \text{with}\ x\equiv \abs{\vb{x}}\,,
\end{equation}
fixed. We are thus in the timelike regime, which is necessary for the commutator to be non-vanishing, and clearly $\kappa$ is the position space analog of $\zeta$.

Let us focus on the contribution to the OPE from the stress-tensor sector. This is the part captured by bulk graviton exchange in holographic theories (see below). We find that this contribution can be determined to be the following:
\begin{equation}
\begin{aligned}
\mathcal{G}_{\rm vac}(t,\kappa)
& =
-2\ms\ii \sin(\pi\Delta)\,\pqty{t^2\, (1-\kappa^2)}^{-\Delta}\,,
\\
\mathcal{G}(t,\kappa \,t)
& = \mathcal{G}_{\rm vac}(t,\kappa) \, F_\kappa(s),
\qquad s \equiv \prn{\frac{t}{\beta}}^d\,,
\\
F_\kappa(s)
& =
1 + \sum_{n = 1}^\infty\,  \lambda_n(\kappa)\,s^n \,.
\end{aligned} \label{eq:ope-ope}
\end{equation}
We are assuming here that $\Delta \notin \mathbb{Z}_{>0}$. The inverse temperature dependence has been subsumed into the dimensionless variable $s$; so the coefficients are in the canonical normalization with $\beta =1$. In terms of the spin-resolved Euclidean thermal
OPE coefficients, the fixed-\(\kappa\) coefficients are (NB: the $J$ sum is in steps of two)
\begin{equation}
\lambda_n(\kappa) =
(-1)^{\frac{n\,d}{2}}
\pqty{1 - \kappa^2}^{\frac{n\,d}{2}} \,
\sum_{J\,\overset{2}{=}\,0}^{2\,n} \,
a_{T^n}(J)\, C_J^{\prn{\Half{d-2}}}\,\pqty{\frac{1}{\sqrt{1-\kappa^2}}}\,.
\label{eq:lambda-ope-coeffs}
\end{equation}
Here $C_J^{(\lambda)}(z)$ is the Gegenbauer polynomial, and
\begin{equation}
a_{T^n}(J)
=
\frac{J!}{2^J\left(\frac{d-2}{2}\right)_J}
\frac{
  f_{\mathcal{O}\mathcal{O} T^n_J}\,
  b_{T^n_J}
}{
  c_{T^n_J}
},
\end{equation}
where $f_{\mathcal{O}\mathcal{O} T^n_J}$ is the OPE coefficient with the multi-stress-tensor, $b_{T^n_J}$ the thermal one-point coefficient of multi-stress-tensor, and
$c_{T^n_J}$ the two-point normalization coefficient of the composite $T^n_J$ and $(x)_a$ the Pochhammer symbol.

The result above is obtained by starting from the Euclidean OPE on $\mathbf{S}^1_\beta\times \mathbb{R}^{d-1}$, which was analyzed in~\cite{Iliesiu:2018fao}, and computing the Lorentzian commutator expectation value by taking the discontinuity across the light-cone branch cut.
The only constraint is that we have restricted our attention to the contribution of the stress tensor. In general other operators would contribute as long as the product of the thermal one-point functions and the OPE coefficients is non-vanishing. In the case of holographic CFTs, we will at the very least have contributions from the double twist fields $\mathcal{O}\, (\partial_\mu \partial^\mu)^n\, \partial_{\mu_1}\cdots \partial_{\mu_J}\, \mathcal{O}$. While double-twist operators contribute to the Euclidean thermal OPE, they don't contribute to the commutator OPE in the classical gravity approximation, as they have vanishing light-cone discontinuity. Therefore, the perturbative expansion of the commutator is entirely fixed by thermal OPE coefficients in multi-stress-tensor sectors.

The stress tensor contribution to the thermal OPE has been analyzed in holographic CFTs by~\cite{Ceplak:2024bja} at zero spatial separation building on the algorithm proposed in~\cite{Fitzpatrick:2019zqz}. The OPE at spacelike separation was analyzed in~\cite{Araya:2026shz}. In the following we perform a similar analysis at timelike separation, which will be related to timelike momentum by a saddle-point relation.

\paragraph{Principal-sheet singularity:} The key point of import is that the OPE coefficients $\lambda_n$ give a series with finite radius of convergence, which is determined by the large order behavior of $\lambda_n$. Generating the OPE coefficients to high orders using the recursion algorithm in~\cite{Buric:2025fye}, we find that $\lambda_n$ has the following large order behavior:
\begin{equation}
\lambda_n(\kappa)
\sim \mathcal{A}(\kappa,\Delta)\, n^{\alpha(\Delta)} \,Y_*(\kappa)^{-n}
\prn{1+ \order{n^{-1}}}\,.
\label{eq:lambdan-large-order}
\end{equation}
The exponent is independent of the fixed ray (viz., $\kappa$) and is consistent with the expression
\begin{equation}
\alpha(\Delta) = 2\,\Delta - \frac{d+1}{2}.
\label{eq:ope-alpha}
\end{equation}

The large-order growth of the OPE coefficients predicts that the stress-tensor sector of the commutator has a singularity on the principal sheet. Specifically, working with the fixed ray specified by $\kappa$ in the variable $t^d$, we predict a singularity at
\begin{equation}
t_*(\kappa)^d = \beta^d \,Y_*(\kappa)\,.
\label{eq:ope-singular-time}
\end{equation}
Equivalently, the function $F_\kappa(s)$  has a principal singularity at
$s = Y_*(\kappa)$, with local behavior given by
\begin{equation}
F_\kappa(s)
\sim \mathcal{A}(\kappa,\Delta)\, \Gamma\pqty{\alpha(\Delta)+1}
\bqty{  1-\frac{s}{Y_*(\kappa)}}^{-\alpha(\Delta) - 1}\,.
\label{eq:ope-local-singularity}
\end{equation}
The action $Y_*(\kappa)$ will be related to a WKB period between AdS boundary and bulk turning points in~\cref{sec:commutator-wkb-periods}.

\paragraph{Non-principal-sheet singularities:} Since at finite $\kappa$ the principal sheet singularity will not be related to the black hole singularity, it is natural to ask if there are still geometric quantities associated with the black hole singularity that can be extracted from the OPE data. The momentum space OPE analysis~\cite{Grassi:2026ekz,Grassi:2026wip} gives a hint: there the principal sheet Borel singularities are associated with boundary-turning-point WKB periods, while non-principal sheet Borel singularities depend on turning-point-singularity periods. In the momentum space context this can be derived precisely from exact WKB analysis, with the non-principal sheet behavior controlled by the DDP formula discussed in~\cref{sec:qperiod-ddp}. Inspired by this, we explore the non-principal sheet singularities in position space OPE. We will indeed find numerical evidence for the appearance of a turning-point-singularity period.

Suppose we analytically continue the OPE data by taking $F_\kappa(s)$ off the principal sheet around the singular locus $s = Y_*(\kappa)$. Denote the resulting function as $F_\kappa^{(1)}(s)$; it is the continuation onto the adjacent Riemann sheet. This continuation can be probed numerically, by adapting the Pad\'e (or Hermite-Pad\'e) approximation to ensure that it is aware of the Riemann sheet of the functions. Let us therefore approximate the function on the principal sheet by
\begin{equation}
\begin{aligned}
Q(u)\, F_\kappa(u) - P(u) - R(u)\,A_*(u)
& = \order{u^N}\,, \qquad
u =\frac{s}{Y_*(\kappa)}\,,
\end{aligned}
\label{eq:ope-sheet-hp}
\end{equation}
with $F_\kappa(u) = F_\kappa\pqty{Y_*(\kappa)\,u}$ as a notational shorthand, and
\begin{equation}
A_*(u) = (1 - u)^{-\alpha(\Delta) -1}\,.
\end{equation}
In this local chart, the singularity has been relocated to $u=1$,
and we can use it to analyze the continuation across the cuts.
Specifically, since $A_*(u)$ rotates by a phase as we pass through the cut to the adjacent sheet (we take $(1-u)^{-p} \to e^{-2\pi\ii p}\, (1-u)^{-p}$ for some $p$ as we go around the cut in a counterclockwise direction), we deduce that
\begin{equation}
F_\kappa^{(1)}(s)
\approx
\frac{P(u)+e^{-2\pi\ii (\alpha(\Delta)+1)}\, R(u)\, A_*(u)}{Q(u)}\,.
\label{eq:ope-sheet-hp-np}
\end{equation}

\begin{figure}[htbp!]
  \centering
  \includegraphics[width=0.84\textwidth]{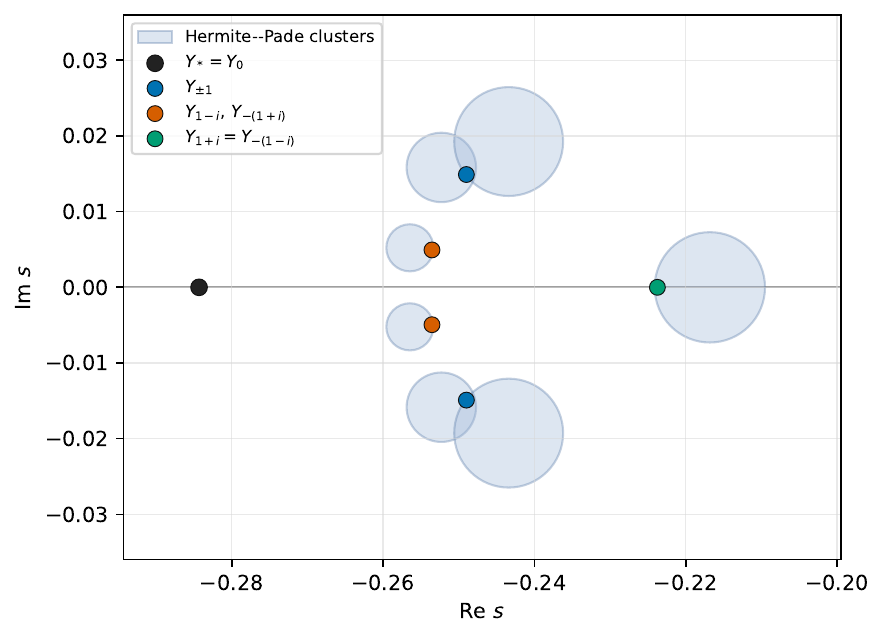}
  \caption{The Hermite--Pad\'e clusters obtained from OPE associated with the non-principal-sheet
    observable $F_\kappa^{(1)}(s)$ defined in~\eqref{eq:ope-sheet-hp-np}. We work in planar \SAdS{5} with a scalar primary of dimension
    $\Delta = \frac{5}{2}$ and have chosen the fixed-ray with $\kappa = \frac{1}{2}$.
    The shaded disks are stable clusters of zeros of  $Q(u)$ in \eqref{eq:ope-sheet-hp}; their radii are the cluster spreads converted to the $s$-plane.
    Solid dots show the WKB action candidates $Y_*=Y_0$,
    $Y_{\pm1}$, $Y_{1 - i}$, $Y_{-(1 + i)}$, and $Y_{1 + i} = Y_{-(1 - i)}$.
  }
  \label{fig:lorentzian-commutator-hp-cluster-halos}
\end{figure}
Given the approximant, we can infer candidate singularities of the
non-principal-sheet function. These are extracted as stable zeros
of $Q(u)$ for which the principal-sheet numerator
$P(u) + R(u)\,A_*(u)$ remains suppressed while the adjacent-sheet numerator
$P(u) + e^{-2\pi \ii (\alpha(\Delta) + 1) }\, R(u)\,A_*(u)$
is non-vanishing.  This test locates candidate singular points,
but does not by itself determine their local singularity type. The stable clusters obtained by this algorithm are shown in~\cref{fig:lorentzian-commutator-hp-cluster-halos}. The analysis was specifically carried out for planar \SAdS{5} black hole where the asymptotic data are available for a scalar primary with $\Delta = \frac{5}{2}$.

\subsection{Relation with WKB periods}
\label{sec:commutator-wkb-periods}

Let us test our OPE result against the WKB actions derived hitherto. This will at the very least help us empirically establish a direct link between the OPE data, and the spectral function asymptotics.

Consider a (relative) contour $\gamma$ and let $V^{\mathrm{cl}}_\gamma$ be the (open) normalized WKB action defined using the leading WKB differential
\begin{equation}
V^{\mathrm{cl}}_\gamma = \frac{2\ii}{\beta\,\rp}\, \int_\gamma\, \lwkb(\zeta)\,.
\end{equation}
We will want to take $\gamma$ to be a combination of paths
from the boundary and singularity to the turning points $\rpr_j$ in the complex $\rpr$ chart introduced in~\cref{sec:examples}. We will use the planar \SAdS{d+1} turning points $\rpr_j$ given in~\eqref{eq:planar-uturn} and thus take $\gamma = \gamma_{\rpr_j, \infty} \cup \gamma_{\rpr_j,0}$. In particular, with our chosen conventions for the cycle $\gamma_{\rstar,\infty}$ running from the distinguished turning point $\rpstar = \rpr_{\frac{d}{2}-1}$ to the boundary (with the same branch and horizon bypass convention used in the preceding period analysis)  we have a direct relation between $V^{\mathrm{cl}}_\gamma$ and $\Iwkb$. Explicitly,
\begin{equation}\label{eq:Pgamma-0inf}
V^{\mathrm{cl}}_{\gamma_{\rstar,\infty}}
=
\frac{\ii}{\beta \,\rp} \, \Iwkb(\zeta)
=\frac{\ii \,e^{\frac{\ii\pi}{d}}}{2\,\sin(\frac{\pi}{d})} \,
\bigg[1-\Pi_d(\zeta)\bigg]\,,
\end{equation}
where $\Pi_d(\zeta)$ is given in~\eqref{eq:Pi-d-general-hypergeometric}.

\begin{figure}[htbp!]
  \centering
  \includegraphics[width=0.90\textwidth]{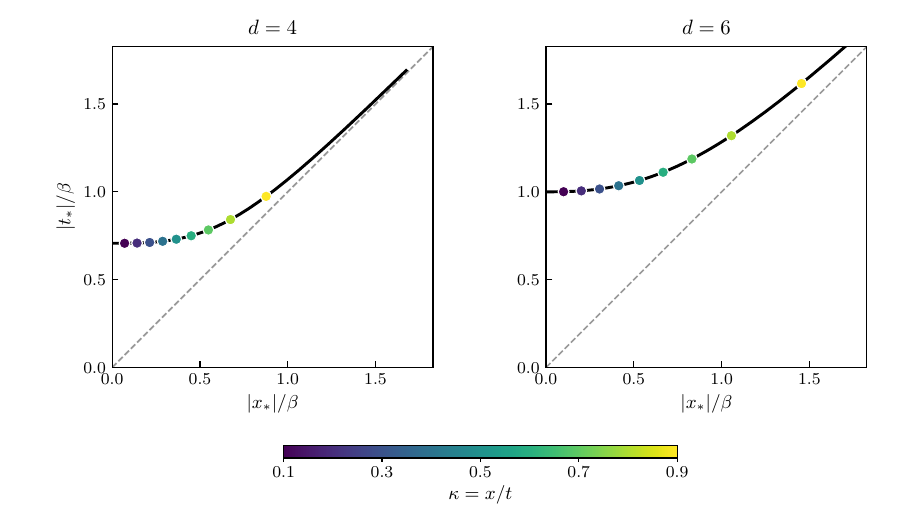}
  \caption{Principal-sheet singularity locus in $x,t$ plane. Black curves give the WKB prediction; colored points are the OPE estimates, with color indicating $\kappa=x/t$.}
  \label{fig:lorentzian-commutator-position-locus}
\end{figure}

We can now use the information from the geodesic analysis~\cref{sec:spgeods} to relate these periods to the spatio-temporal separations on the boundary~\eqref{eq:periodrels1}. With our current normalization conventions, the Lorentzian endpoint separations are
\begin{equation}
(\Delta x)_\gamma(\zeta) = \ii\beta\,\partial_\zeta V^{\mathrm{cl}}_\gamma(\zeta), \qquad
(\Delta t)_\gamma(\zeta)
= -\ii\beta\, \bqty{V^{\mathrm{cl}}_\gamma(\zeta)-\zeta \, \partial_\zeta V^{\mathrm{cl}}_\gamma(\zeta)} \,.
\label{eq:ope-geodesic-endpoints}
\end{equation}
Given this we can predict using the WKB action that the location of the principal sheet singularity for curve $\gamma$ ought to be located at
\begin{equation}
Y_\gamma(\kappa)
= \pqty{\frac{(\Delta t)_\gamma(\zeta_\gamma)}{\beta}}^d
= (-1)^{\frac{d}{2}}\, \bqty{V^{\mathrm{cl}}_\gamma(\zeta_\gamma)-\zeta_\gamma \, \partial_\zeta V^{\mathrm{cl}}_\gamma(\zeta_\gamma)}^d\,.
\end{equation}
The relation between $\zeta$ and $\kappa$, which capture the momentum and position space timelike ratios, respectively, is fixed by the saddle point relation
\begin{equation}\label{eq:kappa-zeta-saddle}
\kappa = \frac{(\Delta x)_\gamma(\zeta_\gamma)}{(\Delta t)_\gamma(\zeta_\gamma)}
= - \frac{\partial_\zeta V^{\mathrm{cl}}_\gamma(\zeta_\gamma)}{V^{\mathrm{cl}}_\gamma(\zeta_\gamma)-\zeta_\gamma \, \partial_\zeta V^{\mathrm{cl}}_\gamma(\zeta_\gamma)}\,.
\end{equation}
Explicitly, at fixed $\kappa$ and for a particular choice of $\gamma$, we solve $\zeta_\gamma = \zeta_\gamma(\kappa)$ using~\cref{eq:kappa-zeta-saddle}. We shall now investigate the principal and non-principal sheet singularities, extracted using the WKB analysis and compare them with those obtained from the OPE. Clearly, the purpose of this section is to demonstrate the agreement between the two, which will indeed transpire to be the case.

\begin{figure}[tbp]
  \centering
  \includegraphics[width=0.88\textwidth]{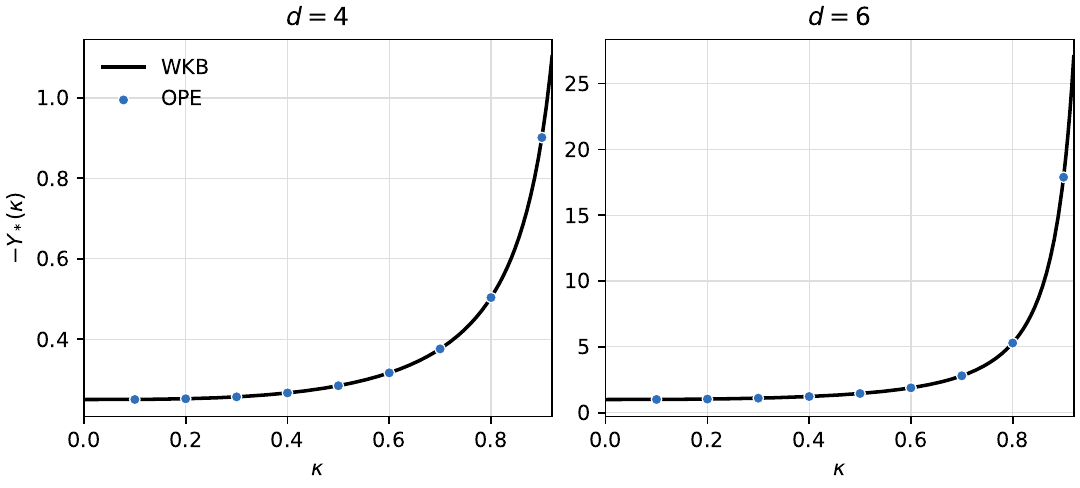}
  \caption{
    Comparison between the WKB prediction and the fixed-ray (specified by $\kappa$) OPE extraction of the timelike commutator action. The solid curves show $-Y_*(\kappa)$ obtained from~\eqref{eq:ope-action-Pi}. The marked (blue) points are obtained
    from the large-order behavior of $\lambda_n(\kappa)$ sampled from the set   $\kappa = \set{0.1,\cdots,0.9}$. We chose $\Delta =\frac{5}{2}$, and $N_{\max} =100$ for $d=4$, and $N_{\max} =80$ for $d=6$. Note that in $d=6$, an operator of dimension $\frac{5}{2}$ corresponds to a scalar field quantized with Neumann boundary conditions.
  }
  \label{fig:lorentzian-commutator-y-kappa}
\end{figure}

\paragraph{The principal-sheet singularity:} Given the principal action $Y_*(\kappa)$ obtained  from the large-order behavior of the OPE coefficients, we can attempt to ascertain the contour $\gamma$ that reproduces it. One finds empirically that the choice
\begin{equation}
\gamma_*
= \gamma_{\rpr_\star,\infty}
\,, \qquad
\zeta_* = \zeta_{\gamma_*}
\,, \qquad
Y_*(\kappa) = Y_{\gamma_*}(\kappa)\,,
\end{equation}
serves the purpose. For the principal sheet singularity, the choice of $\rpr_j$ turns out to be immaterial, so we have chosen it to be our distinguished turning point $\rpstar = \rpr_{\frac{d}{2}-1}$. Doing so, and using~\eqref{eq:Pgamma-0inf} the saddle relation
reduces to
\begin{equation}
\kappa = \frac{\Pi_d'(\zeta_*)}
{1-\Pi_d(\zeta_*)+\zeta_*\Pi_d'(\zeta_*)}. \label{eq:ope-kappa-zeta-Pi}
\end{equation}
As a nice consistency check, we note that for $0 < \kappa < 1$, this primary saddle lies on the real interval $0 < \zeta_*<1$.
The limits $\kappa \to 0$ and $\kappa \to 1$ map to
$\zeta_* \to 0$ and to the light-cone endpoint $\zeta_* \to 1$,
respectively. Computing the associated action we obtain an analytic prediction for the location of the principal sheet singularity
\begin{equation}
Y_*(\kappa) = - \prn{
  \frac{1-\Pi_d(\zeta_*)+\zeta_*\Pi_d'(\zeta_*)}
{2\sin\!\prn{\frac{\pi}{d}}}}^d . \label{eq:ope-action-Pi}
\end{equation}
This result can now be compared against the OPE prediction $Y_*(\kappa)$ obtained from~\eqref{eq:lambdan-large-order}. The two agree quite cleanly as is visible from the plot shown in~\cref{fig:lorentzian-commutator-y-kappa}.

\paragraph{Non-principal-sheet singularities:} We thus see that the principal sheet singularities are accurately predicted by our singularity-anchored WKB representative. While this statement has been known for $\zeta =0$ from the work of~\cite{Ceplak:2024bja} the above analysis shows that it continues to work at finite spatial separation. The non-trivial question is whether the periods $\Aper$ are visible to the OPE data, and, in particular, whether they are aware of the complex turning points. The answer again will be in the affirmative as we now confirm by examining the non-principal sheet singularities.

To organize the non-principal candidates, we start with the $\rstar$ representative and absorb integer combinations of the open cycle contributions from  the segments running from $\rpr_j$ to $0$ into a single label $S$ by introducing
\begin{equation}
V^{\mathrm{cl}}_S(\zeta) = V^{\mathrm{cl}}_{\gamma_{\rstar,\infty}}(\zeta)
+ S\,V^{\mathrm{cl}}_{\gamma_{\rstar,0}}(\zeta), \qquad
S\in \mathbb{Z}[e^{-\frac{2\pi\ii}{d}}] \,.
\label{eq:ope-P-S}
\end{equation}
Specifically, in $d=4$ using relations between the relative and closed cycle periods one can show that
\begin{equation}
V^{\mathrm{cl}}_S(\zeta) = \frac{\ii}{\beta\,\rp} \bqty{\Iwkb(\zeta) +S\, \Aper(\zeta)}\,,
\qquad d=4\,.
\end{equation}
We can view $S$ as abbreviating the integer lattice of open WKB actions
\begin{equation}
S = S(\mathbf m)
= \sum_{j \in \mathbb{Z}_d} m_j e^{-2\pi i j / d}, \qquad \mathbf m
\in \mathbb{Z}^d\,.
\end{equation}

For each $S$ the fixed-ray saddle $\zeta_S=\zeta_S(\kappa)$ is defined
by using $V^{\mathrm{cl}}_S(\zeta_S)$ in~\eqref{eq:kappa-zeta-saddle}, which leads to the action
\begin{equation}
Y_S(\kappa) = (-1)^{\frac{d}{2}}\,
\bigg[V^{\mathrm{cl}}_S(\zeta_S)-\zeta_S\, \partial_{\zeta_S} V^{\mathrm{cl}}_S(\zeta_S)\bigg]^d \,.
\label{eq:ope-action-S}
\end{equation}
We fix the branch by demanding that
\begin{equation}
\lim_{\kappa \to 0}Y_S(\kappa) = Y_*(0).
\end{equation}

We shall examine the features for \SAdS{5}, where the above condition fixes a unique local branch for the nondegenerate
$S$-values. The principal action is denoted as $Y_*(\kappa) = Y_0(\kappa)$. Furthermore, for real $0 < \kappa < 1$, the conjugate fixed-ray saddle branch gives
\begin{equation}
\overline{Y_S(\kappa)} = Y_{-\overline S}(\kappa)\,.
\end{equation}
This means that independent candidates can be related through
$S \sim -\overline S$. Specifically, in $d=4$ the phase is simple and
\begin{equation}
S \in \mathbb{Z}[i], \qquad
S(\mathbf m) =  m_0- \ii m_1-m_2+\ii m_3 \,.
\end{equation}
At the first two period levels $\abs{\mathbf{m}} =1$ and $\abs{\mathbf{m}} =2$, with $\abs{\mathbf{m}} \equiv \sum_{j=0}^3 \,m_j$ the classes are
\begin{equation}
S \in \{\pm1\,, \pm \ii\} \,, \qquad
S \in \{\pm2\,, \pm 2\ii\,, \pm (1 + \ii)\,, \pm (1 - \ii)\}\,,
\end{equation}
respectively, subject to the equivalence condition thanks to the conjugation.
The stable clusters defined in~\cref{sec:commutator-large-order-ope} close to $Y_*(\kappa)$
are indeed identified with some of these WKB candidates, as is clear from~\cref{fig:lorentzian-commutator-hp-cluster-halos}.

\newpage
\part{Exact-WKB analysis: quantum periods and the singularity}
\label{prt:WKB2}

\section{Exact-WKB: the boundary to horizon connection problem}\label{sec:ewkb-invite}

We have hitherto seen that a naive WKB analysis appears to give us access to the geometry in the vicinity of the singularity. However, as noted in that discussion, in each of our examples the propagation from the boundary to the horizon encounters a Stokes line emanating from the complex turning points. When this occurs one is forced to account for the Stokes crossing, which transports the WKB trajectory to the turning point itself first. Therefore, the complex path appears not to reach out directly to the singularity, numerical evidence for our WKB actions notwithstanding.

Our goal now is twofold. First, we need to explain what regions of the spacetime the exact-WKB connection problem accesses naturally. For the physical regime where $\arg \omega =0$ this will turn out to be the exterior region of the black hole with a detour to the complex turning points for $\zeta \in (0,1)$. On the other hand for imaginary momenta $\zeta \in \ii \mathbb{R}$ the WKB path will make it to the real turning point inside the horizon. In neither case, though, will the trajectory get anywhere close to the singularity. Given this we have next to elucidate why it is the case that there is a representative path that accesses the vicinity of the singularity. Specifically, the results of~\cref{prt:WKB1} are obtained by a contour deformation argument, displaying the presence of a `singularity-anchored  representation' of the exact-WKB path. However, there is also a more general story once we allow consideration of general phases $\arg \omega \neq 0$. We explain how the quantum periods can be mined to show that as $\arg \omega$ is varied past some critical value, the exact-WKB trajectory for the boundary to horizon connection problem will have direct access to the singularity.

\paragraph{Setting up the exact-WKB problem:} As elsewhere in the text, we work with $d$ being even. For studying the connection between the boundary $r \to \infty$ and the horizon $r = r_+$ it is useful to work with the projective coordinate $\zpr = \frac{r_+^2}{r^2}$ defined in~\eqref{eq:zproj}.  For concreteness, we will work with the equation of motion of a minimally coupled scalar field of mass squared $\Delta (\Delta -d)$. The wave equation then is of the form
\begin{equation}
\left[
  \frac{\dd^2}{\dd\zpr^2}
  -\mathfrak{w}^2Q_0(\zpr)-Q_2(\zpr)
\right]\psi=0.
\label{eq:wkb-boundary-universal-oper}
\end{equation}
The detailed form of the potentials for the various examples we study is collected in~\cref{app:Q0Q2examples}.

There is one major advantage of working with the projective $\zpr$ coordinate. In it, the leading WKB potential $Q_0$ has a simple pole at the boundary, while $Q_2$ has a double pole. The latter signals that the boundary is a regular singular point of the wave equation. The change in behaviour from a simple pole at leading order to a double pole is well understood in the exact-WKB context, thereby making this choice particularly  efficacious for our purposes.\footnote{In contrast, in the unprojected radial coordinate, $Q_0$ is regular at the boundary, while $Q_2$ still retains the double pole (the regular singular point remains so since power-law projective coordinates generically cannot remove logarithmic ramification loci). The discussion of exact-WKB when a regular point at leading order changes into a singular point at subleading order is more involved and requires a refined analysis near the boundary. }

\paragraph{The connection matrix:} We will be interested in the connection matrix $\mathscr{C}_{\mathrm{bdy},\mathrm{hor}}$ that translates between the Frobenius basis at the boundary and at the horizon, cf.~\eqref{eq:Connect-hor-bdy}. Physically we are asking how the normalizable and non-normalizable modes in the near-boundary expansion are related on-shell with the ingoing and outgoing basis at the outer horizon. Knowledge of the full connection matrix is more than what we need for Green's functions. Nevertheless, we will obtain it in a series of steps, each of which will be intuitive.

Let $\psi_{\mathrm{in}}^{\Gamma}$ and $\psi_{\mathrm{out}}^{\Gamma}$ be the gamma function normalized near horizon basis as in~\cref{sec:RRA}, and let $\psi_{\mathrm{source}}$ and $\psi_{\mathrm{vev}}$ be the corresponding asymptotic basis. The transport we seek can be achieved by
the following steps:
\begin{enumerate}[wide,left=0pt]
  \item  Translate from the asymptotic Frobenius basis $\{\psi_{\mathrm{source}} ,\psi_{\mathrm{vev}}\}$ into the near-boundary WKB basis. This was discussed to leading order in~\cref{sec:nearbdywkb}. We now wish to give an exact-WKB treatment of the same.
  \item Transport the asymptotic WKB basis toward the horizon accounting for Stokes crossings. This step was carried out with a touch of heuristic intuition in~\cref{sec:Iwkb-conj}, which we now seek to remedy. In order to tackle this problem, we will have to allow $\omega$ to pick up a small phase and average over the choice of signs for this phase regulator, denoted $\vartheta$. This is because the $\arg \freq =0$ WKB problem has a saddle Stokes graph (which is clear in~\cref{fig:planar-sads-stokes-cycles,fig:global-SAdS-stokes,fig:rnads5-stokes-cycles}). We will see that the structure gets a bit more intricate with this lateral resolution owing to the horizon being a regular singular point. One has to follow the WKB path to the turning points before reaching the horizon.
  \item Finally, we will need to translate the WKB basis in the near-horizon region to our normalized Frobenius basis.
\end{enumerate}
The connection matrix factorizes across these steps. One should therefore write
\begin{equation}
\begin{aligned}
\mathscr{C}_{\mathrm{bdy},\mathrm{hor}}^{-1}
& =
\pqty{\text{Frobenius to WKB at infinity}} \times
\pqty{\text{Transport to horizon}}
\\
& \qquad  \times
\pqty{\text{WKB to Frobenius  at horizon}}\,.
\end{aligned}
\end{equation}
We will give an explicit parameterization of the factors below, cf.~\eqref{eq:wkb-C-general}. Owing to the factorization one can analyze each step in turn. We shall start with the local properties (steps 1 and 3), highlight some features that are specific to the examples we have already studied, and then move to the question of the actual transport (step 2). We will then discuss various examples, before turning to derive the singularity-anchored representation. Finally, we shall explore the quantum periods' sensitivity to the singularity.

\subsection{Local analysis near singular points}\label{sec:ewkb-local}

The first task is to work out the local behaviour around the regular singular points
of the wave equation, specifically focusing on the AdS boundary, the physical outer
horizon, and the singularity. We will use this analysis to complete the local translation between the WKB and the Frobenius bases in the first two cases.

\subsubsection{Local analysis near the boundary}\label{sec:ewkb-bdy}

In the projective coordinate $\zpr$ the universal near-boundary behaviour of the WKB potentials is
\begin{equation}
Q_0(\zpr)
=
-\frac{1-\zeta^2}{4\,\zpr}+  \order{1},
\qquad
Q_2(\zpr)
=
\frac{\nu^2-1}{4\,\zpr^2}+ \order{\zpr^{-1}},
\qquad
\zpr\, \to\, 0\,.
\label{eq:wkb-boundary-universal-local-data}
\end{equation}
As we have seen before, the leading potential is insensitive to the scaling dimension of the field $\Delta = \frac{d}{2} + \nu$. The subleading piece has the source and expectation value fall-offs.

To compute $\mathscr{C}_{\mathrm{bdy},\mathrm{hor}}$ using exact-WKB analysis, we  first determine the relation between Frobenius basis and exact-WKB basis normalized at the boundary. This is the all-orders version of the statement obtained in~\cref{sec:nearbdywkb}. We will write the general transformation abstractly as
\begin{equation}\label{eq:wkb-boundary-summed-matrix}
\begin{pmatrix}
  \Psi_{+,\bdy}^{(\vartheta)}\\
  \Psi_{-,\bdy}^{(\vartheta)}
\end{pmatrix}
=
\mathcal{B}\subm{bdy}^{(\vartheta)}
\begin{pmatrix}
  \psi_{\mathrm{source}}\\
  \psi_{\mathrm{vev}}
\end{pmatrix},
\end{equation}
where $\Psi_{\pm}^{(\vartheta)}$ are the WKB basis functions for $\arg \freq = \vartheta$.  They are given in terms of the all orders odd momentum
\begin{equation}
\Psi_\pm = \frac{1}{\sqrt{P\subm{odd}}}\, \exp(\pm \int P\subm{odd} \, d\zpr )\,,
\qquad P\subm{odd} \, d\zpr = - \ii \freq \ms \lwkb + \cdots\,.
\end{equation}
We will fix the sheet for the WKB analysis below, by anchoring it on the near-horizon solutions.

The matrix $\mathcal{B}_\bdy^{\vartheta}$ itself factorizes, and can be expressed as a piece that is computable in the pure AdS spacetime and a piece that is sensitive to the black hole.
\begin{equation}
\mathcal{B}\subm{bdy}^{(\vartheta)}
=
\underbrace{
  \frac{1}{\sqrt{\pi}}\,
  \begin{pmatrix}
    \ii e^{-\ii \pi \ms \nu}\, I_\nu(\freq,\zeta)
    &
    \quad \ii e^{\ii \pi \ms \nu}\, I_{-\nu}(\freq,\zeta)
    \\
    I_\nu(\freq,\zeta)
    &
    I_{-\nu}(\freq,\zeta)
  \end{pmatrix}
}_{\text{vacuum AdS contribution}}
\times
\underbrace{
  \begin{pmatrix}
    \left(\mathcal{T}_\bdy^{(\vartheta)}\right)^{-1}
    & 0
    \\
    0 & \mathcal{T}_\bdy^{(\vartheta)}
  \end{pmatrix}
}_{\text{black hole contribution}}\,,
\label{eq:wkb-boundary-exact-undetermined-residual}
\end{equation}
with
\begin{equation}
I_\nu(\freq,\zeta) = e^{\ii \frac{\pi}{2}\ms \nu}\, \Gamma(\nu)
\left(\frac{\freq\sqrt{1-\zeta^2}}2 \right)^{-\nu}\,.
\end{equation}
This factor is the identity contribution in the OPE, since it originates simply from the vacuum AdS asymptotics, see~\cref{sec:nearbdywkb}.

We have taken $\Psi$ to be the exact-WKB basis in one of the two Stokes regions adjacent to the boundary simple pole. The exact-WKB basis in the other region has the boundary normalization  multiplied by the simple pole connection matrix.
Specifically, using the conventions of~\eqref{eq:bdyJbasis} the normalized exact WKB solutions in the chosen Stokes region are
\begin{equation}
\begin{aligned}
\Psi_{+,\bdy}
& =\sqrt{\pi}\,e^{-\ii\frac{\pi}{2}\ms\nu}\sqrt{\zpr}\,
H_\nu^{(2)}(\xi), \\
\Psi_{-,\bdy}
& =\ii\sqrt{\pi}\,e^{\ii\frac{\pi}{2} \ms \nu}\sqrt{\zpr}\,
H_\nu^{(1)}(\xi).
\end{aligned}
\end{equation}
The form of~\eqref{eq:wkb-boundary-exact-undetermined-residual} is justified as follows: the monodromy matrix around the boundary in exact-WKB basis is given by composing a simple pole connection matrix and a branch-cut crossing matrix. In the Frobenius basis, on the other hand, it is given by the standard diagonal form. The form of~\eqref{eq:wkb-boundary-exact-undetermined-residual} is the most general change of basis consistent with the monodromy data. For the local Bessel model corresponding to vacuum AdS, the remainder piece in $\mathcal{T}_\bdy^{(\vartheta)}$ is trivial $\mathcal{T}_\bdy=1$ as we already saw in \eqref{eq:ads-bdy-wkb}.\footnote{The matrix for the vacuum AdS contribution can be inferred from the standard Bessel basis relation
  \[
    H_\nu^{(1)}(\xi) =
    \frac{J_{-\nu}(\xi)-e^{-\ii\pi\nu}J_\nu(\xi)}{\ii\sin(\pi\nu)}\,, \qquad
    H_\nu^{(2)}(\xi)
    =\frac{e^{\ii\pi\nu}J_\nu(\xi)-J_{-\nu}(\xi)} {\ii\sin(\pi\nu)}\,,
  \]
}
The corrections to this arise from the non-trivial fall-offs correcting the leading asymptotic solution.  Hence, it is natural to identify $\mathcal{T}_\bdy$ as the Borel sum of perturbative stress-tensor sector OPE contributions, which are completely fixed by near-boundary data.\footnote{In position space the latter data are precisely the near-boundary expansion analyzed in~\cite{Ceplak:2024bja}. We are working here in momentum space, where the analogous relation has recently been explained in~\cite{Grassi:2026ekz}.} This provides our first link between the exact-WKB and OPE analyses.

The asymptotic perturbative expansion can be computed using the algorithm of~\cite{Afkhami-Jeddi:2025wra} using the boundary aligned Bessel function basis. Denote this expansion as $Z_\bdy$. We will work with the assumption that the Borel sum in the direction specified by $\vartheta$ defines $\mathcal{T}$, viz.,
\begin{equation}\label{eq:wkb-boundary-summed-symbol-assumption}
\mathcal{T}_\bdy^{(\vartheta)}
=
\mathcal{S}_\vartheta[Z_\bdy] =: \mathcal Z_\bdy^{(\vartheta)}\,.
\end{equation}
If $Z_\bdy$ is not Borel-summable in direction $\vartheta$, then it should be computed by median resummation. For real frequency, $\arg \freq = \vartheta=0$, the series $Z_\bdy$ is Borel-summable for even $\frac{d}{2}$, but it is not Borel-summable for odd $\frac{d}{2}$. Singularities of the Borel transform of $Z_\bdy$ are associated with the saddle trajectory reaching the boundary simple pole.

%
\subsubsection{Local analysis near the horizon}\label{sec:ewkb-horizon}

Our next task is to work out the transformation rules from the Frobenius basis at the horizon to the WKB basis. For definiteness, pick the branch where the behavior of the WKB potential near the horizon, which is universal (cf.~\cref{app:Q0Q2examples}), is given by
\begin{equation}
\begin{aligned}
\mathfrak{w}\, \Res_{\zpr=1}
\sqrt{Q_0(\zpr)}\,d\zpr
& =
\frac{\ii\bw}{4\pi}
\,,
\\
\lim_{\zpr\to1}
(\zpr-1)^2\, Q_2(\zpr)
& =
-\frac{1}{4}\,.
\end{aligned}
\label{eq:wkb-horizon-universal-local-data}
\end{equation}
At the horizon we have a WKB basis $\Psi_{\pm,\hor}$, which we use to fix the sheet. The solution $\Psi_{+,\hor}$ is taken to be the outgoing solution, while $\Psi_{-,\hor}$ is the ingoing one. For physical kinematics of the retarded Green's function $\Psi_{-,\hor}$ is the subdominant solution at the future horizon, with $\Im \freq > 0$ defining the retarded rotation.

One can then show that exact-WKB and Frobenius bases at the horizon are related by a matrix $B\subm{hor}$
\begin{equation}\label{eq:wkb-horizon-unit-matching}
\begin{pmatrix}
  \Psi_{+,\hor}\\
  \Psi_{-,\hor}
\end{pmatrix}
=
B\subm{hor}
\begin{pmatrix}
  \psi_{\mathrm{in}}^\Gamma\\
  \psi_{\mathrm{out}}^\Gamma
\end{pmatrix},
\end{equation}
where
\begin{equation}\label{eq:wkb-horizon-unit-matching-matrix}
B\subm{hor}
=
\begin{pmatrix}
  0&
  2^{-\ii \mathbf{w}}
  \left(-\ii \mathbf{w}\right)^{-\frac{1}{2}}
  \Gamma\!\left(1+2\ii \mathbf{w}\right)
  \\[2mm]
  2^{\ii \mathbf{w}}
  \left(-\ii \mathbf{w}\right)^{-\frac{1}{2}}
  \Gamma\!\left(1-2\ii \mathbf{w} \right)&
  0
\end{pmatrix} \,,
\qquad \mathbf{w} = \frac{\bw}{4\pi}\,.
\end{equation}

The main advantage is that the double pole of the regular singular point is already visible in $Q_0(\zpr)$ making the transformation straightforward. We highlight the fact that this matrix needs no Borel summation and is determined perturbatively by indicating it with a regular (non-calligraphic) symbol.

\subsubsection{Local behavior near the singularity}\label{sec:ewkb-sing}

Finally, let us examine the behavior of the potential near the curvature singularity. The  details depend on the background, and therefore have to be analyzed separately for neutral and charged black holes.

\paragraph{Neutral black holes:} For neutral black holes, the singularity is a regular singular point of the differential equation. The local behavior near the singularity is
\begin{equation}\label{eq:wkb-neutral-even-d-black-hole-local-data}
\zpr\to\infty:\qquad
Q_0(\zpr)
\sim
-\frac{\zeta^2}{4\, \pqty{1+K\, \frac{\lads^2}{\rp^2}}\, \zpr^{\frac{d}{2}+1}},
\qquad
Q_2(\zpr)
\sim
-\frac1{4\, \zpr^2}\,.
\end{equation}
Recall that $K$ sets the spatial curvature of the boundary, and $K=0$ for planar black holes. It follows that the order of $Q_0$ at the singularity is
\begin{equation}\label{eq:wkb-neutral-even-d-black-hole-principal-order}
\operatorname{ord}_{\zpr=\infty}
\left(
  Q_0(\zpr)\, d\zpr\otimes d\zpr
\right) =
\frac{d}{2} - 3\,.
\end{equation}
Hence, for $d=4$, the singularity is a simple pole, and the standard simple pole connection formula can be applied for crossing of Stokes lines emanating from the  singularity. For $d>4$, however, $Q_0$ is either regular (in $d=6$) or has zeros at the singularity with $Q_2$ having a double pole. Crossing Stokes lines emanating from the singularity in such cases requires generalization of standard simple zero/pole connection formula.

Fortunately, for the purpose of computing the connection matrix $\mathscr{C}_{\mathrm{bdy},\mathrm{hor}}$, we will not cross Stokes lines emanating from the singularity. At real frequency, all Stokes lines encountered will emanate from simple turning points. Therefore, the connection matrix $\mathscr{C}_{\mathrm{bdy},\mathrm{hor}}$ can be computed using the standard simple zero connection formula, and the behavior at the singularity is not an obstruction.

\paragraph{Charged black holes:} For charged black holes, however, the black hole curvature singularity is not a singular point of the differential equation for minimally coupled scalar fields~\cite{Loganayagam:2022teu} (cf.~\cref{fn:charged-core} for a precise statement in different coordinate charts).  Longitudinal and transverse vector polarizations of conserved currents do, however, have a regular singular point at the curvature singularity, as discussed in~\cref{sec:examples}.

\subsection{The WKB connection matrix}\label{sec:ewkb-connection}

\begin{figure}[bp!]
  \centering
  \begin{subfigure}[t]{\textwidth}
    \centering
    \includegraphics[width=0.98\linewidth]{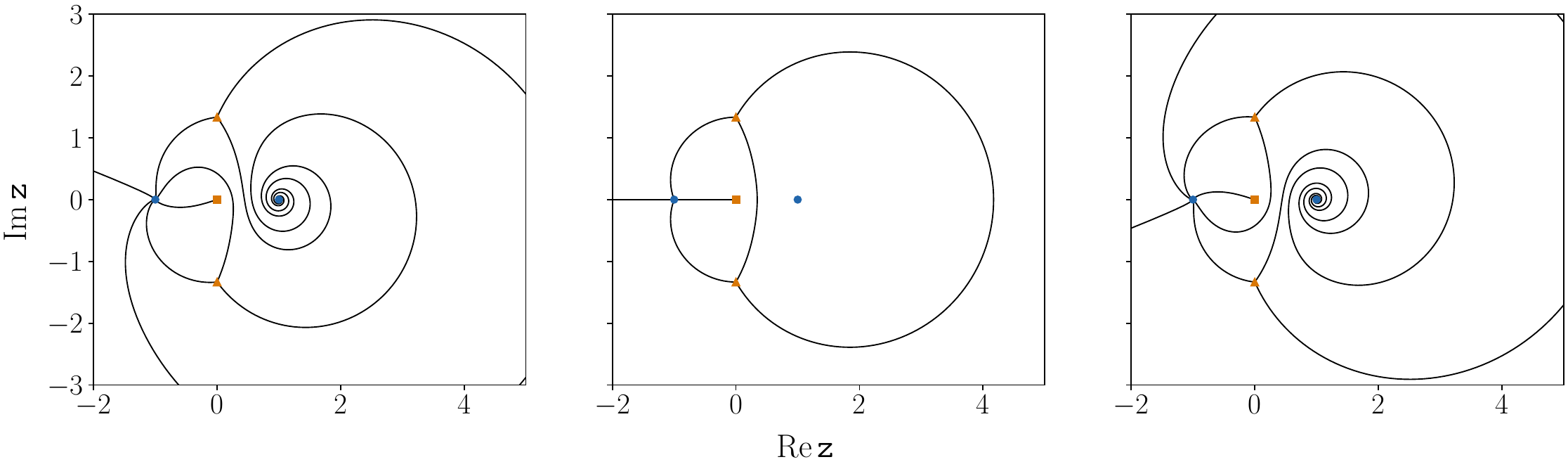}
    \caption{Lateral resolution of Stokes graphs: planar \SAdS{5}}
    \label{fig:sads-lateral-zpr}
  \end{subfigure}
  \par\medskip
  \begin{subfigure}[t]{\textwidth}
    \centering
    \includegraphics[width=0.98\linewidth]{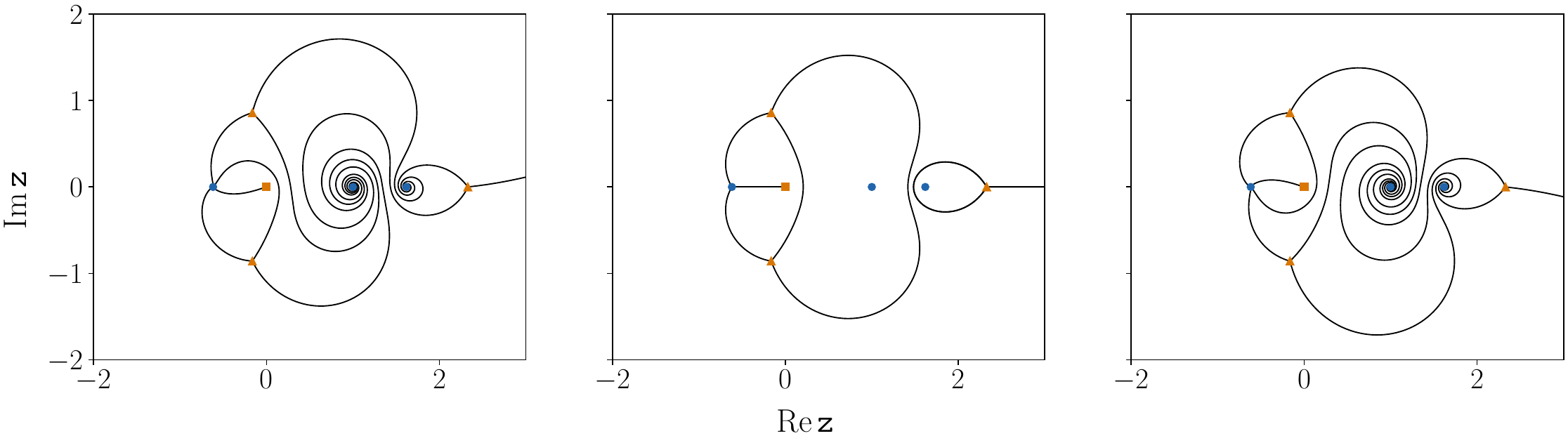}
    \caption{Lateral resolution of planar Stokes graphs: \RNAdS{5}}
    \label{fig:rnads-lateral-zpr}
  \end{subfigure}
  \caption{Examples of Stokes graphs and their lateral resolution for
    (\textsf{a}) neutral and (\textsf{b}) charged black holes.
    The central panel is the saddle graph at critical phase $\vartheta =0$,
    while the left and right are resolutions with negative and positive phases, respectively. The lateral resolution depicts the logarithmic spirals in both cases, and the WKB transport path clearly proceeds from the boundary to the turning point, with no detour toward the singularity. We chose
    $d=4$, fixed $\zeta=\frac{3}{5}$, and picked $Q = 1$ for the charged black hole
  for illustration.}
  \label{fig:stokes-lateral-zpr}
\end{figure}

\begin{figure}[tbhp!]
  \centering
  \begin{subfigure}[t]{\textwidth}
    \includegraphics[width=0.93\linewidth]{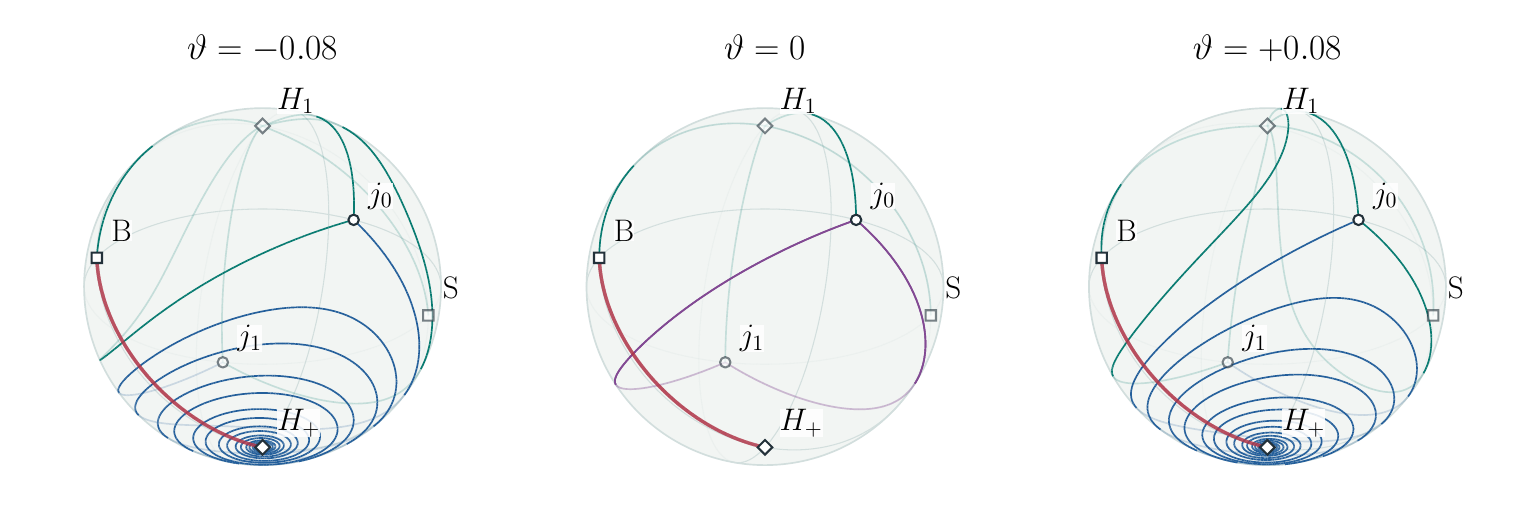}
    \caption{Planar \SAdS{5} with $\zeta=0.65$.}
  \end{subfigure}
  \medskip
  \begin{subfigure}[t]{\textwidth}
    \includegraphics[width=0.93\linewidth]{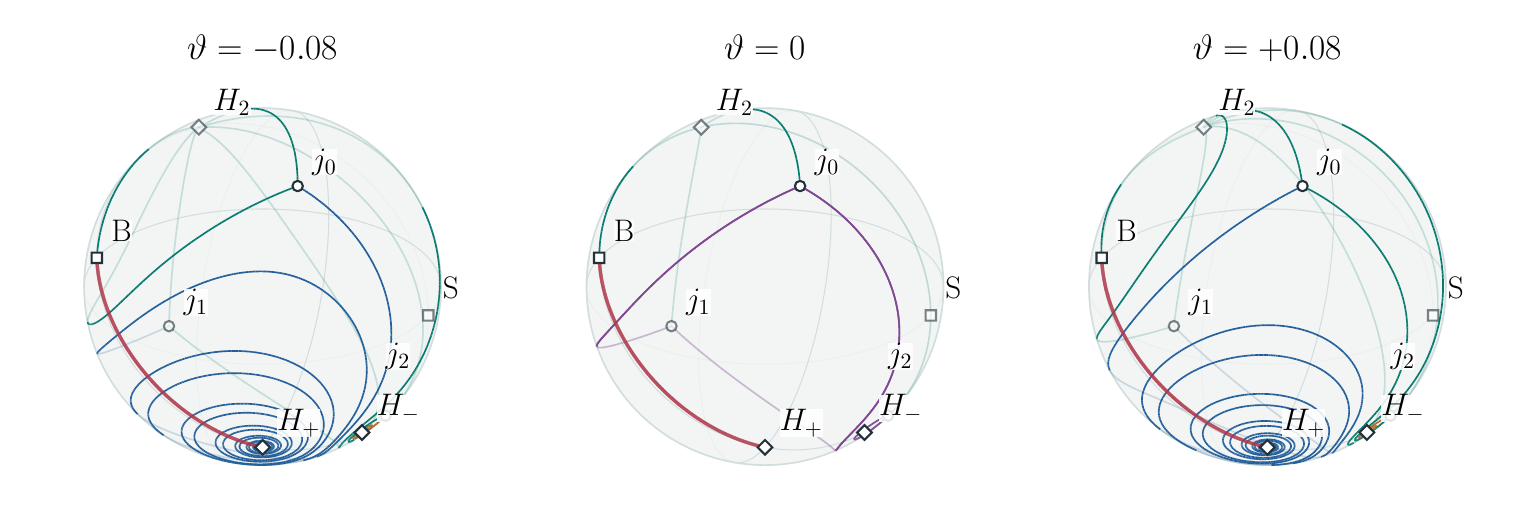}
    \caption{\RNAdS{5} with $\zeta=0.65$ and $\varrho=0.5$}
  \end{subfigure}
  \caption{The Stokes graphs for the examples shown in~\cref{fig:stokes-lateral-zpr} drawn on the Riemann sphere. This spherical projection is helpful for visualization, and for analyzing the crossings along the transport path. A visualizer for the examples studied here can be found at \href{https://mukundrangamani.github.io/stokes-atlas/}{Stokes atlas visualizers}. Here $j_i$ are the turning points, $H_\pm$ the outer/inner horizon, $H_{1,2}$ the complex horizons, while $B$ and $S$ are the boundary and singularity, respectively. }
  \label{fig:planar-ads5-general-sphere}
\end{figure}

We are now ready to complete the analysis of the connection matrix, focusing on the second step in our transport of the solution. To write compact formulae for the Stokes crossings, let us introduce some notation for $\mathrm{SL}(2)$ matrices. Given a parameter $x$ we define upper/lower triangular, and diagonal matrices with entries set by $x$ as
\begin{equation}
\mathsf{U}(x) =
\pmqty{1 & x \\ 0 & 1}\,,
\qquad
\mathsf{L}(x) = \pmqty{1 & 0 \\ x & 1}\,,
\qquad
\mathsf{D}(x) = \mathrm{diag}(x,x^{-1})\,.
\end{equation}
The rationale behind these specific matrices is the fact that we only encounter crossings  of a Stokes line emanating from a simple zero of $Q_0$. Normalize the solution at a simple zero. Then the crossing mixes solutions through multiplication by the  matrices $\mathsf{U}(\ii \eta)$ or $\mathsf{L}(-\ii \eta)$ depending on whether $\Psi_+$ or $\Psi_-$ is dominant. The factor $\eta = \pm1$ encodes the orientation of the crossing: the crossing contributes with
$\eta = +1$ for counterclockwise crossing about the turning point for a $\Psi_+$ dominant path and for clockwise crossing for a $\Psi_-$ dominant path. The $\eta = -1$ captures the opposite orientation in each case. The Stokes graphs for neutral and charged black holes are shown in~\cref{fig:stokes-lateral-zpr} in the $\zpr$ coordinate chart. We have also used a different chart, and depicted the graphs projected onto the Riemann sphere as a visualization aid in~\cref{fig:planar-ads5-general-sphere}. These sphere projections are extremely useful for visualizing the crossings and ascertaining which punctures the lines emanating from the turning points terminate at. For readers interested in exploring this further, we have made a visualizer with examples of neutral and charged black holes available at \href{https://mukundrangamani.github.io/stokes-atlas/}{Stokes atlas visualizers}.

\paragraph{Algorithm for computing the WKB connection:}
The general rule of the exact-WKB analysis is that we start in some Stokes domain and ascertain the solution in a neighbouring domain by mapping the solution basis across domains by a WKB  connection matrix. Along some path which crosses Stokes curves in the order $C_1$, $C_2$, etc., we express the initial basis of solutions in terms of the final basis of solutions through   multiplication by the corresponding Stokes matrices. In our problem we only encounter the aforementioned matrices $\set{\mathsf{U}, \mathsf{L}, \mathsf{D}}$, which makes the analysis straightforward.

As advertised at the outset, transporting WKB solutions from boundary to horizon, exact-WKB analysis in general gives the connection matrix $\mathscr{C}_{\mathrm{bdy},\mathrm{hor}}$ in the form
\begin{equation}\label{eq:wkb-C-general}
\mathscr{C}_{\mathrm{bdy},\mathrm{hor}}^{-1}
= \mathcal{B}\subm{bdy}^{-1} \times  \mathbb{K} \times  \mathsf{D}\pqty{\mathcal{E}_{\bdy,\hor}}
\times B_{\mathrm{hor}}\,.
\end{equation}
We have already encountered the asymptotic and horizon factors $\mathcal{B}\subm{bdy}$ and $B\subm{hor}$. The two new elements are the diagonal matrix with elements fixed as $\mathcal{E}_{\bdy,\hor}$ and the connection factor $\mathbb{K}$. The quantity $\mathcal{E}_{\bdy,\hor}$ is the Borel sum of the open WKB action between the boundary and horizon. It is given as
\begin{equation}
\mathcal{E}_{\bdy,\hor} = \mathcal{S}_\vartheta\, \exp(\int_\hor^\bdy\,P\subm{odd}\, d\zpr)\,.
\end{equation}
The integrand is well-defined to leading order near the boundary, since $Q_0$ only has a simple pole at the boundary. However, since $Q_2$ has a double pole the integral requires regularization. Thankfully, we will not need the details of how to do so, since $\mathcal{E}_{\bdy,\hor}$  drops out in the observables of interest. To evaluate the integral we do have to  account for the fact that the horizon pole contributes a logarithmic divergence, which needs to be regulated.

The matrix $\mathbb{K}$ is a product of connection matrices, and takes the form
\begin{equation}
\mathbb{K} = \prod_{i}\,  \mathsf{K}(\mathcal{X}_{\gamma_i})\,.
\end{equation}
One should read this as follows: start at the boundary and follow the segment up until the first crossing point, say $p_{\times}$. Then follow the Stokes line from $p_{\times}$ to the turning point from which the line emanates. This prescription fixes the orientation. One then takes the product over all crossing points $p_i$ on the transport path from boundary to the horizon and Stokes lines encountered in the process. We define here the Borel summed Voros symbol for a cycle $\gamma_i$ as
\begin{equation}
\mathcal{X}_{\gamma_i} = \mathcal{S}_\vartheta\, \exp(\oint_{\gamma_i}\, P\subm{odd}\, d\zpr)\,.
\end{equation}
To leading order, this involves exponentiating the period integral of $\lwkb$ around $\gamma_i$, viz.,
\begin{equation}
\log \mathcal{X}_{\gamma_i} \sim -\ii \freq \ms \oint_{\gamma_i} \lwkb\,.
\end{equation}
Notice that the ratio of Voros symbols nicely simplifies into a relative period:
\begin{equation}
\mathcal{X}_{ij} := \frac{\mathcal{X}_{\gamma_i}}{\mathcal{X}_{\gamma_j}}
= \mathcal{S}_{\vartheta}\, \exp(\oint_{\gamma_{ij}}\, P\subm{odd}\, d\zpr)\,,
\qquad \gamma_{ij} = \gamma_i - \gamma_j\,.
\end{equation}

\paragraph{The connection for \AdS{} black hole opers:}
As presaged, in all examples we analyze, at real frequency, the Stokes crossings are given by a simple zero Airy connection matrix. Therefore, this matrix is either upper or lower triangular, i.e.,
\begin{equation}\label{eq:wkb-airy-rule}
\mathsf{K}(\mathcal{X})
=
\begin{dcases}
\mathsf{U}\pqty{\ii\eta\, \mathcal{X} }\,,
& \qquad \Psi_+ \ \text{dominant},
\\[1ex]
\mathsf{L}\pqty{-\ii\eta \,\mathcal{X}^{-1} }\,,
& \qquad \Psi_- \ \text{dominant}\,.
\end{dcases}
\end{equation}
We have kept track of the orientation of the crossing lines using $\eta$, and folded in the change involved in translating the normalization point to the boundary. The latter is  captured by the Voros symbol factor $\mathcal{X}$.

We need to know which cycles to use for computing the Voros symbols. The cycle $\gamma_i$ of interest is the lift to the WKB curve of the following path:
start at the boundary, where we have a simple pole $\zpr = 0$, and proceed toward the crossing point $p_i$ along the transport path. At $p_i$ owing to the crossing proceed on toward the turning point along the Stokes line that emanates from the relevant simple zero.
Consequently, the natural cycles that appear in the process of the WKB transport from the boundary toward the horizon are not the cycles that encircle the singularity, but
rather cycles that encircle the boundary and some turning point
\begin{equation}\label{eq:bdy-tp-cycle}
\gamma_j \sim
\set{0, \zpr_j} \,, \qquad \zpr_j \in \set{\zpr\subm{tp}:
Q_0 (\zpr\subm{tp}) = 0}\,.
\end{equation}

The physical problem at real frequency is a critical phase of WKB, as can be seen from the central panels of~\cref{fig:stokes-lateral-zpr}. This is true for both neutral and charged black holes, and is also insensitive to the spatial curvature of the horizon.
Ergo, the exact-WKB transport computation needs to be computed at lateral phases $\vartheta = 0^\pm$. A  universal feature at lateral phases is that the Stokes lines  logarithmically spiral around the horizon, which we recall is a double pole of $Q_0$. As a consequence, the WKB path toward the horizon crosses infinitely many Stokes lines. Nevertheless, the full contribution remains computable and beautifully resums into a statistical Bose-Einstein factor. This was recently pointed out in an analysis of black hole quasinormal modes using exact-WKB~\cite{Miyachi:2025ptm}.

The spiralling behaviour can be inferred from the universal local form of the potential discussed in~\eqref{eq:wkb-horizon-universal-local-data}. The leading contribution to the odd part of the WKB momentum is
\begin{equation}
P_{\mathrm{odd}} \,d\zpr \sim -\frac{\ii \bw}{4\pi}\, \frac{d\zpr}{1-\zpr}\,.
\end{equation}
The local Stokes condition is then
\begin{equation}
\Im\pqty{\frac{\ii \bw}{4\pi}\, \log(1-\zpr)} = \text{constant}
\;\; \Longrightarrow \;\;
\log\abs{1-\zpr} = \tan (\arg \freq) \, \arg (1-\zpr) + \text{constant} \,.
\end{equation}
In fact, this analysis shows quite generally the origin of the universal logarithmic spiral towards a double pole on the Stokes graph.

There are some useful observations that one can make before we proceed with the computation. Starting at a turning point $\zpr_j$ the integral
$\int_{\zpr_j}^{\zpr} P\subm{odd}(\zpr')\, d\zpr'$ is real and monotone along the Stokes trajectory in all examples we analyze. Moreover, if it is positive, it picks out $\Psi_+$, the outgoing mode near the horizon as the dominant solution on every segment of the spiral. This means that all the spiral crossings have Stokes matrices given by $\mathsf{U}$ evaluated on suitable Voros symbols. Since the ingoing solution is subdominant, it remains unaffected as we transport it through the region where the Stokes lines spiral into the horizon.

Finally, if a curve winds once around the future horizon between the crossing locus and the turning point, then one picks up a residue factor from the cycle that encircles the pole from the horizon.  We can write this as
\begin{equation}
\mathcal{X}_{\gamma_j} \mapsto e^{-\beta \omega}\, \mathcal{X}_{\gamma_j}\,,
\end{equation}
with the Boltzmann factor appearing because of the residues at the horizon locus~\eqref{eq:wkb-horizon-universal-local-data}.

When we cross the spiral $n$-times, we pick up a factor of $e^{-n\ms \bw}$ relative to the simple crossing. Suppose that for $\vartheta = 0^+$ the spiral crossings are governed by the $\mathsf{U}$ Stokes factor. Then the overall contributions commute and lead to a contribution
\begin{equation}\label{eq:U-spirals}
\prod_{n \geq 0}\, \mathsf{U}(\ii \eta \, e^{-n\ms \bw} \, \mathcal{X})
= \mathsf{U}\pqty{\ii \eta\, \frac{\mathcal{X}}{1-e^{-\bw}}}\,.
\end{equation}
A similar result applies if the crossings are governed by $\mathsf{L}$ Stokes factor, with reversed cycle orientations. Explicitly,
\begin{equation}
\prod_{n \geq 0}\, \mathsf{L}(-\ii \eta \, e^{-n\ms \bw} \, \mathcal{X}^{-1})
= \mathsf{L}\pqty{-\ii \eta\, \frac{\mathcal{X}^{-1}}{1-e^{-\bw}}}\,.
\end{equation}
The choice is dictated by the lateral resolution and will become clear in examples, cf.~\cref{sec:ewkb-examples}.

It is extremely curious that the exact-WKB problem generates the Bose-Einstein factor by resumming the turning points cycles' Voros symbol contributions. Typically, the statistical factor is argued for using the regularity of the solutions in Euclidean time domain, where the horizon locus is a fixed point of a compact Killing field. This is equivalent to imposing the KMS condition~\cite{Herzog:2002pc}. Alternately, one can work directly in Lorentz signature and use the gravitational Schwinger-Keldysh saddle of~\cite{Glorioso:2018mmw} (see also~\cite{Jana:2020vyx}). In contrast, here we are not focusing on the ingoing solution alone, but rather attempting to build up the full connection matrix. Nevertheless, the laterally resolved Stokes graphs assemble the full picture, by geometrically resumming the Boltzmann factors.

In the examples we analyzed explicitly, the spirals into the horizon originate from at least two turning points. The quantity $\mathcal{X}$ entering~\eqref{eq:U-spirals} will end up being a linear combination of the Voros symbols associated with cycles from the boundary to a subset of turning points. In addition, there may be an ultimate non-spiralling path prior to reaching the horizon. Schematically, therefore, one could write the general formula in terms of the matrix $\mathsf{K}$, taking care to replace it with the appropriate Stokes matrix. To wit,
\begin{equation}
\begin{aligned}
\prod^{\infty}_{n=0} \mathsf{K}_U(e^{-n\beta\omega}\mathcal{X})
&=
\mathsf{K}_U\prn{\frac{\mathcal{X}}{1-e^{-\beta\omega}}}\,,
\\
\prod^{\infty}_{n=0} \mathsf{K}_L(e^{n\beta\omega}\mathcal{X})
&=
\mathsf{K}_L\prn{(1-e^{-\beta\omega})\,\mathcal{X}}.
\end{aligned}
\label{eq:wkb-thermal-sum}
\end{equation}
%

\section{Implications of the exact-WKB connection formula}\label{sec:ewkb-connection-imp}

Let us extract some implications of the exact-WKB expression for the connection matrix~\eqref{eq:wkb-C-general}. Since we have translated the Frobenius basis at the horizon onto the Frobenius basis at AdS asymptopia, we have immediate access to the retarded Green's function. Hence, using this result we can immediately extract the quantities of interest using the definitions given in~\cref{sec:RRA}, the composite monodromy trace~\eqref{eq:monoeq1} and the phase of the retarded Green's function~\eqref{eq:monoratio}.

We display a schematic form of the Green's function and use it to demonstrate that the spectral function cleanly factorizes into components that can be ascribed to OPE and non-perturbative data. The discussion will be somewhat abstract, but will be justified by explicit examples in~\cref{sec:ewkb-examples}. The goal here is to write down compact formulae in terms of Voros symbols, which will prove helpful in understanding the connection to the naive representation of~\cref{sec:specasym}.

\subsection{Thermal correlators}\label{sec:wkb-correlators}

Once we have ascertained the crossings from the Stokes graph, we can work out the entries of the connection matrix $\mathscr{C}_{\bdy,\hor}$. Using this we can immediately write down the boundary correlation functions. The details depend on the explicit form of the differential operator, specifically, the number and location of the turning points. One can nevertheless write down the structural form of the answer, which will be useful for extracting the WKB action.

Assuming for the moment we have derived the matrix $\mathbb{K}$, we can immediately write down the Green's function as
\begin{equation}
\Gret = -\frac{B_i}{A_i} =
\Gret^{\mathrm{vac}}\, \mathcal{T}_\bdy^2\,
\frac{(\mathbb{K}^{-1})_{22} + \ii e^{\ii\pi\nu}\, (\mathbb{K}^{-1})_{21}
}{(\mathbb{K}^{-1})_{22} +\ii  e^{-\ii\pi\nu} \, (\mathbb{K}^{-1})_{21}}\,.
\end{equation}
The vacuum Green's function is given in~\eqref{eq:GR-vac}.
Thus, knowledge of $\mathbb{K}$, which is obtained by multiplying crossing connection matrices, is all we need for the physical answer. Notice that only the lower row of $\mathbb{K}^{-1}$ enters the final result. This is because the retarded Green's function picks out the ingoing mode at the horizon.

As noted earlier, the physical problem of interest is at $\arg\freq = 0$, which is a critical phase of the WKB problem. To resolve this we rotate the frequency by a phase parameterized by $\vartheta$ and take the limits $\vartheta \to 0^\pm$. The laterally rotated matrices are denoted as $\mathbb{K}_\pm$. The matrices $\mathbb{K}_\pm$ can be shown to take the schematic form (the rotations of the Voros symbols are not indicated for simplicity)
\begin{equation}\label{eq:Kgen-pm}
\begin{aligned}
\mathbb{K}_+
&=
\mathsf{U}(-\ii \mathcal{X}_f) \,
\mathsf{L}(\ii \mathcal{X}_\star^{-1})\, \mathsf{U}(-\ii \mathfrak{s}(\mathcal{X}))\,,
\\
\mathbb{K}_-
&=
\mathsf{L}(\ii \mathcal{X}_f)\,
\mathsf{U}(-\ii \mathcal{X}_\star^{-1})\, \mathsf{L}(\ii \mathfrak{s}(\mathcal{X}))\,.
\end{aligned}
\end{equation}
The three factors arise as follows: the first matrix converts between the near-boundary WKB solutions and the intermediate WKB basis before horizon crossings, the second arises  from the transport between the boundary and the distinguished turning point, and the final factor is the contribution from the horizon spirals. The first factor is not present in all examples, but occurs when there is an additional Stokes line that crosses the transport path before the spirals into the horizon. Multiplying the factors out we have
\begin{equation}\label{eq:Kbb-gen}
\mathbb{K}_+^{-1} =
\begin{pmatrix}
  1 + \mathfrak{s}(\mathcal{X})\, \mathcal{X}_\star^{-1}
  & \quad \ii \mathfrak{s}(\mathcal{X}) + \ii \mathcal{X}_f \pqty{1 + \mathfrak{s}(\mathcal{X})\, \mathcal{X}_\star^{-1}}
  \\
  - \ii \mathcal{X}_\star^{-1}
  & 1 + \mathcal{X}_f\, \mathcal{X}_\star^{-1}
\end{pmatrix}\,.
\end{equation}
Here $\mathcal{X}_\star$ is the  Voros symbol for the cycle between the boundary  and the distinguished turning point $\zpr_\star$ singled out in~\cref{sec:perconv}. The factor $\mathfrak{s}(\mathcal{X})$ comprises relative cycle periods between $\zpr_\star$ and other active turning points (schematically indicated by the subscript) in the Stokes graph. It is schematically of the form
\begin{equation}
\mathfrak{s}(\mathcal{X}) = \mathcal{X}_\star \frac{\sum_{j}\, e^{-n_j\ms \bw} \, \mathcal{X}_{j\star}}{1-e^{-\bw}}\,.
\end{equation}
Finally, $\mathcal{X}_f$ is a Voros symbol associated with what might be termed as a final factor. It doesn't always occur, but when it does, it is an additional contribution from a final Stokes line which crosses the transport path before the horizon spirals. This line emanates from one of the turning points and supplies another triangular matrix. Technically, we should decorate the Voros symbols with the lateral rotation sign ($+$ in the case under consideration).  For the other lateral rotation we would have the Voros symbols rotated in the opposite direction; the horizon spirals change directions. The net result is that heuristically the rows are exchanged between $\mathbb{K}_\pm$ (modulo some sign changes).

With the matrix $\mathbb{K}$ explicitly evaluated in terms of the Voros symbols, and the knowledge of the Borel resummed data $\mathcal{T}_\bdy$, $\mathcal{E}_{\bdy,\hor}$, we can work out the connection matrix entries $A_i, A_o$ and $B_i, B_o$. One can then write down the boundary Green's functions in terms of these data. We have
\begin{equation}\label{eq:ewkb-Gret}
\begin{aligned}
\Gret
&=  \Gret^{\mathrm{vac}}\, \mathcal{T}_\bdy^2\,
\frac{(\mathbb{K}^{-1})_{22} +\ii e^{\ii \pi\ms \nu}\, (\mathbb{K}^{-1})_{21}}
{(\mathbb{K}^{-1})_{22} +\ii e^{-\ii \pi\ms \nu}\, (\mathbb{K}^{-1})_{21}}
\\
&=
\Gret^{\mathrm{vac}}\, \mathcal{T}_\bdy^2\,
\frac{1+ \pqty{e^{\ii \pi \nu} + \mathcal{X}_f}\, \mathcal{X}_\star^{-1}}{1+ \pqty{e^{-\ii \pi \nu}  + \mathcal{X}_f}\mathcal{X}_\star^{-1}}\,.
\end{aligned}
\end{equation}
One can extract the phase of the retarded Green's function and obtain its asymptotics on the $\arg \freq = 0$ physical locus. Working to leading order one would get the contribution from the distinguished turning point alone. All other factors drop out and the Green's function simplifies considerably to give\footnote{We are assuming that the median prescription keeps  $\mathcal{T}_\bdy$ real. Otherwise, provided the positive and negative lateral resolutions lead to conjugate boundary factors, and the median factor is their geometric mean, we should include a factor of $\pqty{\frac{\mathcal{T}^+_\bdy}{\mathcal{T}^-_\bdy}}^2$ in~\eqref{eq:ewkb-RRA}. }
\begin{equation}\label{eq:ewkb-RRA}
\exp(2\ii \phGR)
= e^{-2\pi \ii \ms \nu}
\, \frac{1+ \pqty{e^{\ii \pi \nu} + \mathcal{X}_f}\, \mathcal{X}_\star^{-1}}{1+ \pqty{e^{-\ii \pi \nu}  + \mathcal{X}_f}\mathcal{X}_\star^{-1}}
\,
\frac{1+ \pqty{e^{\ii \pi \nu} + \mathcal{X}_f^*}\, \pqty{\mathcal{X}_\star^{-1}}^*}{1+ \pqty{e^{-\ii \pi \nu}  + \mathcal{X}_f^*} \pqty{\mathcal{X}_\star^{-1}}^*}\,.
\end{equation}
Note that the open WKB action between the boundary and the horizon, $\mathcal{E}_{\bdy,\hor}$, cancels out in the ratio of connection coefficients. Thus, the two-point functions are insensitive to this data, as promised. We have written this for one particular lateral resolution $\vartheta \to 0^+$, and will explain how to obtain the median summation in explicit examples below.

\subsection{Factorization of the spectral function from WKB}\label{sec:rho-factorization}

In~\cref{sec:rho-Z-factors} we demonstrated a factorization property of the spectral function for integer conformal dimensions. However, features of the exact-WKB analysis suggest a much more general picture.
The structure of the WKB connection matrix~\eqref{eq:wkb-C-general} itself implies a clean separation between perturbative and non-perturbative contributions to the thermal two-point functions. In fact, the spectral function is forced to take the general form
\begin{equation}\label{eq:rho-factor-WKB}
\rho
=
\rho\subm{vac}\,
\rho\subM{OPE}\,
\rho\subm{np}\,,
\end{equation}
where the factors are associated with different elements of the WKB connection matrix.
One can pretty much read this off directly from~\eqref{eq:ewkb-Gret}, and attribute the contribution from the different parts of the geometry to the three pieces of the spectral function in the above decomposition.

\begin{itemize}[wide, left=0pt]
  \item First, the vacuum contribution is the piece that is captured by the pure AdS geometry without a black hole. This is therefore the part that is given by the matrix with entries $I_{\nu}(\freq,\zeta)$ in~\eqref{eq:wkb-boundary-exact-undetermined-residual}, and is responsible for the $\nu$ dependent ratio in~\eqref{eq:ewkb-Gret}. We may therefore write
    \begin{equation}\label{eq:wkb-vacuum-spectral-factor}
    \rho\subm{vac}
    =
    -2\, \Im\Gret^{\mathrm{vac}}
    =
    \frac{2\pi}{\nu\,\Gamma(\nu)^2} \, \pqty{\frac{\omega^2 - k^2}{4\,\rp^2}}^\nu \,, \qquad \omega > k \geq 0 \,.
    \end{equation}
    This can be directly obtained from~\eqref{eq:GR-vac}.

  \item Second, the near-boundary analysis, as noted around~\eqref{eq:wkb-boundary-summed-symbol-assumption} relates the Borel summed contribution to the OPE data, leading to the identification
    \begin{equation}\label{eq:wkb-ope-factor}
    \rho\subM{OPE}
    =
    \pqty{\mathcal{Z}_\bdy^{\mathrm{med}}}^2.
    \end{equation}
  \item Finally, the non-perturbative factor $\rho\subm{np}$ arises from the actual boundary to horizon transport. This piece depends on the Borel sum of Voros symbols between the boundary and the turning points and should be reconstructed from $\mathbb{K}$. Schematically, it takes the form
    \begin{equation}\label{eq:wkb-non-perturbative-factor}
    \rho\subm{np}
    =
    \mathcal{F}\subm{np}(\mathcal{X}_\star)
    = -\frac{1}{\sin(\pi\ms\nu)}\, \Im\bqty{
      e^{-\ii\pi\ms\nu}\, \pqty{\frac{\mathcal{T}^{+}_\bdy}{\mathcal{Z}^{\mathrm{med}}_{\bdy}}}^2\,  \frac{1+ \pqty{e^{\ii \pi \nu} + \mathcal{X}_f^+}\, \pqty{\mathcal{X}_\star^+}^{-1}}{1+ \pqty{e^{-\ii \pi \nu}  + \mathcal{X}_f^+}\,\pqty{\mathcal{X}_\star^+}^{-1}}
    }
    \end{equation}
    We have kept the positive lateral boundary factor $\mathcal{T}_\bdy^+$ explicit in the retarded Green's function and let $\mathcal Z_\bdy^{\mathrm{med}}$ be the real median
    sum of the boundary OPE factor. This expression is obtained initially for non-integral values of $\nu$; at resonant values the correlator is defined through its contact subtracted limit. To give a  closed expression in median symbols we have to obtain the
    corresponding boundary jump relation. We will give explicit expressions for this function $\mathcal{F}\subm{np}$ below, and show that it actually takes the form suggested in~\eqref{eq:rho-monodromy}. Note that if there is no boundary crossing $\mathcal{X}_f =0$ then the formula simplifies significantly.
\end{itemize}

\section{Exact-WKB data for AdS black holes}\label{sec:ewkb-examples}

We give explicit forms of $\mathbb{K}$ and $\mathcal{F}\subm{np}$ for a subset of examples analyzed hitherto in~\cref{sec:examples}. The bulk of the analysis will be restricted to $\zeta \in (0,1)$. In addition, we unify the neutral black hole discussions treating both the planar and global solutions in one compact setting. In order to do so, we restrict attention to the case where
$\zeta < \zeta\subm{crit}$ so that we do not encounter trapped orbits outside the black hole.

\subsection{Neutral black holes in \AdS{5}}\label{sec:neutral-d4}

%
\begin{figure}[htbp]
  \centering
  \includegraphics[width=0.98\linewidth]{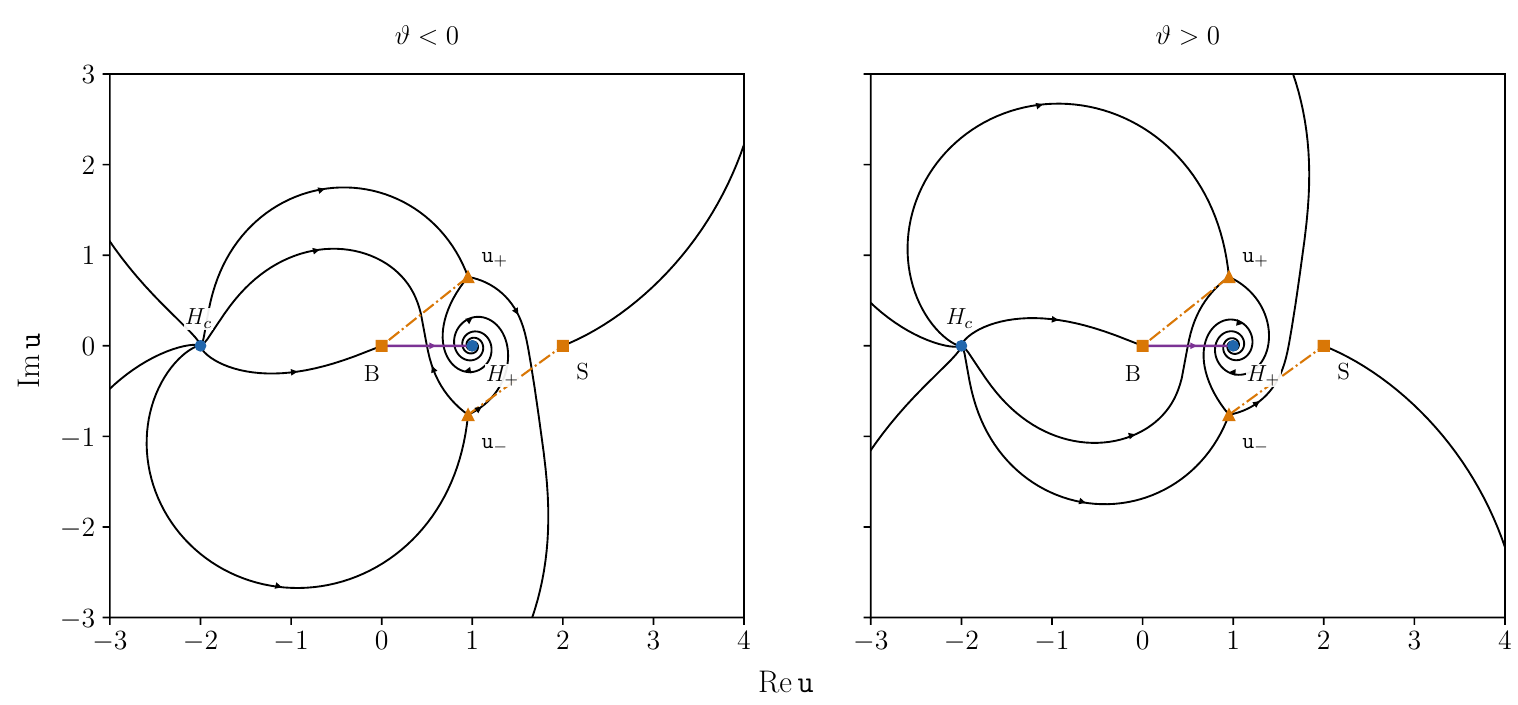}
  \caption{Laterally resolved Stokes graphs for neutral black holes in $d=4$. There are two turning points, which are complex conjugates of each other. The graphs are plotted in the $\upr$-coordinate defined in~\eqref{eq:u-coordinate}. Horizons (outer $H_+$ and complex $H_c$) are indicated as blue circles, orange triangles denote the turning points $\upr_\pm$, and the singularity (S) is the orange square. The transport path from the boundary (B) to the outer horizon ($H_+$) is the purple line, and the orange dashed lines denote the choice of branch cuts.}
  \label{fig:ewkb-d4-neutral}
\end{figure}

In $d=4$ the black hole singularity is a simple pole and as noted in~\cref{sec:pschwads} is a branch point of the WKB spectral curve. It is convenient to bring all the branch points to finite coordinate locations, so we further redefine $\zpr$ into a new coordinate $\upr$ through
\begin{equation}
\upr = \frac{2 \, \zpr}{1 + \zpr}\,.
\label{eq:u-coordinate}
\end{equation}
There are four branch points in all: two turning points and the points at the boundary and singularity. In the parameter range of interest, we have two complex conjugate turning points located at
\begin{equation}
\upr_\pm
=
\frac{ 2+ \pqty{\frac{K}{\rp^2} -2} \, \zeta^2
  \mathbin{\pm}
  \ii\zeta
\sqrt{4\,\pqty{1+\frac{K}{\rp^2}}- \pqty{\frac{K}{\rp^2}  + 2}^2\, \zeta^2}}{
1+2\,\frac{K}{\rp^2} \,\zeta^2},
\qquad
\upr_+ = \upr^*_-.
\label{eq:wkb-d4-tps}
\end{equation}
The boundary to turning point cycles which are natural to work with are then
\begin{equation}
A=\gamma_+ \sim \{0, \upr_+\}, \qquad
B=\gamma_- \sim \{0, \upr_-\}\,.
\label{eq:periods-neutral-d4}
\end{equation}
We have relabeled these cycles as $A$ and $B$ cycles to disambiguate the turning point loci from the phase sign choices.

The Stokes graph for lateral resolution is depicted in~\cref{fig:ewkb-d4-neutral}. Notice that for the positive phase resolution the crossings are indeed governed by the Stokes matrix $\mathsf{U}$. The logarithmic spirals cause the detours to alternate to the two conjugate turning points. There is a single final detour, which is not a spiralling ray, and contributes a final factor. Therefore, for the phase choice $\vartheta = 0^\pm$, the lateral representations of the full connection matrix are given by
\begin{equation}
\mathbb{K}_\pm =
\mathsf{K}_{\pm,\mathrm{final}} \,
\prod_{n=0}^\infty \prn {\mathsf{K}_{\pm,n}^{(1)}\, \mathsf{K}_{\pm,n}^{(2)}}\,.
\label{eq:ewkb-neutral-Kpm}
\end{equation}
We have acknowledged with the notation the binary detour choice (labeled by $(i)$ with $i=1,2$), reserving the sign labels for the phase choice.

The order of operation and the associated factors can be read off from the Stokes graph~\cref{fig:ewkb-d4-neutral}.
Let $\mathcal{X}_{A\pm} = \mathcal{X}_A^{(0^\pm)}$ and $ \mathcal{X}_{B\pm} = \mathcal{X}_B^{(0^\pm)}$ be the Voros symbols for the periods around the A and B cycles after rotation by $\vartheta = 0^\pm$. The orientations are fixed by the reality condition
$\mathcal{X}_{B-} = \mathcal{X}_{A+}^*$ for these lateral symbols. Then we find
\begin{equation}\label{eq:Kmat-d4}
\begin{aligned}
\mathsf{K}_{+,n}^{(1)}
& =
\mathsf{U}\pqty{-\ii e^{-\bw\ms n}\, \mathcal{X}_{B+}^{-1}}\,,
& \qquad
\mathsf{K}_{+,n}^{(2)}
& =
\mathsf{U}\pqty{-\ii e^{-\bw\ms (n+1)}\, \mathcal{X}_{A+}}\,,
\\
\mathsf{K}_{-,n}^{(1)}
& =
\mathsf{L}\pqty{\ii e^{-\bw\ms n}\, \mathcal{X}_{A-}^{-1}}\,,
& \qquad
\mathsf{K}_{-,n}^{(2)}
& =
\mathsf{L}\pqty{\ii e^{-\bw\ms (n+1)}\, \mathcal{X}_{B-}}\,,
\\
\mathsf{K}_{+,\mathrm{final}}
& =
\mathsf{L}\pqty{\ii \mathcal{X}_{A+}^{-1}}\,,
& \qquad
\mathsf{K}_{-,\mathrm{final}}
& =
\mathsf{U}\pqty{-\ii \mathcal{X}_{B-}^{-1}}\,,
\end{aligned}
\end{equation}
The product over windings can be performed as in~\eqref{eq:wkb-thermal-sum}. One finds, for instance
\begin{equation}
\mathbb{K}_+ = \mathsf{L}(\ii \mathcal{X}_{A+}^{-1})\, \mathsf{U}(-\ii \mathfrak{s}(\mathcal{X}_{A+},\mathcal{X}_{B+}))\,,
\qquad
\mathfrak{s}(\mathcal{X}_{A+},\mathcal{X}_{B+})
= \mathcal{X}_{B+}^{-1} \, \frac{1+ e^{-\bw}\, \widetilde{\mathcal{X}}_{AB+}}{1-e^{-\bw}}\,.
\end{equation}
We used here the fact that the product of Voros symbols associated with different cycles concatenates into that for the period over the sum of cycles, viz.,
\begin{equation}\label{eq:ewkb-relcycle}
\mathcal{X}_{A+} \, \mathcal{X}_{B+} = \widetilde{\mathcal{X}}_{AB+}\,,
\end{equation}

To obtain a manifestly real representation of $\mathcal F_{\mathrm{np}}$, it will be convenient to use median resummation of Voros symbols. In the present case, they are given by
\begin{equation}
\begin{aligned}
\mathcal{X}_{A,\mathrm{med}}
& =
\mathcal{X}_{A+}\,
\sqrt{\frac{1+ \mathcal{X}_{A+}\,\mathcal{X}_{B+}\, e^{-\bw}}{
1+ \pqty{\mathcal{X}_{A+}\, \mathcal{X}_{B+}}^{-1}}}
=
\mathcal{X}_{A-}\,
\sqrt{\frac{1+\pqty{\mathcal{X}_{A-}\,\mathcal{X}_{B-}}^{-1}}{
1+ \mathcal{X}_{A-}\,\mathcal{X}_{B-}\, e^{-\bw}}}\,,
\\
\mathcal{X}_{B,\mathrm{med}}
& =
\mathcal{X}_{B+}\,
\sqrt{\frac{1+\pqty{\mathcal{X}_{A+}\, \mathcal{X}_{B+}}^{-1}}{
1+ \mathcal{X}_{A+}\, \mathcal{X}_{B+}\, e^{-\bw}}}
=
\mathcal{X}_{B-}\,
\sqrt{\frac{1+ \mathcal{X}_{A-}\, \mathcal{X}_{B-}\, e^{-\bw}}{
1+ \pqty{\mathcal{X}_{A-}\,\mathcal{X}_{B-}}^{-1}}}\,.
\end{aligned}
\label{eq:ewkb-d4--median}
\end{equation}
Using these the non-perturbative part of the spectral function can be expressed as
\begin{equation}
\mathcal F_{\mathrm{np}}
=
\sinh(\hbw)\, \mathfrak{X}_4^{-1}
\end{equation}
which, as promised, is of the form~\eqref{eq:rho-monodromy} with
\begin{equation}
\begin{aligned}
\mathfrak{X}_4
&=
\cosh(\hbw) +
e^{\hbw}
\abs{\mathcal{X}_{A,\mathrm{med}}}^{-2}
\\
& \qquad +
e^{\hbw}
\sqrt{
  \left(
    1+
    \left\lvert
    \mathcal{X}_{A,\mathrm{med}}
    \right\rvert^{-2}
  \right)
  \left(
    1+
    e^{-\beta\omega}
    \left\lvert
    \mathcal{X}_{A,\mathrm{med}}
    \right\rvert^2
  \right)
}
\Re\bqty{
  e^{-\pi\ii\nu}
  \left(
    \mathcal{X}_{A,\mathrm{med}}
\right)^{-1}}\,.
\end{aligned}
\label{eq:ewkb-sd4-Xfn}
\end{equation}
At integer $\nu$ for planar black holes this reduces to the result obtained  in~\cite{Jia:2026ryl}, providing a nice cross-check of our results. This can be shown to reduce to the form quoted in~\eqref{eq:wkb-non-perturbative-factor}, which was written in terms of the positive lateral rotation. To express the latter in terms of the median symbol one should use the jump relation $\mathcal{X}_{A+}^{-1} = \mathcal{X}_{A,\mathrm{med}}^{-1}\, \sqrt{\frac{1+e^{-\bw}\, \abs{\mathcal{X}_{A,\mathrm{med}}}^2}{1+\abs{\mathcal{X}_{A,\mathrm{med}}}^{-2}}}$, and then substitute $\mathcal{X}_\star^{-1} = \mathcal{X}_{A +}^{-1}$ into the result.

\begin{figure}[tbp]
  \centering
  \includegraphics[width=0.98\linewidth]{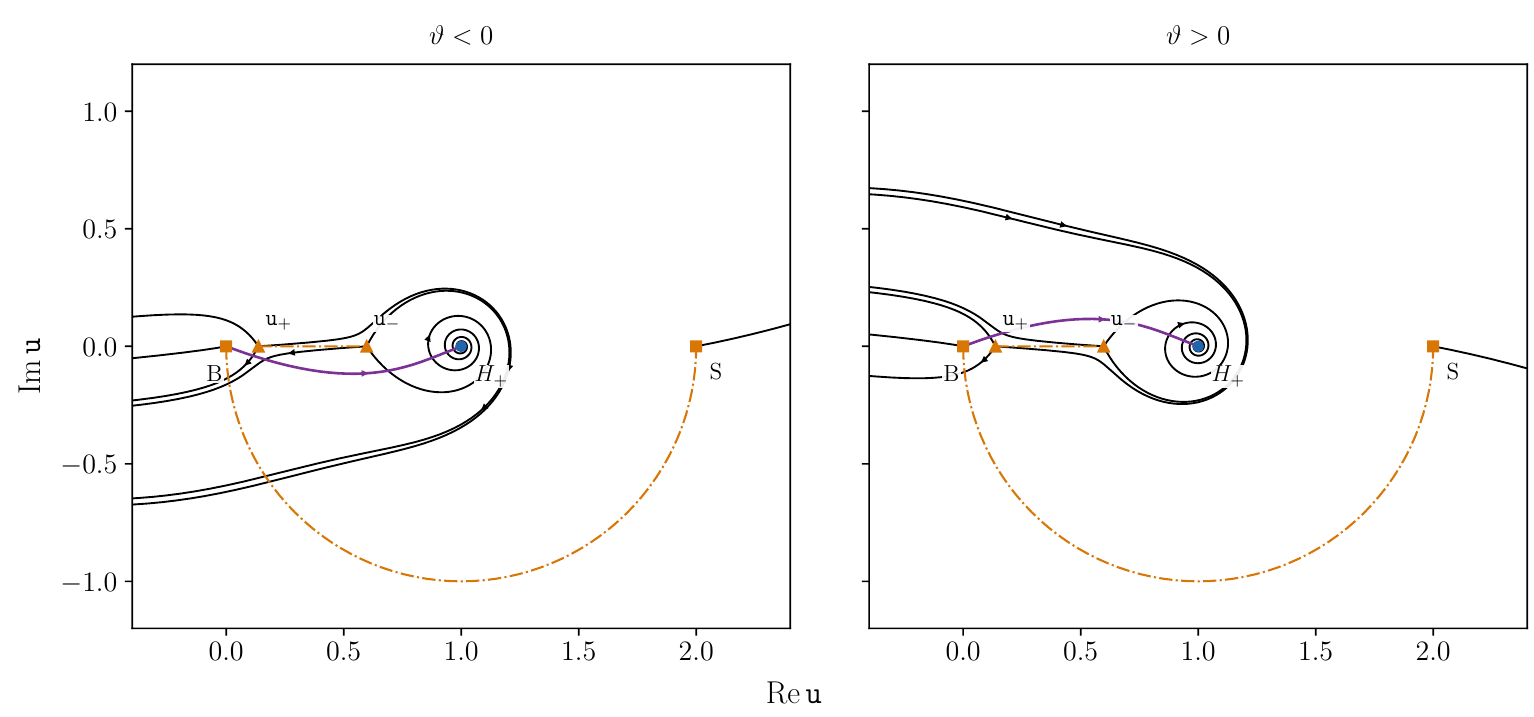}
  \caption{Laterally resolved Stokes graphs for neutral global \SAdS{5} with $\rp= \lads=1$ and $\zeta=97/100$, in the $\upr$ coordinate. The rest of the conventions are as indicated in~\cref{fig:ewkb-d4-neutral}.}
  \label{fig:ewkb-d4-global-exterior}
\end{figure}

\paragraph{Global black hole with exterior turning points:} In the global black hole case for $\zeta > \zeta_{\mathrm{crit}}$ the two turning points lie between the horizon and the boundary. Analyzing the Stokes graph, one finds that only the turning point closest to the horizon has a series of Stokes trajectories that spiral into the horizon singular point. The turning point closer to the boundary does not connect through such spirals. Therefore, the exact-WKB problem is much simpler in this case and more or less works out along lines of  the simpler picture presented in~\cref{sec:globalSAdS} insofar as the bound state eigenenergies are concerned. The decay rate which picks up the contribution from the cycle enclosing the turning points does cut through the spirals, which contribute therefore to the lifetime of the bound state.
The corresponding lateral graphs are shown in~\cref{fig:ewkb-d4-global-exterior}.

There are two closed cycles of interest.
Let $\gamma_{\mathrm{w}}$ be the cycle around the boundary and the turning point closest to the boundary $\upr_+$; this is the bound state well. Additionally, let $\gamma_{\mathrm{b}}$ be the cycle around the two turning points $\upr_\pm$; this is the barrier cycle. We denote their Borel summed Voros symbols by
\begin{equation}
\mathcal{X}_{\mathrm w}
=\mathcal{S}_{0,\mathrm{med}}\!\left[e^{V_{\gamma_{\mathrm w}}}\right],
\qquad
\mathcal{X}_{\mathrm b}
=\mathcal{S}_{0^\pm}\!\left[e^{V_{\gamma_{\mathrm b}}}\right].
\end{equation}
The orientations are chosen such that $\Im V_{\gamma_{\rm w}}>0$ to leading order, and $\mathcal{X}_{\mathrm b}$ is exponentially small. The full exact WKB result for retarded two-point function reads
\begin{equation}
G_{\mathrm R}
=
\Gret^{\mathrm{vac}}\,
\mathcal T_{\mathrm{bdy}}^{\,2}
\frac{
  \sqrt{1+\mathcal{X}_{\mathrm b}}
  +e^{i\pi\nu}\mathcal{X}_{\mathrm w}
  \sqrt{1+e^{-\beta\omega}\mathcal{X}_{\mathrm b}}
}{
  \sqrt{1+\mathcal{X}_{\mathrm b}}
  +e^{-i\pi\nu}\mathcal{X}_{\mathrm w}
  \sqrt{1+e^{-\beta\omega}\mathcal{X}_{\mathrm b}}
}.
\end{equation}
The non-perturbative spectral function factor is
\begin{equation}
\mathcal F_{\mathrm{np}}
=
\frac{
  \mathcal{X}_{\mathrm b}(1-e^{-\beta\omega})
}{
  2+\mathcal{X}_{\mathrm b}(1+e^{-\beta\omega})
  +2\sqrt{(1+\mathcal{X}_{\mathrm b})
  (1+e^{-\beta\omega}\mathcal{X}_{\mathrm b})}
  \,\operatorname{Re}\!\left(e^{-i\pi\nu}\mathcal{X}_{\mathrm w}\right)
}.
\end{equation}

\subsection{Neutral black holes in \AdS{7}}\label{sec:neutral-d6}

\begin{figure}[htbp]
  \centering
  \includegraphics[width=0.98\linewidth]{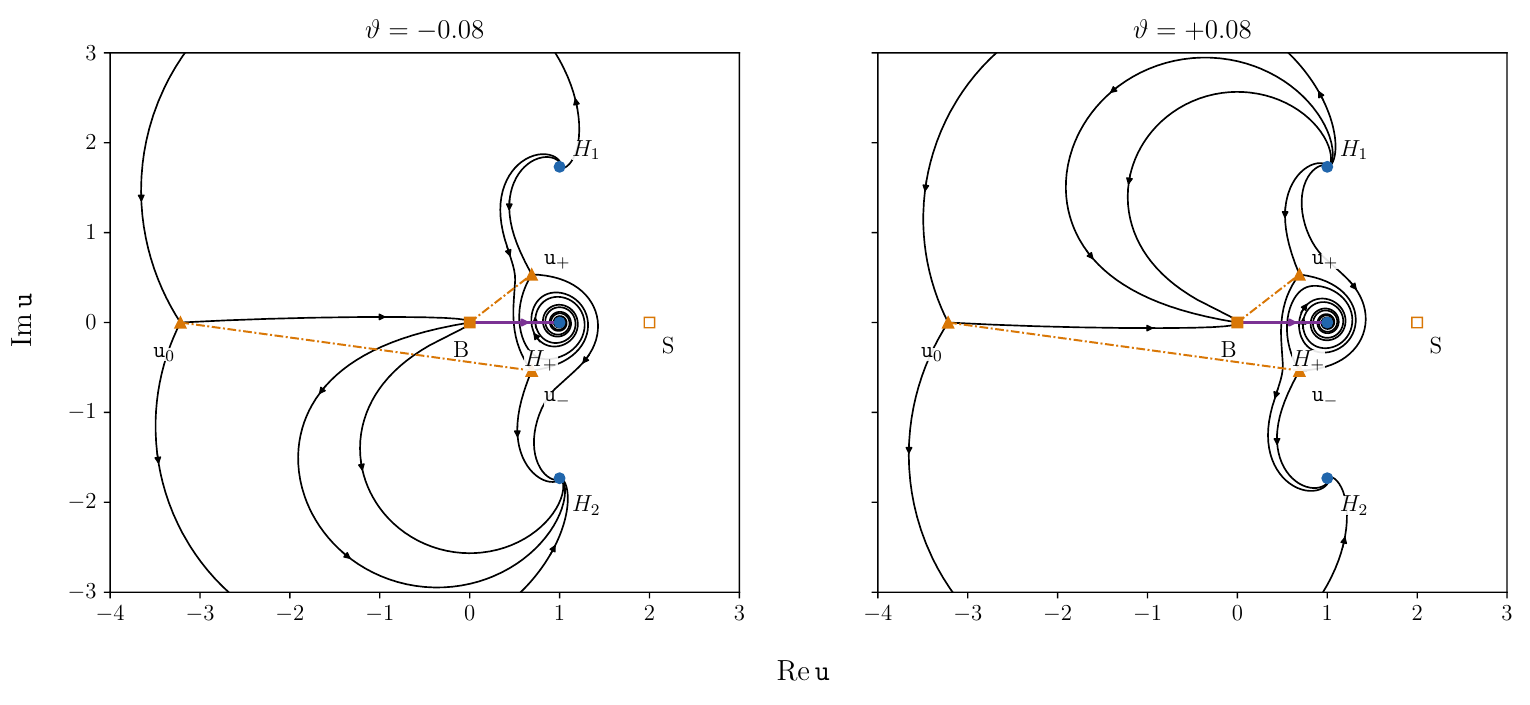}
  \caption{Laterally resolved Stokes graphs for neutral black holes in $d=6$  in the $\upr$ coordinate. The singularity is now an ordinary point of the principal differential and has no cut running out to it. It is therefore denoted as an open orange square; the rest of the conventions are as indicated in~\cref{fig:ewkb-d4-neutral}. }
  \label{fig:ewkb-d6-neutral}
\end{figure}

In $d=6$ both planar and global \SAdS{7} black holes have three turning points; two  of these lie at complex $\upr_\pm$, while the third is real and is located at $\upr_0$. The coordinate values can be deduced in the planar case from~\eqref{eq:planar-uturn}. We will need to consider three distinct cycles
\begin{equation}
A = \gamma_+ \sim \{0,\upr_+\}, \qquad
B = \gamma_- \sim \{0, \upr_-\}, \qquad
C = \gamma_0 \sim \{0,\upr_0 \}\,.
\end{equation}
The third cycle, $\gamma_0$,  is a saddle cycle at $\vartheta=0$ containing the boundary singular point. Based on our earlier discussion in~\cref{sec:examples} this is a general feature for odd $\frac{d}{2}$. This boundary cycle ($C$), however, does not participate in any horizon spiral winding, and can be accounted for as a final connection matrix, see~\cref{fig:ewkb-d6-neutral}.

The full connection matrix for the transport from the boundary to the horizon is given by
\begin{equation}
\mathbb{K}_\pm
=
\mathsf{K}_{\pm,C} \, \mathsf{K}_{\pm,\mathrm{final}}\,
\prod_{n=0}^\infty
\pqty{\mathsf{K}_{\pm,n}^{(1)}\, \mathsf{K}_{\pm,n}^{(2)}}\,.
\label{eq:wkb-d6-Kpm}
\end{equation}
The connection matrices composed of the Stokes crossings are in turn given by
\begin{equation}
\mathsf{K}_{+,C}
= \mathsf{U}\pqty{-\ii \mathcal{X}_{C}} \,,
\qquad
\mathsf{K}_{-,C}
= \mathsf{L}\pqty{\ii \mathcal{X}_{C}} \,.
\label{eq:ewkb-d6-bdyfactor}
\end{equation}
The only difference from $d=4$ is the additional boundary Stokes line crossing owing to the third active turning point $\upr_0$ at the end of the transport.
The non-perturbative factor can be worked out along the same lines. One obtains
\begin{equation}
\mathcal F_{\mathrm{np}}
= \sinh(\hbw)\, \mathfrak{X}_6^{-1}\,,
\end{equation}
where now the denominator factor is
\begin{equation}
\begin{aligned}
\mathfrak{X}_6
& ={}
\sqrt{1+2\cos(\pi\nu)\, \mathcal{X}_C  + \mathcal{X}_C^2}\,
\bqty{\cosh(\hbw) +
\frac{e^{\hbw}}{\abs{\mathcal{X}_{A,\mathrm{med}}}^2}}
\\
& \qquad{}
+
\frac{e^{\hbw}\, \sqrt{
    \pqty{ 1+\abs{\mathcal{X}_{A,\mathrm{med}}}^2}
\pqty{ 1+e^{-\bw}\, \abs{\mathcal{X}_{A,\mathrm{med}}}^2}}}{
\abs{\mathcal{X}_{A,\mathrm{med}}}^3}\,
\Re\bqty{(e^{\ii \pi \ms\nu} + \mathcal{X}_C)\, \mathcal{X}_{A,\mathrm{med}}}\,.
\end{aligned}
\label{eq:ewkb-sd6-Xfn}
\end{equation}
This can be checked to conform to the similar structural form given in~\eqref{eq:wkb-non-perturbative-factor}.

\subsection{Charged black hole in \AdS{5}}\label{sec:charged-d4}

The \RNAdS{5} geometry has three roots for $f(r) =0$ at the outer horizon $H_+: \upr =1$, the inner horizon $H_-: \upr = \frac{2}{1+\varrho}$, and a complex horizon $H_c: \upr =  -\frac{2}{\varrho}$ using~\eqref{eq:Qvarrho} to parameterize the charge. The leading WKB potential has three roots. One of these is real, at $\upr_0$. It lies between the singularity at $\upr = 2$ and the inner horizon at $\upr = \frac{2}{1+\varrho}$. The other two are complex conjugates at $\upr_\pm$. The cycles of interest are again
\begin{equation}
A = \gamma_+ \sim \{0,\upr_+\}, \qquad
B = \gamma_- \sim \{0, \upr_-\} \,.
\end{equation}
There is in fact no direct transport path from the boundary at $\upr =0$ to the real turning point at $\upr_0$. In particular, notice also that the exact-WKB path goes nowhere near the inner horizon. Since it does not cross the inner horizon, the  $\zeta \neq 0$ transport never moves towards any of the future boundaries on the Penrose diagram~\cref{fig:ads-SRN-pd}. This is the primary reason we refrained from depicting timelike trajectories, whose envelope would be a pair of null geodesics, connecting different asymptotic boundaries one to the causal future of the other.

\begin{figure}[htbp]
  \centering
  \includegraphics[width=0.98\linewidth]{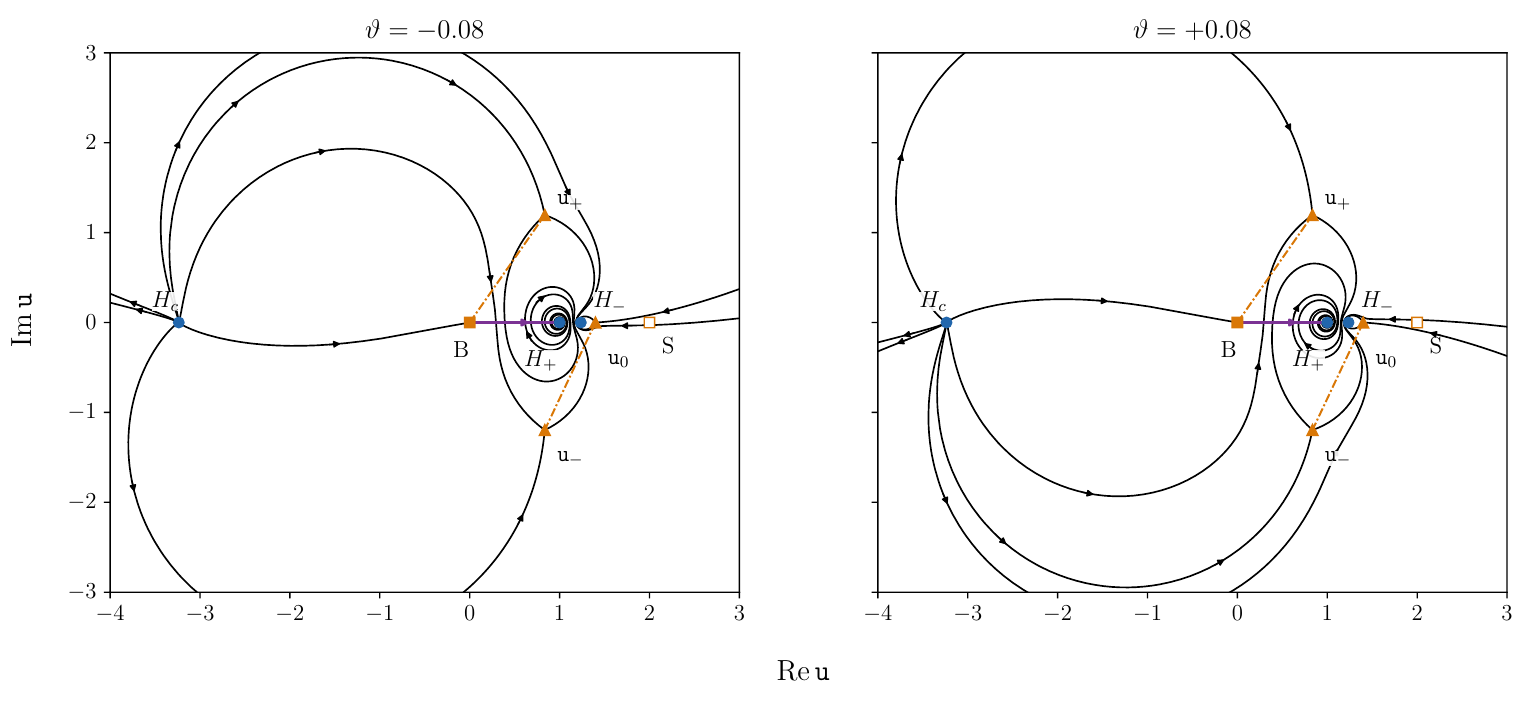}
  \caption{Laterally resolved Stokes graphs for \RNAdS{5} black holes in the $\upr$ coordinate. We chose $\zeta =0.6$ and $Q=1$. The conventions are as indicated in~\cref{fig:ewkb-d6-neutral}, for again the singularity is an ordinary point of the WKB differential. The real turning point does not emit a Stokes trajectory that crosses the transport path. }
  \label{fig:ewkb-d4-charged}
\end{figure}

The order of multiplication of the Stokes factors can be read off from the Stokes graph~\cref{fig:ewkb-d4-charged}. As before define the lateral Voros symbols $\mathcal{X}_{A\pm}$ and
$\mathcal{X}_{B\pm}$ after rotation by $\vartheta = 0^\pm$.  There are again two spiral families into the outer horizon and a direct transport detour to one of the complex turning points. The factors can be read off to be
\begin{equation}\label{eq:Kmat-d4-charge}
\begin{aligned}
\mathsf{K}_{+,n}^{(1)}
& =
\mathsf{U}\pqty{-\ii e^{-\bw\ms n}\, \mathcal{X}_{B+}}\,,
& \qquad
\mathsf{K}_{+,n}^{(2)}
& =
\mathsf{U}\pqty{-\ii e^{-\bw\ms (n+1)}\, \mathcal{X}_{A+}}\,,
\\
\mathsf{K}_{-,n}^{(1)}
& =
\mathsf{L}\pqty{\ii e^{-\bw\ms n}\, \mathcal{X}_{A-}^{-1}}\,,
& \qquad
\mathsf{K}_{-,n}^{(2)}
& =
\mathsf{L}\pqty{\ii e^{-\bw\ms (n+1)}\, \mathcal{X}_{B-}^{-1}}\,,
\\
\mathsf{K}_{+,\mathrm{final}}
& =
\mathsf{L}\pqty{\ii \mathcal{X}_{A+}^{-1}}\,,
& \qquad
\mathsf{K}_{-,\mathrm{final}}
& =
\mathsf{U}\pqty{-\ii \mathcal{X}_{B-}}\,,
\end{aligned}
\end{equation}
The connection matrix $\mathbb{K}$ can be obtained to be
\begin{equation}
\begin{aligned}
\mathbb{K}_+
&= \mathsf{L}\pqty{\ii \mathcal{X}_{A+}^{-1}}\,
\mathsf{U}\pqty{-\ii  \mathfrak{s}_+(\mathcal{X}_{A+},\mathcal{X}_{B+})}\,,
&\qquad
\mathfrak{s}_+
&=
\frac{\mathcal{X}_{B+}\, + e^{-\bw}\, \mathcal{X}_{A+}}{1-e^{-\bw}}
\\
\mathbb{K}_-
&= \mathsf{U}\pqty{-\ii \mathcal{X}_{B-}}\,
\mathsf{L}\pqty{\ii  \mathfrak{s}_-(\mathcal{X}_{A-},\mathcal{X}_{B-})}\,,
&\qquad
\mathfrak{s}_-
&=  \frac{\mathcal{X}_{A-}^{-1} + e^{-\bw}\, \mathcal{X}_{B-}^{-1}}{1-e^{-\bw}}
\end{aligned}
\end{equation}

The median values of the Voros symbols are (NB: now $\mathcal{X}_{AB} = \mathcal{X}_{A}/\mathcal{X}_{B}$)
\begin{equation}\label{eq:rnads5-Xmed}
\mathcal{X}_{A,\mathrm{med}}
= \mathcal{X}_{A+}\, \sqrt{\frac{1+ e^{-\bw}\, \mathcal{X}_{AB+}}{1+ \mathcal{X}_{AB+}^{-1}}}
= \mathcal{X}_{A-}\, \sqrt{\frac{1+ \mathcal{X}_{AB-}^{-1}}{1+ e^{-\bw}\,
\mathcal{X}_{AB-}}}
\end{equation}
Using this we can work out the non-perturbative factor defined in~\eqref{eq:wkb-non-perturbative-factor} to be
\begin{equation}
\begin{aligned}
\mathcal{F}_{\mathrm{np}}
= \sinh(\hbw)\, \mathfrak{X}_{4,c}^{-1}\,.
\end{aligned}
\end{equation}
The function $\mathfrak{X}_{4,c}$ is given by the same expression as for the neutral black holes~\eqref{eq:ewkb-sd4-Xfn}, with the replacement of $\mathcal{X}_{A,\mathrm{med}}$ by the \RNAdS{5} analog given in~\eqref{eq:rnads5-Xmed}.

The only spirals we encounter in the above calculation are around the outer horizon. Nevertheless, the turning points and the Voros periods associated with them depend on all three horizons. The leading order answer, which depends on $\mathcal{X}_{A,\mathrm{med}}$ cares about transport from the boundary to  $\upr_+$. But this contour can easily be deformed to an open contour to the singularity, whence it picks up an explicit contribution linked with each horizon.

\section{Accessing the black hole singularity}\label{sec:ewkb-bhsing}

The exact-WKB analysis for real timelike momentum makes it clear that the cycles of interest which determine the non-perturbative corrections are either
(a) the boundary to turning point cycles, or (b) the relative cycle enclosing two turning points. The latter can be recast in terms of the former owing to~\eqref{eq:ewkb-relcycle}. All in all, the WKB analysis is controlled by turning points and seems to have nothing to do with the singularity, which in all cases analyzed in~\cref{sec:examples} is not a turning point controlling the physical transport path (for $\zeta \in (0,1)$ or for $\zeta \in \ii \mathbb{R}$).

Given that we were able to write the leading WKB action in~\cref{prt:WKB1} in terms of a period involving the singularity, we need to understand how that representative arises out of the exact-WKB answer. The clue is already present in our discussion of the charged \RNAdS{5} black hole where the singularity at $r=0$ was in no way distinguished in the wave equation, but one could nevertheless deform the contour to be anchored at this locus. We explain in~\cref{sec:singrep} how the representation heuristically motivated in~\cref{sec:specasym} can be obtained from our exact connection data. Subsequently, in~\cref{sec:qperiod-ddp} we argue that analytic continuation in $\freq$ provides a more direct access, thanks to the quantum periods being aware of the analytic structure of the full spectral curve.

\subsection{Recovering the singularity anchored representative}\label{sec:singrep}

Consider the expression for the phase of the retarded Green's function~\eqref{eq:ewkb-RRA}. We have expressed this in terms of the all orders boundary to turning point cycle period $\mathcal{X}_\star$. With complex turning points appearing in pairs, the periods for cycles enclosing conjugate turning points are related along the physical WKB path, $\arg \freq =0$. The median resummation respects this conjugation, cf.~\eqref{eq:ewkb-d4--median}, and so we can always focus on one particular turning point $\upr_\star$. Now consider expanding the result in~\eqref{eq:ewkb-RRA} in a series with $\abs{\mathcal{X}_\star^{-1}}$ taken to be small. One finds
\begin{equation}
\exp(2\ii \phGR) =
e^{-2\pi \ii \ms \nu} \bqty{1 + 4\ii\ms \sin(\pi\ms\nu) \, \Re \mathcal{X}_\star^{-1} +
\order{\abs{\mathcal{X}_\star^{-1}}^2}}
\end{equation}
We can read off  $\phGR + \pi \nu $ from this expression to compare with~\eqref{eq:argGR-asymptotics}, leading to
\begin{equation}
\log \mathcal{X}_\star^{-1} = \ii \freq\, V_\star\,, \qquad \Iwkb = V_\star.
\end{equation}
Here $V_\star$ is the leading classical period as defined in~\eqref{eq:Vmone}. The cycle is the boundary to the distinguished turning point cycle, and the integrand is the universal WKB differential. Hence, with $\gamma_\star \sim \set{0, \upr_\star}$ we have
\begin{equation}
V_\star = \oint_{\gamma_\star} \lwkb = 2\, \int_0^{\upr_\star} \lwkb\,.
\end{equation}
In the last equality we evaluated the contour integral that encircles both $\upr =0$ and $\upr = \upr_\star$ by a direct integral of the WKB differential.

This integral can be evaluated in any coordinate chart, and moreover, one is free to deform the integration path. Let us do so by taking a detour to the singularity $\upr = 2$ in the coordinate~\eqref{eq:u-coordinate} and then pass onto the turning point. This deformed path runs along the real $\upr$ axis through horizons onto the singularity, and goes up to the turning point on the primary sheet. The return leg is on the second sheet. We have to employ a principal value prescription to account for the pole at the horizon which picks up a half residue contribution. No other singularities or cuts are crossed in the deformation. Therefore,
\begin{equation}
\begin{aligned}
2\int_0^{\upr_\star} \lwkb
& = 2\,\mathrm{PV}\int_0^2 \lwkb + 2\,\int_2^{\upr_\star} \lwkb + 2 \pi\,\ii \sum_{k}\, \eta_k\, \Res_k \lwkb\,.
\end{aligned}
\end{equation}
This is precisely the decomposition obtained in~\eqref{eq:Iwkb}.

One could argue that this had to work out given the numerical evidence at hand. Nevertheless, understanding how the singularity-anchored representation arises is helpful. The essential point of the above argument is the contour deformation. One could simply say that the physical WKB path which picks out the cycle $\gamma_\star$ is homologous to the path used in writing~\eqref{eq:Iwkb}. The latter is anchored at the singularity by fiat, but there is no compelling reason to single out the black hole singularity. One could just as well have picked any other point in the spacetime which gives a decomposition in the same homology class. One is therefore led to conclude that the access to the singularity from the spectral function asymptotics is a mirage on the real frequency ray at fixed $\zeta$ (independent of $\arg \zeta$).

\subsection{Quantum periods and black hole singularity}\label{sec:qperiod-ddp}

We now understand that the WKB periods which naturally appear in exact-WKB analysis of thermal two-point functions at finite $\zeta$ are the ones surrounding the boundary and some turning point. The two-point function therefore appears to have no direct relation with the black hole singularity. This is also in line with our geodesic considerations in~\cref{sec:spgeod-mom}.

However, unlike the classical geodesic problem, the two-point function is defined by the full Schr\"odinger problem and depends on quantum periods rather than only classical periods. The exact-WKB boundary to turning point period contains information about the full quantum singularity to turning point period through the DDP-like jump formula~\cite{Delabaere:1997srq}. A neutral black hole in $d+1$ dimensions has $\frac{d}{2}$ turning points. At the black hole singularity the quadratic differential has order $\frac{d}{2} -3$  in the projective coordinate chart. Odd orders of the differential produce ramifications of the double cover. In $d=4$ we have a simple pole, an ordinary point in $d=6$, and higher order in $d\geq 8$.  Since the required jump formula is known for  branch points of order unity, we will focus on $d=4$. Analyzing higher dimensional examples with $d = 2n >4$ requires one to generalize the results for higher order of the quadratic differential.

\paragraph{The DDP jump formula:} The basic idea behind the DDP jump formula is the following. Say we have a finite Stokes trajectory $\ell$ connecting two simple turning points, and let $\gamma_\ell$ be the cycle on the spectral curve. Due to its presence the Borel summation of Voros symbols associated with other cycles $\gamma$ will have two-sided limits $\mathcal{X}^\pm_{\gamma, \vartheta_c}$. We have allowed for the possibility that the Borel singularity appears along the direction $\vartheta_c$. The two limiting values are related by the jump which depends on $\mathcal{X}_{\gamma_\ell, \vartheta_c}$ as
\begin{equation}
\mathcal{X}^-_{\gamma, \vartheta_c}
= \mathcal{X}^+_{\gamma, \vartheta_c}\, (1 + \mathcal{X}_{\gamma_\ell, \vartheta_c})^{-\langle \gamma_\ell, \gamma \rangle}\,.
\end{equation}
The multiplication by the factor involving the Voros symbol of the saddle trajectory $\ell$ is the action of a Stokes automorphism, and is captured as
\begin{equation}
\mathfrak{S}_{\gamma_\ell} : \mathcal{X}_{\gamma, \vartheta_c}
\mapsto \mathcal{X}_{\gamma, \vartheta_c}  (1 + \mathcal{X}_{\gamma_\ell, \vartheta_c})^{-\langle \gamma_\ell, \gamma \rangle}\,,
\end{equation}
implying that the jump formula is equivalent to the composition rule for the Borel sums
\begin{equation}
\mathcal{S}_{\vartheta_c^-} = \mathcal{S}_{\vartheta_c^+} \circ \mathfrak{S}_{\gamma_\ell}
\end{equation}

\paragraph{The quantum double well analogy:} Let us unpack the jump formula in a simple setting before proceeding with the black hole problem. The quantum mechanical double well potential provides a useful example to capture the information in the DDP formula. The classical motion in the double well potential is of course oblivious to the tunneling region under the potential barrier. However, the complete quantum mechanical treatment should encode the tunneling effect, which is closely related to the DDP formula. Let $V_{\mathrm{pert}}$ be the perturbative period in a classically allowed region (this is the cycle $\gamma$), and $V_{\mathrm{np}}$ be the non-perturbative period in the tunneling region (associated now with the finite saddle cycle $\gamma_\ell$). The DDP formula gives the jump of quantum perturbative period at $\vartheta = \arg \hbar =0$ in terms of quantum non-perturbative period. Assuming the intersection between the cycles is unity, we have
\begin{equation}
\frac{\mathcal{X}^{(0^-)}_{\mathrm{pert}}}{\mathcal{X}^{(0^+)}_{\mathrm{pert}}} = \prn{1 + \mathcal{X}^{-1}_{\mathrm{np}}}^{-1}.
\label{eq:wkb-double-well-ddp}
\end{equation}

The analogy to the black hole singularity problem should be clear. Classical geodesics do not enter the vicinity of the black hole singularity at finite $\zeta$. In the quantum WKB analysis of the black hole wave equation, the boundary to turning point period is analogous to the perturbative period in the classically allowed region. The singularity to turning point period is the analog of the WKB  analysis being sensitive to a tunneling region through the non-perturbative period associated with a finite saddle trajectory.

\begin{figure}[thbp!]
  \centering
  \begin{subfigure}[t]{\textwidth}
    \centering
    \includegraphics[width=0.7\linewidth]{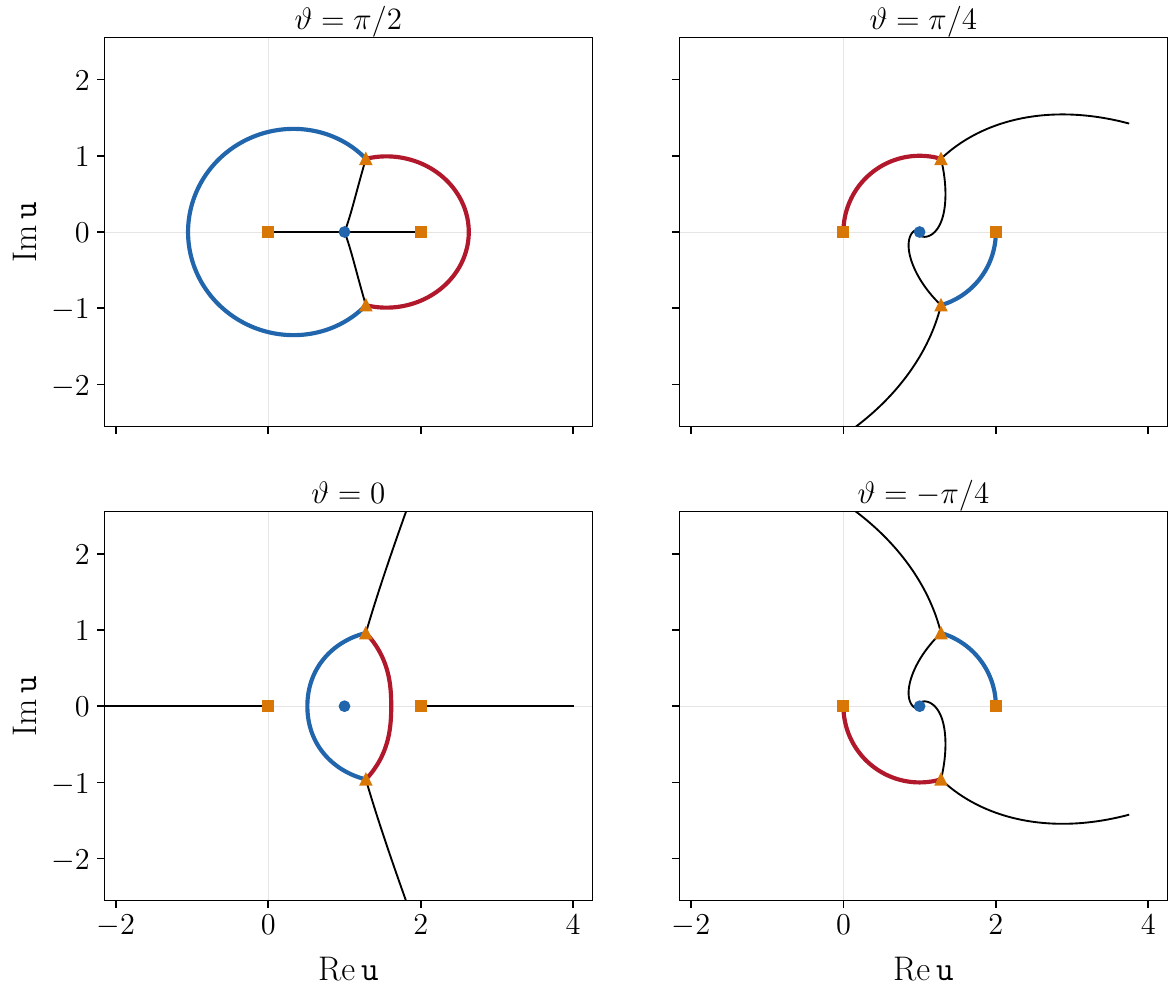}
    \caption{Critical phases of planar \SAdS{5}}
  \end{subfigure}
  \par\medskip
  \begin{subfigure}[t]{\textwidth}
    \centering
    \includegraphics[width=0.98\linewidth]{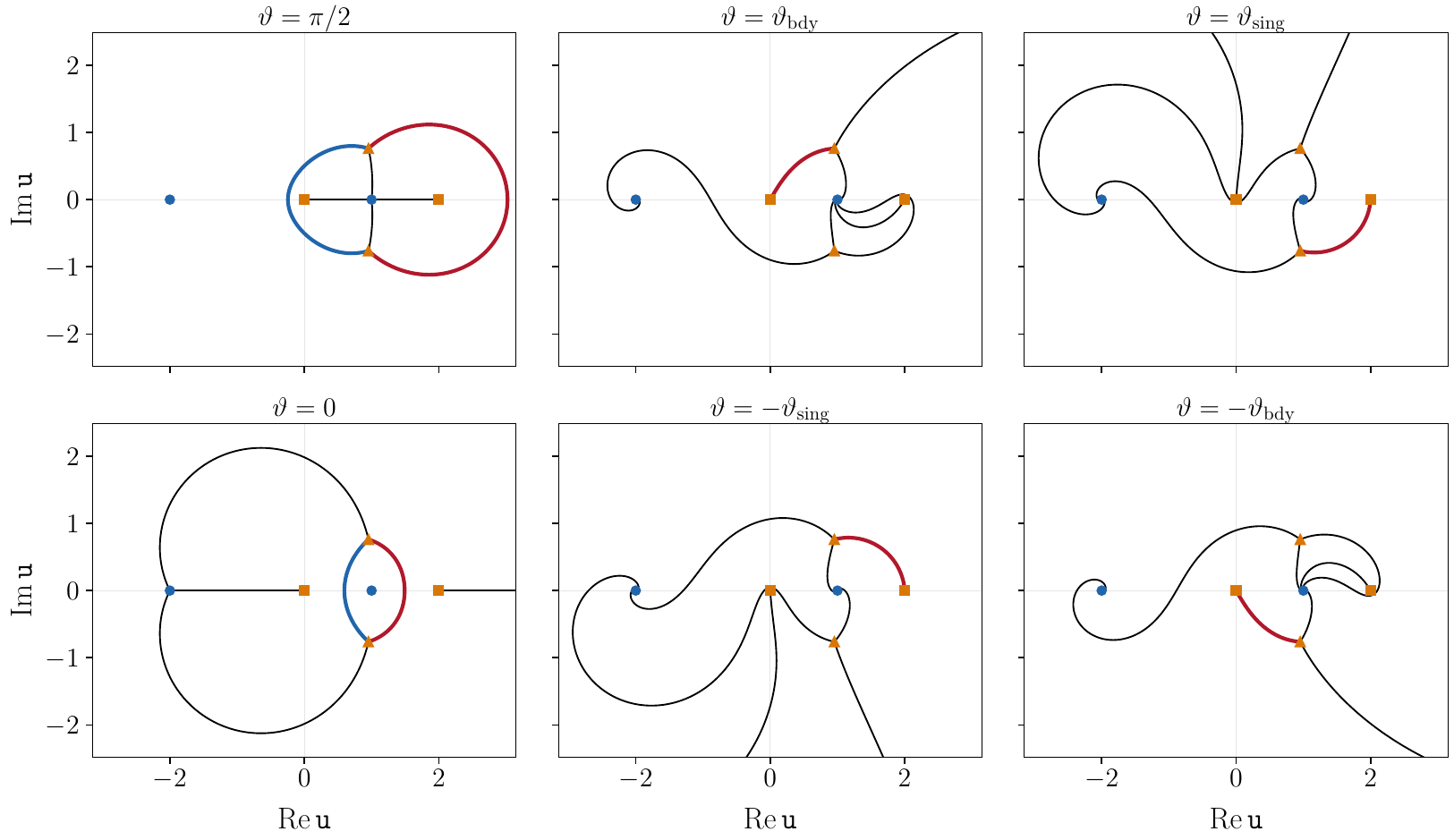}
    \caption{Critical phases of global \SAdS{5} with $\rp = 1$}
  \end{subfigure}
  \caption{The Stokes graphs drawn at the critical phases for neutral black holes in $d=4$. We have fixed  $\zeta= \frac{3}{5}$ so that both turning points are complex in the planar and global black holes. As we dial $\vartheta = \arg \freq$ we encounter different finite saddle Stokes lines, connecting the turning points to the boundary as expected, but also connecting them to the singularity. The graphs are drawn in the $\upr$ coordinate of~\eqref{eq:u-coordinate}. }
  \label{fig:neutral-d4-critical-graphs}
\end{figure}

\paragraph{The DDP phases for \SAdS{5} black holes:} To understand the contribution from the jump formula consider the Stokes graphs at all critical phases as we vary $\vartheta = \arg \freq$ for both planar and global \SAdS{5} black holes. The result is depicted in~\cref{fig:neutral-d4-critical-graphs}.

Of particular relevance for us are the critical phases
\begin{equation}
\vartheta = \pm\vartheta_{\mathrm{sing}}
\end{equation}
where the singularity to turning point period corresponds to a saddle cycle and therefore enters into the jump formula for the boundary to turning point period.
The planar black holes reside at a degenerate limit where $
\vartheta_{\mathrm{sing}}  = \vartheta_{\bdy} = \frac{\pi}{4}$. This is not the case for the global black hole where the critical phases are differentiated between the singularity and the boundary cleanly.

To write down the jump formula, we recall the boundary to horizon connection problem involves the periods associated with the boundary to turning point cycles $A$ and $B$, defined in~\eqref{eq:periods-neutral-d4}. Working in the $\upr$ coordinate let us denote the singularity to turning point cycles as
\begin{equation}
A^\prime = \{\upr_-, 2 \}, \qquad B^\prime = \{\upr_+, 2 \}.
\end{equation}

For a spherical black hole, the boundary $A$ cycle has non-zero intersection with the saddle singularity $B^\prime$ cycle at critical phase $\vartheta = -\vartheta_{\mathrm{sing}}$. Explicitly the critical phase is given by
\begin{equation}
\vartheta_{\mathrm{sing}}
=\frac{\pi}{2}+\arg V_{A_s}.
\end{equation}
This leads to the jump formula
\begin{equation}
\frac{\mathcal{X}^{(-\vartheta_{\mathrm{sing}})^{-}}_{A}}{
\mathcal{X}^{(-\vartheta_{\mathrm{sing}})^{+}}_{A}}
= \prn{1 + \prn{\mathcal{X}^{(-\vartheta_{\mathrm{sing}})}_{B^\prime} }^{-1} }^{2}\,.
\end{equation}
Consequently, the median resummed value of $\mathcal{X}^{(-\vartheta_{\mathrm{sing}})}_{A,\mathrm{med}}$, which enters the connection coefficients, carries a signal of the singularity through the $B^\prime$ cycle period. We should emphasize that this is qualitatively different from our singularity anchored representation. There we shoehorned the singularity  anchored cycle into the $\vartheta = 0$ periods, which was a homology representative choice. In contrast, the statement here is that if we examine the asymptotics of the spectral function at $\arg \freq = -\vartheta_{\mathrm{sing}}$ then there is necessarily information from the vicinity of the singularity.

For the planar black holes, the situation as mentioned above is degenerate. The boundary $A$ cycle has non-zero intersection with both saddle singularity $B^\prime$ cycle and saddle boundary $B$ cycle at critical phase $\vartheta = -\frac{\pi}{4}$. This leads to a jump formula, which mixes both contributions. One finds
\begin{equation}
\frac{\mathcal{X}_{A}^{(-\pi/4)^{-}}}{\mathcal{X}_{A}^{(-\pi/4)^{+}}}
=
\frac{
  \left[
    1+
    \left(\mathcal{X}_{B'}^{(-\pi/4)}\right)^{-1}
  \right]^2
}{
  \left[
    1+e^{\pi\ii\nu}
    \left(\mathcal{X}_{B}^{(-\pi/4)}\right)^{-1}
  \right]
  \left[
    1+e^{-\pi\ii\nu}
    \left(\mathcal{X}_{B}^{(-\pi/4)}\right)^{-1}
  \right]
}.
\end{equation}
While more involved, it nevertheless still makes clear that for $\arg \freq = -\frac{\pi}{4}$ one has direct information about the black hole singularity through an analog of a quantum tunneling mechanism.

\newpage
\part{A core-scaling limit: toward the singularity}
\label{prt:WKB3}

\section{The core-scaling regime}\label{sec:tripscale-core}

The results of~\cref{prt:WKB2} make it clear that
in the linear-scaling limit the WKB trajectory does not get to the vicinity of the singularity.
One can, of course, attempt to use the singularity-anchored representative as in~\cref{prt:WKB1} or use the quantum tunneling argument for $\arg\omega \neq 0$. In both cases, we are making an assumption that the spacetime is analytic to permit these excursions, and the signature of the black hole singularity is somewhat indirect.
One could instead work at high frequency, but keep momentum finite. This would effectively set $\zeta =0$ in the WKB analysis. While this has been analyzed hitherto in~\cite{Afkhami-Jeddi:2025wra} by working out the transseries coefficients numerically, a complete exact-WKB analysis of the problem has not yet been undertaken.
The reason for this was explained in~\cite[Section 8]{Jia:2026ryl}: at $\zeta=0$ the turning points merge with the black hole singularity. As a consequence there is a competition between the leading and subleading terms of the WKB potential $Q_0$ and $Q_2$, with $\omega^2 \, Q_0 \sim Q_2$ in some neighborhood of $r =0$. This implies that one has to treat the core region of the geometry differently. In~\cite{Jia:2026ryl} this was achieved by expanding the exact solution in the vicinity of the singularity and matching to the exterior WKB solution. This matching strategy is guaranteed to work, but one does require control over the core region solution.

We now take inspiration from the analysis of~\cite{Afkhami-Jeddi:2025wra} and argue that the exact-WKB analysis of~\cref{prt:WKB2} can be adapted to the case of finite momentum. In fact, we study a slight variation: we take $\omega, \abs{\vb{k}} \to \infty$, but at different rates keeping a new combination $\trs$ fixed. The momentum will be scaled sub-linearly with frequency to achieve this.  We refer to this choice of parameter scaling as the \emph{core-scaling limit}. In this scaling regime, the turning points are confined to the vicinity of the singularity and lie at distinct loci; at the very least one has access to the  part of the geometry with high curvatures. The crux of the argument is realizing that the Stokes crossings encountered in the transport path from the boundary, the spirals connecting the black hole horizon and the turning points are no longer simply of the Airy type. Rather, they are given in terms of the reflection amplitude from the singularity~\cite{Afkhami-Jeddi:2025wra}. The analysis in this case can be done straightforwardly for any $\arg \zeta$, so we generalize our considerations and study the problem for arbitrary complex phase of the momentum. In fact, as $\arg \zeta$ changes, the turning points rotate around the origin. The WKB path makes it down to them and reflects off of the singularity, which leads to the modification to the Stokes crossing formula.

We begin with a description of the scaling regime and the core, near-singularity region equation that arises in the limit. The second step is to account for the reflection amplitude and modify the Stokes crossing factor. With these two results at hand we can write down the exact-WKB connection matrix. The end result will be similar to our analysis in~\cref{prt:WKB2}, with the transport matrix generalized from the Airy form.

\subsection{The core geometry: zooming into the singularity}\label{sec:core-scale}

Let us first define the core-scaling regime of interest. Consider the contribution from
the leading WKB potential $Q_0(r)$ and the subleading piece $Q_2(r)$. The former is universal and is given by~\eqref{eq:Q0wkb}. The part we need from $Q_2(r)$ is the simple fact that for neutral black holes the locus $r=0$ is a regular singular point of the wave equation.
Therefore, examining the two pieces to the potential in the original radial chart, we have as $r\to 0$
\begin{equation}
Q_0 \sim -\frac{r^{2\,d-4}}{\rp^{2\,d}} - \zeta^2\, \frac{r^{d-4}}{\rp^{d}}\,, \qquad
\omega^{-2}\, Q_2\propto -\frac{1}{\omega^2\, r^2}\,.
\end{equation}
As noted in~\cite{Jia:2026ryl} the core region of the spacetime where $Q_2$ becomes comparable to $Q_0$ is when
\begin{equation}\label{eq:core-r-scale}
\omega^2\, r^2 \sim r^{4-2\,d}   \;\; \Longrightarrow \;\;
r\sim \omega^{-\frac{1}{d-1}} \,.
\end{equation}
In this domain we are no longer able to neglect the fact that the full potential harbors a  regular singular point. Technically, this means that we have a hard Bessel type edge to the spectral curve, and the scaling~\eqref{eq:core-r-scale} homogenizes it.
Zooming into this region at fixed $\zeta$ would, of course, not be interesting, since the physical turning points lie at $r\sim \zeta^{\frac{2}{d}}$. The only way to have spatial resolution while accessing the hard edge core is to scale
\begin{equation}
\zeta \sim r^{\frac{d}{2}} \sim \omega^{-\frac{d}{2\,(d-1)}}\,.
\end{equation}
With this scaling the turning points are consigned to lie in the core region, and one can thus access the strong curvature region, while continuing to disambiguate the turning point loci from the black hole singularity.  The fractional powers appearing in the transseries~\cite{Afkhami-Jeddi:2025wra} originate from the scaling into the core region, cf.~\cite{Jia:2026ryl}.

Therefore, define the \emph{core-scaling limit} as
\begin{equation}\label{eq:triple-scale}
\boxed{
  \omega, \abs{\vb{k}}\to \infty \,,  \qquad
  \abs{\vb{k}} \sim \omega^{\frac{d-2}{2\,(d-1)}}\,, \qquad
  \qquad \trs \equiv \zeta \, \freq^{\frac{d}{2\,(d-1)}} = \text{fixed}\,.
}
\end{equation}
Note that $\abs{\vb{k}} \to \infty$ but at a different rate than in the linear-scaling regime (where it was linear).

The above discussion holds for the case of \SAdS{d+1} black holes in $d\geq 4$. For charged
\RNAdS{5}, as noted in~\cref{sec:ewkb-sing} the curvature singularity is a regular point of the
differential operator for minimally coupled fields in the projected coordinate chart.\footnote{The coefficient of the putative double pole at $\rpr=0$ in \RNAdS{d+1} is proportional to $d(d-4)$. This makes it clear that in $d=4$ the singularity locus is an ordinary point. In the unprojected $r$ coordinate chart the coefficient is $\frac{(d-1)\,(d-3)}{4}$. However, the wavefunction is free of a logarithmic branch and thus in general dimensions $r=0$ is an apparent singular point~\cite{Loganayagam:2022teu}. Note that even ordinary or apparent singular points in the original radial coordinate $r$ can end up with the wavefunctions having branch cuts in the projective coordinate $\rpr$ or $\zpr$. \label{fn:charged-core}}  The dynamics of transverse vector and longitudinal scalar modes of photons and gravitons has a regular singular point at the curvature singularity. However, the scaling into the core region  is different. In general if $f(r) \sim r^{-\gamma}$ as $r\to 0$  the scaling we need is
\begin{equation}
r \sim \omega^{\frac{1}{1-\gamma}}\, \qquad \zeta \sim r^{\frac{\gamma}{2}}\,.
\end{equation}
For \RNAdS{d+1} black holes $\gamma = 2(d-1)$.

Having defined the scalings we wish to consider, we now proceed with the exact-WKB analysis.
Steps 1 and 3 of our timelike $\zeta$ exact-WKB treatment are unmodified, for the
solution bases at the boundary and the horizon are unaffected by the scaling of momentum. The crux is to understand Step 2, transport from the boundary to the horizon with the turning points now lying in the core region.

There are two equivalent ways to understand the transport. One involves zooming into the core region by using a coordinate chart adapted to the scaling. In this chart, we end up with a core oper with a hard edge. However, a new feature is that the WKB turning points encountered earlier disappear, and are replaced by a set of
core turning points. The latter are effective resolutions of the singularity owing to our use of rescaled coordinates. They are directly helpful to work out the core region Stokes graph topology, and dictate the connections possible between the roots.\footnote{We use the Stokes graph terminology even though in the core region oper there will be no obvious large WKB parameter. The nature of what the crossings mean will become clear in due course.}

An alternate strategy is to use the original WKB coordinate $\int \lwkb$, which absolves one of having to locate the core region turning points. This action coordinate chart flattens the transport and is particularly effective for the $\zeta = 0$ analysis, when the turning points have coalesced into the singularity. For non-vanishing $\zeta$, however, the action chart is branched at the real turning points. The end result for the transport is the same in either chart, for the connection data are covariant.

As both approaches have their merits, we therefore split our discussion into two parts. We first analyze the core region differential equation after a suitable scaling of coordinates. After explaining how we can obtain the solutions in the core region, we show how to account for the Stokes crossings. The upshot will be a modification of our transport matrix $\mathbb{K}$ from~\cref{sec:ewkb-connection}. We then turn to a detailed analysis  using the action chart~\cref{sec:core-wkb-borel}, which illustrates how to recover the same result without resolving the core region.

\subsection{The core region oper: the scaling chart}\label{sec:core-scaling}

Since we want to work in the vicinity of the singularity, it is more efficacious to use a projective coordinate chart where the black hole singularity is at a finite locus.\footnote{One could just work in the $\upr$ coordinate introduced in~\eqref{eq:u-coordinate}, which has the advantage that all the interesting loci, boundary, horizon, and singularity are at finite positions. We will use this chart to draw Stokes graphs later. For now the local analysis is simpler to explore in the $\rpr$ coordinate.} Let us revert to  $\rpr = \frac{r^2}{\rp^2}$. Furthermore,  $\zeta$ is defined as before~\eqref{eq:zetadef}. For our present analysis instead of restricting it to the open unit interval or to being purely imaginary, we assume it to be small and let its  phase be arbitrary. Then inspired by~\eqref{eq:triple-scale} introduce rescaled coordinates
\begin{equation}\label{eq:core-chart}
\rtr = \freq^{\frac{2}{d-1}}\, \rpr\,, \qquad
\trs = \zeta\, \freq^{\frac{d}{2\,(d-1)}}\,.
\end{equation}
In these coordinates, the outer horizon is pushed off to $\rtr = \freq^{\frac{2}{d-1}}$, and the physical region of interest between it and the boundary is the domain
$\rtr \in  (\freq^{\frac{2}{d-1}}, \infty)$. As long as we take $\freq \gg 1$ we do not immediately resolve this region, so having understood the properties in the core region, we should splice that data back into the physical domain.

Substituting into our scalar wave oper, we find the oper
\begin{equation}\label{eq:core-oper}
\mathfrak{D}_{d}[\trs] :=
\dv[2]{}{\rtr} + \frac{1}{4} \bqty{\rtr^{d-3} + \trs^2\, \rtr^{\frac{d}{2} -3} + \frac{1}{\rtr^2}} \,.
\end{equation}
This form of the leading oper is valid for neutral black holes in all even dimensions $d\geq 4$.\footnote{In the global \SAdS{d+1} case the coefficients of the potential are modified by $\rp$ dependence. Set $M_+ = \frac{\rp^2}{1+ \rp^2}$ (recall $\lads =1$). The potential of the core oper is modified to
$M_+^2 \,\rtr^{d-3} + M_+ \,\trs^2\, \rtr^{\frac{d}{2}-3} + \rtr^{-2}$, which reduces to the planar case for $\rp \gg 1$. Nevertheless,  we can treat this case in parallel for we can absorb the $\rp$ factors into $\rtr$ and $\trs$. Redefining $\rtr \to M_+^{\frac{2}{d-1}}\, \rtr$ and $\trs \to M_+^{\frac{1}{2\,(d-1)}}\trs$, we bring the global core oper to the form given  in~\eqref{eq:core-oper}.\label{fn:global-core-caveat}}
As written, there is no obvious large parameter for this core region; the leading dependence on $\freq$ is scaled out. There are corrections to this oper in inverse fractional powers of $\freq$, which comprise an infinite series. For planar black holes the corrections start at
$\order{\freq^{-\frac{d}{d-1}}}$, while for global black holes they originate at $\order{\freq^{-\frac{d-2}{d-1}}}$. One set of corrections arises, for example, from expanding $f(r)$ which for planar \SAdS{d+1} would have a factor $\sim 1- \freq^{- \frac{d}{d-1}}\, \rtr^{\frac{d}{2}}$ in the denominator of the potential.  These are dealt with later when we discuss explicit examples, cf.~\eqref{eq:deltaQ-core} and~\eqref{eq:rnads-deltaQ}. In particular, they provide perturbative corrections to the leading solution we discuss here.

In our earlier discussion we encountered WKB turning points at the roots of $\rtr^{\frac{d}{2}} = -\trs^2$, where we are approximating the solutions described in~\eqref{eq:planar-uturn}. Equivalently, these are the roots obtained by balancing the first two terms in the oper~\eqref{eq:core-oper} (and thus do not distinguish the horizon spatial curvature). These were  separated from the singularity located at $\rtr = 0$, but are approaching it in the limit $\trs \to 0$. This vanishing momentum limit (which would also result from keeping momentum finite in the high frequency regime) is well-studied, and forms the anchor for our discussion.

\begin{figure}[th!]
  \centering
  \includegraphics[width=0.9\linewidth]{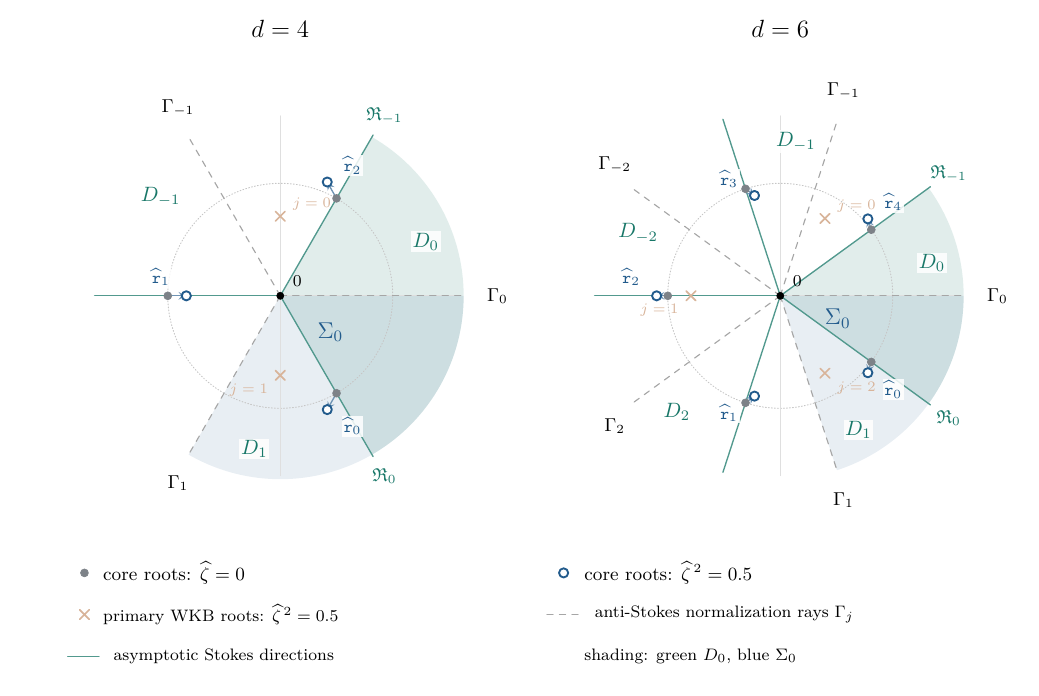}
  \caption{The core region in the $\rtr$ coordinate chart for  planar \SAdS{d+1} in $d=4$ and $d=6$.
    Grey dots mark the $d-1$ core roots at $\trs=0$ on the dotted unit circle, while
    blue open circles mark their continuations to $\trs^{\,2}=0.5$, with (short) arrows
    indicating how the roots follow the (real) momentum deformation. Pale orange crosses are the $\frac{d}{2}$ primary WKB roots at the same value of $\trs$.
    We distinguish the root sets by their indices: primary WKB roots are labeled by $j$, while the blue core root indices are $a$. All the primary WKB roots
    coalesce with the black hole singularity at the origin at zero momentum, whereas the
    core roots remain separated. The former are not germane for the core analysis, but have been indicated here for reference. The plot also shows the Stokes and anti-Stokes lines: the latter are the dashed rays  $\Gamma_a$ where we normalize the asymptotic oscillatory
  solutions, and the former are the lines across which we pick up the core Stokes multipliers. The two shaded wedges are the Stokes domains denoted $D_a$ and the domains   $\Sigma_a$ between adjacent anti-Stokes lines. The figure should be read as displaying the roots and the relevant sectors in the core region.}
  \label{fig:scaling-core-regions}
\end{figure}

However, as we noted above the core region does not have a WKB parameter. The oper~\eqref{eq:core-oper} should be treated exactly. It is nevertheless useful to ask what the turning points of its potential are. These lie at the loci which are the roots of the polynomial
\begin{equation}\label{eq:core-poly}
\rtr^{d-1} + \trs^2 \, \rtr^{\frac{d}{2} - 1 } + 1 = 0\,.
\end{equation}
These form a set of $d-1$ roots, unlike the primary WKB turning points, which were fewer, e.g., the planar and global black holes have $\frac{d}{2}$ turning points.

To get a sense for these roots notice that at $\trs = 0$ they lie on the unit circle in our rescaled core chart:
\begin{equation}\label{eq:core-turningpts}
\rtr_{a}(0) = \exp(-\frac{\pi \ii}{d-1}\, (2\,a+1))\,, \quad a \in \set{0, 1, \cdots d-2}\,,
\qquad \text{for}\ \trs = 0\,.
\end{equation}
So while the primary WKB turning points would have completely coalesced into the singularity in this limit, zooming into the core region with the new coordinate chart, `resolves' the singularity into new core turning points. The oper still has a regular singular point at $\rpr =0$, but the potential vanishes on the unit circle. These core turning points are at finite $\rtr$ locations, but really they lie at a distance of $\order{\freq^{-\frac{2}{d-1}}}$  from the origin in the $\rpr$ coordinate. So the external problem still sees the transport as reaching all the way to the singularity; the chart~\eqref{eq:core-chart} just zooms into this region to separate the turning points from the double pole.

Turning on $\trs$ will still retain these $d-1$ roots, and all of them will lie away from the origin. We follow them from $\trs =0$ along a specified path avoiding discriminant loci (for example, the polynomial in $d=4$ has double roots when $4\,\trs^6 + 27 =0$). The first correction is $\rtr_a(\trs) = \rtr_a(0) - \frac{\trs^2}{d-1}\, \bqty{\rtr_a(0)}^{1-\frac{d}{2}} + \order{\trs^4}$.  On the other hand, the WKB turning points inherited from our finite momentum analysis are, as noted above, only $\frac{d}{2}$ in number (restricting to planar \SAdS{d+1} black holes). Their locations can be read off from~\eqref{eq:planar-uturn} to be
\begin{equation}\label{eq:core-inherited}
\rtr_{j} = \trs^{\frac{4}{d}}\, \exp(\frac{2\pi \ii}{d}\, (2\,j+1))\,,
\qquad j \in \{0, 1,\cdots \frac{d}{2}-1\}\,.
\end{equation}
So we have $\frac{d}{2}-1$ new roots. The relation between them is particularly transparent when $\trs \gg 1$, for then we have $\frac{d}{2}$ roots which scale as $\trs^{\frac{4}{d}}$ and $\frac{d}{2}-1$ roots which scale as $\trs^{-\frac{4}{d-2}}$. The latter disappear into the singularity for large $\trs$ but are happily resolved for finite $\trs$. The linear-scaling limit, or indeed finite momentum problem, fails to resolve the singularity since it always ends up scaling parameters to obfuscate the core region. In~\cref{fig:scaling-core-regions} we have depicted the core roots and the unresolved primary roots for some small fixed $\trs$. The relevance of domains and Stokes/anti-Stokes lines indicated in the figure will become clear in the following analysis.

\subsection{The core Stokes multiplier}\label{sec:core-multiplier}

We now work out the details of the solutions of the core oper. Absent a large parameter, we have to solve this differential equation exactly. Fortunately, at $\trs =0$ this is straightforward in terms of Bessel functions. The general $\trs$ case can then be handled iteratively; one can argue that this iteration is convergent.

With the solutions at hand we can ask how the core solutions mix as we traverse the $\rtr$ plane around the origin. This is again, happily, informed by the Bessel connection formula at $\trs = 0$, which then can be upgraded for $\trs \neq 0$. We will record the result as giving a core reflection amplitude, through a function $\mathfrak{R}(\trs)$. One important caveat: this is only the leading order contribution from the core. One should further correct the core factor in inverse fractional powers of $\freq$, incorporating the terms we dropped in writing the core oper~\eqref{eq:core-oper}. These can also be dealt with by the  iterative method, albeit as a formal series in inverse (fractional) powers of the frequency.

\paragraph{The zero momentum case:}
Consider~\eqref{eq:core-oper} and specialize to
$\trs = 0$. The oper $\mathfrak{D}_{d}[0]$ is the standard Bessel operator at order-$0$. The differential equation
\begin{equation}
\dv[2]{\psi}{\rtr} + \frac{1}{4}\pqty{\rtr^{d-3} + \frac{1}{\rtr^2}} \psi = 0\,,
\end{equation}
admits the following solution basis anchored around $\rtr =0$
\begin{equation}\label{eq:core-J0}
\begin{aligned}
\psi\subm{core}^{(1)}(\rtr;\trs=0)
& = \sqrt{\rtr}\,  J_0 \pqty{\frac{\rtr^{\frac{d-1}{2}}}{d-1}}\,,
\\
\psi\subm{core}^{(2)}(\rtr,\trs =0)
& = \sqrt{\rtr}\,  \bqty{
  \frac{\pi}{d-1} \, Y_0 \pqty{\frac{\rtr^{\frac{d-1}{2}}}{d-1}}
  - \frac{2\,\gamma\subM{E} - 2\,\log(2\, (d-1))}{d-1}\,
J_0 \pqty{\frac{\rtr^{\frac{d-1}{2}}}{d-1}}}\,.
\end{aligned}
\end{equation}
The coefficients were chosen to ensure that the Wronskian is unity. We subtracted the $J_0$ piece in the second solution so that the resonant logarithm is canonical; near $\rtr =0$, $\psi\subm{core}^{(1)} = \sqrt{\rtr} + \cdots$ and $\psi\subm{core}^{(2)} = \psi\subm{core}^{(1)}\, \log \rtr + \order{\rtr^{d - \frac{1}{2}}}$. This makes the monodromy canonical in Jordan normal form.

We are however interested in the WKB solutions, for which the natural basis is around $\rtr \gg1$. These are better captured in terms of Hankel functions
\begin{equation}\label{eq:core-Hankel}
\begin{aligned}
\psi\subm{core}^{\pm}(\rtr;\trs=0)
& =
\sqrt{\frac{\pi}{d-1}}\, e^{\pm \ii \frac{\pi}{4}}\, \sqrt{\rtr}\,
H_0^{(1,2)}\pqty{\frac{\rtr^{\frac{d-1}{2}}}{d-1}}\,,
\\
\psi\subm{core}^{\pm}(\rtr;\trs=0)
& \to \sqrt{2}\, \rtr^{-\frac{d-3}{4}}\, \exp(\pm \ii\frac{\rtr^{\frac{d-1}{2}}}{d-1})\,, \qquad \rtr \gg 1.
\end{aligned}
\end{equation}
The Wronskian of these solutions is $W(\psi\subm{core}^+, \psi\subm{core}^- ) = -2\ms\ii$. Notice
that $\rtr \to \infty$ is the internal infinity of the core region. We will have to match the WKB solutions fixed in the Hankel basis above with the basis transported from the AdS boundary.

As an aside, we can generalize our considerations and define a leading oper $\mathfrak{D}_{d,g}[\trs]$, by replacing the coefficient of the double pole at $\rtr =0$ to be $\frac{g}{4}$. At $\trs =0$ this is still a Bessel operator, albeit of order $\frac{1}{d-1}\, \sqrt{1-g}$. Many of the above statements  can immediately be generalized to this case. This will be relevant for opers associated with polarizations of conserved currents. There is also one special case: $g =0$ in $d=4$ is the canonical Airy oper, which including subleading corrections was analyzed in~\cite{Jia:2026ryl}. Of course, we have already used its properties for our local crossing analysis at the turning points in~\cref{prt:WKB1,prt:WKB2}.

Having understood the turning points, let us turn to an important question: in which direction of the core coordinate chart should we normalize the solutions for large $\rtr$?  Fortunately, at $\trs = 0$, we can exploit known properties of the asymptotic Hankel basis~\eqref{eq:core-Hankel}. The Hankel solutions $H_0^{(1)}(z)$ and $H_0^{(2)}(z)$, whose asymptotic behavior is $e^{\pm \ii z}$, are recessive in the upper and lower half $z$ planes, respectively. The Hankel continuation identities of import are
\begin{equation}\label{eq:Hankelcontinue}
H_0^{(1)}(e^{\ii\pi}\,z) =  -H_0^{(2)}(z)\,,
\qquad
H_0^{(2)}(e^{\ii\pi}\,z) = H_0^{(1)}(z)  +2\, H_0^{(2)}(z) \,.
\end{equation}

Accounting for the argument of the Hankel solutions being $\rtr^{\frac{d-1}{2}}$ in the $\rtr$ coordinate chart, the boundaries of the oscillatory behavior are oriented along the rays
\begin{equation}
\Gamma_a: \arg \rtr = - \frac{2\pi}{d-1}\, a\,, \qquad a \in \set{0,1,\cdots d-2}\,.
\end{equation}
These are the anti-Stokes lines drawn in~\cref{fig:scaling-core-regions} on which the imaginary part of the WKB phase of $\psi\subm{core}^\pm$, which is $\Im(\rtr^{\frac{d-1}{2}})$, vanishes. We have ordered these clockwise.  The Stokes lines, on the other hand, are where the $\Re(\rtr^{\frac{d-1}{2}})$ vanishes. The Stokes domains are therefore
\begin{equation}
D_a: -\frac{(2\,a+1)\,\pi}{d-1}\, <\arg \rtr < - \frac{(2\,a-1)\, \pi}{d-1} \,.
\end{equation}
What we are after is the Stokes factor relating solutions across adjacent domains $D_a$. The domains between the anti-Stokes rays, $\Gamma_{a}$ and $\Gamma_{a+1}$, denoted $\Sigma_a$, which contain a single turning point $\rtr_a$ from the set~\eqref{eq:core-turningpts}, are useful to compare the bases associated with $D_a$ and $D_{a+1}$.

Let us fix the reference solution basis along the ray $\Gamma_0$, as the $\psi\subm{core}^\pm (\rtr;\trs)$ such that they  asymptote as stated in~\eqref{eq:core-Hankel}.
To obtain the solution basis along $\Gamma_1$ we can use the fact that the oper~\eqref{eq:core-oper} is homogeneous of degree $-2$ under discrete phase rotations,
$\rtr \to \lambda^{-1}\, \rtr$ and $\trs \to \lambda^{\frac{d-2}{4}}\, \trs$, with the phase $\lambda$ given below. This immediately gives a solution basis along $\Gamma_1$ as
\begin{equation}\label{eq:core-homogeneity}
\psi\subm{core}^\pm(\rtr; \trs) \bigg|_{\Gamma_1}
= \ii \lambda^{\frac{1}{2}}\, \psi\subm{core}^\mp(\lambda^{-1}\, \rtr; \lambda^{\frac{d-2}{4}}\, \trs)\bigg|_{\Gamma_0} \,,
\qquad \lambda = e^{-\frac{2\pi \ii}{d-1}}\,.
\end{equation}

Consider the wedge region $\Sigma_0$ bounded by the rays $\Gamma_0$ and $\Gamma_1$.
We continue the solutions normalized on $\Gamma_0$ and $\Gamma_1$ to compare them at a point in $\Sigma_0$.  Working through the phases we find
\begin{equation}
\psi\subm{core}^-(\rtr; 0) \bigg|_{\Gamma_1} = \psi\subm{core}^-(\rtr; 0) \bigg|_{\Gamma_0}\,,
\qquad
\psi\subm{core}^+(\rtr; 0) \bigg|_{\Gamma_1} = \psi\subm{core}^+(\rtr; 0) \bigg|_{\Gamma_0} + 2\ms\ii \psi\subm{core}^-(\rtr; 0) \bigg|_{\Gamma_0} \,.
\end{equation}
Thus, at $\trs = 0$ using the properties of the Hankel solutions we learn that $\psi\subm{core}^-(\rtr;0)$ is the recessive solution throughout the wedge. Moreover, the Hankel identity~\eqref{eq:Hankelcontinue} tells us how the dominant solution transforms. We can write
\begin{equation}
\pmqty{\psi\subm{core}^+(\rtr; 0) \\ \psi\subm{core}^-(\rtr;0)}_{\Gamma_0 \in D_0}
=  \pmqty{1 & -2\ms\ii \\ 0& 1} \, \pmqty{\psi\subm{core}^+(\rtr; 0) \\ \psi\subm{core}^-(\rtr;0)}_{\Gamma_1 \in D_1}
\end{equation}
The matrix relating these two solution bases is the canonical sectorial connection matrix of the core oper (at $\trs =0$). This is an exact connection matrix for the core oper~\eqref{eq:core-oper}. The core connection matrix
at $\trs =0$ is therefore $\mathsf{U}(-2\ii)$ replacing $\mathsf{U}(-\ii)$ of the Airy solutions.

\paragraph{The general case with $\trs \neq 0$:}

Once we have an analytic solution basis for $\mathfrak{D}_{d}[0]$ we can obtain the corresponding basis for $\mathfrak{D}_{d}[\trs]$ using the standard variation of parameters method. The solutions are given as
\begin{equation}
\psi\subm{core}^{\pm}(\rtr;\trs)
=
\psi\subm{core}^{\pm}(\rtr;0) + \frac{\trs^2}{4}\,
\int_{\rtr}^\infty\, \mathfrak{G}(\rtr,\rtr')\, \rtr'^{\frac{d}{2}-3}\,
\psi\subm{core}^{\pm}(\rtr';\trs)\, d\rtr'\,,
\end{equation}
where the Green's function is simply (the Wronskian factor is explicitly included)
\begin{equation}
\mathfrak{G}(\rtr,\rtr') = \frac{\ii}{2} \pqty{
  \psi\subm{core}^{-}(\rtr;0)\, \psi\subm{core}^{+}(\rtr';0)
  -
  \psi\subm{core}^{+}(\rtr;0)\, \psi\subm{core}^{-}(\rtr';0)
}\,.
\end{equation}
This can be solved iteratively as a Born series. The contribution after $n$ steps of iterated integration can be shown to be bounded for $\rtr \ge \rtr_c > 0$ (for some fixed $\rtr_c$) and so the series converges uniformly locally. In particular, at every step one ends up with a finite order polynomial in $\trs$. After factoring out the leading oscillatory amplitudes, the ordered
$n$-fold integrals are bounded by
$\bigl(C_c\,\abs{\trs}^{2}\bigr)^n/\Gamma(n+1)$,
uniformly on compact momentum sets.
Continuing the solution along a fixed path then gives solutions (and therefore Wronskians which we need below) whose dependence on the momentum is entire. Therefore, we can, in particular,  freely rotate the phase of $\zeta$ in this limit, allowing us to analyze the behavior as a function of $\arg\trs$ in one shot.

Once again the basis of solutions, $\psi\subm{core}^{\pm}(\rtr,\trs)$, is anchored on the ray $\arg \rtr =0$. We now require the core connection matrix, so that we can compare canonically normalized solution bases associated with adjacent normalization rays (after continuing both sets to a common domain). The core connection matrix is expected again to be upper triangular, since the recessive solution remains invariant under continuation; only the dominant solution picks up a contribution along the recessive one. The factor of import is the  \emph{core Stokes multiplier} $\mathfrak{R}(\trs)$; it replaces the $2\ms\ii$ of the Hankel connection formula.

The argument given for $\trs =0$ can be generalized to $\trs \neq 0$ and the core connection matrix takes the form
\begin{equation}\label{eq:core-conn-matrix}
\pmqty{\psi\subm{core}^+(\rtr; \trs) \\ \psi\subm{core}^-(\rtr;\trs)}_{D_0}
=  \pmqty{1 & \ii \mathfrak{R}_0(\trs) \\ 0& 1} \, \pmqty{\psi\subm{core}^+(\rtr; \trs) \\ \psi\subm{core}^-(\rtr;\trs)}_{D_1}\,.
\end{equation}
Once again the idea is that the recessive solution is common, but the dominant solution differs by a piece involving the recessive one. Given this, the core multiplier $\mathfrak{R}_0(\trs)$ across domains $D_0$ and $D_1$ may itself be computed directly in terms of the Wronskian between the two sets of dominant solutions on the anti-Stokes rays
\begin{equation}\label{eq:core-multiplier}
\mathfrak{R}_0(\trs)
=  -\frac{1}{2}\, \operatorname{\text{Wr}}(\psi\subm{core}^+\big|_{\Gamma_0},
\psi\subm{core}^+\big|_{\Gamma_1}) \,.
\end{equation}

While we worked out the relation between the two neighboring rays $\Gamma_0$ and $\Gamma_1$, we can readily generalize this to any pair $\Gamma_j$ and $\Gamma_{j+1}$ then the core Stokes multiplier between domains $D_j$ and $D_{j+1}$ to be given by
\begin{equation}\label{eq:core-conn-j}
\begin{aligned}
\pmqty{\psi\subm{core}^+(\rtr; \trs) \\ \psi\subm{core}^-(\rtr;\trs)}_{D_j}
=
\begin{dcases}
\mathsf{U}\pqty{\ii \mathfrak{R}_j(\trs)} \,
\pmqty{\psi\subm{core}^+(\rtr; \trs) \\ \psi\subm{core}^-(\rtr;\trs)}_{D_{j+1}} \,,
&\qquad \text{for}\ j \ \text{even}\,, \\
\mathsf{L}\pqty{\ii \mathfrak{R}_j(\trs)} \,  \pmqty{\psi\subm{core}^+(\rtr; \trs) \\ \psi\subm{core}^-(\rtr;\trs)}_{D_{j+1}}  ,
&
\qquad \text{for}\ j \ \text{odd}\,.
\end{dcases}
\end{aligned}
\end{equation}
The core multiplier can be deduced to be
\begin{equation}\label{eq:trs-j}
\mathfrak{R}_j(\trs) = \mathfrak{R}_0(\trs_j)\,, \qquad
\trs_{j} = e^{-\frac{\pi \ii}{2} \ms j \ms a_d} \, \trs \equiv
e^{-\frac{\pi \ii j\ms (d-2)}{2\,(d-1)}}\, \trs \,,
\end{equation}
in terms of $\trs_j$ the phase rotated core momentum parameter. The subscript $j$ on $\trs_j$
exists to emphasize that we should rotate $\trs$ if we are keeping $\zeta$ fixed.
The rotation of $\trs$ follows from the homogeneity of the oper under simultaneous scaling of both the coordinate and the momentum parameter~\eqref{eq:core-homogeneity}.
In writing the above we have introduced a shorthand for the $d$ dependence by introducing $a_d$. Going forward, expressions are succinct if we adopt the notation
\begin{equation}\label{eq:abd-def}
a_d =  \frac{d-2}{d-1}
\,, \qquad
b_d = \frac{d}{d-1} = 2-a_d\,,
\end{equation}

\paragraph{Frequency corrections:}  As indicated above, this core Stokes multiplier factor has corrections starting with terms of order $\freq^{-\alpha}$ for  $\alpha = \frac{d}{d-1}$ for planar  (and $\alpha = \frac{d-2}{d-1}$ for global) \SAdS{d+1} black holes.

These corrections can also be worked out using the very same variation of parameters method. In particular, we can extend~\eqref{eq:core-multiplier} to include frequency corrections. Let $\delta_\freq Q$ be the corrections to the core oper~\eqref{eq:core-oper}. Explicitly, for neutral planar black holes
\begin{equation}\label{eq:deltaQ-core}
\delta_\freq Q = \sum_{n=1}^\infty\,
\frac{\freq^{-n\ms b_d}}{4}\, \bqty{
  (n+1)\, \rtr^{\frac{(n+2)\ms d}{2}-3}
  +\trs^{2}\, \rtr^{\frac{(n+1)\ms d}{2} -3}
  +\left(\nu^2+\frac{n\, d^2}{4}\right)
\rtr^{\frac{n\ms d}{2}-2}} \,.
\end{equation}
Given this we solve for the corrections to $\psi\subm{core}^\pm$ starting from the seed solution to~\eqref{eq:core-oper} using variation of parameters. Working in the domain $\Sigma_0$ the solutions are normalized along the rays $\Gamma_0$ and $\Gamma_1$ to have no additional constant in the asymptotic exponent (the WKB momentum), and have unit leading WKB coefficient.  This fixes the integration constants for the solutions we work with. Denote these solutions as $\psi\subm{core}^\pm(\rtr; \trs,\freq)$ with a formal power series expansion
\begin{equation}\label{eq:core-psipm-freq}
\left.\psi\subm{core}^\pm(\rtr; \trs,\freq) \right|_{\Gamma_j}=
\left.\psi\subm{core}^\pm(\rtr; \trs) \right|_{\Gamma_j}+
\sum_{n=1}^\infty \, \freq^{-n\ms b_d} \; \left.\delta_n\psi\subm{core}^\pm(\rtr; \trs)\right|_{\Gamma_j}\,.
\end{equation}

Plugging the core solutions thus obtained into~\eqref{eq:core-multiplier} gives us a formal power series
\begin{equation}\label{eq:core-Rformal-Sigma0}
\mathfrak{R}_0(\trs, \freq) = \mathfrak{R}_0(\trs) + \sum_{n=1}^\infty\, \freq^{-n\,b_d}\, \delta_n \mathfrak{R}_0(\trs)\,.
\end{equation}
This can be worked out to any desired order iteratively. For instance, the first variation of the core Stokes multiplier as a function of frequency is then given to be
\begin{equation}\label{eq:core-R1}
\delta_1\mathfrak{R}_0(\trs)
=
-\frac{1}{2}\, \operatorname{FP}\,
\int_{\infty_0}^{\infty_1}
\operatorname{Coeff}\bqty{\freq^{-b_d}} \Big(\delta_{\freq} Q \Big)\,
\left.\psi_{\rm core}^{+}\right|_{\Gamma_0}
\left.\psi_{\rm core}^{+}\right|_{\Gamma_1}
\,d\rtr\,.
\end{equation}
Here we are imagining the integration path coming in from infinity along $\Gamma_0$ and exiting back to infinity along $\Gamma_1$. The prescription for taking the finite part (FP) uses the normalized solutions  with no constants in the asymptotic WKB primitives along the specified rays. Having understood what happens across the primary Stokes domains $D_0$ and $D_1$ (whose normalized solutions were analytically continued to a common comparison wedge $\Sigma_0$), we can extend this discussion to any adjacent pair of rays $\Gamma_j$ and $\Gamma_{j+1}$. The only difference is that in addition to rotating the momentum we should also rotate the frequency corrections,

\begin{equation}\label{eq:core-Rformal-Sigmaj}
\mathfrak{R}_j(\trs, \freq) = \mathfrak{R}_0(\trs_j) + \sum_{n=1}^\infty\, \freq^{-n\,b_d}\, e^{-\ii\pi \ms j\ms n\ms b_d}\, \delta_n \mathfrak{R}_0(\trs_j)\,.
\end{equation}

Let us assume that this has been achieved to any given desired order in the inverse frequency expansion, at least as a formal power series in $\freq$. We reserve $\mathfrak{R}(\trs,\freq)$ for the resulting core multiplier when obtained as this formal power series. One could further Borel resum the $\freq$ series, and the resulting multiplier will be denoted as $\mathcal{R}^{(\vartheta)}(\trs,\freq)$. To wit,
\begin{equation}
\mathcal{R}_j^{(\vartheta)}(\trs,\freq) = \mathcal{S}_{\vartheta} \bqty{\mathfrak{R}_j(\trs,\freq)}\,.
\end{equation}
We will analyze the Borel series more explicitly in~\cref{sec:core-wkb-borel}.

Once we have the core Stokes multiplier we are essentially done.  The core problem will then map dominant and recessive solutions across domains by factors of
\begin{equation}
\mathsf{U}(\ii\mathcal{R}_j^{(\vartheta)}(\trs,\freq))
\,, \qquad
\mathsf{L}(\ii\mathcal{R}_j^{(\vartheta)}(\trs,\freq))\,,
\end{equation}
depending on the parity of $j$, respectively.

\subsection{The transport through the core}\label{sec:core-transport}

\begin{figure}[ht!]
  \centering
  \includegraphics[width=0.9\linewidth]{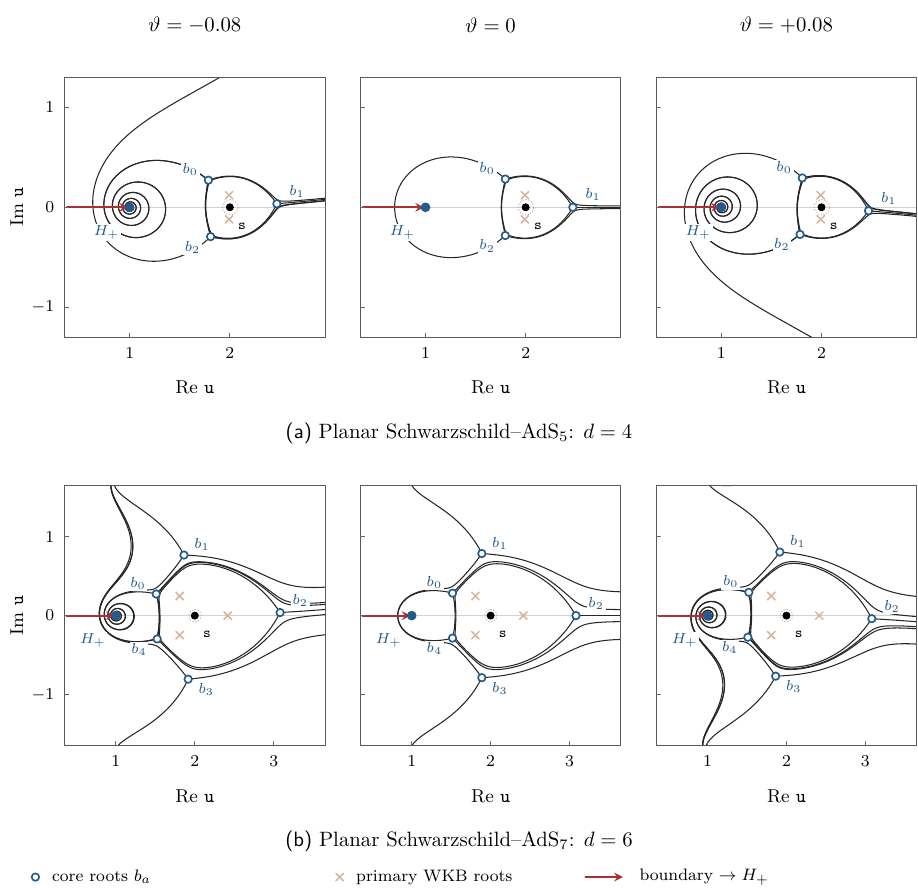}
  \caption{Lateral resolution of the full finite-frequency Stokes graph in the
    core regime, in the physical $\mathtt u$-chart for planar \SAdS{d+1} with $d=4,6$. We show a region that comprises the horizon and singularity.
    We chose $\freq=12\, e^{\ii\vartheta}$, $\nu= 0.75$, and held $\zeta$
  fixed with $\trs^{\,2}=0.1$ at $\vartheta=0$. The phase rotated panels with negative phase on left and  positive phase on the right therefore have  $\trs_\vartheta^{\,2}=0.1e^{id\vartheta/(d-1)}$. Our choice of parameters makes the core turning points (blue open circles, labeled $b_i$ for simplicity) visible explicitly. In reality the entire core region will shrink as $\freq \gg 1$, and the Stokes lines will emanate from close to the singularity at $\upr = 2$. The primary turning points are still indicated as orange crosses for reference. The Stokes lines circulate around $\upr =2$, and spiral into the horizon (for $\vartheta \neq 0$). The zoomed in perspective does not capture Stokes lines from the core turning point that cross the transport path close to the boundary. This does not happen in $d=4$, but in $d=6$ a line from $b_2$ intersects the transport path.}
  \label{fig:core-stokes-laterals}
\end{figure}

We are finally in a position to assemble all the pieces and construct the analog of~\eqref{eq:wkb-C-general}. There are four factors in that transport formula, of which  only $\mathbb{K}$ is modified owing to the presence of the core region. The factors
$\mathcal{B}_\bdy$ and $B_\hor$ are as given in~\eqref{eq:wkb-boundary-exact-undetermined-residual} and ~\eqref{eq:wkb-horizon-unit-matching-matrix}, respectively. We once again will not need to dwell on the factor $\mathsf{D}\pqty{\mathcal{E}_{\bdy,\hor}}$ which will drop out of physical observables. All that is left is the computation of the matrix $\mathbb{K}$ for transport through the core.

\begin{figure}[tbhp!]
  \centering
  \begin{subfigure}[t]{\textwidth}
    \includegraphics[width=0.93\linewidth]{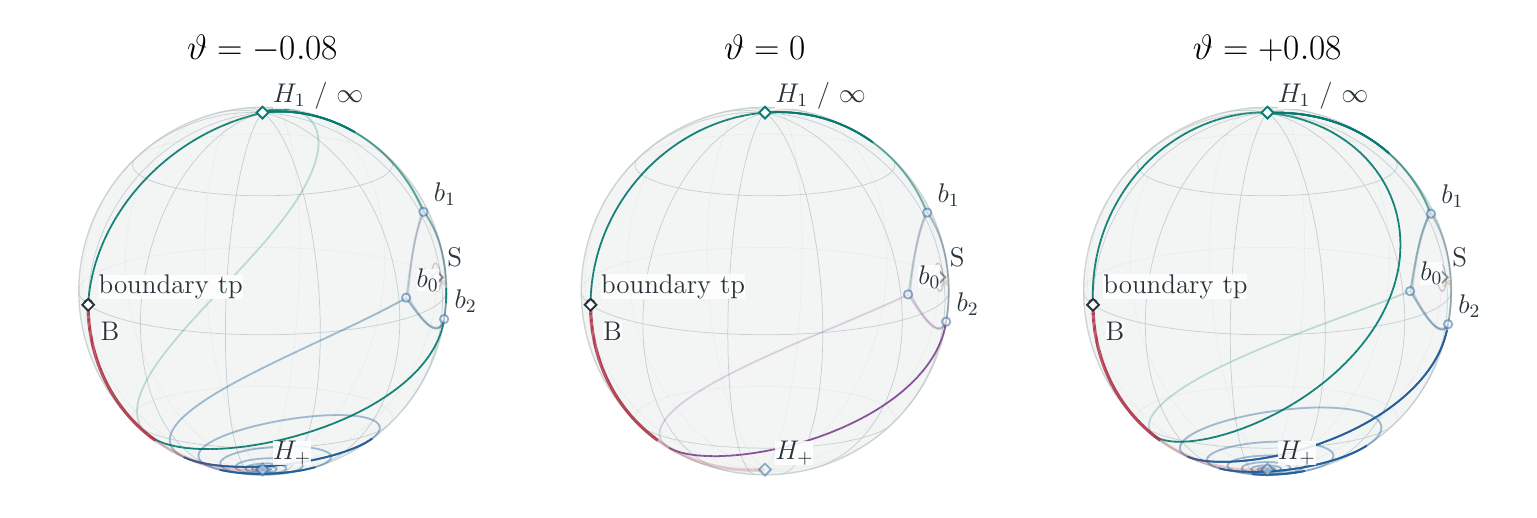}
  \end{subfigure}
  \begin{subfigure}[t]{\textwidth}
    \includegraphics[width=0.93\linewidth]{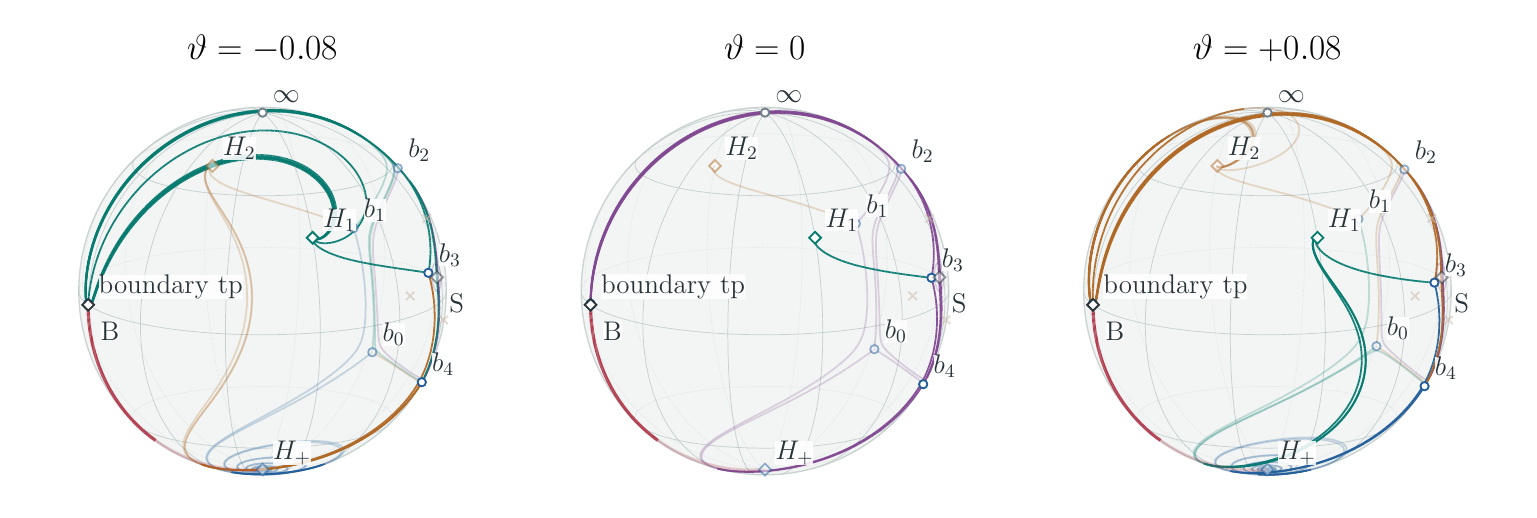}
  \end{subfigure}
  \caption{The Stokes graphs for \SAdS{5} and \SAdS{7} in the core-scaling regime drawn on the Riemann sphere. We use the resolved core coordinate and project it onto the sphere. The parameters chosen are $\abs{\freq}=12$,  $\nu=0.75$, and $\trs_0^2=0.1$. The lines ending on the near boundary turning-point are now clear in the \SAdS{7} example. See \href{https://mukundrangamani.github.io/stokes-atlas/}{Stokes atlas visualizer} for an interactive version.}
  \label{fig:planar-sads5-core-sphere}
\end{figure}

The matrix $\mathbb{K}$ still has the feature that it is a product of connection matrices.  This can be immediately understood if we draw the Stokes graph with the core retained, see~\cref{fig:core-stokes-laterals} and~\cref{fig:planar-sads5-core-sphere}.  Following the transport path from the boundary to the horizon we again encounter Stokes crossings, which we follow to their emitting turning point. In this case, with our core resolution, these are the core turning points $\rtr_a$ given as roots of~\eqref{eq:core-poly}, which has now been translated into the $\upr$ chart so that the horizon can be made visible.\footnote{In the graphs we have denoted them as $b_a$ to avoid cluttering the figure.} As before at $\arg \freq =0$ we have a saddle graph, so~\cref{fig:core-stokes-laterals} also displays the lateral resolutions for small $\arg \freq =  \vartheta$.

When implementing the lateral resolution we keep $\zeta$ fixed, and therefore have to transform $\trs$. Furthermore, the phase rotation resolves the double pole of the oper at the horizon into spirals, but it does not do so for the singularity (at $\upr =2$ in this chart). This is because the coefficient of the inverse-square term at the singularity is fixed at $\frac{1}{4}$ and is frequency independent. So its Stokes contours are closed loops encircling the double pole, limiting to a self-returning loop anchored on one of the core roots. In $d=4$ (resp.~$d=6$) this loop runs from root $\rtr_{a=1}$ (resp.~$\rtr_{a=2}$) back to itself, but in doing so passes very close to the other roots. The dynamic range in the plots makes this hard to disambiguate. Each core root is a simple turning point, so it can have no more than three lines connecting it to other loci. For comparison to the local analysis of ~\cref{sec:core-multiplier}, let us note that the principal rays $\Gamma_0$ and $\Gamma_1$ used there, are emitted from $b_0$. What is not explicit in~\cref{fig:core-stokes-laterals} is the fact that there is a Stokes line in $d=6$ from $b_2$ that crosses the transport path close to the boundary. The lines of interest are $\Gamma_{-3}$ and $\Gamma_{-2}$ delimiting sectors with fixed normalized WKB solutions, so we need to rotate the core-multiplier accordingly. We can now assemble the pieces necessary to compute $\mathbb{K}$.

\paragraph{The transport to the core:} First, assume that the core Stokes multiplier is known, and the core connection matrix is given in terms of it as $\mathsf{U}(\ii\eta\ms\mathcal{R}^{(\vartheta)}(\trs,\freq))$ or $\mathsf{L}(\ii\eta\ms\mathcal{R}^{(\vartheta)}(\trs,\freq))$ depending on the lateral rotation and the domain in question.  Next, let the transport from the boundary to the singularity be determined in terms of the leading WKB potential as in~\cref{prt:WKB1}. Further, let the transport from the boundary to the core be captured by the Voros symbol
\begin{equation}\label{eq:Xcore}
\mathcal{X}\subm{core}
= \exp(2\ii \freq\, \int_{\mathrm{sing}}^\bdy \lwkb(\zeta=0) )
= e^{i\,\freq \, \Iwkb(\zeta =0)}\,,
\end{equation}
which we recognized to be the WKB action~\eqref{eq:Iwkb} evaluated at $\zeta =0$.
We will examine this closely in~\cref{sec:core-data-transport}.   This will give a diagonal factor $\mathsf{D}(\mathcal{X}\subm{core}^{\frac{1}{2}})$ in the transport.   We combine this factor with the hitherto obtained core connection matrix for the crossings.
This entails computing $ \mathsf{D}(\mathcal{X}\subm{core}^{\frac{1}{2}})^{-1}\,
\mathsf{M}(\ii\eta\ms \mathcal{R}^{(\vartheta)} )\,  \mathsf{D}(\mathcal{X}\subm{core}^{\frac{1}{2}})$, where $\mathsf{M} \in \set{\mathsf{U}, \mathsf{L}}$ and ends up producing factors
\begin{equation}\label{eq:wkb-core-rule}
\mathsf{K}\subm{bdy,core}
=
\begin{dcases}
\mathsf{U}\pqty{\ii\eta\ms \mathcal{X}\subm{core}^{-1} \, \mathcal{R}(\trs,\freq)}
& \qquad \Psi_+ \ \text{dominant},
\\[1ex]
\mathsf{L}\pqty{\ii\eta\ms \mathcal{X}\subm{core} \, \mathcal{R}(\trs,\freq) }\,,
& \qquad \Psi_- \ \text{dominant}\,.
\end{dcases}
\end{equation}
Thus, the basic transport from the boundary to the core is dictated precisely by our naive WKB action at $\zeta =0$ (when our intuitive explanation in~\cref{prt:WKB1} would have obviously held). The factor will enter our considerations as one of the Stokes lines from the core turning points will cross the transport path.
This line is unfortunately not visible in Stokes graphs displayed in~\cref{fig:core-stokes-laterals}, but can be seen in~\cref{fig:planar-sads5-core-sphere} and will be made explicit in examples below.

\paragraph{The horizon spirals:} The part that is visible is where the horizon spirals intersect the transport path. These require us to transport from the crossing to the core turning point, and account as before, for the horizon winding. The bare transport to the turning point will result in a factor
$\mathcal{X}_a$, and the winding gives a thermal factor, all in all leading to factors
\begin{equation}
\prod_{n\geq 0}\, \mathsf{U}(\ii \eta\ms e^{-n\ms\bw} \, \mathcal{X}_a) = \mathsf{U}\pqty{
\ii \eta\ms \frac{\mathcal{X}_a}{1-e^{-\bw}}}\,.
\end{equation}
However, now we have to give a prescription to compute $\mathcal{X}_a$, for it unlike the symbols appearing in~\cref{sec:ewkb-connection} has to be computed exactly in the core region. We shall do so explicitly in~\cref{sec:core-data-spirals}.

\paragraph{Assembling $\mathbb{K}\subm{core}$:}
With the Stokes crossings accounted for, we can then argue that  upon assembling the pieces we will end up with a simple modification of~\eqref{eq:Kbb-gen}. The replacement rule we need is simply\footnote{We have not yet established that we have determined the core WKB solutions with exponential accuracy. Technically, our statement should be read as a formal asymptotic equality. We explore the Borel structure of the core region in~\cref{sec:core-wkb-borel} and defer these considerations until then.}
\begin{equation}\label{eq:Xstar-core-rep}
\mathcal{X}_\star^{-1} \mapsto - \mathcal{X}\subm{core} \, \mathcal{R}(\trs,\freq) \,.
\end{equation}
Accounting for the lateral rotation and ordering we use in the core analysis we find that~\eqref{eq:Kgen-pm} holds with~\eqref{eq:Xstar-core-rep}. In particular, one can obtain
\begin{equation}\label{eq:Kbb-core}
\mathbb{K}\subm{+,core}^{-1} =
\begin{pmatrix}
  1 - \mathfrak{s}(\mathcal{X})\, \mathcal{X}\subm{core}\, \mathcal{R}(\trs,\freq)\,
  \,
  & \quad \ii \mathfrak{s}(\mathcal{X}) + \ii \, \mathcal{X}_f\, \pqty{
  1- \mathfrak{s}(\mathcal{X})\, \mathcal{X}\subm{core}\, \mathcal{R}(\trs,\freq)}
  \\
  \ii\mathcal{X}\subm{core}\,  \mathcal{R}(\trs,\freq)
  &1 - \mathcal{X}_f\, \mathcal{X}\subm{core}\,  \mathcal{R}(\trs,\freq)
\end{pmatrix}\,.
\end{equation}
The factor $\mathfrak{s}(\mathcal{X}_a)$ captures the contribution from the transport to the core turning points from all the spiral crossings. Schematically, it takes the form
\begin{equation}
\mathfrak{s}(\mathcal{X}) = \frac{1}{1-e^{-\bw}} \, \sum_a\,\mathcal{X}_a\,.
\end{equation}
We emphasize  that this factor essentially involves transport all the way to the singularity insofar as the exterior region is concerned.  The additional factor of $\mathcal{X}_f$ is the contribution from a Stokes line that intersects the transport path but does not spiral into the horizon.

\paragraph{The boundary Green's function:}
Given this expression and the remaining factors, we can immediately write down the analog of~\eqref{eq:ewkb-Gret}, which fortunately, only involves the lower row of the above. We have
\begin{equation}
\Gret = \Gret^{\mathrm{vac}} \, \mathcal{T}\subm{bdy}^2\,
\frac{1 - \pqty{e^{\ii\pi\,\nu} + \mathcal{X}_f}\, \mathcal{R}(\trs,\freq)\, \mathcal{X}\subm{core}}{1 - \pqty{ e^{-\ii\pi\,\nu} + \mathcal{X}_f}\, \mathcal{R}(\trs,\freq)\, \mathcal{X}\subm{core}}
\end{equation}
In this schematic expression it appears that the horizon factors and the terms proportional to $\mathcal{X}_a$ drop out. This will be true for one sign of lateral rotation but not for the other. Therefore, the median prescription will carry information about the spiral crossings and transport contained in $\mathfrak{s}(\mathcal{X})$. We will find that the factors entering the Green's function from the transport will be
\begin{equation}
\mathcal{R}^{(+)}(\trs,\freq)\, \mathcal{X}^+\subm{core} \,,
\qquad
\frac{\mathfrak{s}(\mathcal{X}^-)}{1+ \mathfrak{s}(\mathcal{X}^-)\, \mathcal{R}^{(-)}(\trs,\freq)\, \mathcal{X}^-\subm{core}}\,,
\end{equation}
for the two choices of phase rotation. The point of import is that the leading order answer will be independent of the $\mathcal{X}_a$.

\section{Core-scaling regime of AdS black holes}\label{sec:core-examples}

We now examine the transseries obtained in the core-scaling regime. We will first work out the core Stokes multiplier for neutral black holes, and then examine specific cases in detail to extract the median answer for $\mathbb{K}$ and the non-perturbative part of the spectral function.

Some aspects of the core region data necessary for the exact-WKB analysis can
be worked out universally. We will therefore compute these first, before specializing to Voros symbols that depend on the backgrounds in question.

\subsection{Transport to the core region}\label{sec:core-data-transport}

There is one useful simplification in the computation, which is worth understanding at the outset. The core transport factor $\mathcal{X}\subm{core}$ is actually universal, and is precisely given by~\eqref{eq:Xcore}. The key argument is the contour deformation described in~\cref{fig:planar-Pphys-deformation},
along with a few other observations.

First, note that the leading WKB differential $\lwkb$ defined in~\eqref{eq:lambdaWKB} which is obtained from the Schr\"odinger potential $Q_0$ has a simple pole (for non-extremal black holes being considered here) at the horizon loci, i.e., at roots of $f(r)$, say $r = r_k$.  This translates to a statement about the all orders odd momentum\footnote{The residues are computed with the branch choice $P_{\mathrm{odd}}\,dr = -i\,\omega\, \lwkb$ with $\lwkb$ taken to be positive outside the horizon.}
\begin{equation}
\Res_{r_k} P_{\mathrm{odd}}\, dr = -\frac{\ii \beta_k\ms \omega}{4\pi} \,.
\end{equation}
Second, for even-dimensional static black holes, the metric function is even. Third, expanding around $\zeta =0$, we remove all the turning points by Taylor expanding.
The full potential then expands in even powers of the radial coordinate $r$. Integrating from the singularity to the boundary termwise one finds that the finite part vanishes. The statement follows once one checks that the horizon residues vanish for the subleading odd momenta $P_n$ with $n\geq 1$. This implies that along any path $C: \mathrm{sing} \to \bdy$ the finite part of the odd momentum contribution  collapses to the integral of the tortoise coordinate differential $dr_*$, viz.,
\begin{equation}
\frac{1}{2}\,V\subm{core}
= \ii \operatorname{\text{FP}}\int_C\, \frac{1}{\freq} \,P_{\mathrm{odd}}\, dr
= \rp\, \int_C \, \frac{dr}{r^2\,f(r)}
\equiv
\rp\,  \int_C \, dr_* \,.
\end{equation}
The integral itself can be evaluated by an analog of the sum rule~\eqref{eq:Psumrule}. Extend the range of the integral which is over the half-line $r=0$ to $r =\infty$ to the full real axis and close the contour over the upper half-plane. We pass below each real horizon locus by a semicircular deformation (note orientation is reversed relative to the convention in~\cref{fig:planar-Pphys-deformation}).   Each complex horizon in the upper half-plane gives a residue term -- this is precisely what the principal value integral for $\Pphys$ computes at $\zeta =0$. Each semicircular arc picks up half a residue. Since real horizon loci come in pairs, the contributions add, and it suffices to look at those on the positive axis. This leads us to
\begin{equation}
\rp\, \int_{0}^\infty \, \frac{dr}{r^2\,f(r)}
= \ii \sum_{\Im r_k > 0} \, \frac{\rp\,\beta_k}{4} + \ii
\sum_{r_j > 0}\, \frac{\eta_j\, \rp\, \beta_j}{4}\,.
\end{equation}
Here $\eta_j$ as before is $\pm 1$ for passage below or above the real pole.

Therefore, the transport to the core is fixed completely by the leading order WKB action, and the lateral rotations are trivial to account for by conjugation.
To wit,
\begin{equation}\label{eq:Xcore-lateral}
\mathcal{X}\subm{core}^{+}
= e^{\ii \freq \ms V\subm{core} } \,, \qquad
\mathcal{X}\subm{core}^{-}
= e^{-\ii \freq \ms V^*\subm{core} }
\,,
\qquad V\subm{core} \equiv \Iwkb(\zeta =0)\,.
\end{equation}

\paragraph{Examples:}
We can immediately write down the factors appearing for various examples:
\begin{itemize}[wide,left=0pt]
  \item  For planar \SAdS{d+1} black holes with $d$ even we have
    $r_k = r_+\, e^{\frac{2\pi \ms\ii k}{d}}$ and  $\beta_k = \frac{4\pi}{d\, r_k}$, with $k =0, \cdots, d-1$ leading to the familiar result~\cite{Ceplak:2024bja,Dodelson:2025jff}
    \begin{equation}
    \frac{1}{2}\, \Iwkb(\zeta = 0)
    = \frac{\ii \pi}{d} + \frac{\ii \pi}{d} \,
    \sum_{k=1}^{\frac{d}{2}-1}\, e^{-\frac{2\pi \ms\ii k}{d}} =  \frac{\pi}{d} \pqty{\ii + \cot\frac{\pi}{d}} \,.
    \end{equation}
  \item For global \SAdS{d+1} black holes we need the roots of $r^2 + 1 - \frac{\rp^{d-2}}{r^{d-2}}\, (1+\rp^2)$. We find
    \begin{equation}
    \frac{1}{2}\,\Iwkb(\zeta =0)
    =
    \begin{dcases}
    \frac{\pi \ms\rp}{2\, (2\,\rp^2 +1)} \, \bqty{\ii \rp + \sqrt{1+\rp^2}}
    &\qquad d=4\,,\\
    \frac{\pi\ms\rp}{2\,(3\,\rp^2 +2)}
    \bqty{\ii \rp + \frac{2\, (1+\rp^2) + \rp \, \sqrt{1+\rp^2}}{\sqrt{1+ \rp^2 + 2\, \rp\, \sqrt{1+ \rp^2}}}}, &\qquad d =6\,.
    \end{dcases}
    \end{equation}
    For general even $d$ one can write a convergent series expansion (for $\abs{M} > M_c(d)$ with $M_c(d) <1$ which holds for black holes dominating the canonical ensemble)
    in the inverse mass $M = \rp^{d-2}\, (1+\rp^2)$ as
    \begin{equation}
    \frac{1}{2}\,\Iwkb(\zeta =0)
    =
    \frac{e^{\frac{\ii \pi}{d}}\,\rp}{d\, M^{\frac{1}{d}}}\,
    \sum_{n=0}^\infty\, \frac{\Gamma\pqty{\frac{2\,n+1}{d}}\, \Gamma\pqty{n + 1 - \frac{2\,n +1}{d}}}{\Gamma(n+1)}\, \pqty{-\frac{e^{\frac{\pi\ms \ii}{d}}}{M^{\frac{1}{d}}}}^{2n}\,.
    \end{equation}
  \item For \RNAdS{5} we have the inner and outer horizons, which are real loci $r = r_\pm$ and a complex horizon at $r = i\sqrt{r_+^2 + r_-^2}$.  Using the
    inverse temperatures  given in~\eqref{eq:rnads5-residue-horizon-periods}, we  obtain
    \begin{equation}
    \Iwkb(\zeta =0) = \frac{r_+}{2}\,
    \bqty{\beta_c + \ii (\beta - \beta_{\mathrm{inner}})} \,,
    \end{equation}
    which is the limiting value of~\eqref{eq:rnads5-relative-action-explicit} at vanishing momentum.
  \item Notice that the product of laterally resolved core transport factors satisfies a simple relation
    \begin{equation}\label{eq:Xcore-neutral-identity}
    \mathcal{X}\subm{core}^+ \, \mathcal{X}\subm{core}^- = e^{-\bw + \beta_{\mathrm{inner}}\ms\omega}\,.
    \end{equation}
    This follows from~\eqref{eq:Xcore-lateral} and the explicit expressions above. The inner horizon contribution is absent for neutral black holes. This result is valid for real frequencies.
\end{itemize}

\subsection{Horizon spirals}\label{sec:core-data-spirals}

Having understood the core transport factor, we should next account for the contribution from the horizon windings. A priori one should expect that it is largely determined in terms of $\mathcal{X}\subm{core}$ and the core Stokes multiplier. We will indeed confirm this to be the case.

Across a spiral crossing the dominant solution picks up a contribution of the recessive one. Since we are heading to the horizon the dominant solution in our retarded convention is the outgoing mode, while the recessive one is the ingoing mode. The symbol $\mathcal{X}_a$ heading toward a core root $\rtr_a$ is defined by
\begin{equation}
\Psi\subm{out}^{\mathrm{before}} = \Psi\subm{out}^{\mathrm{after}} - \ii \mathcal{X}_a\, \Psi\subm{in}
\end{equation}
So once again we can extract $\mathcal{X}_a$ by invoking the Wronskian between the solution bases before and after the crossing
\begin{equation}\label{eq:Xa-core-Wronskian}
\mathcal{X}_a = -\ii \frac{\operatorname{\text{Wr}}\pqty{\Psi\subm{out}^{\mathrm{before}}, \Psi\subm{out}^{\mathrm{after}}} }{\operatorname{\text{Wr}}\pqty{\Psi\subm{out}^{\mathrm{after}},  \Psi\subm{in}}}\,.
\end{equation}

The problem of computing the Wronskians can now be factorized. Assume that we have started with the WKB basis at the boundary. After crossing the Stokes lines that are not involved in the horizon spirals, we have a WKB basis $\Psi_\pm^{\mathrm{int}}$. The two bases are related by a factor $\mathsf{U}(-\ii \mathcal{X}_f)$, which we should account for in due course. The remaining piece is the direct comparison of the ingoing and outgoing modes across a Stokes line. This can be done at the first two Stokes crossings, noting that the two crossings head to different core turning points and are oppositely oriented. The local analysis here is similar to the general $\zeta$ exact-WKB, and the aforementioned difference is visible say in~\eqref{eq:Kmat-d4}.

With this understanding, we can ask how the ingoing and outgoing waves at the horizon behave in the core region. This is sufficient to evaluate the Wronskians explicitly. Consider two consecutive domains $\Sigma_0$ between $\Gamma_0$ and $\Gamma_1$, and $\Sigma_{-1}$ between $\Gamma_{-1}$ and $\Gamma_{0}$; cf.~\cref{fig:scaling-core-regions}. We can use the core Stokes multiplier, which relates solutions as
\begin{equation}
\begin{aligned}
\left.\psi\subm{core}^+\right|_{\Gamma_1}
&=
\left.\psi\subm{core}^+\right|_{\Gamma_0} - \ii \mathfrak{R}_0(\trs,\freq)\,\left.\psi\subm{core}^-\right|_{\Gamma_0} \,,
\\
\left.\psi\subm{core}^-\right|_{\Gamma_{-1}}
&=
\left.\psi\subm{core}^-\right|_{\Gamma_0} + \ii \mathfrak{R}_{-1}(\trs,\freq)\,\left.\psi\subm{core}^+\right|_{\Gamma_0} \,.
\end{aligned}
\end{equation}
Take $\psi\subm{core}^+$ to be the reference ingoing solution, common to these three rays. The outgoing representatives are determined by picking representatives along each ray, transporting them all to the same reference point, and normalizing them to have Wronskian $2\ii$ with our reference ingoing solution. Our choice of outgoing solutions obtained thus is
\begin{equation}
\left.\psi\subm{core}^-\right|_{\Gamma_{-1}} \,,\qquad
\frac{1}{-\ii \mathfrak{R}_0(\trs,\freq)}\, \left.\psi\subm{core}^+\right|_{\Gamma_{1}}\,,\qquad
\left.\psi\subm{core}^-\right|_{\Gamma_0}\,,
\end{equation}
respectively. The ordering matters here: these are three successive outgoing solutions. So the Wronskians we want to compute should be between successive pairs, but this is now easily done since we have fixed all the normalizations. Moreover, it is clear that the result will only depend on the core reflection factor, and the transport to the core.
For positive lateral rotation, we then find the first crossing involves the first two  outgoing solutions in the above, while the second crossing involves the second two. Denote the two transport factors as $\mathcal{X}_{1,2}$.

Should there be no Stokes line from a core root prior to the horizon spirals, viz., when $\mathcal{X}_f = 0$ we find the following simple expression for the factors
\begin{equation}\label{eq:Xa-core-factors}
\begin{aligned}
\mathcal{X}_1^{+}
&\sim
\mathcal{X}\subm{core}^{-} \bqty{-\mathfrak{R}_{-1}(\trs,\freq) + \frac{1}{\mathfrak{R}_0(\trs,\freq)}} \\
\mathcal{X}_2^{+}
&\sim
- \mathcal{X}\subm{core}^{-} \; \frac{1}{\mathfrak{R}_0(\trs,\freq)} \,.
\end{aligned}
\end{equation}
When we do have a Stokes line crossing the transport path we should include its contribution. The factor $\mathcal{X}_f$ itself can be determined using the same strategy. We will defer it to the examples below.

\subsection{Neutral black holes}\label{sec:core-neutral}

We now turn to assembling the pieces for neutral black holes. We will largely focus on planar black holes, and indicate changes necessary at the end for the global ones.

\subsubsection{The core Stokes multipliers}\label{sec:core-data-multiplier}

Having established what the transport factor to the core geometry is, we can proceed with the transport to the core region. Therefore,  we turn next to the core Stokes multiplier. Our goal is to determine it as a function of $\trs$ to leading order in the frequency expansion. We claim that one can obtain the following series expansion
\begin{equation}\label{eq:core-neutralr-mom}
\mathfrak{R}_0(\trs)
= -2 - \sum_{n=1}^\infty\, \mathfrak{r}_{n}\,
e^{-\frac{\ii\pi}{2}\, \frac{d-2}{d-1}\, n }\, \trs^{2\,n}
\end{equation}
The first few coefficients can be explicitly obtained for planar \SAdS{d+1} black holes to be
\begin{equation}
\begin{aligned}
\mathfrak{r}_1
& =
4\, \sin^2\pqty{\frac{\pi\ms a_d}{2}} \, \mathfrak{I}_1
\\
\mathfrak{r}_2
&=
2\, \sin^2(\pi\, a_d)\, \mathfrak{I}_2 - 4\, \sin^2\pqty{\frac{\pi\ms a_d}{2}} \,
\cos\pqty{\pi\, a_d} \, \mathfrak{I}_1^2
\\
\mathfrak{r}_3
&=
\frac{4}{3}\, \sin^2\pqty{\frac{3\pi\ms a_d}{2}}\, \mathfrak{I}_3
- 4\, \sin\pqty{\frac{\pi\ms a_d}{2}} \,
\sin\pqty{\pi\, a_d} \, \cos\pqty{\frac{3\pi\ms a_d}{2}}  \mathfrak{I}_1 \, \mathfrak{I}_2
\\
& \qquad
- \frac{8}{3}\, \sin^3\pqty{\frac{\pi\ms a_d}{2}} \,
\sin\pqty{\frac{3 \pi\ms a_d}{2}}  \mathfrak{I}_1^3\,.
\end{aligned}
\end{equation}
The Born series integrals involve modified Bessel functions and are given as
\begin{equation}\label{eq:Ij-Rcore}
\begin{aligned}
\mathfrak{I}_1
&= (d-1)^{a_d -2}\,
\int_0^\infty\, dt\, t^{a_d - 1}\, I_0(t)\, K_0(t)
=
\Big(2\, (d-1)\Big)^{a_d -2}\,
\frac{\Gamma\pqty{\frac{a_d}{2}}^2\, \Gamma(1-a_d)}{\Gamma\pqty{1-\frac{a_d}{2}}^2}\,,
\\
\mathfrak{I}_2
&=
2\, (d-1)^{2(a_d - 2)} \,
\int_0^\infty\,dt\,  t^{a_d -1 }\, K_0(t)^2 \;
\int_0^t\, du\, u^{a_d -1 }\, I_0(u)^2\,,
\\
\mathfrak{I}_3
&=
6\, \pqty{d-1}^{3\,(a_d-2)} \,
\int_0^\infty\,dt\,  t^{a_d -1 }\, K_0(t)^2 \,
\int_0^t\, du\, u^{a_d -1}\, I_0(u)\, K_0(u) \,\int_0^u\, dv\, v^{a_d -1}\, I_0(v)^2 \,.
\end{aligned}
\end{equation}
Higher coefficients can be systematically extracted similarly.  Recall that $a_d$ and $b_d$ are introduced in~\eqref{eq:abd-def}. The derivation for these expressions is given in~\cref{app:determinants}.

While one can obtain this by using the Green's function obtained using variation of parameters and computing Wronskians as explained in~\cref{sec:core-multiplier}, it turns out to be efficient to relate this coefficient to a ratio of determinants of Schr\"odinger operators. We explain how to do this in~\cref{app:determinants}.
Explicitly, to the first three orders we find
the following series in $d=4$ and $d=6$, respectively:
\begin{equation}
\begin{aligned}
d=4: & \qquad
\mathfrak{R}_0(\trs) =
-2 - 2.8851\, e^{- \frac{\pi\ii}{3}}\, \trs^2 -
1.6856\, e^{-\frac{2\pi\ii}{3}}\, \trs^{4} + 0.5737\, \trs^6 + \cdots\,, \\
d=6: &\qquad
\mathfrak{R}_0(\trs) =
-2 - 2.3251\, e^{- \frac{2\pi\ii}{5}}\, \trs^2
- 1.2398\, e^{-\frac{4\pi\ii}{5}}\, \trs^{4}
- 0.4136\, e^{-\frac{6\pi\ii}{5}}\, \trs^6 + \cdots \,.
\end{aligned}
\end{equation}

\paragraph{Leading frequency correction:} The next step is to obtain the frequency dependent corrections to the core Stokes multiplier. To do so analytically, we work at $\trs =0$ and use our solutions~\eqref{eq:core-Hankel}. The leading frequency correction can be computed at vanishing $\trs$ using~\eqref{eq:core-R1}. The explicit expressions for the potential correction and the wavefunctions being known we evaluate this to be
\begin{equation}\label{eq:core-neutralR-freq}
\begin{aligned}
\mathfrak{R}_0(0,\freq)
&= -2
- C_d\, e^{\frac{\ii\pi}{2}\,a_d}\, \pqty{\nu^2 - \frac{d^2}{4\, (2\,d-1)}}\, \freq^{-b_d }+ \order{\freq^{-2\,b_d}}\,,
\\
C_d
&=
\frac{2\sqrt{\pi}\, \cos(\frac{\pi}{2\,(d-1)})\, (d-1)^{1-a_d}\,
\Gamma\pqty{\frac{b_d}{2}}\,\Gamma\pqty{\frac{1+a_d}{2}}}{\Gamma\pqty{\frac{a_d}{2}}^2}\,.
\end{aligned}
\end{equation}
Evaluating these explicitly we find
\begin{equation}
\begin{aligned}
d=4: & \qquad
\mathfrak{R}_0(0,\freq) =
-2 - 0.943\, e^{\frac{\pi\ii}{3}}\, \pqty{\nu^2 - \frac{4}{7}}\, \freq^{-\frac{4}{3}} + \order{\freq^{-\frac{8}{3}}} \\
d=6: &\qquad
\mathfrak{R}_0(0,\freq) =
-2 - 1.505\, e^{ \frac{2\pi\ii}{5}}\,  \pqty{\nu^2 - \frac{9}{11}}\, \freq^{-\frac{6}{5}} + \order{\freq^{-\frac{12}{5}}} \,.
\end{aligned}
\end{equation}

\paragraph{The core transport factors:} The last pieces we need to determine are the factors $\mathcal{X}_a$ that are associated with the core turning points.  For \SAdS{5} black holes we don't have much to do, since there is no Stokes line intersecting the transport path that does not spiral into the horizon, viz., $\mathcal{X}_f  =0$.  The above data then immediately determine $\mathcal{X}_a$.

This is not so in $d=6$ where we do have a crossing and $\mathcal{X}_f \neq 0$. Working out the details we find the Stokes line emanates from $b_2$. The laterally resolved factors are
\begin{equation}
\mathcal{X}_f^+ = -e^{-\bw}\, \mathcal{R}^{(+)}_{-3}(\trs,\freq)\,, \qquad
\mathcal{X}_f^- = -e^{-\bw}\, \mathcal{R}^{(-)}_{2}(\trs,\freq)\,.
\end{equation}

\paragraph{Generalization to global black holes:}
The above results can also be lifted to give the multiplier for global \SAdS{d+1} black holes. For this we need the identification mentioned in~\cref{fn:global-core-caveat}, which says that we should adjust $\trs$ accordingly. In fact, setting it to be
\begin{equation}
\trs^2 = \ell\,(\ell + d-2)\, \rp^{-2\,a_d}\, (1+\rp^2)^{-\frac{1}{d-1}}\, \freq^{-a_d}\,,
\end{equation}
allows us to use the same expressions as the planar case. The frequency corrections appear at different orders, but the leading term can be computed as in the planar case to be
\begin{equation}\label{eq:core-globalR-freq}
\mathfrak{R}_0^{\rm global}(0,\freq)
=-2-
\frac{M_+^{\frac{1}{d-1}}}{\rp^2}\,
\frac{(d-1)\, (d-2)^2}{2\, (2\,d-3)}
\,\mathfrak{r}_1 \,e^{-\frac{\ii\pi}{2} \, a_d}\,
\freq^{-a_d} + \order{\freq^{-2\,a_d}, \freq^{-b_d}}\,.
\end{equation}
%

\subsubsection{The  non-perturbative spectral function}\label{sec:core-np-spectral}

We now have all the pieces assembled to compute the median answer for the physical thermal observables.  For the positive frequency rotation  we can adopt the conventions from~\cref{sec:neutral-d4}. Let the  oriented cycles participating in the spirals be $A$ and $B$. Then, in both $d=4$ and $d=6$ we have
\begin{equation}
\begin{aligned}
\mathcal{X}_{\star +} &= \mathcal{X}_{A+}\,,
&\qquad
\mathcal{X}_{1 +} &= \mathcal{X}_{B+}^{-1}
&\qquad
\mathcal{X}_{2 +} &= e^{-\bw}\, \mathcal{X}_{A+} \,,
\\
\mathcal{X}_{\star -} &= \mathcal{X}_{B-}\,,
&\qquad
\mathcal{X}_{1 -} &= \mathcal{X}_{A-}^{-1}
&\qquad
\mathcal{X}_{2 -} &= e^{-\bw}\, \mathcal{X}_{B-}  \,.
\end{aligned}
\end{equation}
For the non-spiralling Stokes crossing close to the boundary in $d=6$
\begin{equation}
\mathcal{X}_{C+} = \mathcal{X}_{f+} \,, \qquad
\mathcal{X}_{C-} = \mathcal{X}_{f-} \,.
\end{equation}
Translating our computations of the core data into this notation, we have
\begin{equation}
\begin{aligned}
\mathcal{X}_{A+}^{-1}
&=
- \mathcal{X}\subm{core}^+\, \mathcal{R}_0^{(+)}(\trs, \freq) \,,
\\
\mathcal{X}_{B+}^{-1}
&=
\mathcal{X}\subm{core}^-\, \bqty{-\mathcal{R}_{-1}^{(+)}(\trs, \freq) +
\frac{1}{\mathcal{R}^{(+)}_0(\trs, \freq) }}\,,
\\
\mathcal{X}_{C+}
&=
- e^{-\bw}\, \mathcal{R}^{(+)}_{-3}(\trs, \freq) \,.
\end{aligned}
\end{equation}
We can plug these expressions directly into the formulae derived in~\cref{sec:neutral-d4} and~\cref{sec:neutral-d6} and obtain the non-perturbative piece of the spectral function. Explicitly,
\begin{itemize}[wide,left=0pt]
  \item  For planar \SAdS{5} we find with $\theta_4 = \hbw - \pi\nu$
    \small{
      \begin{equation}\label{eq:psads5-rhonp-freq}
      \begin{aligned}
      \rho\subm{np}(\omega,0)
      &= 1- 2\,e^{-\hbw}\,\bqty{
        2\,  \cos\theta_4 +  C_4\, \freq^{-\frac{4}{3}}\,
        \pqty{\nu^2-\frac{4}{7}}\,
      \cos(\ms\theta_4 + \frac{\pi}{3}) + \order{\freq^{-\frac{8}{3}}}}
      \\
      & \;
      + 8\,e^{-\bw}\, \bqty{\cos(2\ms\theta_4) + C_4\, \freq^{-\frac{4}{3}}\,
        \pqty{\nu^2-\frac{4}{7}}\,
      \cos(2\ms\theta_4 + \frac{\pi}{3}) + \order{\freq^{-\frac{8}{3}}}} +  \order{e^{-\frac{3}{2}\ms\bw}}\,,
      \end{aligned}
    \end{equation}}
    with $C_4 = 0.943$ obtained from the core Stokes multiplier factor~\eqref{eq:core-neutralR-freq}. Our leading answer agrees with the results of~\cite{Afkhami-Jeddi:2025wra}. In~\cref{sec:core-wkb-borel} we will show that the subleading term of order $\freq^{-\frac{8}{3}}$ also agrees with their result.
  \item Similarly, for planar \SAdS{7}, one has with $\theta_6 = \frac{\sqrt{3}\, \beta\ms\omega}{2} - \pi\ms\nu$
    \small{
      \begin{equation}
      \begin{aligned}
      \rho\subm{np}(\omega,0)
      &= 1- 2\,e^{-\hbw}\,\bigg[
        2\, \cos \theta_6 + C_6\, \freq^{-\frac{6}{5}}\,
        \pqty{\nu^2-\frac{9}{11}}\,
        \cos(\theta_6 + \frac{2\pi}{5})
      + \order{\freq^{-\frac{12}{5}}}\bigg]
      \\
      & \quad
      + \,e^{-\bw}\,\bigg[8\,\cos(2 \ms\theta_6) -2\,\cos(\pi \nu)
        + C_6\, \freq^{-\frac{6}{5}}\, \pqty{\nu^2-\frac{9}{11}}
        \pqty{8\,\cos(2 \ms\theta_6 + \frac{2\pi}{5}) -\cos(\pi \nu) }
        \\
        & \qquad \qquad  \qquad
        + \order{\freq^{-\frac{12}{5}}}
      \bigg]
      + \order{e^{-\frac{3}{2}\ms\bw}}\,,
      \end{aligned}
    \end{equation}}
    where evaluating~\eqref{eq:core-neutralR-freq} we find $C_6 = 1.505$.
  \item For global \SAdS{d+1} we have instead
    \begin{equation}
    \begin{aligned}
    d=4: &\quad
    &\rho\subm{np}(\omega,0)
    &=
    1- e^{-\hbw}\,\bigg[4\, \cos\theta_4 + c_4\, M^{-\frac{1}{3}}\, \omega^{-\frac{2}{3}}\,
    \cos(\theta_4 - \frac{\pi}{3}) \bigg]+ \cdots \,,
    \\
    d=6: &\quad
    & \rho\subm{np}(\omega,0)
    &=
    1- e^{-\hbw}\,\bigg[4\, \cos\theta_6 + c_6\, M^{-\frac{1}{5}}\, \omega^{-\frac{4}{5}}\,
      \cos(\theta_6 - \frac{2\pi}{5})
      \\
      &&& \qquad \qquad
    + 2\,c'_6\, \pqty{\nu^2 - \frac{9}{11}} M^{\frac{1}{5}}\, \omega^{-\frac{6}{5}} \, \cos(\theta_6 + \frac{2\pi}{5})\bigg]+ \cdots \,,
    \end{aligned}
    \end{equation}
    with
    \begin{equation}
    \theta_d = 2\,\freq\,  \Pphys^{(d)}(0) - \pi \ms\nu \,, \qquad c_4 = 6.924\,,
    \qquad c_6 = 20.668\,, \qquad c_6' = 1.5045\,.
    \end{equation}
    These coefficients, which are indexed by the dimension $d$, have been determined from~\eqref{eq:core-globalR-freq} for $d=4,6$.
    In $d=4$ determining the next correction at $\order{\omega^{-\frac{4}{3}}}$ necessitates evaluating the second correction due to spatial curvature, but this is not the case in $d=6$; $c_6'$ quoted above can be deduced from the planar black hole.
\end{itemize}

\subsection{Charged black holes}

Thus far, we have examined neutral black holes, for which the locus $r=0$ is a regular singular point. This motivated our core-scaling regime. As already indicated in~\cref{sec:core-scale} the situation for scalar fields in charged black hole backgrounds is different for the locus $r=0$ is at best an apparent singular point of the  Schr\"odinger oper, cf.~\cref{fn:charged-core}. In $d=4$ the situation is even simpler as $\rpr=0$ is just an ordinary point in the projected coordinate chart; there is no singularity in $Q_2$ as $\rpr\to 0$. Specifically, the wave equation is
\begin{equation}
\dv[2]{\psi}{\rpr} +
\bqty{\freq^2 \frac{\rpr^3 - \zeta^2 \, (\rpr-1) (\rpr -\varrho) (\rpr  + 1 + \varrho)}{4\, (\rpr-1)^2 (\rpr -\varrho)^2 (\rpr  + 1 + \varrho)^2} + \order{\freq^0}}\, \psi = 0\,,
\end{equation}
with the coefficient of the $\order{\freq^0}$ term being finite at $\rpr \to 0$.

We can nevertheless ask what happens as we scale momentum to zero, and whether there is a scaling regime that can be tackled. The turning points which are solutions of~\eqref{eq:rn-rturnp}  collapse to the singularity, so whether we can access the region close to it is still germane. From the oper we see that such a scaling requires
\begin{equation}
\rpr^5\, \freq^2 \sim \order{1}
\;\; \Longrightarrow \;\;
\rtr = \frac{1}{Q^{\frac{4}{5}}}\,
\rpr\, \freq^{\frac{2}{5}}\,.
\end{equation}
We defined the core chart by absorbing the constant $Q^2 = \varrho\,(1+\varrho)$ to simplify the expressions below. The scaling of $\zeta$ then follows
\begin{equation}
\trs^2 = \frac{1}{Q^{\frac{2}{5}}}\, \zeta^2\, \freq^{\frac{6}{5}} \,.
\end{equation}

In these coordinates the core oper becomes
\begin{equation}
\dv[2]{\psi}{\rtr} + \frac{1}{4} \bqty{\rtr^3 - \trs^2} \psi = 0\,.
\end{equation}
This oper has corrections from the $\order{\freq^0}$ terms that we have not explicitly written out. What is of import is that the potential is analytic at $\rtr =0$, and the core roots are resolved to
\begin{equation}
\rtr_a = \trs^{\frac{2}{3}}\, \exp(-\frac{2\pi \ii}{3}\, a) \,, \quad a \in \set{0,1,2}
\,, \qquad \text{for}\ \trs \neq 0 \,.
\end{equation}
At $\trs =0$ the equation can be solved in terms of Bessel functions with
\begin{equation}
\psi\subm{core}^\pm(\rtr;0)
= \sqrt{\frac{\pi}{5}}\, e^{\pm \frac{7\pi \ii}{20}}\, \sqrt{\rtr}\, H_{\frac{1}{5}}^{(1,2)}\pqty{\frac{\rtr^{\frac{5}{2}}}{5}} \,.
\end{equation}
Once we have this solution we can exploit the same strategy as in the case of neutral black holes to define rays $\Gamma_j$ and the domains $\Sigma_j$. The Stokes multipliers for the normalized solution upon crossing domains at $\trs = 0$ are determined by the Bessel connection formulae.  Dependence on $\trs$ can be worked out order by order in $\trs$ perturbation theory, as can the subleading frequency part (the latter as a formal power series).

To work out the details we should compute the corrections to the potential. These turn out to be
\begin{equation}\label{eq:rnads-deltaQ}
\delta_\freq Q = \frac{1+ Q^2}{4\, Q^{\frac{6}{5}}\, \freq^{\frac{2}{5}}}\, \rtr \pqty{2\, \rtr^3 - \trs^2} + \frac{(1+Q^2)^2}{4\, Q^{\frac{12}{5}}\, \freq^{\frac{4}{5}}} \, \bqty{\rtr^2 \, \pqty{3\,\rtr^3 - \trs^2 } +1} + \order{\freq^{-\frac{6}{5}}} \,.
\end{equation}
Using the arguments given in~\cref{sec:core-multiplier} we then deduce the analog of~\eqref{eq:core-R1} to be\footnote{This relies on using the zero momentum action coordinate to simplify the correction to the potential.}
\begin{equation}
\begin{aligned}
\mathfrak{R}_0(\trs,\freq)
&=
\mathfrak{R}_0(\trs) -\frac{(1+Q^2)\trs^{\,2}}{56\, Q^{\frac{6}{5}}\, \freq^{\frac{2}{5}}}
\, \operatorname{FP}\int_{\infty_0}^{\infty_1} \, d\rtr\, \rtr\,
\left.\psi_{\rm core}^{+}\right|_{\Gamma_0}
\left.\psi_{\rm core}^{+}\right|_{\Gamma_1} + \order{\freq^{-\frac{4}{5}}}\,.
\end{aligned}
\end{equation}

The core Stokes multiplier at leading order in frequency turns out to have the following momentum expansion:
\begin{equation}
\mathfrak{R}_0(\trs)
= - 2 \,\cos(\frac{\pi}{5}) + (0.617 - 1.899\ms \ii)\, \trs^2  +
(0.892 + 0.648\ms\ii) \, \trs^4 + (-0.304 + 0.221 \ms\ii) \,\trs^6 + \cdots
\end{equation}
Computing the corrections in frequency, starting at $\trs = 0$ we instead find
\begin{equation}
\begin{aligned}
\mathfrak{R}_0(0,\freq)
&=
-2\, \cos(\frac{\pi}{5})
- \frac{(1+Q^2)^2}{21\,Q^{\frac{12}{5}}}\, \frac{
\Gamma\pqty{\frac{1}{5}}^2\, e^{-\frac{2\pi \ii}{5}}}{2^{\frac{6}{5}}\, 5^{\frac{7}{10}}\, \Gamma\pqty{\frac{3}{5}}}\, \freq^{-\frac{4}{5}}
\\
&\qquad
+ \frac{2^{\frac{6}{5}}\, 5^{\frac{7}{10}}\, \Gamma\pqty{\frac{4}{5}}^2 \, e^{\frac{2\pi \ii}{5}}}{\Gamma\pqty{\frac{2}{5}}}\, \bqty{\frac{Q^{\frac{2}{5}}}{4}\, \pqty{\nu^2 - \frac{8}{11}}
+ \frac{9\, (1+Q^2)^3}{1078\, Q^{\frac{18}{5}}}}\, \freq^{-\frac{6}{5}}+ \order{\freq^{-\frac{8}{5}}}\,.
\end{aligned}
\end{equation}
A priori there should have been a correction proportional to $\freq^{-\frac{2}{5}}$ but around zero momentum this piece vanishes. The reason can be understood analytically by working in the action coordinate chart. For the purposes of our analysis, it suffices that we can compute the next-to-leading-order corrections.

\begin{figure}[htbp!]
  \centering
  \includegraphics[width=0.9\linewidth]{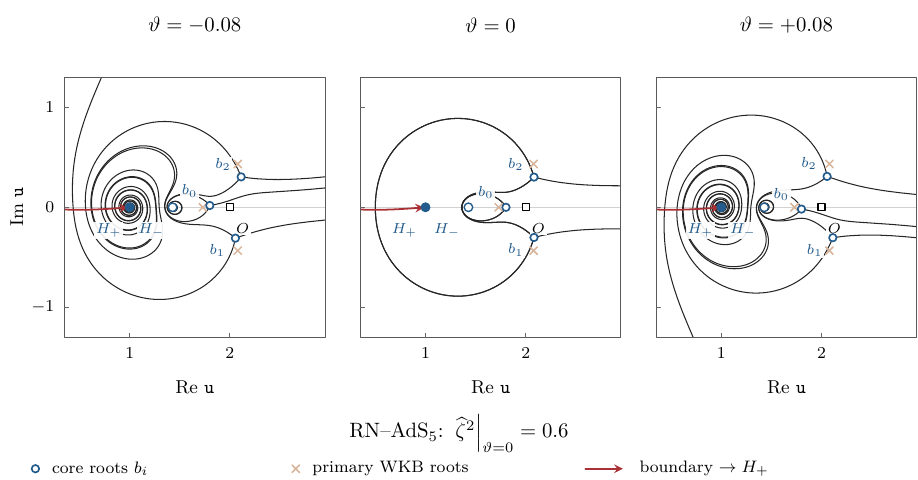}
  \caption{Lateral resolution of the full finite-frequency Stokes graphs in the core regime for planar \RNAdS{5}. We use the $\upr$ coordinate, and have chosen $\varrho = 0.4$, $\zeta = 0.1085845$ and $\nu = 1.3$. The frequency is set to be  $\freq =24\, e^{\ii\vartheta}$ and the lateral rotation holds $\zeta$ fixed.
  The conventions are as in~\cref{fig:core-stokes-laterals}. The core turning points are indicated as the blue open circles and the transport path comes into the outer horizon from the boundary (not shown). Notice that the Stokes lines crossing the transport path head toward $b_1$ and $b_2$ which are the complex core roots, and not to the real core root $b_0$. A Stokes line heading toward the inner-horizon at $H_-$ does not cross the transport path.  }
  \label{fig:rn-core-stokes-laterals}
\end{figure}

\begin{figure}[hbtp!]
  \centering
  \includegraphics[width=0.93\linewidth]{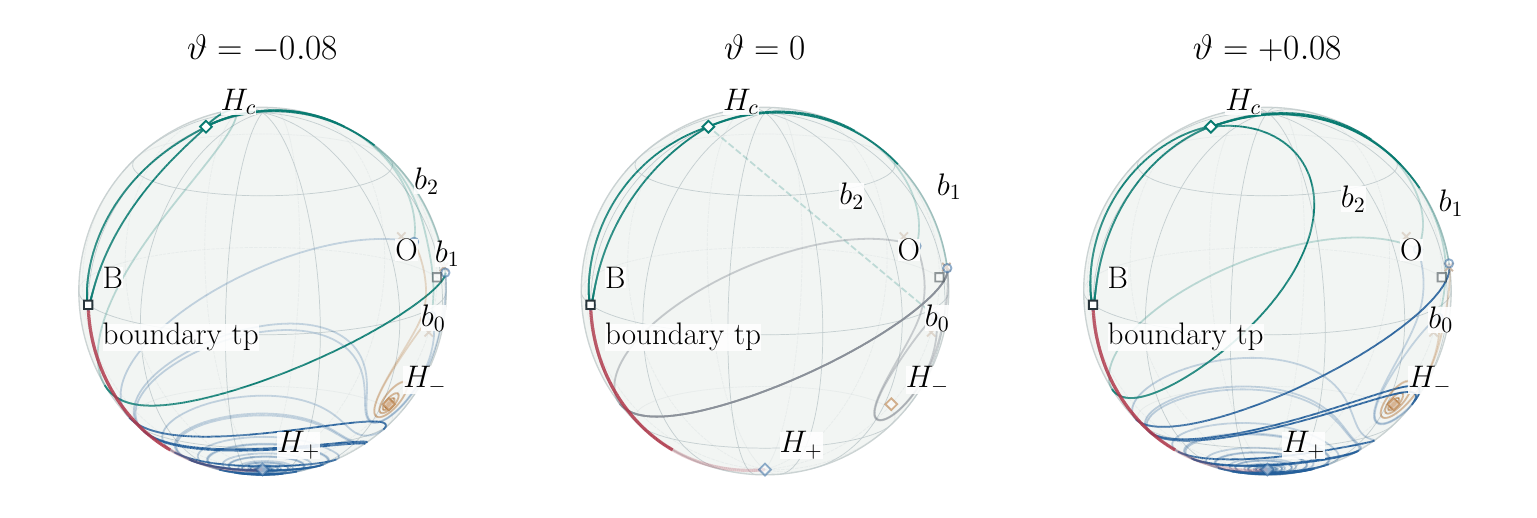}
  \caption{The Stokes graphs for \RNAdS{5} in the core-scaling regime drawn on the Riemann sphere. The resolved core coordinate chart is projected onto the sphere, and we chose $\abs{\freq}=12$, with $\nu=0.75$ and $\trs_0^2=0.1$. See \href{https://mukundrangamani.github.io/stokes-atlas/}{Stokes atlas visualizer} for an interactive version.}
  \label{fig:planar-rnads5-core-sphere}
\end{figure}

The Stokes graph for the core region with the core turning-points resolved is shown in~\cref{fig:rn-core-stokes-laterals} and~\cref{fig:planar-rnads5-core-sphere}. The key point we wish to highlight is that no Stokes line crossing the transport path emanates from the inner horizon $H_-$. The main contribution is already the piece contained in $\mathcal{X}\subm{core}$ which involves the inner horizon temperature from the period contribution.

Given the graph, we can map the calculation in the core back to the general discussion of~\cref{sec:charged-d4}. We need the identifications
\begin{equation}
\mathcal{X}_{1+} = \mathcal{X}_{B+}\,,
\qquad
\mathcal{X}_{2+} = e^{-\bw}\,\mathcal{X}_{A+}\,,
\end{equation}
and there is no non-spiralling line crossing the transport path. Note that since the spirals all end at the outer horizon $H_+$ we only encounter the standard thermal factor.
The main point we need to know is which Stokes sector is involved for the line exiting the core region. This turns out to be the multiplier between $D_1$ and $D_2$, viz., $\mathfrak{R}_1(0,\freq)$. Working through the details, the non-perturbative piece of the spectral function is (NB: $\theta_{\mathrm{RN}} = \frac{\beta_c\ms\omega}{2} - \pi \nu$)
\small{
  \begin{equation}
  \begin{aligned}
  \rho\subm{np}(\omega,0)
  &=
  1 - e^{-\frac{1}{2}\ms(\beta-\beta_{\mathrm{inner}})\ms\omega}
  \bigg[4\, \cos(\frac{\pi}{5})\, \cos(\theta_{\mathrm{RN}} )
    - 0.19\, \frac{(1+Q^2)^2}{Q^{\frac{12}{5}}} \,\freq^{-\frac{4}{5}}\, \cos(\theta_{\mathrm{RN}}  - \frac{\pi}{5})
    \\
    &\quad
    + 8.662 \bqty{\frac{Q^{\frac{2}{5}}}{4}\, \pqty{\nu^2 - \frac{8}{11}}
    + \frac{9\, (1+Q^2)^3}{1078\, Q^{\frac{18}{5}}}}\, \cos(\theta_{\mathrm{RN}} + \frac{\pi}{5}) \freq^{-\frac{6}{5}}
  + \order{\freq^{-\frac{8}{5}}}\bigg]
  + \order{e^{-(\beta-\beta_{\mathrm{inner}})\ms\omega}}\,.
  \end{aligned}
\end{equation}}
This formula has not yet been explicitly derived by a careful median prescription, but agrees with our numerical checks. It also agrees to leading orders with the result recently quoted in~\cite{Chakravarty:2026fdm}. However, their expression, which was obtained by a numerical fitting procedure appears to miss the $\freq^{-\frac{6}{5}}$ term. The analytic calculation predicts this contribution with a non-vanishing coefficient; it is plausible that a numerical fit in a finite frequency window is unable to resolve the contribution.

\section{Toward an exact-WKB treatment}\label{sec:core-wkb-borel}

In exact WKB analysis, one considers the Borel sum of formal WKB solutions, and connection formulae of WKB solutions are understood as jumps of Borel sums due to Borel singularities. Generically, the locations of the Borel singularities are controlled by classical WKB actions, with Stokes multipliers determined from the discontinuity across a Borel singularity. To complement our discussion of~\cref{sec:tripscale-core}, we now apply this paradigm directly to the core-scaling regime. In order to do so, we motivate first a different coordinate chart, one which directly uses the WKB action.

\subsection{The WKB action chart}\label{sec:core-wkb-chart}

The above discussion exploited a coordinate chart that zoomed into the core region and resolved the black hole singularity into core turning points. We could alternately choose to work in a chart which does not resolve the core region, but is somewhat more natural from the perspective of the exact WKB problem.
We can also decide to work with an alternate chart, using the leading order potential at large frequency. Recall that the scalar wave equation takes the form
\begin{equation}
\dv[2]{\psi}{\zpr} - \bqty{\freq^2\, Q_0(\zpr) + Q_2(\zpr)} \psi(\zpr) = 0\,.
\end{equation}
We define a new coordinate $y$ from the WKB potential, and further redefine the wavefunction so that the operator retains its Schr\"odinger form. Set
\begin{equation}
y(\zpr) = \int^\zpr \, \sqrt{-Q_0(\zpr')}\, d\zpr' \,, \qquad
\psi = \frac{1}{(-Q_0(\zpr))^{\frac{1}{4}}}\, \Psi(y)\,.
\end{equation}
The new wavefunction satisfies
\begin{equation}
\dv[2]{\Psi}{y} + \pqty{\freq^2 + \frac{Q_2}{Q_0} + \frac{1}{4\,Q_0^2}\, \dv[2]{Q_0}{\zpr}
- \frac{5}{16\, Q_0^3}\, \pqty{\dv{Q_0}{\zpr}}^2 } \Psi(y) = 0\,.
\end{equation}
We have left it implicit that one should solve for the potential in terms of $y$ after inverting the coordinate transform.

Notice that the location of the singularity $\zpr \to \infty$ maps to a finite position in the $y$ coordinate. Moreover, if we anchor ourselves at the boundary we would have
\begin{equation}
\lim_{\zpr \to \infty} y(\zpr) = \int_0^\infty\, \sqrt{-Q_0(\zpr')} \, d\zpr'
\end{equation}
whose principal value is just the $\Pphys$ defined in~\eqref{eq:Pphys-def}. The passage across the horizon will also pick up the residue contribution. This should not be surprising, since the chart uses the action as the coordinate.

In the neighborhood of the primary turning points~\eqref{eq:planar-uturn}  $Q_0$ vanishes. If $Q_0 \sim c_j\, (\zpr - \zpr_j)$ is the local behavior, then  $y -y_j\sim \frac{2}{3}\, \sqrt{-c_j}\, (\zpr - \zpr_j)^{\frac{3}{2}}$. The Schr\"odinger potential now inherits a local double pole $\frac{5}{36}\,(y-y_j)^{-2}$ whose coefficient specifies the local branching of the coordinate. The local exponents are fixed by the residue to be $\frac{1}{6}$ and $\frac{5}{6}$, respectively. These data have to be kept track of, since the coordinate
$y$ is not globally invertible. This is not an obstacle per se, but makes for more bookkeeping.

Thus far, nothing about the change to the action coordinate chart was specific to the triple scaling regime. Let us further specialize to this limit, and use the action coordinate
\begin{equation}\label{eq:y-action-coord}
d\widehat{y} = \frac{1}{2}\, \sqrt{\rtr^{\frac{d}{2}-3}\, \pqty{\rtr^{\frac{d}{2}} + \trs^2}}\, d\rtr
\;\; \Longrightarrow \;\;
\widehat{y} \sim \frac{\rtr^{\frac{d-1}{2}}}{d-1}\,.
\end{equation}
One can also work out the local behavior in the core should we wish to retain non-zero $\trs$ by working in the regime $\rtr \ll \trs^{\frac{4}{d}}$; in this case the action coordinate becomes $\widehat{y} \sim  \frac{2\,}{d-2} \, \trs \, \rtr^{\frac{d-2}{4}}$.
We will use the $\trs=0$ action coordinate in the following. The resulting wave equation takes a universal Bessel form (after a suitable transformation of the wavefunction)
\begin{equation}
\dv[2]{\Psi}{\widehat{y}} +  \bqty{1 + \frac{1}{4\, \widehat{y}^2} +
\trs^2\, \pqty{(d-1)\, \widehat{y}}^{-\frac{d}{d-1}} }\Psi = 0\,.
\end{equation}
We have expanded the result for small $\trs$ to emphasize the universal behavior. The potential now has a series of corrections in $\trs$ and $\freq^{-\frac{d}{d-1}}$, whose coordinate dependence can be extracted straightforwardly. The former will lead to the connection coefficient $\mathfrak{R}(\trs)$, in this approach.

The main point to emphasize is that in this action coordinate chart the core turning points are no longer resolved. The transport appears to be unhindered all the way onto the singularity. Nevertheless, the local Bessel nature of the equation changes the Stokes multiplier (already for $\trs = 0$) from the canonical Airy form. Keeping track of $\trs$ dependence is also possible by anchoring oneself on this leading Bessel solution.

The overall strategy is to work out the Borel sums of the full problem starting from this solution. While~\cref{sec:core-multiplier} did this by placing the onus on the core multiplier for an equation bereft of any WKB parameter, we will now attempt to do so directly in the framework of exact-WKB.

\subsection{Analysis at vanishing momentum}\label{sec:borel-wkb-zero}

As noted above the WKB chart is simplest when the action coordinate does not branch. To understand the set-up of exact-WKB it is therefore useful to start from the simpler setting where $\zeta = 0$. In this case we can work out the action coordinate analytically, and the wave equation becomes quite tractable. We will show that the connection formula derived using the Borel analysis reproduces the core Stokes multiplier for $\zeta =0$.

At $\zeta =0$, the leading order potential $Q_0(r) =- r^{-4}\, f(r)^{-2}$, which implies that the action coordinate is related directly to the tortoise coordinate $r_*$. For definiteness, and to work with exponentially growing/damped solutions, rotate $\widehat{y} = \ii \freq\,x$ relative to~\cref{sec:core-wkb-chart} at $\zeta =0$. Explicitly,  define
\begin{equation}\label{eq:xaction-coord}
x(r) = \ii \rp \, \int_0^r\, \frac{dr'}{r'^2\,f(r')}
= \ii\rp \Big(r_*(r) - r_*(0)\Big) \,.
\end{equation}
Changing variables the wave equation takes the form
\begin{equation}\label{eq:borel-wkb-zero-oper}
\dv[2]{\Psi}{x} - \pqty{\freq^2 - \frac{1}{4\,x^2}  +V(x)} \Psi(x) =0\,,
\end{equation}
where we note that we have not scaled the coordinate by factors of frequency. We will first work with $x>0$ so that $\Psi_+$ is the dominant solution. The potential $V(x)$ has a series expansion in $x$. For instance, in the case of neutral planar \SAdS{d+1} black holes it takes the form:\footnote{We will focus primarily  on planar neutral black holes; the corrections in frequency are different for global black holes, as already noted in~\cref{sec:core-scaling}.}
\begin{equation}
V(x) = \sum_{n\geq 1} \, v_n\, x^{n\, b_d -2} \,.
\end{equation}

Our goal is to solve this equation to all orders in $\freq$ using WKB methods.
To anchor the discussion note that the solutions behave as $e^{\pm \freq x}$ at large $x$, and map to the large $\rtr$ asymptotics of the core region solutions
$\psi\subm{core}^\mp(\rtr;0)$. Suppose we denote the formal WKB solutions as $\Psi_\pm(x,\freq)$, where we assume that we have obtained the aforesaid solution.
Our convention is $\Psi_\pm(x,\freq) \leftrightarrow \psi\subm{core}^\mp(\rtr;0)$ in the WKB regime.

\subsubsection{Deriving the core connection formula}\label{sec:borel-core-connect0}

Consider the Borel transform of $\Psi_\pm(x,\freq)$ with $\tau$ being the Borel variable conjugate to $\freq$. The Borel transformed functions $\widehat{\Psi}(x,\tau)$ satisfy
\begin{equation}\label{eq:borel-wkb-waveeq}
\pqty{\pdv[2]{x} - \pdv[2]{\tau} + \frac{1}{4\,x^2}}\, \widehat{\Psi}
= V(x)\, \widehat{\Psi}\,.
\end{equation}
We recognize this to be a physical wave equation of the Klein-Gordon form, at imaginary tortoise coordinate $x$ and a time coordinate $\tau$. Given that near the singularity the role of spatial and temporal coordinates defined in the exterior swaps in the canonical chart~\eqref{eq:staticbh}, this is quite natural.

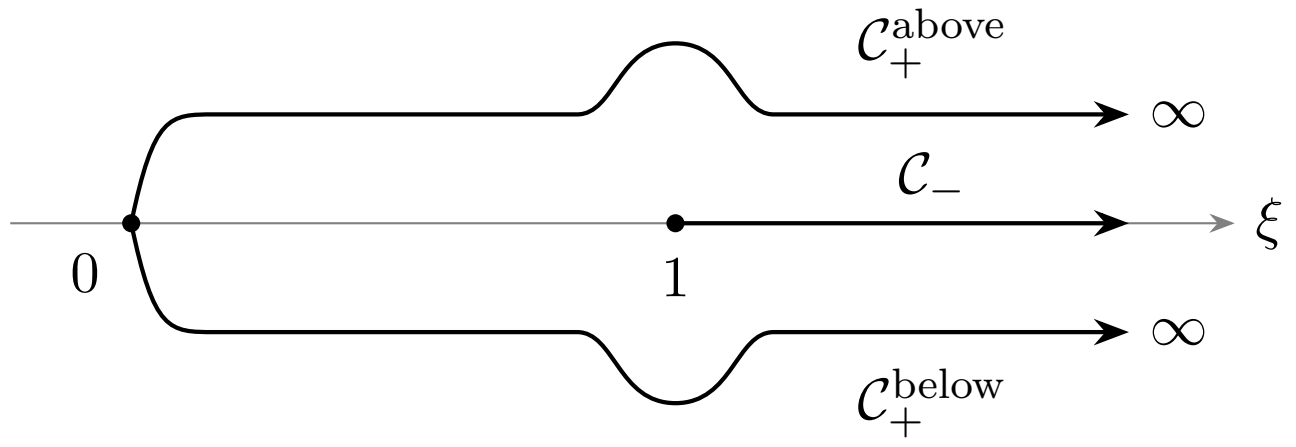
\begin{figure}[thbp!]
  \centering
  \begin{tikzpicture}[x=1cm,y=1cm,>=Stealth]
    \draw[->,gray] (-3,0)--(5.1,0) node[right,black] {$\xi$};
    \draw[->,thick] (-2.2,0)
    .. controls (-2.05,.72) and (-1.95,.72) .. (-1.65,.72)
    --(.75,.72)
    .. controls (1.0,.72) and (1.02,1.19) .. (1.4,1.19)
    .. controls (1.78,1.19) and (1.8,.72) .. (2.05,.72)--(4.4,.72);
    \draw[->,thick] (-2.2,0)
    .. controls (-2.05,-.72) and (-1.95,-.72) .. (-1.65,-.72)
    --(.75,-.72)
    .. controls (1.0,-.72) and (1.02,-1.19) .. (1.4,-1.19)
    .. controls (1.78,-1.19) and (1.8,-.72) .. (2.05,-.72)--(4.4,-.72);
    \draw[->,thick] (1.4,0)--(4.4,0);
    \fill (-2.2,0) circle (1.7pt) node[below left=3pt] {$0$};
    \fill (1.4,0) circle (1.7pt) node[below=3pt] {$1$};
    \node[above=3pt] at (3.1,.72) {$\mathcal{C}_+^{\mathrm{above}}$};
    \node[below=3pt] at (3.1,-.72) {$\mathcal{C}_+^{\mathrm{below}}$};
    \node[above=2pt,inner sep=1pt] at (3.1,0) {$\mathcal{C}_-$};
    \node[right] at (4.4,.72) {$\infty$};
    \node[right] at (4.4,-.72) {$\infty$};
  \end{tikzpicture}
  \caption{Contours for the Laplace representation~\eqref{eq:borel-wkb-laplace-representation}. The Borel singularity at $\xi=1$ leads to the Stokes connection formula for the dominant solution $\Psi_+$, with the Stokes
  multiplier determined by the behavior near the Borel singularity.}
  \label{fig:borel-wkb-contours}
\end{figure}

This wave equation can now be solved perturbatively in $x$ treating the potential as a correction to the leading order problem defined by the l.h.s.~of~\eqref{eq:borel-wkb-waveeq}. Specifically, we solve the problem with both $\tau$ and $x$ taken to be small, but keeping their ratio fixed. Define
\begin{equation}\label{eq:borel-xidef}
\xi = \frac{\tau +x}{2\,x} \,,
\end{equation}
and consider a perturbative ansatz
\begin{equation}\label{eq:borel-pert-ansatz}
\widehat{\Psi}_\pm\Big(x, x\, (2\,\xi -1)\Big)
= \frac{1}{\sqrt{x}}\, \sum_{n\geq 0}\, x^{n\,b_d}\, \psi^\pm_n(\xi) \,.
\end{equation}
The exact WKB solutions are given by the Borel sum
\begin{equation}\label{eq:borel-wkb-laplace-representation}
\Psi_\pm(x,\freq)
=
2\,x\,\int_{\mathcal{C}_\pm}
e^{-\freq \,x\, (2\ms\xi-1)}\,
\widehat{\Psi}_\pm\bigl(x,x\ms(2\ms\xi-1)\bigr)\,d\xi\,.
\end{equation}
The contours along which the integral is taken are shown in~\cref{fig:borel-wkb-contours}.

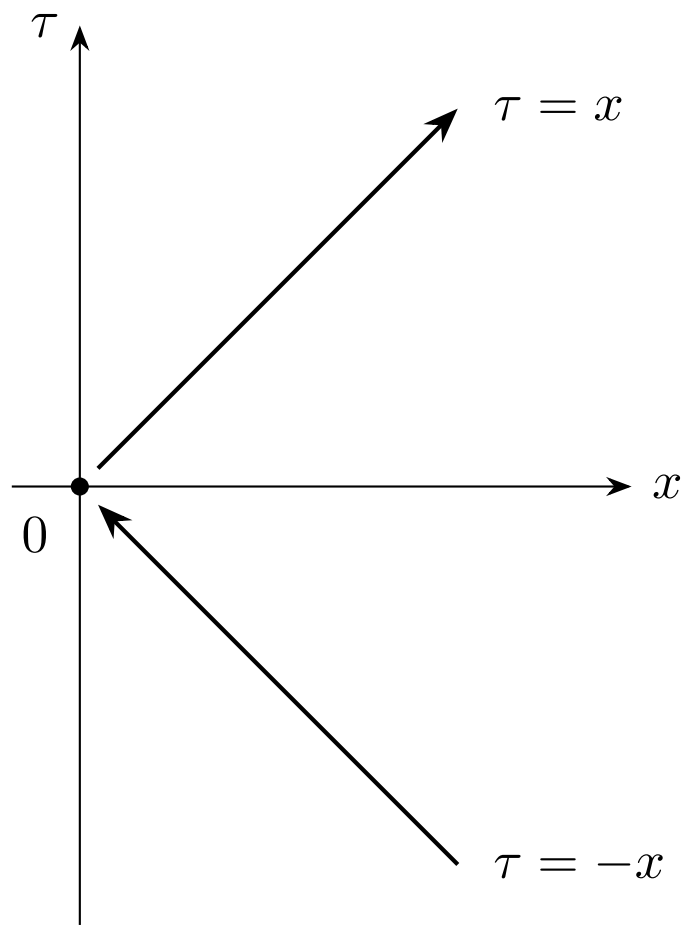
\begin{figure}[htbp!]
  \centering
  \begin{tikzpicture}[>=Stealth, font=\small, x=1cm, y=1cm]
    \draw[->,thin] (-.45,0)--(3.65,0) node[right] {$x$};
    \draw[->,thin] (0,-2.9)--(0,3.05) node[left] {$\tau$};
    \draw[black,thick,->] (2.5,-2.5)--(.12,-.12);
    \draw[black,thick,->] (.12,.12)--(2.5,2.5);
    \fill (0,0) circle (1.7pt) node[below left=3pt] {$0$};
    \node[anchor=west] at (2.6,-2.5) {$\tau=-x$};
    \node[anchor=west] at (2.6,2.5) {$\tau=x$};
  \end{tikzpicture}
  \caption{The wave characteristics for the auxiliary Borel equation~\eqref{eq:borel-wkb-waveeq}. The wave is incoming from the past, scatters off the center, and propagates to the future. The connection problem is tantamount to computing the reflection coefficient.}
  \label{fig:borel-wkb-characteristics}
\end{figure}

Since we have a wave equation, it is natural to view the perturbative approach in the language of scattering, cf.~\cite{Afkhami-Jeddi:2025wra}. Indeed, in this action chart the WKB wavefunctions are exponential waves
$e^{\pm \freq\ms x}$ that are suitably normalized, cf.~\eqref{eq:borel-pert-ansatz}.  In the Borel conjugate variable, we can think of an incoming wave from the past, traveling along $\tau = -x$ scattering off the origin, and propagating to the future along $\tau = +x$. The past characteristic corresponds to $\xi = 0$, and the future characteristic to $\xi =1$, respectively.

Plugging the perturbative ansatz~\eqref{eq:borel-pert-ansatz} into the auxiliary Borel wave equation~\eqref{eq:borel-wkb-waveeq} we find a set of recursion relations
\begin{equation}\label{eq:borel-zero-pertrel}
\begin{aligned}
\bqty{
  \xi\, (1-\xi)\, \pdv[2]{\xi}
  +\pqty{\frac{3}{2} - n\, b_d - (3- 2\, n\, b_d)\, \xi} \, \pdv{\xi}
-(1-n\, b_d)^2 } \psi^\pm_n
=
- \sum_{j=1}^{n}\, v_j\, \psi^\pm_{n-j} \,.
\end{aligned}
\end{equation}
Recall that our aim is to solve this system anchored on the WKB waves. Using the WKB data we can write down the profiles for the perturbative coefficients
$\psi^\pm_n$ as an expansion in the asymptotic wave region, viz., near the past
or future characteristics.
In particular, the leading order result is
\begin{equation}\label{eq:borel-zero-bessel}
\begin{aligned}
\psi^+_{0}(\xi)
&=
\frac{1}{\sqrt{2\pi}}\, \frac{1}{\sqrt{\xi\, (1-\xi)}}\,,
&& \qquad 0<\xi<1\,,
\\
\psi^-_{0}(\xi)
&=
\frac{1}{\sqrt{2\pi}}\, \frac{1}{\sqrt{\xi\, (\xi-1)}}\,,
&&\qquad \xi>1\,.
\end{aligned}
\end{equation}
These are the characteristics of the Bessel functions, which we have encountered earlier in~\eqref{eq:core-Hankel} and can be verified using the translation in~\eqref{eq:y-action-coord} and the translation $\widehat{y} = \ii\freq x$. In fact,~\eqref{eq:borel-zero-pertrel} has a second independent solution, $\frac{\arcsin\sqrt{\xi}}{\sqrt{\xi\, (1-\xi)}}$ which is regular at $\xi =0$. We have chosen normalization such that the incoming solution along the past characteristics does not include this piece.  At higher orders we find
\begin{equation}\label{eq:borel-zero-endpoint}
\begin{aligned}
\psi^+_n(\xi)
&=
\sum_{k\geq0}
\frac{\psi^+_{n,k}}{\Gamma\pqty{k+\frac{1}{2}}}\,
(2\,\xi)^{k-\frac{1}{2}}\,,
&& \quad \xi\to 0\,,
\\
\psi^-_{n}(\xi)
&=
\sum_{k\geq0}
\frac{\psi^-_{n,k}}{\Gamma\pqty{k+\frac{1}{2}}}\,
\pqty{2\, (\xi-1)}^{k-\frac{1}{2}}\,,
&&\quad \xi\to1\,.
\end{aligned}
\end{equation}

As in the core-scaling chart~\cref{sec:tripscale-core} the variation of parameters method can be exploited to systematically construct $\psi^\pm_{n}$ from the leading term~\cref{eq:borel-zero-bessel}. To make progress, we make a few assumptions: we assume that the coefficient functions
$\psi^\pm_n(\xi)$ only have singularities at $\xi =0$ and $\xi =1$. This supposes that the location of Borel singularities is still given by the WKB action at  $\tau=\pm x$. In general, we anticipate the part with the discontinuity, which does not holomorphically extend through $\xi =1$,  takes the form:
\begin{equation}\label{eq:borel-series-refl}
\begin{aligned}
\psi^+_n(\xi)
&=
\sum_{m=0}^{n}\,
\sum_{k\geq0}\,
\frac{\mathfrak R_{-1,m}(0)}{2\,\sin(\pi(m\,b_d- \frac{1}{2}))}\,
\frac{
\psi^+_{n-m,k}\, \pqty{2\, (1-\xi)}^{m\,b_d+k-\frac{1}{2}}}{
\Gamma\pqty{m\, b_d + k+ \frac{1}{2}}}
\\
&\qquad \qquad +\; \text{holomorphic at }\xi=1\,.
\end{aligned}
\end{equation}
In the exact WKB approach, we therefore define the all-orders Stokes multiplier $\mathfrak R_{-1,m}$ as the coefficient of non-holomorphic terms at $\xi=1$ via a connection problem between $\xi=0$ and $\xi=1$. The reason for this definition is that it is the non-holomorphic term that will contribute to the jump of the Borel sum of $\psi_+$ from the Borel singularity at $\xi=1$. The leading Stokes multiplier can be obtained from the  Bessel profiles. It is given by the previously computed answer
\begin{equation}\label{eq:borel-bessel-reflection}
\mathfrak R_{-1,0}(0) = -2\,,
\end{equation}
where we have used a shorthand $\mathfrak R_{j,n}(0)=\delta_n\mathfrak R_j(0)$. We can obtain this directly from~\eqref{eq:borel-zero-bessel}, since
\begin{equation}
\operatorname{\text{Disc}} \bqty{2\, (1-\xi)}^\alpha = 2\ii\sin(\pi \ms\alpha)\, \bqty{2\, (\xi-1)}^\alpha\,,
\quad \xi > 1\,.
\end{equation}
We defined the discontinuity via $\operatorname{\text{Disc}}f(\xi) = f(\xi - \ii 0) - f(\xi + \ii 0)$ with $\xi >1$.  The $m^{\mathrm{th}}$ mode $\mathfrak R_{-1,m}(0)$ folds in the factor of $\freq^{-m\, b_d}$. Therefore, the discontinuity across the branch cut $\xi>1$ from the Borel singularity $\xi=1$ is then given by:
\begin{equation}\label{eq:borel-zero-discontinuity}
\begin{aligned}
\operatorname{Disc} \psi^+_{n}(\xi)
&=
-2 \ms\ii \psi^-_n(\xi)
+ \ii \sum_{m=1}^{n}\, \frac{2^{m\, b_d}\, \mathfrak R_{-1,m}(0)}{\Gamma(m\,b_d)}\,
\int_1^\xi\, dv\,
(\xi-v)^{m\,b_d-1}\, \psi^-_{n-m}(v)\,.
\end{aligned}
\end{equation}
Here we have used the parity relation between WKB coefficients $(-1)^k\psi^+_{n-m,k}
=\psi^-_{n-m,k}$ and the identity
\begin{equation}
\begin{aligned}
\frac{2^{m\, b_d}}{\Gamma(m\,b_d)}
\int_1^\xi\, dv\,
(\xi-v)^{m\,b_d - 1}
\frac{[2\, (v-1)]^{k-\frac{1}{2}}}
{\Gamma\pqty{k+\frac{1}{2}}}
=
\frac{[2\,(\xi-1)]^{m\,b_d +k-\frac{1}{2}}}
{\Gamma(mb_d+k+\frac12)}\,.
\end{aligned}
\end{equation}

If we further assume that the Laplace transform of $\widehat{\Psi}$ commutes with the small $x$ expansion, then we deduce the connection formula of import.  In the $x$ chart we have the relation
\begin{equation}
\begin{aligned}
\Psi_+^{D_{-1},\vartheta}
-\Psi_+^{D_0,\vartheta}
&=
2\,x\,\int_{\mathcal{C}_-}\,
d\xi\, e^{-\freq \ms x\ms(2\,\xi-1)}
\operatorname{Disc}\widehat{\Psi}_+
\bigl(x,x\ms(2\ms\xi-1)\bigr)
\\
&=
\ii\,\mathcal{R}_{-1}^{(\vartheta)}(0,\freq)\, \Psi_-^{D_0,\vartheta}.
\end{aligned}
\end{equation}
Here $\mathcal{R}_{-1}^{(\vartheta)}(0,\freq)$ denotes the Borel sum of the all-orders Stokes multiplier. It refers to the crossing $D_0\to D_{-1}$ in the convention of~\cref{sec:core-multiplier}, cf.~\cref{fig:scaling-core-regions} for a pictorial depiction.

\begin{figure}[hbt!]
  \centering
  \includegraphics[width=0.9\linewidth]{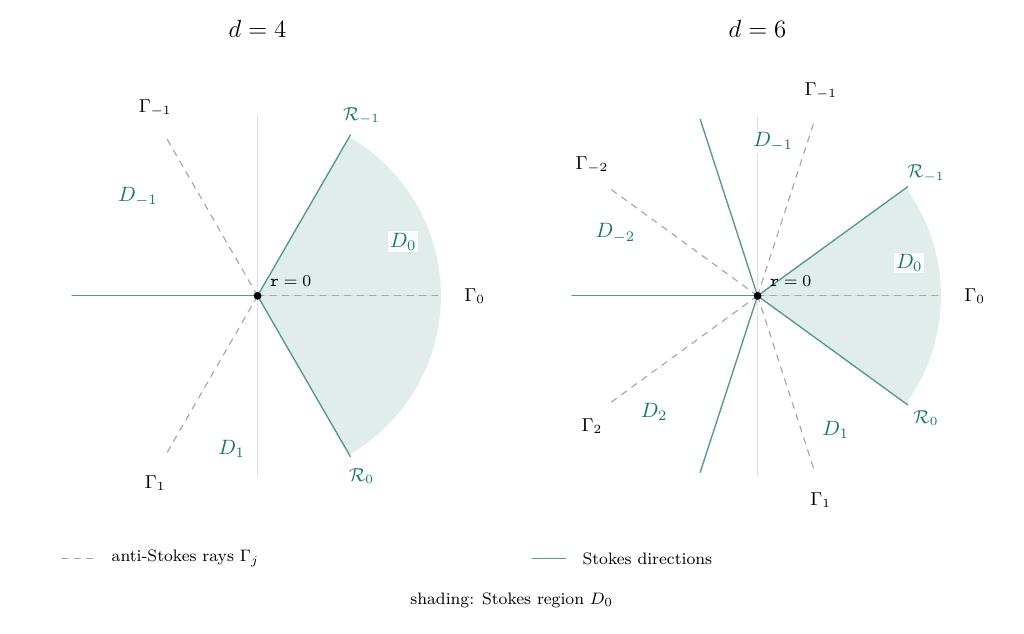}
  \caption{The local Stokes geometry near $\rpr =0$, with the Stokes regions $D_j$ shown. The Stokes lines are solid olive green, while the anti-Stokes lines are dashed. Crossing the Stokes line involves the multipliers $\mathcal{R}_j$.  This is a variant of~\cref{fig:scaling-core-regions} drawn in the $\rpr$ coordinate chart, with the Stokes lines emanating from the singularity, and the core turning points remaining unresolved.  }
  \label{fig:borel-core-regions}
\end{figure}

We are now in a position to  go back to the $\rpr$ coordinate. To do so, we need to recall the relation $x(\rpr) \sim -\frac{\ii}{d-1}\, \rpr^{\frac{d-1}{2}}$ which implies as in~\cref{sec:core-multiplier} the $\rpr$ plane  is divided into $(d-1)$ Stokes domains $D_j$. We already presaged this in our connection formula above, where the $x$ chart captures the domains $D_{0}$ and $D_{-1}$. In fact, the anti-Stokes lines used in~\cref{sec:tripscale-core}, viz., the rays $\Gamma_j$ and $\Gamma_{j-1}$ in the $\rpr$ coordinate chart with $j=0$ refer here to the incoming and outgoing waves.

Under $x \to e^{-i\pi}x$, the $\rpr$ coordinate rotates as $\rpr \to e^{-\frac{2\pi \ii}{d-1}}\, \rpr$.  We then let $\mathfrak R_{j-1}(0,\freq)$ denote the Stokes multiplier for connecting the domains $D_j$ and $D_{j-1}$.  There are two phase rotations to keep track of. First, under  $x \to e^{-i\pi}x$ the coefficients in the expansion of the potential rotate as $v_n \to e^{-\ii\pi \ms n\ms b_d} \,v_n$. Second, we need the translation to the overall Stokes multiplier associated with crossing between Stokes regions $D_j$ and $D_{j-1}$. This is modified from the answer deduced in the $x$ chart by  $\mathfrak R_{-1,n}(0) \to e^{-i \pi j n b_d}\,  \mathfrak R_{-1,n}(0)$:
\begin{equation}\label{eq:borel-wkb-zero-rotated-series}
\mathfrak R_{j-1}(0,\freq)
=
-2+\sum_{n\ge1}
e^{-i\pi jn b_d}\mathfrak R_{-1,n}(0)\freq^{-n b_d}.
\end{equation}
The general core connection formula is then given by
\begin{equation}\label{eq:borel-wkb-zero-regional-connection}
\begin{pmatrix}
  \Psi_+^{D_{j-1},\vartheta}\\
  \Psi_-^{D_{j-1},\vartheta}
\end{pmatrix}
=
\begin{cases}
\begin{pmatrix}
  1& \ii \mathcal{R}_{j-1}^{(\vartheta)}(0,\freq)\\
  0&1
\end{pmatrix}
\begin{pmatrix}
  \Psi_+^{D_j,\vartheta}\\
  \Psi_-^{D_j,\vartheta}
\end{pmatrix},
&j\ \text{even},\\[12pt]
\begin{pmatrix}
  1&0\\
  \ii \mathcal{R}_{j-1}^{(\vartheta)}(0,\freq)&1
\end{pmatrix}
\begin{pmatrix}
  \Psi_+^{D_j,\vartheta}\\
  \Psi_-^{D_j,\vartheta}
\end{pmatrix},
&j\ \text{odd}.
\end{cases}
\end{equation}
This completes the derivation of the core multipliers, in a coordinate chart where the singularity remains unresolved and is the only turning point (at $\zeta =0$). It should not be a surprise that the calculation using the Borel sums results in the same expression as that obtained earlier. Both are attempting to solve the core problem, albeit in different local charts, but the Stokes domains and charts  relevant for the physical problem are identical.

\subsubsection{Reflection coefficients for planar \SAdS{d+1} black holes}\label{sec:borel-wkb-coefficients}

The strategy for finding the reflection coefficients is no different from the earlier analysis in~\cref{sec:core-data-multiplier}. The only difference is that our current ordering of domains picks a related Stokes multiplier. To keep the discussion self-contained we give here the main results for planar black holes (neutral).  The first two reflection coefficients we need are given by
\begin{equation}\label{eq:borel-wkb-first-reflection-coefficients}
\begin{aligned}
\mathfrak{R}_{-1,1}(0)
&=
4\sin^2\left(\frac{\pi\ms b_d}{2}\right) \mathsf{A}_1(b_d)\,v_1,
\\[4pt]
\mathfrak{R}_{-1,2}(0)
&=
4\sin^2(\pi \ms b_d)
\left[ \mathsf{A}_1(2b_d)\,v_2 +\mathsf{A}_2(b_d)\,v_1^2
\right]
- 4\, \sin^2\!\left(\frac{\pi b_d}{2}\right) \mathsf{A}_1(b_d)^2\,v_1^2.
\end{aligned}
\end{equation}
where
\begin{equation}\label{eq:borel-wkb-amplitude-functions}
\begin{aligned}
\mathsf{A}_1(b_d)
&=
\frac{2^{b_d-2}\, \Gamma\pqty{\frac{b_d}{2}}^2\, \Gamma(1-b_d)}{\Gamma\pqty{1- \frac{b_d}{2}}^2}
=  - \frac{(d-1)^{a_d}}{4\,\sin^2\pqty{\frac{\pi\ms a_d}{2}}} \, C_d\,,
\\
\mathsf{A}_2(b_d)
&=
\frac{2^{2\,b_d-2}\Gamma(b_d)^3\,\Gamma(1-b_d)}
{16\,(b_d-1)^3}
\Biggl[
  (5\ms b_d^2-14\ms b_d+10)
  \frac{\Gamma(3-2\ms b_d)}{\Gamma(b_d)\, \Gamma(2-b_d)^3}
  \\
  &\hspace{4cm} +
  \frac{(b_d-2)^3}{b_d}
  \,{}_6F_5\!\left(
    \begin{matrix}
      b_d-1,b_d-1,b_d-1,b_d-1,\frac{b_d+1}{2},\frac{b_d-2}{2}\\
      1,1,1,\frac{b_d-1}{2},\frac{b_d+2}{2}
  \end{matrix};1\right)
\Biggr].
\end{aligned}
\end{equation}
The leading answer is written in terms of $C_d$ introduced in~\eqref{eq:core-neutralR-freq} to make direct contact with the previous analysis (the identity uses $a_d + b_d =2$, cf.~\eqref{eq:abd-def}).  The other difference is that we should shift the frequency to the neighboring Stokes domain;  $\mathfrak{R}_{-1,1}(0) = e^{\ii \pi \ms b_d}\, \delta_1\, \mathfrak{R}_0(0)$. The potential coefficients are
\begin{equation}
\begin{aligned}
v_1
&=
\frac{e^{-\frac{\ii \pi \ms a_d}{2}}}{(d-1)^{a_d}}\, \bqty{\nu^2 - \frac{d^2}{4\, (2\,d -1)}}
\\
v_2
&=
e^{-\ii\pi \ms a_d}\, (d-1)^{b_d - a_d}\, \frac{d+1}{2\,d -1}\, \bqty{\nu^2 +
\frac{d^2\, (d^2 - 5\,d +2)}{4\, (2\,d -1)\, (3\,d -1)}}\,.
\end{aligned}
\end{equation}
%

\subsubsection{Non-perturbative spectral function for \SAdS{5}}\label{sec:borel-wkb-planar-four}

With the connection formula~\cref{eq:borel-wkb-zero-regional-connection}  obtained thus, we may complete the exact-WKB analysis as in the preceding sections. We will present here the result for the
transseries expression for the non-perturbative part of the spectral function. We  have already obtained the leading result in~\cref{sec:core-np-spectral}. The main point of this exercise is to demonstrate that the calculation done in the action chart reproduces the same answer.

Specializing to $d=4$, and using $\freq = \frac{\omega}{r_+} = \frac{\beta\omega}{\pi}$ with the latter following from $\beta = \frac{\pi}{r_+}$, we find the exact WKB expression for $\rho_{\mathrm{np}}$ to be given by
\begin{equation}\label{eq:borel-wkb-planar-four-spectral-factor}
\rho_{\mathrm{np}}
=
\frac{1-e^{-\bw}}{
  \abs{
    \sqrt{1+
    \left(\abs{\mathcal{R}_{-1}^{\mathrm{med}}\pqty{0,\freq}}^2-1\right) e^{-\bw}}
    - \mathcal{R}_{-1}^{\mathrm{med}}\pqty{0,\freq}\,
e^{\ii\pi\ms \nu-\frac{1+\ii}{2}\, \bw}}^2 }\,.
\end{equation}
The full transseries expansion can then be obtained as
\begin{equation}\label{eq:borel-wkb-planar-four-transseries}
\begin{aligned}
\rho_{\mathrm{np}}
&\sim
1+2\, \Re \sum_{n\geq1}\, \sum_{m\geq0}\,
\frac{(-1)^m \, \pqty{\frac{n}{2}}_m}{m!}\,
\left[\mathfrak{R}_{-1}\!\left(0,\freq\right)\right]^n
\left[\abs{\mathfrak{R}_{-1}\!\left(0,\freq\right)}^2-1 \right]^m\\
&\qquad \qquad
\times
\exp(-\frac{n+2\,m}{2}\, \bw  +\ii n\left(\pi\nu- \hbw\right) )\,.
\end{aligned}
\end{equation}
The leading non-perturbative sector is the one with $(n,m)=(1,0)$:
\begin{equation}\label{eq:borel-wkb-planar-four-leading-sector}
\begin{aligned}
\rho_{\mathrm{np}}\bigg|_{(1,0)}
=
2\, e^{-\hbw}\, \Re \left[
e^{\ii (\pi\ms \nu- \hbw)}\,  \mathfrak{R}_{-1}\pqty{0,\freq} \right].
\end{aligned}
\end{equation}
We have also  matched the leading answer given in~\eqref{eq:psads5-rhonp-freq} with the result obtained in~\cite{Afkhami-Jeddi:2025wra}. Evaluating the second coefficient, we find that our analytic answer also agrees with their numerically determined fit.  Explicitly, specializing to a scalar primary with $\Delta =4$, i.e., a massless scalar field in the bulk, we obtain
\small{
  \begin{equation}
  \begin{aligned}
  \rho_{\mathrm{np}}(\omega,0)
  &=
  1 - 2\,e^{-\hbw} \Bigg[
    2\,\cos\theta_4 +\frac{24\,C_4}{7}\, \freq^{-\frac{4}{3}}
    \cos(\theta_4+\frac{\pi}{3})
    + \frac{C_4'}{2}\,\freq^{-\frac{8}{3}} \,  \cos(\theta_4+\frac{2\pi}{3})
  +\order{\freq^{-4}} \Bigg]
  \\
  &\quad
  + 8\, e^{-\bw} \Bigg[ \cos(2\theta_4) +\frac{24\,C_4}{7}\,\freq^{-\frac{4}{3}}
    \cos(2\theta_4+\frac{\pi}{3})
    +\left(\frac{C_4'}{2}+\frac{144\,C_4^2}{49}\right)
    \freq^{-\frac{8}{3}}
    \cos\left(2\theta_4+\frac{2\pi}{3}\right)
    \\
    & \qquad\qquad  \quad
  + \order{\freq^{-4}} \Bigg] + \order{e^{-\frac{3}{2}\,\bw}}\,.
  \end{aligned}
\end{equation}}
The first coefficient and phase of the second coefficient agree analytically with~\cite{Afkhami-Jeddi:2025wra} as noted around~\eqref{eq:psads5-rhonp-freq}.  For $\nu =2$, we have the coefficient
$\frac{48\,C_4}{7} = 6.46639$ which agrees with them.
The norm of our second coefficient is
\begin{equation}\label{eq:borel-wkb-accm-second-coefficient}
\begin{aligned}
C_4' = 2\, \abs{\mathfrak{R}_{-1,2}(0)}
&=6^{2/3}\Bigg[
  \frac{864}{49}
  \frac{\Gamma(\tfrac{2}{3})^6}{\Gamma(\tfrac{1}{3})^4}
  +\frac{428}{539}
  \frac{\Gamma(\tfrac{1}{3})^3}{\Gamma(\tfrac{2}{3})^2} -
  \frac{16}{49}\Gamma(\tfrac{1}{3})^3\Gamma(\tfrac{2}{3})
  \,{}_6F_5\!\left(
    \begin{matrix}
      \frac{1}{3},\frac{1}{3},\frac{1}{3},\frac{1}{3},\frac{7}{6},-\frac{1}{3}\\
      1,1,1,\frac{1}{6},\frac{5}{3}
    \end{matrix};1
  \right)
\Bigg]\\
&\simeq 6.939713
\end{aligned}
\end{equation}
This is close to their numerical fit obtained in~\cite{Afkhami-Jeddi:2025wra} who find
$C_4' \simeq 6.93972$.

\subsection{Analysis at non-zero momentum}\label{sec:borel-wkb-scaled}

While we focused on $\zeta=0$ thus far for the action chart analysis of the Borel sums, we can readily generalize the discussion to non-vanishing rescaled momentum $\trs$. We now outline the salient features of this analysis to illustrate that it agrees with the multiplier obtained in the core-scaling chart.

\subsubsection{Connection formula}\label{sec:borel-wkb-scaled-connection}

As noted at the outset in~\cref{sec:tripscale-core} we can adapt a chart to the action coordinate obtained from $Q_0(\trs)$. However, since the action coordinate for $\trs =0$ was unbranched, it is efficacious to continue using it and treat momentum dependence as providing corrections. Per se, this is similar in spirit to how we used the core-scaling chart.

The wave equation in action coordinate $x$ introduced in~\eqref{eq:xaction-coord} is
\begin{equation}\label{eq:borel-wkb-scaled-action-equation}
\left[
  \partial_x^2 - \freq^2 + \frac{1}{4\,x^2} -w_0(\trs)\, \freq^{a_d}\,
x^{-b_d} \right]\Psi =
\left[
  V(x) +\freq^{a_d}\, W(x;\trs)
\right]\Psi\,.
\end{equation}
Here the potential $V(x)$ is unchanged, but there is now a contribution from the momentum dependent part. These terms of the  potential have an expansion
\begin{equation}\label{eq:borel-wkb-scaled-potential-series}
\begin{aligned}
w_0(\trs)
&=
\trs^2\, \pqty{e^{\frac{\ii\pi }{2}}\,(d-1)}^{-b_d},\\
W(x;\trs)
&=
\sum_{j\geq 1} \, w_j(\trs)\, x^{(j-1)\, b_d},
\end{aligned}
\end{equation}
with all $w_j$ proportional to $\trs^2$.

The explicit fractional powers of frequency in the $\trs$ dependent part of the potential require use of fractional derivatives in the Borel time $\tau$. At zero momentum ($\trs= 0$) fractional inverse powers of frequency arose instead from the local expansion of the potential. If we wish to obtain a result as a function of $\trs$, then we should learn how to account for such fractional powers directly in the Borel sum. Multiplication by the fractional frequency power becomes a fractional derivative in Borel time. The fractional derivative operator itself can be defined by
\begin{equation}\label{eq:borel-wkb-fractional-derivative}
\mathcal{D}_\pm^\alpha g(x,\tau)
=
\frac{\partial}{\partial\tau}
\left[ \frac{1}{\Gamma(1-\alpha)} \, \int_{\mp x}^{\tau}\, du\,
(\tau-u)^{-\alpha}g(x,u) \right], \qquad
0<\alpha<1\,.
\end{equation}
The Borel transforms of $\Psi_\pm$ now obey the (fractional) differential equation
\begin{equation}\label{eq:borel-wkb-scaled-borel-equation}
\begin{aligned}
\left[ \partial_x^2-\partial_\tau^2+\frac{1}{4x^2} - w_0(\trs)\,x^{-b_d}\, \mathcal{D}_\pm^{a_d}
\right]\widehat{\Psi}_\pm
=
\left[ V(x) + W(x;\trs) \,\mathcal{D}_\pm^{a_d} \right]\widehat{\Psi}_\pm\,.
\end{aligned}
\end{equation}

To solve the equation~\eqref{eq:borel-wkb-scaled-borel-equation} one generalizes the perturbative ansatz~\cref{eq:borel-pert-ansatz} to
\begin{equation}\label{eq:borel-wkb-scaled-homogeneous-expansion}
\widehat{\Psi}_\pm\bigl(x,x\,(2\ms\xi-1);\trs\bigr)
=x^{-\frac{1}{2}}\, \sum_{n\geq0}\, x^{n\, b_d}\,\psi_n^\pm(\xi;\trs)\,.
\end{equation}
Plugging in we find that the  perturbative profiles $\psi^\pm_n(\xi;\trs)$ now satisfy
\begin{equation}\label{eq:borel-wkb-scaled-hierarchy}
\begin{aligned}
&
\Biggl[ \xi\, (1-\xi)\, \partial_\xi^2
  +\left(\frac{3}{2}-n\, b_d - (3-2\, n\, b_d)\,\xi\right)\partial_\xi
  -(1-n\, b_d)^2 +\frac{w_0}{2^{a_d}}\,  \mathcal{D}_{\xi,\pm}^{a_d}
\Biggr]\psi_n^\pm(\xi;\trs)
\\
&\qquad
= -\sum_{j=1}^{n} \left[ v_j\, \psi_{n-j}^\pm(\xi;\trs)
  +\frac{w_j}{2^{a_d}}\, \mathcal{D}_{\xi,\pm}^{a_d}\,\psi_{n-j}^\pm(\xi;\trs)
\right], \qquad n\geq0\,.
\end{aligned}
\end{equation}

In the $\xi$ coordinate the fractional derivative operators are explicitly given as
\begin{equation}
\begin{aligned}
\mathcal{D}_{\xi,+}^{\alpha}g(\xi)
&=
\frac{1}{\Gamma(1-\alpha)} \, \dv{\xi} \int_0^\xi\, dv\, (\xi-v)^{-\alpha}\, g(v)\,,
\\
\mathcal{D}_{\xi,-}^{\alpha}g(\xi)
&=
\frac{1}{\Gamma(1-\alpha)} \, \dv{\xi}
\int_1^\xi\, dv\, (\xi-v)^{-\alpha}\, g(v)\,, \qquad 0<\alpha<1\,.
\end{aligned}
\end{equation}
Already at $n=0$, the profile depends on $\trs$ through $w_0$; it reduces
to the Bessel profile~\eqref{eq:borel-zero-bessel} at zero momentum.
The fractional frequency expansion generalizes~\eqref{eq:borel-zero-endpoint} to
\begin{equation}\label{eq:borel-wkb-scaled-endpoint-series}
\begin{aligned}
\psi_n^+(\xi;\trs)
&=
\sum_{k\geq0} \, \frac{\psi_{n,k}^+(\trs)}
{\Gamma\!\left(\frac{k}{d-1}+\frac{1}{2}\right)} (2\ms\xi)^{\frac{k}{d-1}-\frac{1}{2}},
&&\qquad
\xi\to0,\\
\psi_n^-(\xi;\trs)
&=
\sum_{k\geq0} \, \frac{\psi_{n,k}^-(\trs)}
{\Gamma\!\left(\frac{k}{d-1}+\frac{1}{2}\right)}
\bigl[2(\xi-1)\bigr]^{\frac{k}{d-1}-\frac{1}{2}},
&&\qquad
\xi\to1\,.
\end{aligned}
\end{equation}
The idea is to continue to use WKB recursion to fix the coefficients  $\psi^\pm_{n,k}$.

As in~\eqref{eq:borel-series-refl}, assuming the following connection structure, we define the all-orders Stokes multiplier $\mathfrak R_{-1,n}(\trs)$ by a connection problem between $\xi=0$ and $\xi=1$, i.e.,
\begin{equation}\label{eq:borel-wkb-scaled-reflection-definition}
\begin{aligned}
\psi_n^+(\xi;\trs)
&= \sum_{m=0}^{n}\sum_{k\geq0}
\frac{ \mathfrak{R}_{-1,m}(\trs)\, \psi_{n-m,k}^-(\trs)
[2\, (1-\xi)]^{m\, b_d+\frac{k}{d-1}-\frac{1}{2}} }{
  2\sin\!\left[\pi\left(m\, b_d+\frac{k}{d-1}-\frac{1}{2}\right)\right]
\Gamma\!\left(m\, b_d+\frac{k}{d-1}+\frac{1}{2}\right) }
\\
&\qquad
+ \text{terms holomorphic at }\xi=1\,.
\end{aligned}
\end{equation}
The remaining step is essentially the same as in the vanishing momentum case. By the same manipulation leading to~\eqref{eq:borel-zero-discontinuity}, we have
\begin{equation}\label{eq:borel-wkb-scaled-discontinuity}
\begin{aligned}
\operatorname{Disc}\psi_n^+(\xi;\trs)
= \ii \mathfrak{R}_{-1,0}(\trs)\, \psi_n^-(\xi;\trs)
+\ii\sum_{m=1}^{n}\,
\frac{2^{m \,b_d}\, \mathfrak{R}_{-1,m}(\trs)}{\Gamma(m \,b_d)}
\int_1^\xi\, dv\,  (\xi-v)^{m\, b_d-1}\, \psi_{n-m}^-(v;\trs)\,.
\end{aligned}
\end{equation}

Using the same assumption as for vanishing momentum that one can exchange the order of the Laplace transform and small $x$ expansion, we obtain the connection formula as
\begin{equation}\label{eq:borel-wkb-scaled-summed-connection}
\begin{aligned}
\Psi_+^{D_{-1},\vartheta}-\Psi_+^{D_0,\vartheta}
&=
\ii \mathcal{R}_{-1}^{(\vartheta)}(\trs,\freq)\, \Psi_-^{D_0,\vartheta}\,,
\\
\Psi_-^{D_{-1},\vartheta}
&=
\Psi_-^{D_0,\vartheta}\,.
\end{aligned}
\end{equation}
For Stokes multipliers associated with other Stokes lines (see~\cref{fig:borel-core-regions}), under the clockwise rotation
$x\mapsto e^{-\ii\pi \ms j}x$, the momentum $\trs$ also needs to be phase rotated as explained in~\cref{eq:trs-j}. The phase-rotated Stokes multiplier~\cref{eq:borel-wkb-zero-rotated-series} is then generalized to
\begin{equation}\label{eq:borel-wkb-scaled-rotated-series}
\begin{aligned}
\mathfrak{R}_{j-1}(\trs,\freq)
&=
\mathfrak{R}_{-1,0}(\trs_j) +\sum_{n\geq1}\, e^{-\ii\pi \ms j\ms n \ms b_d}\,
\mathfrak{R}_{-1,n}(\trs_j)\, \freq^{-n b_d}.
\end{aligned}
\end{equation}
This leads to the connection formula, the analog of~\eqref{eq:core-conn-j} adapted to our present conventions, which reads
\begin{equation}\label{eq:borel-wkb-scaled-regional-connection}
\begin{pmatrix}
  \Psi_+^{D_{j-1},\vartheta}\\
  \Psi_-^{D_{j-1},\vartheta}
\end{pmatrix}
=
\begin{dcases}
\begin{pmatrix}
  1& \ii \mathcal{R}_{j-1}^{(\vartheta)}(\trs,\freq)\\
  0&1
\end{pmatrix}
\begin{pmatrix}
  \Psi_+^{D_j,\vartheta}\\
  \Psi_-^{D_j,\vartheta}
\end{pmatrix},
&j\ \text{even},
\\
\begin{pmatrix}
  1&0\\
  \ii\mathcal{R}_{j-1}^{(\vartheta)}(\trs,\freq)&1
\end{pmatrix}
\begin{pmatrix}
  \Psi_+^{D_j,\vartheta}\\
  \Psi_-^{D_j,\vartheta}
\end{pmatrix},
&j\ \text{odd}\,.
\end{dcases}
\end{equation}
Once we have this core connection data, we can assemble the boundary to horizon connection matrix using the same methods described earlier.

\subsubsection{The planar \SAdS{5} spectral function}\label{sec:borel-wkb-scaled-coefficients}

\begin{figure}[hbtp!]
  \centering
  \includegraphics[width=0.98\linewidth]{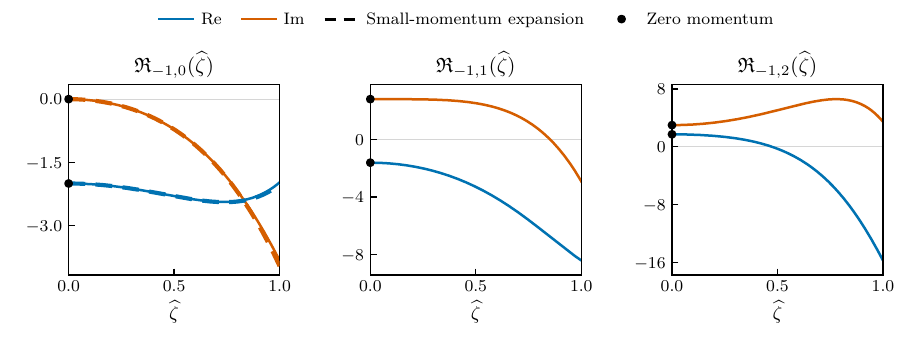}
  \caption{Real (blue) and imaginary (orange) parts of $\mathfrak R_{-1,n}(\trs)$ for $d=4$, $\nu=2$, and $n=0,1,2$. Solid curves are numerical evaluations at finite $\trs$. Black markers at $\trs=0$ give the analytic values~\cref{eq:borel-wkb-first-reflection-coefficients}. Dashed curves in the leading panel show the sixth-order small-momentum truncation of~\cref{eq:core-neutralr-mom}, converted to the $j=-1$ sector.}
  \label{fig:borel-wkb-scaled-coefficients}
\end{figure}

We illustrate the general momentum dependence in the core-scaling regime with an example, supplementing our determination in $\trs$ perturbation theory described in~\cref{sec:core-neutral}. The reflection coefficients can be computed numerically at finite $\trs$ or analytically at small $\trs$ as in~\cref{eq:core-neutralr-mom}. The first three coefficients for a massless scalar in a $d=4$ neutral planar black hole are shown in~\cref{fig:borel-wkb-scaled-coefficients}.

Once we have the reflection coefficients, we generalize the  exact-WKB results of~\cref{sec:borel-wkb-planar-four} straightforwardly to account for finite $\trs$. We find the non-perturbative spectral function to be given by~\eqref{eq:borel-wkb-planar-four-spectral-factor} with the median Stokes multiplier upgraded to account for the $\trs$ dependence. Noting that $\freq = \frac{\bw}{\pi}$ we have
\begin{equation}\label{eq:borel-wkb-scaled-planar-four-spectral-factor}
\rho_{\mathrm{np}}
= \frac{1-e^{-\bw}}{
  \abs{
    \sqrt{1+
    \left( \abs{\mathcal{R}_{-1}^{\mathrm{med}}(\trs,\freq)}^2-1\right)e^{-\bw}}
- \mathcal{R}_{-1}^{\mathrm{med}}(\trs,\freq) \, e^{\ii\pi\nu-\frac{1+\ii}{2}\bw}}^2}.
\end{equation}
The full formal transseries is
\begin{equation}\label{eq:borel-wkb-scaled-planar-four-transseries}
\begin{aligned}
\rho_{\mathrm{np}}
&\sim
1+2\, \Re \sum_{n\geq1}\,\sum_{m\geq0}\,
\frac{(-1)^m\, \pqty{\frac{n}{2}}_m}{m!}\,
\left[\mathfrak{R}_{-1}\!\left(\trs,\freq\right)\right]^n
\left[ \abs{\mathfrak{R}_{-1}\!\left(\trs,\freq\right)}^2-1
\right]^m
\\
&\qquad \qquad
\times\exp[
-\frac{n+2\,m}{2}\,\bw + \ii n\left(\pi\ms\nu-\hbw\right) ]\,.
\end{aligned}
\end{equation}
In particular, for a massless scalar field, dual to a primary with  $\nu=2$, the leading non-perturbative sector gives
\begin{equation}\label{eq:borel-wkb-scaled-planar-four-leading-sector}
\begin{aligned}
\rho_{\mathrm{np}}\bigg|_{(1,0)}
\sim 2\, e^{-\hbw}\,  \Re\Bigg\{ e^{-\ii\hbw}\Bigg[
    \mathfrak{R}_{-1,0}(\trs) + \mathfrak{R}_{-1,1}(\trs)
    \left(\frac{\pi}{\bw}\right)^{\frac{4}{3}}
    + \mathfrak{R}_{-1,2}(\trs) \left(\frac{\pi}{\bw}\right)^{\frac{8}{3}}
+ \cdots\Bigg]\Bigg\}.
\end{aligned}
\end{equation}
The first three coefficients are shown in~\cref{fig:borel-wkb-scaled-coefficients}. Setting $\trs=0$ recovers~\cref{eq:borel-wkb-planar-four-spectral-factor,eq:borel-wkb-planar-four-transseries,eq:borel-wkb-planar-four-leading-sector}, to the same leading order. This generalizes the analysis of~\cite{Afkhami-Jeddi:2025wra} to include momentum corrections in the core.

\newpage
\part{Future prospects}\label{prt:conclude}

\section{Discussion}\label{sec:discuss}

In the course of our discussion, we have explored several features of thermal two-point functions in planar holographic field theories. Specifically, we argued that the OPE and the non-perturbative contributions to the spectral function cleanly factorize; this follows immediately from a WKB analysis of the bulk wave equations.  Moreover, the leading exponential rate of the latter is universal and is insensitive to the quantum numbers of the operators in question.  This particular feature had earlier been observed in some specific cases, for example~\cite{Afkhami-Jeddi:2025wra} noted similar behavior for massless scalar and conserved  currents, but our statement is valid generally.

The leading WKB action in the high frequency limit controls (a) the non-perturbative contribution to the spectral function (b) the asymptotic highly damped quasinormal mode spectrum, and (c) the asymptotics of thermal OPE coefficients. Again some of these statements have been known in some specific kinematic regimes, typically at zero spatial momentum (or in some cases spatial separation). Our observation is that these hold for arbitrary timelike momentum taken to scale linearly with the frequency. The asymptotics of quasinormal modes along fixed timelike momentum rays (i.e., $\frac{k}{\omega}$ fixed) are  still determined by classical period integrals on the WKB spectral curve.
In addition to understanding high frequency non-perturbative features of the thermal observables,
the analysis also sheds light on how these probe the black hole singularity in holographic CFTs.

\looseness=-1
The finite timelike momentum problem has non-trivial WKB turning points, which are complex radial positions. So we not only get to analyze spatial resolution of the field theory observables, but we also get to ask about the black hole singularity's imprint. The exact-WKB problem says that the relative WKB action between the boundary and the complex turning point completely controls the non-perturbative transseries at large real frequency.   Nevertheless, the leading action has a singularity-anchored representative. This is a trajectory that follows the bouncing null geodesic of~\cite{Fidkowski:2003nf} but then detours to the complex turning point and picks up the corresponding classical period. This representative is, however, just a particular choice that one can make, and by no means a distinguished one.

Furthermore, in some situations, like spherically symmetric black holes, the turning points end up being real and in the exterior of the black hole along the transport path. This occurs above a threshold angular momentum, and corresponds to closed null orbits in the black hole geometry. The WKB trajectory in such cases does not even make it to the horizon, as it encounters a turning point en route. The result is a bound state like spectrum, seen as nearly undamped quasinormal modes. Such bound states are a signal of bulk-cone singularities. The global black hole example serves to demonstrate that the bouncing geodesic singularity of~\cite{Fidkowski:2003nf} and the bulk-cone singularity of~\cite{Hubeny:2006yu} are manifestations of the same phenomenon (which was the original intuition for the introduction of the latter), albeit seen in different parameter ranges.

There are two regimes where the singularity's imprint is clean. The first is when we explore the WKB asymptotics along rays with $\arg \omega \neq 0$. As we dial the phase of the frequency, we pick up contributions from the cycle between the complex turning points and the singularity. This is visible both in the exact-WKB analysis through the DDP jump formula, and also in the non-principal-sheet behavior of the OPE coefficients. Note that these effects involve subleading WKB corrections, and are thus dependent on the observable in question, and more importantly, assume the analytic structure of the bulk wave operator is known.

The second regime is the core-scaling regime where we scale momentum at a sublinear rate with frequency. In this limit the turning points lie close to the singularity and modify the subleading WKB problem considerably. Exact-WKB techniques can fail if the singularity structure of the subleading potential terms is much worse than that of the leading potential, especially when the turning points of the latter approach the singular locus of the former.  We have extended here the exact-WKB analysis to such situations. The leading order WKB approximation needs to be corrected to account for the singular behaviour of the subleading potential. The core-scaling was deduced by balancing the two contributions and treating the near-singularity region exactly. The change in the WKB analysis is that the core-region supplies a non-trivial Stokes multiplier, which can be seen as a reflection coefficient for scattering off the singularity, as emphasized in~\cite{Afkhami-Jeddi:2025wra}. This analysis not only closes a gap from our previous investigation~\cite{Jia:2026ryl}, but also provides a hint for probing more general near-singularity behavior (outside the remit of static thermal ensembles). We comment on this and some broader implications of our analysis below.

\paragraph{A caveat regarding analytic continuation:} The WKB analysis crucially relies on analytic properties of the bulk wave equations, which we use to argue for specific features of the dual CFT's thermal correlators. It is worth recording its limitations, especially in the context of employing the singularity-anchored representative to declare that the leading WKB answer involves transport all the way down to the black hole singularity. Consider the case of general timelike momentum for simplicity. The physical problem is the determination of a causal observable, retarded/advanced thermal Green's function. This observable is characterized by a wave operator specified in the black hole exterior, and has suitable boundary conditions imposed at the AdS boundary and the outer horizon. Suppose we replace the core of the black hole spacetime by a smooth singularity free region in a $C^\infty$  manner, replacing $f(r)$ for planar black holes by a smooth function $f_{\mathrm{reg}}(r)$. A simple example is to replace $r_+^d \to r_+^d\, \sigma(r)$ for some sigmoid function $\sigma(r)$ which smoothly grows from $0$ to $1$ in a region close to the singularity (say from $r=0$ to $r= \alpha\, \rp$ with $\alpha < \frac{1}{2}$).  The exterior opers for the two geometries will be the same, so one ends up with the same WKB data because one is analytically continuing this exterior geometry, and it uniquely lands us on the original black hole solution. The geometry with the near-singularity region modified by $\sigma(r)$ is not analytic. The data necessary for WKB such as turning points, Stokes graphs etc., are ambiguous without additional information being supplied. Our point is simply that one has to exercise care in extending causal observables, which are  only cognizant about the domain of outer communication, to infer information about causally inaccessible regions. This point has been made in several works hitherto, notably in~\cite{Marolf:2004fy} which emphasized the limitations of using analytic spacetimes to make inferences of bulk geometric structures.

\paragraph{The charged black hole interior:} Investigations of the imprint of the timelike singularity of charged \RNAdS{} black holes have suggested that the relevant bouncing geodesic is a limit of a sequence of timelike geodesics that connect two asymptotic regions, one to the future of the other via a passage through the Cauchy (inner) horizon~\cite{Ceplak:2025dds,Dodelson:2025jff,AliAhmad:2026wem,Chakravarty:2026fdm}. We would like to argue that this interpretation is too quick. For a fixed timelike momentum ray, this is immediate: the complex turning points relevant for exact-WKB show no inkling of the inner horizon, even if one can find a singularity-anchored representative to capture the leading WKB asymptotics. The situation in the core-scaling limit might appear more promising, for now the transport path makes its way all the way to the singularity. The real radial projection of the core region near the singularity does indeed lie beyond the Cauchy horizon. Nevertheless, one cannot interpret the complex continuation of the transport path as connecting boundaries which are causally stacked. The observable is still the causal retarded Green's function, but it gets contributions from complex geodesics. The endpoints of the correlator remain anchored in one thermofield double chamber of the Penrose diagram. The complex WKB path should not, in our opinion, be used to argue for a displacement of the insertions in the Lorentzian section.

\paragraph{The core-region connection problem \& Kasner epochs:} The core-scaling analysis we explored has interesting implications for recent investigations examining the imprint of Kasner epochs in the asymptotic behavior of quasinormal frequencies~\cite{Hartnoll:2026vhu,Xiao:2026pir} or OPE coefficients~\cite{Ceplak:2026ard}. We can immediately translate their observations to local modifications in the computation of the non-perturbative spectral function. Additionally, one can infer how a chaotic BKL like series of transitions~\cite{Belinsky:1970ew} can be modeled directly in our framework.

We need the following simple observation to reinterpret our core analysis. The Kasner geometry near a spacelike singularity is characterized by anisotropic exponents. An ansatz for the local geometry is
\begin{equation}
ds^2 = -d u^2 + u^{2\,p_t}\, dt^2 + \sum_i\, u^{2\, p_i}\, dx_i^2 \,, \qquad
p_t + \sum_i\, p_i = 1\,,
\end{equation}
with the coordinate $u$ measuring proper time from the singularity.
Realize that the neighborhood of the spacelike singularity  of \SAdS{d+1} black holes is exactly of this form, with  $p_t = \frac{2}{d} -1$ and $p_i = \frac{2}{d}$. Moreover, the anisotropy between the asymptotic temporal and spatial directions is what is responsible for the different $\omega$ scaling of the corrections to the core Stokes multiplier. The momentum corrections appeared as $\omega^{-a_d}$, with $a_d = \frac{d-2}{d-1}$, but the frequency corrections are at $\freq^{-b_d}$, with
$b_d = \frac{d}{d-1}$. In the above local model the subleading WKB corrections from $p_i$ and $p_t$ appear with exponents $a_i = \frac{2\,(1-p_i)}{1-p_t}$ and $b_t = \frac{2}{1-p_t}$, respectively.
It is also immediately clear that for such a geometry, the only ingredient necessary for the exact-WKB problem is the core Stokes multiplier $\mathcal{R}^{(\vartheta)}$, which should be obtained as a function of frequency (conjugate to $t$) and momenta $k_i$ (conjugate to $x_i$).

Several generalizations are immediate. We can model multi-epoch scenarios by modifying the core oper suitably. A simple model which mimics the example considered in~\cite{Hartnoll:2026vhu} is to consider the following wave operator in the projective $\rpr$ coordinate: $\dv[2]{\rpr} + \freq^2\, \rpr^a\, (\rpr+ \epsilon)^{1-a} + \freq^2\, \zeta^2\, \rpr^{-1} + \frac{1}{4}\,\rpr^{-2}$ for some suitable parameters $a,\epsilon$. The core scalings are different for small and large values of $\epsilon$, but can be analyzed with the same techniques described herein. One can for example check that $\delta \rho_{\mathrm{np}} = 2\, \Re\bqty{e^{-\ii \pi \nu}\, \mathcal{X}_{\mathrm{core}}\, \delta \mathcal{R}} + \cdots$ will be sensitive to the $\epsilon$ dependence of $\mathcal{R}$. The upshot is that the subleading corrections of the spectral functions capture the scaling exponents of the different epochs. It should also be clear that BKL transitions between epochs can also be handled by this local model, by composition of the core Stokes multipliers. To carry out the exercise explicitly, one requires the ordered connection matrices, a common normalization to compute Wronskians, knowledge of the transition and the associated propagation data, and finally the temporal extent of each epoch.
The essential point to appreciate is that the WKB problem is completely factorized: the core part can be solved independently of the transport problem, and the two can be spliced together to obtain the complete picture. It would be interesting to flesh out this picture.

\newpage
\section*{Acknowledgements}
It is a pleasure to thank Matthew Dodelson, Alba Grassi, Veronika Hubeny, Don Marolf, Andrei Parnachev, Steve Shenker, Zhenbin Yang, and Sasha Zhiboedov for illuminating discussions. We would also like to thank Andrei Parnachev for comments on a draft version of the manuscript.
H.F.J.~and M.R.~would like to thank the Hamilton Mathematics Institute, Trinity College, Dublin, and CERN for hospitality during the project.
In addition, M.R.~also would like to acknowledge the Aspen Center for Physics, which is supported by National Science Foundation grant PHY-2210452, for providing a stimulating environment during the course of this project. H.F.J.~acknowledges support from the Tsinghua Shuimu Scholar program. M.R.~was supported by U.S. Department
of Energy grant DE-SC0009999 and by funds from the University of California.

\noindent
\textbf{Statement of AI use:} We acknowledge the use of GPT Codex 5.5/6.0/6.1 (OpenAI), and Claude Opus 4.8/5.0/5.1, Fable 5.0 (Anthropic) for development of some of these ideas, and for assistance with computations. The authors take sole responsibility for the accuracy of this work.
M.R.~also acknowledges support from OpenAI through the ChatGPT for Academic Researchers program, which provided complimentary access to the OpenAI models used in this work.

\appendix

\part{Appendices}\label{prt:appendix}

\section{Scalar wave equations and their opers}\label[appendix]{app:Q0Q2examples}

The wave equations for minimally coupled scalar fields in \AdS{d+1} black hole backgrounds can be put in the form~\eqref{eq:wkb-boundary-universal-oper}. We collect here the explicit form of $Q_0$ and $Q_2$ for black holes in $d=4, 6, 8$ for reference.

\paragraph{Black holes in $d=4$:} We consider both neutral and charged asymptotically \AdS{5} black holes. In the former case we examine both the planar and global solutions, while all charged solutions analyzed herein are planar.

\begin{itemize}[wide,left=0pt]
  \item For the planar \SAdS{5} solution, we have in the projective $\zpr$ coordinate
    \begin{equation}\label{eq:scalar-psads5}
    Q_0(\zpr)
    =
    - \frac{1-\zeta^2+\zeta^2\, \zpr^2}{4\, \zpr\, (1-\zpr^2)^2}\,, \qquad
    Q_2(\zpr)
    = -\frac{(1+\zpr^2)^2 + \nu^2 \, (\zpr^2 -1)}{4\,\zpr^2 \, (1-\zpr^2)^2} \,.
    \end{equation}
    There are two turning points located at
    \begin{equation}
    \zpr = \mp i\, \frac{\sqrt{1-\zeta^2}}{\zeta}\,.
    \end{equation}
  \item In the case of the global black hole set $\mathfrak{h}_4(\zpr) = r_+^2 + (1+r_+^2)\, \zpr$ in units where $\lads = 1$. Then,
    \begin{equation}\label{eq:scalar-gsads5}
    \begin{aligned}
    Q_0(\zpr)
    &=
    - \frac{r_+^2\, \bqty{r_+^2-\zeta^2\, (1-\zpr) \, \mathfrak{h}_4(\zpr)}}
    {4\, \zpr\, (1-\zpr)^2\, \mathfrak{h}_4(\zpr)^2}\,,
    \\
    \qquad Q_2(\zpr)
    &=
    -\frac{\bqty{r_+^2 +(1+r_+^2)\, \zpr^2}^2
    +\nu^2\, r_+^2\, (\zpr-1) \mathfrak{h}_4(\zpr)}
    {4\,\zpr^2\, (\zpr-1)^2 \, \mathfrak{h}_4(\zpr)^2}\,.
    \end{aligned}
    \end{equation}
    There is a complex horizon at $\zpr_c = - \frac{r_+^2}{1+r_+^2}$ and there are two turning points (whose location depends on $\zeta$ as discussed in the main text).
  \item For the $\RNAdS{5}$ solution, we parameterize the charge using $\varrho$ defined in~\eqref{eq:Qvarrho}. The potential comprises the two terms
    \begin{equation}\label{eq:scalar-rnads5}
    \begin{aligned}
    Q_0(\zpr)
    &=
    - \frac{1-\zeta^2+\zeta^2(1+\varrho+\varrho^2)\,\zpr^2
    -\zeta^2\,\varrho(1+\varrho)\,\zpr^3}{4\, \zpr\, (1-\zpr)^2\, (1-\varrho \,\zpr)^2\, \pqty{1+(1+\varrho)\, \zpr}^2}\,,
    \\
    Q_2(\zpr)
    &=
    \frac{\nu^2}{4\, \zpr^2\, (1-\zpr)\, (1-\varrho \,\zpr)\, \pqty{1+(1+\varrho)\, \zpr}}
    \\
    &\qquad
    - \frac{1+2\, (1+\varrho+\varrho^2)\,\zpr^2
      -8\,\varrho(1+\varrho)\, \zpr^3
    +(1+\varrho+\varrho^2)^2\, \zpr^4 }{4\,\zpr^2\,(1-\zpr)^2\, (1-\varrho \,\zpr)^2\, \pqty{1+(1+\varrho)\, \zpr}^2}\,.
    \end{aligned}
    \end{equation}
    The oper has regular singular points at
    \begin{equation}
    H_+: \zpr =1\,, \qquad H_-: \zpr = \varrho^{-1}\,,
    \qquad
    H_3: \zpr=-(1+\varrho)^{-1}\,,
    \end{equation}
    which are the physical outer and inner horizons, and the complex horizon.  There are three turning points, which are the roots of the cubic in the numerator of $Q_0$.
\end{itemize}

\paragraph{Neutral black holes in $d=6$:} We only consider neutral black holes now; in $d=6$ their spectral curves end up being elliptic.
\begin{itemize}[wide,left=0pt]
  \item  The scalar oper for a planar \SAdS{7} black hole is built out of potentials
    \begin{equation}\label{eq:finite-interval-planar-d6-oper}
    \begin{aligned}
    Q_0(\zpr)
    & =
    - \frac{1-\zeta^2+\zeta^2\,\zpr^3}
    {4\, \zpr\, (1-\zpr^3)^2},
    \\
    Q_2(\zpr)
    & =
    \frac{\nu^2(1-\zpr^3)\,  -\pqty{1+7\,\zpr^3+ \zpr^6}}
    {4\,\zpr^2\, (1-\zpr^3)^2}\,.
    \end{aligned}
    \end{equation}
  \item For the  global \SAdS{7}, we define $\mathfrak{h}_6(\zpr) =
    \rp^2 + \zpr -(1+ \rp^2)\,\zpr^3$ in terms of which
    \begin{equation}
    \begin{aligned}
    Q_0(\zpr)
    & =
    - \frac{r_+^2\, \bqty{r_+^2- \zeta^2 \, \mathfrak{h}_6(\zpr)}}
    {4\, \zpr\, \mathfrak{h}_6(\zpr)^2}\,,
    \\
    Q_2(\zpr)
    & =
    \frac{r_+^2\, \nu^2}{4\,\zpr^2\, \mathfrak{h}_6(\zpr)}
    - \frac{\rp^4 -3\, \rp^2\, \zpr -3\, \zpr^2+7\, \rp^2\, (1+\rp^2)\,\zpr^3 + 6\,(1+\rp^2)\, \zpr^4+(1+\rp^2)^2\, \zpr^6 }{4\,\zpr^2\, \mathfrak{h}_6(\zpr)^2}\,.
    \end{aligned}
    \label{eq:finite-interval-global-d6-oper}
    \end{equation}
    The three horizon locations lie at the roots of $\mathfrak{h}_6(\zpr)$, and the turning points lie at the roots of $r_+^2- \zeta^2 \, \mathfrak{h}_6(\zpr)$.
\end{itemize}

\paragraph{Black holes in $d=8$:} In this case we only focus on
planar \SAdS{9}, for which
\begin{equation}
\begin{aligned}
Q_0(\zpr)
& =
- \frac{1-\zeta^2+\zeta^2\, \zpr^4}
{4\, \zpr\, (1-\zpr^4)^2}\,, \\
Q_2(\zpr)
& =
-\frac{\nu^2\, \zpr^4-\nu^2+\zpr^8+14\,\zpr^4+1}
{4\,\zpr^2\,(1-\zpr^4)^2}\,.
\label{eq:finite-interval-planar-d8-oper}
\end{aligned}
\end{equation}
There are a total of four turning points in this case, although two of them do not send out critical Stokes rays towards the outer horizon locus at $\zpr =1$.

\section{Imaginary momentum spectral curves and asymptotics }\label[appendix]{app:imaginary-mom}

In this appendix we explore the leading asymptotics of the spectral function at fixed imaginary momentum ratio $\zeta = \ii \chi$ with $\chi \in \mathbb{R}$.
The analysis is meant to complement the discussion of~\cref{sec:examples}. We primarily highlight features that are novel for imaginary momentum. In addition, we verify that the leading prediction for the WKB action agrees with direct numerical solution of the differential equations for the composite monodromy trace and the phase of the retarded Green's function.

\subsection{Planar \SAdS{} black holes}
\label{sec:planar-chi}

The analysis for imaginary momentum parallels that in~\cref{sec:pschwads}. The main change is that
when $\zeta = \ii \chi$ with $\chi \in \mathbb{R}_+$ the turning points are at
\begin{equation}
\rpr_k= \pqty{\frac{\chi^2}{1+\chi^2}}^{\frac{2}{d}}\, \exp(\frac{4\pi\ii}{d}\, k)\,, \qquad k \in \set{0,1,\cdots, \frac{d}{2} -1}\,.
\end{equation}
Now $\rpr_0$ is a real turning point and lies between the singularity and the
horizon $0 < \rpr_0< 1$. This turning point is real for all even $d$. It still
remains true that for $d \equiv 2 \pmod{4}$ the singularity is a regular point
of the leading WKB spectral curve. Since we now have a real turning point, the ingoing
WKB trajectory along $\arg \rpr = 0$ will first encounter $\rpbull = \rpr_0 > 0$, and
not venture toward the singularity. Hence, the canonical representative for the
physical phase picked up by the wavefunction is simply
\begin{equation}
\Pphys = \rp\, \mathrm{PV}\, \int_{\rpbull}^\infty\, \lwkb
\end{equation}
Evaluating the integral we obtain
\begin{equation}
\Pphys =  \frac{\pi}{d}\, \cot \frac{\pi}{d} \, \pqty{
1 + e^{-\ii\pi (1- \frac{1}{d})}\, \sec \frac{\pi}{d}\, \Pi_d(\ii \chi)}\,.
\end{equation}
In addition, since
we cross the outer horizon to get to the turning point we have a half thermal
period contribution.

While the black hole singularity was not relevant for the above, we can nevertheless extend the domain of integration down to $\rpr = 0$ and include the compact cycle contribution between $\rpr = 0$ and $\rpbull$.
For $d \equiv 0 \pmod{4}$ the compact cycle would be the analog of the $A_s$
cycle. For $d \equiv 2 \pmod{4}$ one could simply extend the open cycle to
include the regular point at $\rpr = 0$. In either case, doing so has no effect on the physical answer; the additional contribution simply cancels between the two
terms. However, it does offer the advantage that we can write the expression for
the WKB action uniformly for either choice of $\arg \zeta$.  The leading WKB action can then be shown to be
\begin{equation}
\Iwkb = \frac{\beta\,r_+}{2}\, \bqty{\ii + \cot\frac{\pi}{d}\, + e^{-\ii \pi(1-\frac{1}{d})} \, \csc\frac{\pi}{d}\, \Pi_d(\ii\chi) }\,.
\end{equation}

\begin{figure}[htp!]
  \centering
  \includegraphics[width=0.9\linewidth]{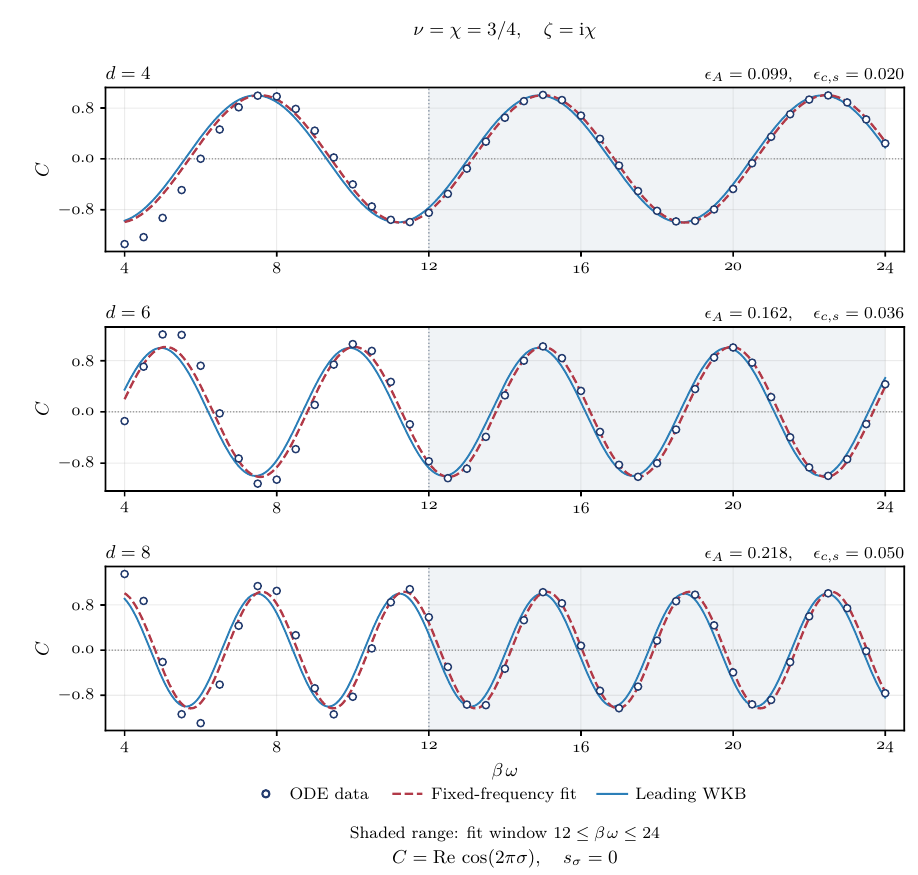}
  \caption{
  The real part of the composite monodromy trace $C=\Re\cmtrace$ for planar \SAdS{d+1} at imaginary momentum, $\zeta=\frac{3}{4}\ms\ii$, with $\nu= 0.75$, for $d=4,6,8$. The exponential envelope is predicted to be absent, so we display a single panel for the numerical data (there being nothing to strip off to expose the oscillations). The plot uses the same set of conventions as~\cref{fig:planar-phase-oscillations}.  The shaded region indicates the data that were used for fitting (the residuals are quoted for  this region). We again fit to the  ansatz $h_c\cos(\xi_d\bw)+h_s\sin(\xi_d\bw)$ (red dashed curve), or a cosine alone, setting $h_s =0$ (blue curve). The fitted curve is extrapolated to the unshaded region, where we see it does reasonably well.}
  \label{fig:imaginary-zeta-planar-trace}
\end{figure}

\begin{figure}[hbpt!]
  \centering
  \includegraphics[width=0.9\linewidth]{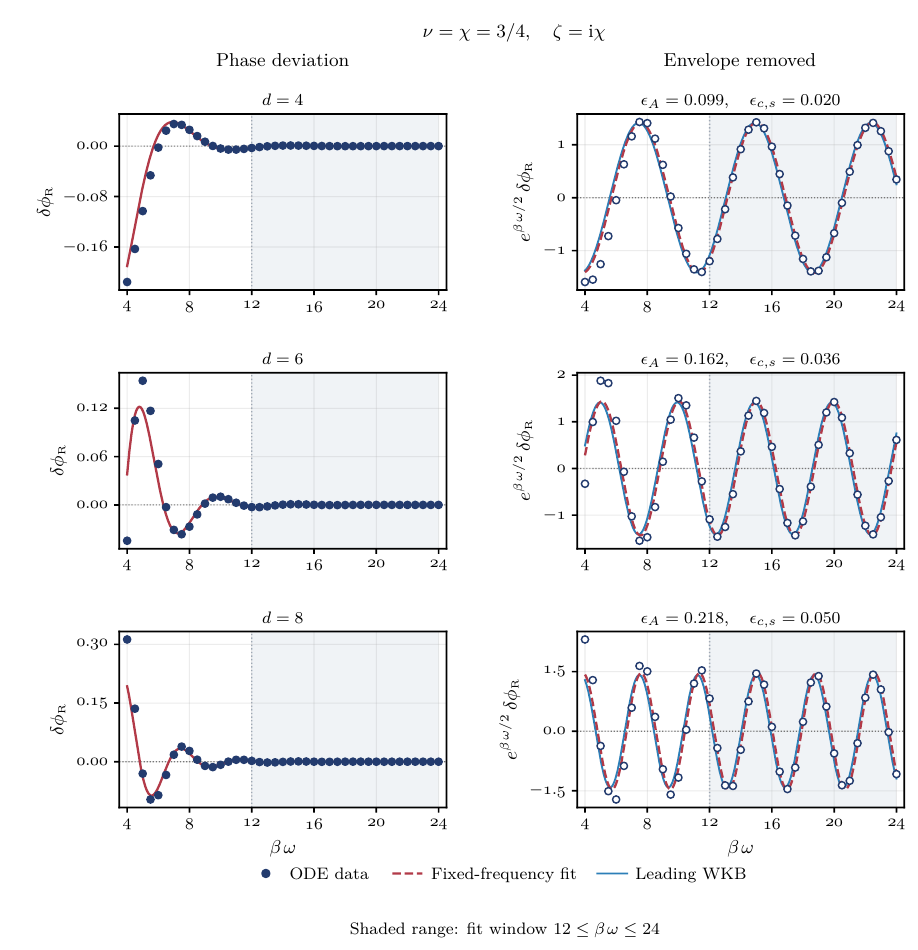}
  \caption{
    The phase of the retarded Green's function for planar \SAdS{d+1} at imaginary momentum, $\zeta=\frac{3}{4}\ms\ii$, with $\nu= 0.75$. Now the left panels show the phase $\delphR$ directly and the right ones strip off the exponential.
    $\delphR$ is as defined in~\eqref{eq:lead-phiGR}; the right panels remove its damping by multiplying by $e^{\bw/2}$.
    The conventions are as in~\cref{fig:planar-phase-oscillations}, and the fit uses the data from the shaded region at higher frequency. This fit is extrapolated to  smaller frequencies as in~\cref{fig:imaginary-zeta-planar-trace}.
  }
  \label{fig:imaginary-zeta-planar-phase}
\end{figure}

The results for the composite monodromy and the retarded Green's function phase
follow directly from this expression. Computing these numerically we again find
perfect agreement as can be seen
from~\cref{fig:imaginary-zeta-planar-trace} and~\cref{fig:imaginary-zeta-planar-phase},
respectively. There are, however, some qualitative differences that are worth highlighting:
\begin{itemize}[wide,left=0pt]
  \item  Since the turning point is real-valued, the integral $\Pphys$ is likewise real. So the leading order result for the composite monodromy is purely oscillatory with
    \begin{equation}
    \cos(2\pi \sigma) \sim \cos(2\,\Pphys\, \freq) - \cos(\pi\ms \nu)\, e^{-\hbw}\,.
    \end{equation}
    The second term is the contribution isolated in~\eqref{eq:sigma-asymptotics}.
  \item Likewise, the retarded Green's function phase has a damping set by the thermal period alone with no leading order contribution to this rate from any compact cycle. Explicitly,
    \begin{equation}
    \delphR \sim 2\, \sin(\pi\ms\nu)\, e^{-\frac{\beta\,r_+}{2}\, \freq}\, \cos(2\,\Pphys\, \freq)\,.
    \end{equation}
\end{itemize}

\subsection{Global \SAdS{} black holes}\label{sec:global-immom}

\begin{figure}[bh!]
  \centering
  \includegraphics[width=0.9\linewidth]{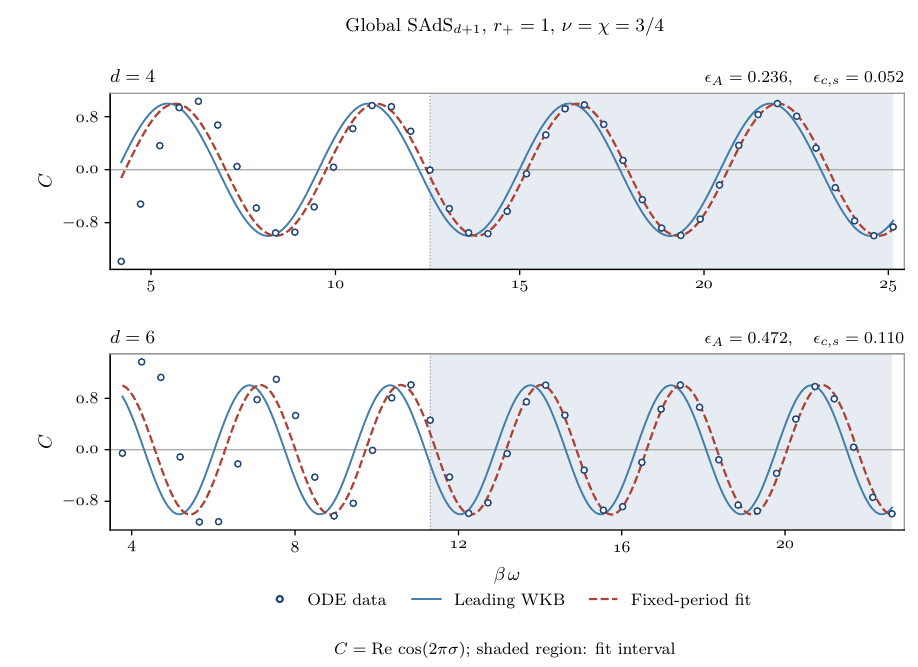}
  \caption{
    The real part of the composite monodromy trace for global \SAdS{d+1} ($d=4,6$), with $\rp=1$, $\zeta=\ii\ms \frac{3}{4}$ and $\nu=0.75$. The conventions are as in~\cref{fig:imaginary-zeta-planar-trace}.
  }
  \label{fig:imaginary-zeta-global-trace}
\end{figure}

\begin{figure}[th!]
  \centering
  \includegraphics[width=0.9\linewidth]{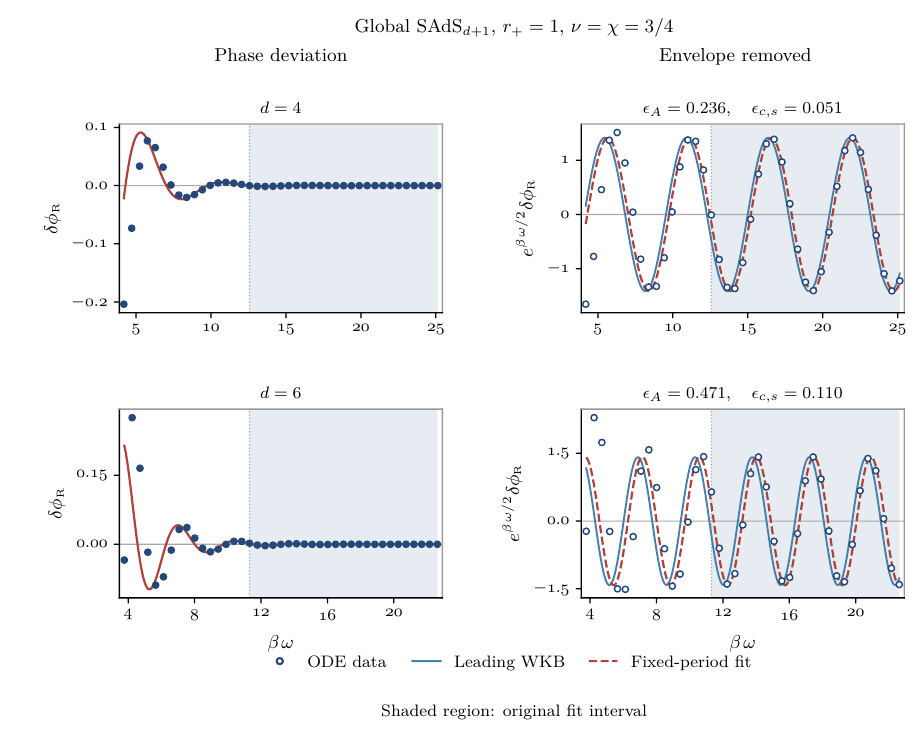}
  \caption{
    The retarded Green's function phase for the same set of parameters
    as~\cref{fig:imaginary-zeta-global-trace}, using the conventions outlined there.
  }
  \label{fig:imaginary-zeta-global-phase}
\end{figure}

For imaginary momentum $\zeta = \ii \chi$ the story largely parallels the discussion for planar \SAdS{d+1} black holes. The turning point equation is
\begin{equation}
(1+\chi^2)\, \rpr^{\frac{d}{2}} + \chi^2\, \rpr^{\frac{d}{2}-1} - \chi^2\, \rp^{d-2}\, (1+\rp^2) = 0\,,
\end{equation}
which can always be seen to have a real root $\rpbull$ between the horizon and the singularity. So the WKB path reaches this turning point along the real radial trajectory, and simply turns around. One never encounters a turning point outside the horizon, and thus there are no bound states in the spectrum either. The WKB action is obtained again as the contribution of the free propagation to the turning point, modulo a thermal half-period contribution from the horizon crossing. To wit,
\begin{equation}
\Iwkb = 2\, \Pphys + i\, \frac{\beta\, r_+}{2} \,,
\qquad
\Pphys = \rp\, \mathrm{PV} \int_{\rpbull}^\infty\, \lwkb(\zeta = \ii\chi)\,.
\end{equation}
The integral can again be evaluated in terms of elliptic integrals, but the explicit expressions are not illuminating in themselves.

The composite trace is again purely oscillatory, with no exponential envelope as is clear from the numerical data displayed in~\cref{fig:imaginary-zeta-global-trace}. The retarded Green's function phase has a damping, but this is set by the thermal half-period, and is depicted in~\cref{fig:imaginary-zeta-global-phase}.

\subsection{Planar \RNAdS{5} black hole}\label{sec:rn-immom}

\begin{figure}[bh!]
  \centering
  \includegraphics[width=0.9\linewidth]{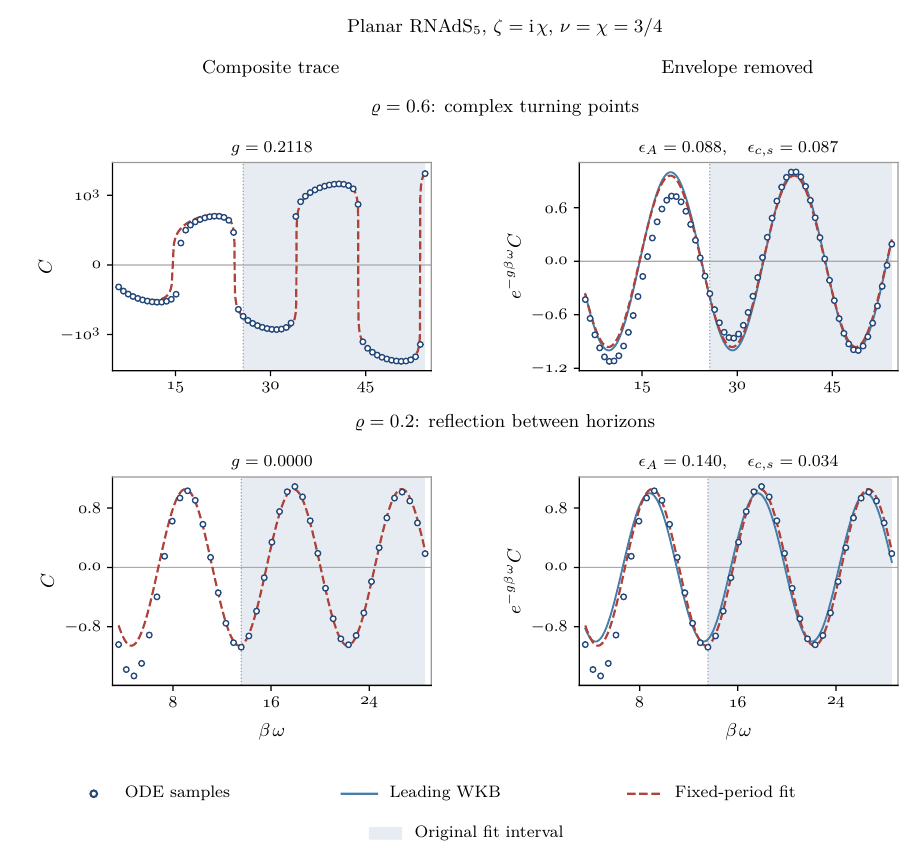}
  \caption{
    The real part of the composite monodromy trace for  planar \RNAdS{5}, with
    $\zeta=\ii\ms \frac{3}{4}$ and $\nu=0.75$. The top row is for
    $\varrho=0.6$, when we have a pair of complex turning points and a real turning point at $\rpbull<0$. The bottom row picks $\varrho=0.2$ to illustrate the situation where we have a real turning point between inner and outer horizons.
    We use the predicted exponent $g=\half-\Im\Iwkb/(\beta\rp)$. The conventions are as in earlier plots (only the top left panel uses a symmetric
    logarithmic scale).
  }
  \label{fig:imaginary-zeta-rn-trace}
\end{figure}

Unlike the neutral black holes discussed hitherto, the \RNAdS{5} example has some new features when we take the momentum to be imaginary $\zeta \in \ii \mathbb{R}_+$. This can be seen from the turning point equation, which reads with $\zeta = \ii \chi$
\begin{equation}
(1+\chi^2)\,\rpr^3 - \chi^2 (1+\varrho + \varrho^2)\, \rpr + \chi^2\, \varrho(1+\varrho) =0\,.
\end{equation}
The roots of the cubic change character as we vary $\chi$.

We find a critical case for $\chi = \chi_c$ whence there is a double turning point between the inner and outer horizons, at
\begin{equation}
\rpr_c = \frac{3}{2}\, \frac{\varrho\, (1+\varrho)}{1+ \varrho + \varrho^2}\,.
\end{equation}
This occurs for
\begin{equation}
\chi_c = \frac{3\,\sqrt{3}\, \varrho\, (1+\varrho)}{(1-\varrho)\, (2+\varrho)\, (1 + 2\,\varrho)}\,.
\end{equation}

\begin{figure}[tp!]
  \centering
  \includegraphics[width=0.9\linewidth]{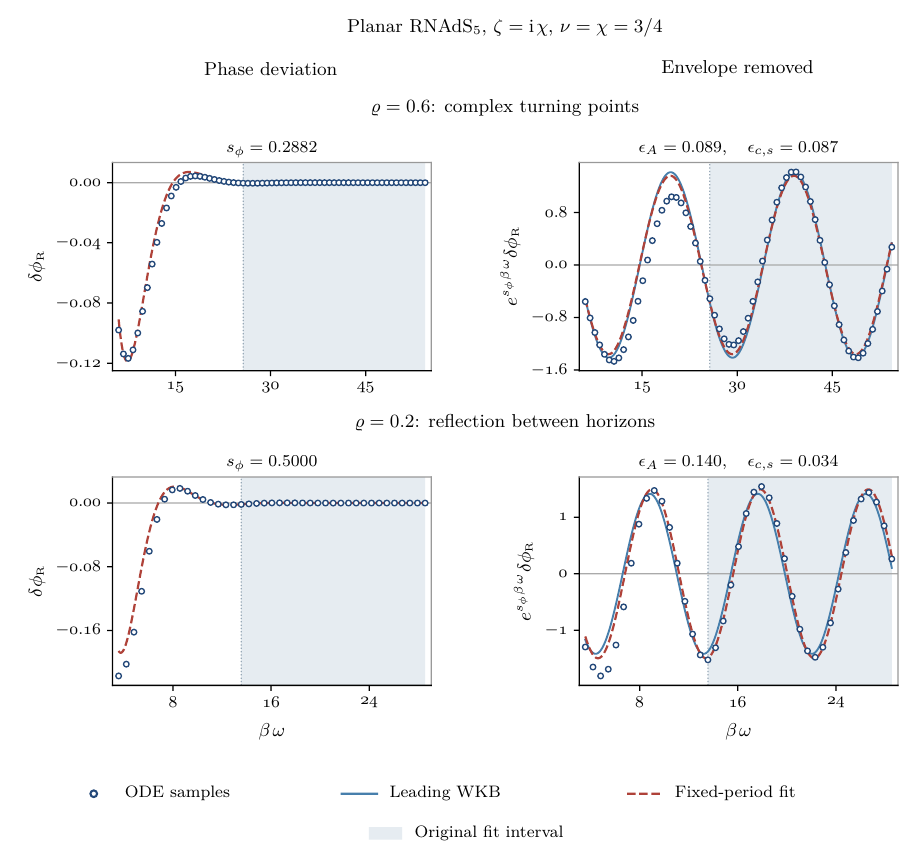}
  \caption{
    The phase of the retarded Green's function for the same set of parameters as~\cref{fig:imaginary-zeta-rn-trace}.  We use the predicted exponent $s_\phi=\Im\Iwkb/(\beta\rp)$ in the right panels. The conventions are as detailed in earlier plots.
  }
  \label{fig:imaginary-zeta-rn-phase}
\end{figure}

For $0<\chi < \chi_c$ there is a single real root $\rpbull < 0$ and a pair of complex conjugate roots, entirely analogous to the timelike momentum case. However, for $\chi > \chi_c$ all three turning points are at real $\rpr$, ordered as
$\rpr_0 < 0 < \varrho < \rpr_1 < \rpr_2 < 1$. So the relevant turning point is $\rpbull = \rpr_2$, and it lies between the outer and inner horizons above the critical imaginary momentum. The picture here is analogous to the global black hole with overcritical timelike momentum, albeit with a twist: at imaginary momentum the charged black hole's real turning point requires crossing the outer horizon.

While the analytic expressions for the periods and WKB action are obtainable in closed form, the expressions are not illuminating in themselves. Therefore, we simply demonstrate the validity of our ansatz by comparing the results for the composite monodromy trace and the retarded Green's function phase against numerical ODE integration in~\cref{fig:imaginary-zeta-rn-trace} and~\cref{fig:imaginary-zeta-rn-phase}, respectively. We show this for different values of the charges chosen to illustrate the two kinds of behavior discussed above.

\section{Determination of core multipliers}\label[appendix]{app:determinants}

We outline a strategy to compute the core Stokes multiplier $\mathfrak{R}(\trs, \freq)$ without explicitly working out the Wronskians. Consider the core oper~\eqref{eq:core-oper} with a new rotated coordinate $ x = e^{\frac{\ii\pi}{d-1}}\,\rtr$. This is to work with exponentials rather than oscillatory solutions (and explains the presence of the modified Bessel functions in~\eqref{eq:Ij-Rcore}). Write the resulting equation as an eigenvalue problem for $f(x) = \psi(e^{-\frac{\ii\pi}{d-1}}\,x)$
\begin{equation}
\mathscr{L} f(x) = E\,v(x) f(x)\,,
\qquad
\mathscr{L} = -\dv[2]{x} + \frac{1}{4}\, x^{d-3} - \frac{1}{4\,x^2} \,,
\qquad
\qquad v(x) =  \frac{1}{4}\,x^{\frac{d}{2}-3}\,.
\end{equation}
At $x=0$ we will impose the boundary condition that picks the logarithm-free solution  $\frac{f(x)}{\sqrt{x}} \to 1$ for the eigenvalue problem. This Friedrichs condition picks the spectral problem to compute the connection.
Of course for the wave equation  of interest we want to set $E= e^{-\frac{\ii \pi}{2}\,a_d}\trs^2$, but shall do so in due course.

Consider the determinant of the following operator built from the inverse kinetic operator convolved with the perturbing potential
\begin{equation}
\mathscr{B} = v(x)^{\frac{1}{2}}\,
\mathscr{L}^{-1}\,
v(x)^{\frac{1}{2}}\,.
\end{equation}
This positive weighted operator is trace class. We can consider its determinant, viz.,
\begin{equation}
\mathscr{D}(E) = \det\pqty{1 - E\, \mathscr{B}} \,,
\qquad
\log \mathscr{D}(E) = -\sum_{n=1}^\infty\, \frac{\mathfrak{b}_n}{n}\, E^{n}\,,
\end{equation}
where we also recognized that the determinant has a series expansion as well
(here $\mathfrak{b}_n=\Tr(\mathscr{B}^n)$). The series for $\log \mathscr{D}$ is valid for $\abs{E} < \abs{\mathscr{B}}^{-1}$ with the operator norm.
The core Stokes multiplier can be obtained directly from the determinant
\begin{equation}\label{eq:core-ref-det}
\mathfrak{R}(\trs) =
- \frac{\mathscr{D}\pqty{e^{-\frac{3\ii\pi}{2}\, a_d} \,\trs^2}+ \mathscr{D}\pqty{e^{\frac{\ii\pi}{2}\, a_d} \,\trs^2} }{\mathscr{D}\pqty{e^{-\frac{\ii\pi}{2}\, a_d} \,\trs^2}}\,.
\end{equation}
Using the expansion of the determinant one recovers the necessary expansion of the core Stokes multiplier.

The derivation of the expansion above proceeds by using some simple identities involving the Wronskian. Suppose that $f_E$ is the solution that is regular at the origin $x=0$ and $g_E$ is the solution that is exponentially decaying at large $x$.
For example, at  $E = 0$ such  wavefunctions are
\begin{equation}
\sqrt{x} \, I_0\pqty{\frac{x^{\frac{d-1}{2}}}{d-1}}\,, \qquad
\sqrt{x}\, K_0\pqty{\frac{x^{\frac{d-1}{2}}}{d-1}}\,.
\end{equation}
The former is indeed regular at the origin, and the latter is decaying at large $x$. To use this in the determinant identity we fix normalizations by requiring
\begin{equation}
\frac{f_E(x)}{\sqrt{x}} \to 1\,, \quad \text{as } x\to 0\,,
\qquad
\frac{g_E(x)}{\sqrt{x}\,K_0\pqty{\frac{x^{\frac{d-1}{2}}}{d-1} }} \to 1
\,, \quad \text{as } x\to \infty\,.
\end{equation}

Let $G_E(x,y)$ be the Green's function of $\mathscr{L} - E\, v(x)$, which given the solutions above can be assembled as
\begin{equation}
G_E(x,y) = - \frac{f_E(x_<)\, g_E(x_>)}{\operatorname{\text{Wr}}(f_E,g_E)}\,, \qquad
x_<  = \min(x,y), \quad x_> = \max(x,y) \,.
\end{equation}
It satisfies
\begin{equation}\label{eq:greenid}
G_E(x,y) = G(x,y) + E\, \int_0^\infty\, dz\, G(x,z) \,v(z)\, G_E(z,y)\,,
\end{equation}
where $G(x,y)$ is the Green's function of $\mathscr{L}$, which given our basis is simply
\begin{equation}
G(x,y) = \frac{2\,\sqrt{x\,y}}{d-1}\, I_0\pqty{\frac{x_<^{\frac{d-1}{2}}}{d-1}}\, K_0\pqty{\frac{x_>^{\frac{d-1}{2}}}{d-1}}\,,
\end{equation}
This follows from the kernel identity for $(\mathscr{L}-E\,v)^{-1}$. Moreover, multiplying this kernel by $\sqrt{v(x)}$ and  $\sqrt{v(y)}$  on the two sides one finds the  relation
\begin{equation}
v^{\frac{1}{2}}\, \pqty{\mathscr{L} - E\, v}^{-1}\,v^{\frac{1}{2}}
=  \pqty{1- E\, \mathscr{B}}^{-1}\, \mathscr{B} \,.
\end{equation}
Taking the trace and using the kernel on the diagonal we end up learning that
\begin{equation}
\int_0^\infty\, v(x)\, G_E(x,x)\, dx =
\Tr\bqty{\pqty{1- E\, \mathscr{B}}^{-1}\, \mathscr{B}}\,.
\end{equation}

However, we also have the relation
\begin{equation}
\pdv{E}\,\log\operatorname{\text{Wr}}(f_E,g_E)
= -
\int_0^\infty\, dx\,v(x)\, G_E(x,x) \,.
\end{equation}
To see this, we use
\begin{equation}
\pdv{x}\, \operatorname{\text{Wr}} \pqty{\pdv{E}f_E,g_E}
= v(x)\, f_E(x)\,g_E(x) \,, \qquad
\pdv{x}\, \operatorname{\text{Wr}} \pqty{f_E,\pdv{E}g_E}
= -v(x)\, f_E(x)\,g_E(x) \,.
\end{equation}
Therefore, we conclude
\begin{equation}
\pdv{E}\,\log\operatorname{\text{Wr}}(f_E,g_E)
= - \Tr\bqty{\pqty{1- E\, \mathscr{B}}^{-1}\, \mathscr{B}}
=  \pdv{E}\, \log\,\mathscr{D}(E)\,.
\end{equation}
This can be integrated to give an expression for the ratio of Wronskians, i.e.,
\begin{equation}
\mathscr{D}(E) =
\frac{\operatorname{\text{Wr}} \pqty{f_E,g_E} }{\operatorname{\text{Wr}} \pqty{f_0,g_0} }\,.
\end{equation}

The final useful step here is to write the regular solution at the origin using the WKB basis.
Suppose $\psi_0$ is such a solution of the original wave equation. It can be expanded in the WKB basis along the rays $\Gamma_0$ and $\Gamma_1$ independently. Focus first on the basis along $\Gamma_0$ and expand
$\psi_0(\rtr; \trs)
= A_+(\trs) \, \psi\subm{core}^+(\rtr; \trs) + A_-(\trs) \, \psi\subm{core}^-(\rtr; \trs)$.
One can check that
\begin{equation}
\begin{aligned}
\frac{A_+(\trs)}{A_+(0)}
&=
\frac{\operatorname{\text{Wr}} \pqty{\psi_0(\trs),\psi\subm{core}^-(\trs)\big|_{\Gamma_0}} }{\operatorname{\text{Wr}} \pqty{\psi_0(0),\psi\subm{core}^-(0)\big|_{\Gamma_0} }} =
\mathscr{D}(e^{-\frac{\ii\pi}{2}\,a_d}\, \trs^2)
\\
\frac{A_-(\trs)}{A_-(0)}
&=
\frac{\operatorname{\text{Wr}} \pqty{\psi_0(\trs), \psi\subm{core}^+(\trs)\big|_{\Gamma_0}} }{\operatorname{\text{Wr}} \pqty{\psi_0(0),\psi\subm{core}^+(0)\big|_{\Gamma_0} }} =
\mathscr{D}(e^{\frac{\ii\pi}{2}\,a_d}\, \trs^2)\,.
\end{aligned}
\end{equation}
The final step is to realize that the core Stokes multiplier relates the bases on $\Gamma_0$ and $\Gamma_1$. One has
\begin{equation}
\pmqty{A_+(\trs)\\ A_-(\trs)}_{\Gamma_1}
= \pmqty{1 & 0 \\ \ii \mathfrak{R}(\trs) & 1} \, \pmqty{A_+(\trs)\\ A_-(\trs)}_{\Gamma_0}
= -\ii
\pmqty{A_-(e^{-\frac{\ii\pi}{2}\,a_d}\, \trs)\\ A_+(e^{-\frac{\ii\pi}{2}\,a_d}\trs)}_{\Gamma_0}
\end{equation}
The first equality is the basis rotation through the core Stokes multiplier adapted to coefficients, and the second exploits homogeneity of the core oper~\eqref{eq:core-homogeneity}.  Finally, one solves for $\mathfrak{R}$ to obtain~\eqref{eq:core-ref-det}, and uses $A_-(0) = \ii A_+(0)$ which follows from Bessel identities. Once this is achieved, we can identify $\mathfrak{b}_n$ in the expansion of the determinant with the integrals~\eqref{eq:Ij-Rcore}.

We have outlined the steps for computing the correction to the core Stokes multiplier in an expansion in $\trs$, but it should be clear that an identical strategy can be employed to obtain the formal power series in $\freq^{-b_d}$. There is one important difference: the perturbation is growing and needs a prescription for its evaluation.

\printbibliography

@article{Oh:2026dky,
    author = "Oh, Hyunwoo",
    title = "{Kasner Exponents from the Bouncing Singularity}",
    eprint = "2610.31699",
    archivePrefix = "arXiv",
    primaryClass = "hep-th",
    month = "10",
    year = "2026"
}

@article{Parisini:2023nbd,
    author = "Parisini, Enrico and Skenderis, Kostas and Withers, Benjamin",
    title = "{The ambient space formalism}",
    eprint = "2312.03820",
    archivePrefix = "arXiv",
    primaryClass = "hep-th",
    doi = "10.1007/JHEP05(2024)296",
    journal = "JHEP",
    volume = "05",
    pages = "296",
    year = "2024"
}

@article{Hatsuda:2026ghx,
    author = "Hatsuda, Yasuyuki and Shiga, Tomohito",
    title = "{Exact WKB and Quantum Periods for Extremal Black Hole Quasinormal Modes}",
    eprint = "2605.01321",
    archivePrefix = "arXiv",
    primaryClass = "hep-th",
    reportNumber = "RUP-26-7",
    month = "5",
    year = "2026"
}

@article{Hatsuda:2026qaw,
    author = "Hatsuda, Yasuyuki",
    title = "{Systematic High Overtone Expansion of Kerr Quasinormal Modes}",
    eprint = "2610.38019",
    archivePrefix = "arXiv",
    primaryClass = "gr-qc",
    month = "10",
    year = "2026"
}

@article{Basha:2018bvi,
    author = "Basha, Roi and Itzhaki, Nissan and Liram, Lior",
    title = "{High Energy Reflection Coefficients of Black Holes and Branes}",
    eprint = "1808.02036",
    archivePrefix = "arXiv",
    primaryClass = "hep-th",
    month = "8",
    year = "2018"
}

@article{Itzhaki:2018rld,
    author = "Itzhaki, Nissan and Liram, Lior",
    title = "{A stringy glimpse into the black hole horizon}",
    eprint = "1801.04939",
    archivePrefix = "arXiv",
    primaryClass = "hep-th",
    doi = "10.1007/JHEP04(2018)018",
    journal = "JHEP",
    volume = "04",
    pages = "018",
    year = "2018"
}

@article{Valach:2026xjk,
    author = "Valach, Samuel and Vrbica, Mile",
    title = "{The Spectral Anatomy of Complex-Time Singularities}",
    eprint = "2611.18520",
    archivePrefix = "arXiv",
    primaryClass = "hep-th",
    month = "11",
    year = "2026"
}

@article{Marolf:2004fy,
    author = "Marolf, Donald",
    title = "{States and boundary terms: Subtleties of Lorentzian AdS / CFT}",
    eprint = "hep-th/0412032",
    archivePrefix = "arXiv",
    doi = "10.1088/1126-6708/2005/05/042",
    journal = "JHEP",
    volume = "05",
    pages = "042",
    year = "2005"
}

@article{Afkhami-Jeddi:2025wra,
  author        = {Afkhami-Jeddi, Nima and Caron-Huot, Simon and Chakravarty, Joydeep and Maloney, Alexander},
  title         = {{Imprint of the black hole singularity on thermal two-point functions}},
  eprint        = {2510.21673},
  archiveprefix = {arXiv},
  primaryclass  = {hep-th},
  year          = {2026},
  doi           = {10.1103/1j3s-fzjl},
  journal       = {Phys. Rev. D},
  volume        = {114},
  number        = {2},
  pages         = {025005},
}

@article{Alday:2009aq,
  author        = {Alday, Luis F. and Gaiotto, Davide and Tachikawa, Yuji},
  title         = {{Liouville Correlation Functions from Four-dimensional Gauge Theories}},
  eprint        = {0906.3219},
  archiveprefix = {arXiv},
  primaryclass  = {hep-th},
  doi           = {10.1007/s11005-010-0369-5},
  journal       = {Lett. Math. Phys.},
  volume        = {91},
  pages         = {167--197},
  year          = {2010}
}

@article{AliAhmad:2026wem,
  author        = {Ali Ahmad, Shadi and Almheiri, Ahmed and Lin, Simon},
  title         = {{Continuing past the inner horizon using WKB}},
  eprint        = {2601.02354},
  archiveprefix = {arXiv},
  primaryclass  = {hep-th},
  month         = {1},
  year          = {2026}
}

@article{Almheiri:2026cpv,
  author        = {Almheiri, Ahmed and Dodelson, Matthew and Komargodski, Zohar},
  title         = {{Quantum gravity on the bulk cone}},
  eprint        = {2610.31610},
  archiveprefix = {arXiv},
  primaryclass  = {hep-th},
  month         = {10},
  year          = {2026}
}

@article{Amado:2008hw,
  author        = {Amado, Irene and Hoyos-Badajoz, Carlos},
  title         = {{AdS black holes as reflecting cavities}},
  eprint        = {0807.2337},
  archiveprefix = {arXiv},
  primaryclass  = {hep-th},
  reportnumber  = {IFT-UAM-CSIC-08-45},
  doi           = {10.1088/1126-6708/2008/09/118},
  journal       = {JHEP},
  volume        = {09},
  pages         = {118},
  year          = {2008}
}

@article{Aminov:2020yma,
  author        = {Aminov, Gleb and Grassi, Alba and Hatsuda, Yasuyuki},
  title         = {{Black Hole Quasinormal Modes and Seiberg{\textendash}Witten Theory}},
  eprint        = {2006.06111},
  archiveprefix = {arXiv},
  primaryclass  = {hep-th},
  reportnumber  = {RUP-20-18},
  doi           = {10.1007/s00023-021-01137-x},
  journal       = {Annales Henri Poincare},
  volume        = {23},
  number        = {6},
  pages         = {1951--1977},
  year          = {2022}
}

@article{Buric:2026sit,
    author = "Buri{\'c}, Ilija and Gusev, Ivan and Parnachev, Andrei",
    title = "{Stringy resolution of bouncing singularities}",
    eprint = "2610.12311",
    archivePrefix = "arXiv",
    primaryClass = "hep-th",
    month = "10",
    year = "2026"
}

@article{Araya:2026shz,
  author        = {Araya, Ignacio J. and Esper, Chantelle and Jia, Yueke and Kulaxizi, Manuela and Parnachev, Andrei},
  title         = {{Bulkcone Singularities and Complex Geodesics}},
  eprint        = {2602.12893},
  archiveprefix = {arXiv},
  primaryclass  = {hep-th},
  month         = {2},
  year          = {2026}
}

@article{Arnaudo:2026axe,
  author        = {Arnaudo, Paolo and Iossa, Cristoforo and Karlsson, Robin and Withers, Benjamin},
  title         = {{OPE = QNM}},
  eprint        = {2607.24909},
  archiveprefix = {arXiv},
  primaryclass  = {hep-th},
  month         = {7},
  year          = {2026}
}

@article{Arnaudo:2026der,
  author        = {Arnaudo, Paolo and Withers, Benjamin},
  title         = {{Analytic structure of holographic thermal correlators from Fourier series}},
  eprint        = {2603.13469},
  archiveprefix = {arXiv},
  primaryclass  = {hep-th},
  year          = {2026},
  doi           = {10.1007/JHEP06(2026)205},
  journal       = {JHEP},
  volume        = {06},
  pages         = {205},
}

@article{Arnaudo:2026tcy,
  author        = {Arnaudo, Paolo and Withers, Benjamin},
  title         = {{Bouncing singularities in Schwarzschild: a geometric origin of the QNM convergence region}},
  eprint        = {2605.16489},
  archiveprefix = {arXiv},
  primaryclass  = {gr-qc},
  month         = {5},
  year          = {2026}
}

@article{Barrat:2026jfg,
  author        = {Barrat, Julien and Bozkurt, Deniz N. and Marchetto, Enrico and Miscioscia, Alessio and Pomoni, Elli},
  title         = {{Analytic thermal bootstrap in momentum space: From thermal OPE to QNMs}},
  eprint        = {2607.24919},
  archiveprefix = {arXiv},
  primaryclass  = {hep-th},
  reportnumber  = {DESY-26-094, YITP-SB-2026-13},
  month         = {7},
  year          = {2026}
}

@article{Batra:2026gbq,
  author        = {Batra, Gauri and Levine, Adam and Shenker, Stephen H.},
  title         = {{A backreacted bouncing geodesic in 2D}},
  eprint        = {2609.32071},
  archiveprefix = {arXiv},
  primaryclass  = {hep-th},
  month         = {9},
  year          = {2026}
}

@article{Batra:2026otu,
  author        = {Batra, Gauri},
  title         = {{A non-prompt null ray cures its own singularity}},
  eprint        = {2610.22478},
  archiveprefix = {arXiv},
  primaryclass  = {hep-th},
  month         = {10},
  year          = {2026}
}

@article{Belinsky:1970ew,
  author  = {Belinsky, V. A. and Khalatnikov, I. M. and Lifshitz, E. M.},
  title   = {{Oscillatory approach to a singular point in the relativistic cosmology}},
  doi     = {10.1080/00018737000101171},
  journal = {Adv. Phys.},
  volume  = {19},
  pages   = {525--573},
  year    = {1970}
}

@article{Bonelli:2022ten,
  author        = {Bonelli, Giulio and Iossa, Cristoforo and Panea Lichtig, Daniel and Tanzini, Alessandro},
  title         = {{Irregular Liouville Correlators and Connection Formulae for Heun Functions}},
  eprint        = {2201.04491},
  archiveprefix = {arXiv},
  primaryclass  = {hep-th},
  doi           = {10.1007/s00220-022-04497-5},
  journal       = {Commun. Math. Phys.},
  volume        = {397},
  number        = {2},
  pages         = {635--727},
  year          = {2023}
}

@article{Brecher:2004gn,
  author        = {Brecher, Dominic and He, Jianyang and Rozali, Moshe},
  title         = {{On charged black holes in anti-de Sitter space}},
  eprint        = {hep-th/0410214},
  archiveprefix = {arXiv},
  doi           = {10.1088/1126-6708/2005/04/004},
  journal       = {JHEP},
  volume        = {04},
  pages         = {004},
  year          = {2005}
}

@article{Buric:2025fye,
  author        = {Buri{\'c}, Ilija and Gusev, Ivan and Parnachev, Andrei},
  title         = {{Holographic Correlators from Thermal Bootstrap}},
  eprint        = {2508.08373},
  archiveprefix = {arXiv},
  primaryclass  = {hep-th},
  year          = {2026},
  doi           = {10.1007/JHEP05(2026)059},
  journal       = {JHEP},
  volume        = {05},
  pages         = {059},
}

@article{Buric:2026qsp,
  author        = {Buri{\'c}, Ilija and Chang, Chi-Ming and Gusev, Ivan and Helfenberger, Elizabeth and Parnachev, Andrei and Rangamani, Mukund},
  title         = {{Thermal two-point functions in SYK and complex-time singularities}},
  eprint        = {2607.05258},
  archiveprefix = {arXiv},
  primaryclass  = {hep-th},
  year          = {2026},
  doi           = {10.1007/JHEP10(2026)228},
  journal       = {JHEP},
  volume        = {10},
  pages         = {228},
}

@article{Caron-Huot:2009ypo,
  author        = {Caron-Huot, S.},
  title         = {{Asymptotics of thermal spectral functions}},
  eprint        = {0903.3958},
  archiveprefix = {arXiv},
  primaryclass  = {hep-ph},
  doi           = {10.1103/PhysRevD.79.125009},
  journal       = {Phys. Rev. D},
  volume        = {79},
  pages         = {125009},
  year          = {2009}
}

@article{Ceplak:2024bja,
  author        = {{\v{C}}eplak, Nejc and Liu, Hong and Parnachev, Andrei and Valach, Samuel},
  title         = {{Black hole singularity from OPE}},
  eprint        = {2404.17286},
  archiveprefix = {arXiv},
  primaryclass  = {hep-th},
  doi           = {10.1007/JHEP10(2024)105},
  journal       = {JHEP},
  volume        = {10},
  pages         = {105},
  year          = {2024}
}

@article{Ceplak:2025dds,
  author        = {{\v{C}}eplak, Nejc and Liu, Hong and Parnachev, Andrei and Valach, Samuel},
  title         = {{Fooling the Censor: Going beyond inner horizons with the OPE}},
  eprint        = {2511.09638},
  archiveprefix = {arXiv},
  primaryclass  = {hep-th},
  year          = {2026},
  doi           = {10.1007/JHEP05(2026)001},
  journal       = {JHEP},
  volume        = {05},
  pages         = {001},
}

@article{Ceplak:2026ard,
  author        = {{\v{C}}eplak, Nejc and Valach, Samuel},
  title         = {{Where's Kasner? OPE Signature of Kasner Exponents}},
  eprint        = {2609.23776},
  archiveprefix = {arXiv},
  primaryclass  = {hep-th},
  month         = {9},
  year          = {2026}
}

@article{Chakravarty:2026fdm,
  author        = {Chakravarty, Joydeep and Zigdon, Yoav},
  title         = {{A signature of the timelike singularity in holography}},
  eprint        = {2610.30134},
  archiveprefix = {arXiv},
  primaryclass  = {hep-th},
  month         = {10},
  year          = {2026}
}

@article{Coviello:2026hdc,
  author        = {Coviello, Chiara and Gupta, Ansh and Hennigar, Robie A. and Shi, Kai and Svesko, Andrew},
  title         = {{Quasinormal modes and quantum black hole interiors}},
  eprint        = {2609.26724},
  archiveprefix = {arXiv},
  primaryclass  = {hep-th},
  month         = {9},
  year          = {2026}
}

@article{Delabaere:1997srq,
  author  = {Delabaere, Eric and Dillinger, Herv{\'e} and Pham, Fr{\'e}d{\'e}ric},
  title   = {{Exact semiclassical expansions for one-dimensional quantum oscillators}},
  doi     = {10.1063/1.532206},
  journal = {J. Math. Phys.},
  volume  = {38},
  number  = {12},
  pages         = {6126--6184},
  year    = {1997}
}

@article{Dodelson:2023nnr,
  author        = {Dodelson, Matthew and Iossa, Cristoforo and Karlsson, Robin and Lupsasca, Alexandru and Zhiboedov, Alexander},
  title         = {{Black hole bulk-cone singularities}},
  eprint        = {2310.15236},
  archiveprefix = {arXiv},
  primaryclass  = {hep-th},
  reportnumber  = {CERN-TH-2023-192},
  doi           = {10.1007/JHEP07(2024)046},
  journal       = {JHEP},
  volume        = {07},
  pages         = {046},
  year          = {2024}
}

@article{Dodelson:2023vrw,
  author        = {Dodelson, Matthew and Iossa, Cristoforo and Karlsson, Robin and Zhiboedov, Alexander},
  title         = {{A thermal product formula}},
  eprint        = {2304.12339},
  archiveprefix = {arXiv},
  primaryclass  = {hep-th},
  reportnumber  = {CERN-TH-2023-062},
  doi           = {10.1007/JHEP01(2024)036},
  journal       = {JHEP},
  volume        = {01},
  pages         = {036},
  year          = {2024}
}

@article{Dodelson:2024atp,
  author        = {Dodelson, Matthew},
  title         = {{Ringdown in the SYK model}},
  eprint        = {2408.05790},
  archiveprefix = {arXiv},
  primaryclass  = {hep-th},
  reportnumber  = {CERN-TH-2024-135},
  doi           = {10.21468/SciPostPhys.19.3.081},
  journal       = {SciPost Phys.},
  volume        = {19},
  number        = {3},
  pages         = {081},
  year          = {2025}
}

@article{Dodelson:2025jff,
  author        = {Dodelson, Matthew and Iossa, Cristoforo and Karlsson, Robin},
  title         = {{Bouncing off a stringy singularity}},
  eprint        = {2511.09616},
  archiveprefix = {arXiv},
  primaryclass  = {hep-th},
  doi           = {10.1007/JHEP07(2026)161},
  journal       = {JHEP},
  volume        = {07},
  pages         = {161},
  year          = {2026}
}

@article{Dodelson:2026gak,
  author        = {Dodelson, Matthew and Gupta, Om and Mezei, M{\'a}rk and Wang, Diandian},
  title         = {{On the temperature dependence of quasinormal modes in SYK and holography}},
  eprint        = {2606.22679},
  archiveprefix = {arXiv},
  primaryclass  = {hep-th},
  month         = {6},
  year          = {2026}
}

@article{Liu:2026luf,
    author = "Liu, Mingxuan and Ruan, Shan-Ming",
    title = "{Smoothing the Bouncing-Geodesic Singularity in AdS/CFT}",
    eprint = "2611.03347",
    archivePrefix = "arXiv",
    primaryClass = "hep-th",
    month = "11",
    year = "2026"
}

@article{Dodelson:2026hjm,
  author        = {Dodelson, Matthew and Lin, Henry W. and Xie, Haitian and Yeh, Jessica},
  title         = {{Matrix scattering and the black hole interior}},
  eprint        = {2610.31700},
  archiveprefix = {arXiv},
  primaryclass  = {hep-th},
  month         = {10},
  year          = {2026}
}

@article{Festuccia:2005pi,
  author        = {Festuccia, Guido and Liu, Hong},
  title         = {{Excursions beyond the horizon: Black hole singularities in Yang-Mills theories. I.}},
  eprint        = {hep-th/0506202},
  archiveprefix = {arXiv},
  reportnumber  = {MIT-CTP-3641},
  doi           = {10.1088/1126-6708/2006/04/044},
  journal       = {JHEP},
  volume        = {04},
  pages         = {044},
  year          = {2006}
}

@article{Festuccia:2008zx,
  author        = {Festuccia, Guido and Liu, Hong},
  title         = {{A Bohr-Sommerfeld quantization formula for quasinormal frequencies of AdS black holes}},
  eprint        = {0811.1033},
  archiveprefix = {arXiv},
  primaryclass  = {gr-qc},
  reportnumber  = {MIT-CTP-3995, SCIPP-08-11},
  doi           = {10.1166/asl.2009.1029},
  journal       = {Adv. Sci. Lett.},
  volume        = {2},
  pages         = {221--235},
  year          = {2009}
}

@article{Fidkowski:2003nf,
  author        = {Fidkowski, Lukasz and Hubeny, Veronika and Kleban, Matthew and Shenker, Stephen},
  title         = {{The black hole singularity in AdS/CFT}},
  eprint        = {hep-th/0306170},
  archiveprefix = {arXiv},
  doi           = {10.1088/1126-6708/2004/02/014},
  journal       = {JHEP},
  volume        = {02},
  pages         = {014},
  year          = {2004}
}

@article{Fitzpatrick:2019zqz,
  author        = {Fitzpatrick, A. Liam and Huang, Kuo-Wei},
  title         = {{Universal Lowest-Twist in CFTs from Holography}},
  eprint        = {1903.05306},
  archiveprefix = {arXiv},
  primaryclass  = {hep-th},
  doi           = {10.1007/JHEP08(2019)138},
  journal       = {JHEP},
  volume        = {08},
  pages         = {138},
  year          = {2019}
}

@article{Gaiotto:2009hg,
  author        = {Gaiotto, Davide and Moore, Gregory W. and Neitzke, Andrew},
  title         = {{Wall-crossing, Hitchin systems, and the WKB approximation}},
  eprint        = {0907.3987},
  archiveprefix = {arXiv},
  primaryclass  = {hep-th},
  doi           = {10.1016/j.aim.2012.09.027},
  journal       = {Adv. Math.},
  volume        = {234},
  pages         = {239--403},
  year          = {2013}
}

@article{Gaiotto:2012rg,
  author        = {Gaiotto, Davide and Moore, Gregory W. and Neitzke, Andrew},
  title         = {{Spectral networks}},
  eprint        = {1204.4824},
  archiveprefix = {arXiv},
  primaryclass  = {hep-th},
  doi           = {10.1007/s00023-013-0239-7},
  journal       = {Annales Henri Poincare},
  volume        = {14},
  pages         = {1643--1731},
  year          = {2013}
}

@article{Ghosh:2019rcj,
  author        = {Ghosh, Animik and Maxfield, Henry and Turiaci, Gustavo J.},
  title         = {{A universal Schwarzian sector in two-dimensional conformal field theories}},
  eprint        = {1912.07654},
  archiveprefix = {arXiv},
  primaryclass  = {hep-th},
  doi           = {10.1007/JHEP05(2020)104},
  journal       = {JHEP},
  volume        = {05},
  pages         = {104},
  year          = {2020}
}

@article{Giombi:2026kdz,
  author        = {Giombi, Simone and Li, Yue-Zhou and Shan, Jieru},
  title         = {{Bouncing singularities and thermal correlators on line defects}},
  eprint        = {2603.11012},
  archiveprefix = {arXiv},
  primaryclass  = {hep-th},
  year          = {2026},
  doi           = {10.1007/JHEP07(2026)170},
  journal       = {JHEP},
  volume        = {07},
  pages         = {170},
}

@article{Glorioso:2018mmw,
  author        = {Glorioso, Paolo and Crossley, Michael and Liu, Hong},
  title         = {{A prescription for holographic Schwinger-Keldysh contour in non-equilibrium systems}},
  eprint        = {1812.08785},
  archiveprefix = {arXiv},
  primaryclass  = {hep-th},
  reportnumber  = {MIT-CTP/5095; EFI-18-21},
  month         = {12},
  year          = {2018}
}

@article{Grassi:2026ekz,
  author        = {Grassi, Alba and Jia, Hewei Frederic},
  title         = {{Resurgence of Black Holes in Thermal OPE}},
  eprint        = {2610.37554},
  archiveprefix = {arXiv},
  primaryclass  = {hep-th},
  reportnumber  = {CERN-TH-2026-262},
  month         = {10},
  year          = {2026}
}

@article{Grozdanov:2026cut,
  author        = {Grozdanov, Sa{\v{s}}o and Valach, Samuel and Vrbica, Mile},
  title         = {{Bouncing geodesics, black hole singularities, and singularities of thermal correlators}},
  eprint        = {2603.15598},
  archiveprefix = {arXiv},
  primaryclass  = {hep-th},
  year          = {2026},
  doi           = {10.1007/JHEP08(2026)022},
  journal       = {JHEP},
  volume        = {08},
  pages         = {022},
}

@article{Gubser:1998bc,
  author        = {Gubser, S. S. and Klebanov, Igor R. and Polyakov, Alexander M.},
  title         = {{Gauge theory correlators from noncritical string theory}},
  eprint        = {hep-th/9802109},
  archiveprefix = {arXiv},
  reportnumber  = {PUPT-1767},
  doi           = {10.1016/S0370-2693(98)00377-3},
  journal       = {Phys. Lett. B},
  volume        = {428},
  pages         = {105--114},
  year          = {1998}
}

@article{Hartnoll:2026vhu,
  author        = {Hartnoll, Sean A. and Zhiboedov, Alexander},
  title         = {{Can one hear the shape of a black hole singularity?}},
  eprint        = {2609.07514},
  archiveprefix = {arXiv},
  primaryclass  = {hep-th},
  reportnumber  = {CERN-TH-2026-217},
  month         = {9},
  year          = {2026}
}

@article{Herzog:2002pc,
  author        = {Herzog, C. P. and Son, D. T.},
  title         = {{Schwinger-Keldysh propagators from AdS/CFT correspondence}},
  eprint        = {hep-th/0212072},
  archiveprefix = {arXiv},
  reportnumber  = {NSF-ITP-02-175, INT-PUB-02-53},
  doi           = {10.1088/1126-6708/2003/03/046},
  journal       = {JHEP},
  volume        = {03},
  pages         = {046},
  year          = {2003}
}

@article{Horowitz:2009wm,
  author        = {Horowitz, Gary and Lawrence, Albion and Silverstein, Eva},
  title         = {{Insightful D-branes}},
  eprint        = {0904.3922},
  archiveprefix = {arXiv},
  primaryclass  = {hep-th},
  reportnumber  = {NSF-KITP-09-53, SLAC-PUB-13584, SU-ITP-09-16, BRX-TH-607},
  doi           = {10.1088/1126-6708/2009/07/057},
  journal       = {JHEP},
  volume        = {07},
  pages         = {057},
  year          = {2009}
}

@article{Hubeny:2006yu,
  author        = {Hubeny, Veronika E and Liu, Hong and Rangamani, Mukund},
  title         = {{Bulk-cone singularities {\&} signatures of horizon formation in AdS/CFT}},
  eprint        = {hep-th/0610041},
  archiveprefix = {arXiv},
  reportnumber  = {DCPT-06-29, MIT-CTP-3775},
  doi           = {10.1088/1126-6708/2007/01/009},
  journal       = {JHEP},
  volume        = {01},
  pages         = {009},
  year          = {2007}
}

@article{Iliesiu:2018fao,
  author        = {Iliesiu, Luca and Kolo{\u{g}}lu, Murat and Mahajan, Raghu and Perlmutter, Eric and Simmons-Duffin, David},
  title         = {{The Conformal Bootstrap at Finite Temperature}},
  eprint        = {1802.10266},
  archiveprefix = {arXiv},
  primaryclass  = {hep-th},
  reportnumber  = {CALT-TH-2018-013, PUPT-2550},
  doi           = {10.1007/JHEP10(2018)070},
  journal       = {JHEP},
  volume        = {10},
  pages         = {070},
  year          = {2018}
}

@article{Iliesiu:2020qvm,
  author        = {Iliesiu, Luca V. and Turiaci, Gustavo J.},
  title         = {{The statistical mechanics of near-extremal black holes}},
  eprint        = {2003.02860},
  archiveprefix = {arXiv},
  primaryclass  = {hep-th},
  doi           = {10.1007/JHEP05(2021)145},
  journal       = {JHEP},
  volume        = {05},
  pages         = {145},
  year          = {2021}
}

@article{Iwaki_2015,
  title     = {Exact WKB Analysis and Cluster Algebras II: Simple Poles, Orbifold Points, and Generalized Cluster Algebras},
  volume    = {2016},
  issn      = {1687-0247},
  doi       = {10.1093/imrn/rnv270},
  number    = {14},
  journal       = {Int. Math. Res. Not.},
  publisher = {Oxford University Press (OUP)},
  author    = {Iwaki, Kohei and Nakanishi, Tomoki},
  year          = {2016},
  pages         = {4375--4417},
  eprint        = {1409.4641},
  archiveprefix = {arXiv},
  primaryclass  = {math.CA},
}

@article{Iwaki:2014vad,
  author        = {Iwaki, Kohei and Nakanishi, Tomoki},
  title         = {{Exact WKB analysis and cluster algebras}},
  eprint        = {1401.7094},
  archiveprefix = {arXiv},
  primaryclass  = {math.CA},
  doi           = {10.1088/1751-8113/47/47/474009},
  journal       = {J. Phys. A},
  volume        = {47},
  number        = {47},
  pages         = {474009},
  year          = {2014}
}

@article{Iwaki:2025tmq,
  author        = {Iwaki, Kohei},
  title         = {{Les Houches Lectures on Exact WKB Analysis and Painlev{\'e} Equations}},
  eprint        = {2512.17599},
  archiveprefix = {arXiv},
  primaryclass  = {math-ph},
  month         = {12},
  year          = {2025}
}

@article{Jana:2020vyx,
  author        = {Jana, Chandan and Loganayagam, R. and Rangamani, Mukund},
  title         = {{Open quantum systems and Schwinger-Keldysh holograms}},
  eprint        = {2004.02888},
  archiveprefix = {arXiv},
  primaryclass  = {hep-th},
  doi           = {10.1007/JHEP07(2020)242},
  journal       = {JHEP},
  volume        = {07},
  pages         = {242},
  year          = {2020}
}

@article{Jia:2024zes,
  author        = {Jia, Hewei Frederic and Rangamani, Mukund},
  title         = {{Holographic thermal correlators and quasinormal modes from semiclassical Virasoro blocks}},
  eprint        = {2408.05208},
  archiveprefix = {arXiv},
  primaryclass  = {hep-th},
  doi           = {10.1007/JHEP12(2024)047},
  journal       = {JHEP},
  volume        = {12},
  pages         = {047},
  year          = {2024}
}

@article{Jia:2025jbi,
  author        = {Jia, Hewei Frederic and Rangamani, Mukund},
  title         = {{Thermal spectral function asymptotics and black hole singularity in holography}},
  eprint        = {2512.15114},
  archiveprefix = {arXiv},
  primaryclass  = {hep-th},
  year          = {2026},
  doi           = {10.1103/tzm2-jjvg},
  journal       = {Phys. Rev. D},
  volume        = {114},
  number        = {4},
  pages         = {046025},
}

@article{Jia:2026pmv,
  author        = {Jia, Yueke and Kulaxizi, Manuela},
  title         = {{Bulk Phase Shift and Singularity}},
  eprint        = {2602.06558},
  archiveprefix = {arXiv},
  primaryclass  = {hep-th},
  month         = {2},
  year          = {2026}
}

@article{Jia:2026ryl,
  author        = {Jia, Hewei Frederic and Rangamani, Mukund},
  title         = {{Exact holographic thermal spectral functions: OPE, non-perturbative corrections, and black hole singularity}},
  eprint        = {2604.10803},
  archiveprefix = {arXiv},
  primaryclass  = {hep-th},
  month         = {4},
  year          = {2026}
}

@book{KawaiTakei2005,
  author     = {Takahiro Kawai and Yoshitsugu Takei},
  title      = {Algebraic Analysis of Singular Perturbation Theory},
  series     = {Translations of Mathematical Monographs},
  volume     = {227},
  publisher  = {American Mathematical Society},
  address    = {Providence, RI},
  year       = {2005},
  isbn       = {978-0-8218-3547-0},
  translator = {Goro Kato},
  note       = {Iwanami Series in Modern Mathematics},
  pagetotal     = {129},
  doi           = {10.1090/mmono/227},
}

@article{Komargodski:2026oob,
  author        = {Komargodski, Zohar},
  title         = {{Fate of Bouncing Singularities at Finite $G_N$}},
  eprint        = {2609.31309},
  archiveprefix = {arXiv},
  primaryclass  = {hep-th},
  month         = {9},
  year          = {2026}
}

@article{Kovtun:2006pf,
  author        = {Kovtun, Pavel and Starinets, Andrei},
  title         = {{Thermal spectral functions of strongly coupled N=4 supersymmetric Yang-Mills theory}},
  eprint        = {hep-th/0602059},
  archiveprefix = {arXiv},
  reportnumber  = {NSF-KITP-06-06},
  doi           = {10.1103/PhysRevLett.96.131601},
  journal       = {Phys. Rev. Lett.},
  volume        = {96},
  pages         = {131601},
  year          = {2006}
}

@article{Lisovyy:2022flm,
  author        = {Lisovyy, O. and Naidiuk, A.},
  title         = {{Perturbative connection formulas for Heun equations}},
  eprint        = {2208.01604},
  archiveprefix = {arXiv},
  primaryclass  = {math-ph},
  doi           = {10.1088/1751-8121/ac9ba7},
  journal       = {J. Phys. A},
  volume        = {55},
  number        = {43},
  pages         = {434005},
  year          = {2022}
}

@article{Liu:2009dm,
  author        = {Liu, Hong and McGreevy, John and Vegh, David},
  title         = {{Non-Fermi liquids from holography}},
  eprint        = {0903.2477},
  archiveprefix = {arXiv},
  primaryclass  = {hep-th},
  reportnumber  = {MIT-CTP-4023},
  doi           = {10.1103/PhysRevD.83.065029},
  journal       = {Phys. Rev. D},
  volume        = {83},
  pages         = {065029},
  year          = {2011}
}

@article{Loganayagam:2022teu,
  author        = {Loganayagam, R. and Rangamani, Mukund and Virrueta, Julio},
  title         = {{Holographic thermal correlators: A tale of Fuchsian ODEs and integration contours}},
  eprint        = {2212.13940},
  archiveprefix = {arXiv},
  primaryclass  = {hep-th},
  doi           = {10.1007/JHEP07(2023)008},
  journal       = {JHEP},
  volume        = {07},
  pages         = {008},
  year          = {2023}
}

@article{Miyachi:2025ptm,
  author        = {Miyachi, Taiga and Namba, Ryo and Omiya, Hidetoshi and Oshita, Naritaka},
  title         = {{Path to an exact WKB analysis of black hole quasinormal modes}},
  eprint        = {2503.17245},
  archiveprefix = {arXiv},
  primaryclass  = {hep-th},
  reportnumber  = {RIKEN-iTHEMS-Report-25},
  doi           = {10.1103/1gmr-9f1g},
  journal       = {Phys. Rev. D},
  volume        = {111},
  number        = {12},
  pages         = {124045},
  year          = {2025}
}

@article{Motl:2003cd,
  author        = {Motl, Lubos and Neitzke, Andrew},
  title         = {{Asymptotic black hole quasinormal frequencies}},
  eprint        = {hep-th/0301173},
  archiveprefix = {arXiv},
  reportnumber  = {HEP-UK-0017, HUTP-03-A005},
  doi           = {10.4310/ATMP.2003.v7.n2.a4},
  journal       = {Adv. Theor. Math. Phys.},
  volume        = {7},
  number        = {2},
  pages         = {307--330},
  year          = {2003}
}

@article{Natario:2004jd,
  author        = {Natario, Jose and Schiappa, Ricardo},
  title         = {{On the classification of asymptotic quasinormal frequencies for d-dimensional black holes and quantum gravity}},
  eprint        = {hep-th/0411267},
  archiveprefix = {arXiv},
  doi           = {10.4310/ATMP.2004.v8.n6.a4},
  journal       = {Adv. Theor. Math. Phys.},
  volume        = {8},
  number        = {6},
  pages         = {1001--1131},
  year          = {2004}
}

@article{Son:2002sd,
  author        = {Son, Dam T. and Starinets, Andrei O.},
  title         = {{Minkowski space correlators in AdS / CFT correspondence: Recipe and applications}},
  eprint        = {hep-th/0205051},
  archiveprefix = {arXiv},
  reportnumber  = {INT-PUB-02-34},
  doi           = {10.1088/1126-6708/2002/09/042},
  journal       = {JHEP},
  volume        = {09},
  pages         = {042},
  year          = {2002}
}

@article{Teaney:2006nc,
  author        = {Teaney, Derek},
  title         = {{Finite temperature spectral densities of momentum and R-charge correlators in N=4 Yang Mills theory}},
  eprint        = {hep-ph/0602044},
  archiveprefix = {arXiv},
  doi           = {10.1103/PhysRevD.74.045025},
  journal       = {Phys. Rev. D},
  volume        = {74},
  pages         = {045025},
  year          = {2006}
}

@article{Voros1983ReturnQuarticOscillator,
  author  = {Voros, Andr{\'e}},
  title   = {The Return of the Quartic Oscillator. The Complex WKB Method},
  journal       = {Ann. Inst. H. Poincar{\'e} Phys. Th{\'e}or.},
  volume  = {39},
  number  = {3},
  pages   = {211--338},
  year    = {1983},
  url     = {https://www.numdam.org/item/AIHPA_1983__39_3_211_0/}
}

@unpublished{Grassi:2026wip,
  author = {Grassi, Alba and Jia, Hewei Frederic},
  year = {2026},
  note   = {In preparation}
}

@unpublished{Caron-Huot:2026unp,
  author = {Caron-Huot, Simon and Simmons-Duffin, David and Stanford, Douglas},
  year = {2026},
  note   = {Unpublished}
}

@article{Witten:1998qj,
  author        = {Witten, Edward},
  title         = {{Anti de Sitter space and holography}},
  eprint        = {hep-th/9802150},
  archiveprefix = {arXiv},
  reportnumber  = {IASSNS-HEP-98-15},
  doi           = {10.4310/ATMP.1998.v2.n2.a2},
  journal       = {Adv. Theor. Math. Phys.},
  volume        = {2},
  pages         = {253--291},
  year          = {1998}
}

@article{Xiao:2026pir,
  author        = {Xiao, Zi-Qing and Nian, Jun and Li, Li},
  title         = {{Channel-Resolved High-Overtone Quasinormal Modes Decode Kasner Scaling inside Black Holes}},
  eprint        = {2609.17426},
  archiveprefix = {arXiv},
  primaryclass  = {hep-th},
  month         = {9},
  year          = {2026}
}

\end{document}